\documentclass[12pt,preprint]{aastex}
\usepackage[]{natbib}
\usepackage{amsmath}
\usepackage{calligra}
\usepackage{epstopdf}
\input epsf

\begin{document}

\numberwithin{equation}{section}
\numberwithin{table}{section}
\numberwithin{figure}{section}

\def\wisk#1{\ifmmode{#1}\else{$#1$}\fi}

\def\lt     {\wisk{<}}
\def\gt     {\wisk{>}}
\def\le     {\wisk{_<\atop^=}}
\def\ge     {\wisk{_>\atop^=}}
\def\lsim   {\wisk{_<\atop^{\sim}}}
\def\gsim   {\wisk{_>\atop^{\sim}}}
\def\kms    {\wisk{{\rm ~km~s^{-1}}}}
\def\Lsun   {\wisk{{\rm L_\odot}}}
\def\Zsun   {\wisk{{\rm Z_\odot}}}
\def\Msun   {\wisk{{\rm M_\odot}}}
\def\um     {$\mu$m}
\def\mic     {\mu{\rm m}}
\def\sig    {\wisk{\sigma}}
\def\etal   {{\sl et~al.\ }}
\def\eg     {{\it e.g.\ }}
 \def\ie     {{\it i.e.\ }}
\def\bsl    {\wisk{\backslash}}
\def\by     {\wisk{\times}}
\def\half {\wisk{\frac{1}{2}}}
\def\third {\wisk{\frac{1}{3}}}
\def\nwm2sr {\wisk{\rm nW/m^2/sr\ }}
\def\nw2m4sr {\wisk{\rm nW^2/m^4/sr\ }}

\title{Measuring bulk motion of X-ray clusters via the kinematic Sunyaev-Zeldovich effect:
summarizing the "dark flow" evidence and its implications}

\author{A. Kashlinsky\altaffilmark{1}, F.  Atrio-Barandela\altaffilmark{2},
 H.  Ebeling\altaffilmark{3}}
\altaffiltext{1}{SSAI and Observational Cosmology Laboratory, Code
665, Goddard Space Flight Center, Greenbelt MD 20771;
alexander.kashlinsky@nasa.gov} \altaffiltext{2}{Fisica Teorica,
  University of Salamanca, 37008 Salamanca, Spain}
   \altaffiltext{3}{Institute for Astronomy, University of Hawaii, 2680
  Woodlawn Drive, Honolulu, HI 96822}

\begin{abstract}

Large-scale peculiar velocities provide a rare opportunity to probe the physics and conditions of the early Universe and to test the completeness of the standard cosmological paradigm. Particularly important is to probe the velocity field on scales $\ga 100$ Mpc, where the amplitude and coherence length of the gravitationally induced velocity field is predicted robustly independently of cosmological parameters or evolution. Both galaxies and galaxy clusters have been used over the last several decades to probe the large-scale velocity field, albeit with different resolution and to different depth. In this review we present a comprehensive discussion of peculiar velocity field measured recently on very large scales with a novel method using X-ray galaxy clusters as tracers. The measurement is based on the kinematic component of the Sunyaev-Zeldovich (KSZ) effect produced by Compton scattering of cosmic microwave background (CMB) photons off the hot intracluster gas, and uses a large catalog of X-ray selected clusters and all-sky CMB maps obtained with the WMAP satellite. The method probes the dipole of the CMB temperature field evaluated at the cluster positions and within the apertures in which the CMB monopole contribution vanishes, thereby isolating the signal remaining from the KSZ effect produced by coherently moving clusters. The detection of a highly significant dipole out to the depth of at least $\sim 800$Mpc casts doubt on the notion that gravitational instability from the observed mass distribution is the sole -- or even dominant -- cause of the detected motions. Rather it appears that the flow may extend across the entire observable Universe. Possible implications include the possibility to constrain the primeval preinflationary structure of space-time and its landscape, and/or the need for modifications of presently known physics (e.g. arising from a higher-dimensional structure of gravity). We review these possibilities in light of the measurements described here and specifically discuss the prospects of future measurements and the issues they should resolve. We address the consistency of these large-scale velocity measurements with those obtained on smaller scales by studies using galaxies as tracers, and resolve the discrepancies with two recent claims based on modified CMB analysis schemes.
\end{abstract}

\newpage
\tableofcontents
\newpage

\section{Introduction}
\label{sec:intro}

At the beginning of the 20th century, the Hubble interpretation of the radial velocity of
galaxies yielded the first observational, and at the time radical, evidence for expanding-Universe models proposed independently by Friedman and Lema\^{i}tre as an immediate consequence of Einstein's field equations\footnote{There is currently a lively discussion whether the observational establishment of the law of expansion known as the {\it Hubble} law, $v=Hr$, should originally be credited to Lema\^{i}tre (see e.g. Way \& Nussbaumer 2011). While taking note of the arguments in favor of Lema\^{i}tre's work preceding Hubble's, we will follow below the common notation referring to the law of expansion as the Hubble law.}. The expansion of the Universe later became well supported by further observations, making it clear that, unless the laws of physics were modified e.g. such as proposed by Hoyle in his steady-state cosmology, the Universe would have been very hot and dense at early times, originating in a singularity a finite time ago. Thus the model was nicknamed the "Big Bang" by its formidable and notable opponent, Sir Fred Hoyle. Gamow and Alpher later developed the hot Big Bang theory, predicting the universal abundance of light elements (primarily $He^4$) and properties of the left-over relic radiation. The discovery of such relic radiation in the form of the Cosmic Microwave Background (CMB) by Pensias \& Wilson in 1964 firmly established the hot Big Bang as the correct model for the post-singularity evolution of the Universe. Later measurements, such as of the abundance of light elements and in particular of deuterium, have further confirmed the validity of the Big Bang model.

The hot Big Bang model received further important verification with the discovery of the CMB anisotropies and measurements of its highly accurate black-body spectrum by the COBE DMR and FIRAS instruments (Smoot et al 1992, Mather et al 1990). The former provided  important information on the large-scale power distribution of the structures at the last scattering surface via the Sachs-Wolfe effect (Sachs \& Wolfe 1967, hereafter SW), while the latter indicated that the Universe had gone through a hot dense phase as required to thermalize the leftover relic radiation.

Observationally, the Universe appeared homogeneous and isotropic on scales causally unconnected at early times within the standard Big Bang model, when the cosmological horizon of the standard Big Bang model was much smaller than their separation distances. This great puzzle was elegantly explained (theoretically) by the inflationary paradigm first proposed independently by Kazanas (1980) and Guth (1981). According to that model the vacuum-like physics at very early times  generated a repulsive equation of state, where the pressure $p=w\rho$ with $w\simeq -1$. This then led to exponential (inflationary) expansion of the initially causally connected patches of space-time which, in a very short time, inflated to scales much greater than the present-day cosmological horizon. Our Universe was then posited to be one such patch thereby resolving the ``horizon problem".  Various plausible mechanisms for the inflationary expansion were later proposed by Gott (1982), Starobinsky (1982), Linde (1983), Albrecht \& Steinhardt (1992), and others to show that this inflationary model is a logical consequence of the physics expected from high-energy vacuum processes (see review by Olive 1990).

It was only at the close of the 20th century, however, that the Universe's rate of expansion was found to be not constant but in fact accelerating, a finding that was instrumental in establishing the current cosmological concordance model. These findings got on firm ground with the measurements of the dimming of distant SNIa (Riess et al 1998, Perlmutter et al 1998, Schmidt et al 1998), but had in fact been proposed by Efstathiou et al (1992) from the observed large-scale power of galaxy clustering \cite{mesl}.

The power spectrum of the matter distribution at the time of last scattering was measured by the COBE satellite to be approximately of the Harrison-Zeldovich form (Harrison 1967, Zeldovich 1972 -- hereafter HZ) consistent with the simplest inflationary predictions (Gorski et al 1994). The post-COBE observations of the CMB structure on arcminute scales, first by balloon experiments (MAXIMA, Hanany et al 2000 and BOOMERANG, Lange et al 2001, de Bernardis et al 2000) and and then by the WMAP satellite (Bennett et al 2003, Spergel et al 2003), revealed the (expected - e.g. Bond \& Efstathiou 1984) presence of the acoustic Doppler peaks in the CMB power spectrum. By measuring the position of these peaks, the observations established the flatness of the Universe as predicted by the inflationary model (Lange et al 2001). The structure of the CMB has by now been measured up to nine(!) Doppler peaks, probing the multipoles to $\ell \sim 3,000$ (Keisler et al 2011). This confirms the validity of the standard cosmological model with data spanning over three orders of magnitude in angular scales.

The first decade of the 21st century marked the remarkable emergence of precision cosmology. Fundamental cosmological parameters have been determined with high precision, the contributions of the basic constituents of the Universe (baryons, dark matter, dark energy, cosmic radiation fields, etc) to its matter and energy budget have been accurately measured, and a standard
cosmological model describing the evolution of structure in the Universe has been established. In addition, our understanding of
high-energy physics became sufficiently advanced to start connecting phenomena on cosmological scales to the quantum physics of the earliest moments of the Big Bang. It is then important to probe possible observational deviations from the standard concordance model and explore their implications for the physics beyond the inflationary model. A way to do this is by designing experiments
that test the completeness and validity of the standard cosmological model and, at the same time, help us obtain a better description of the space time of the pre-inflationary Universe.

In this context, an important observable -- and the subject of this review and our recent work -- is the distribution of large-scale peculiar velocities in the Universe. Peculiar velocities are deviations from the universal expansion, which, in the standard gravitational instability picture, should probe the overall mass distribution of the gravitating matter in the Universe. Inflation makes robust predictions for the mass distribution on sufficiently large scales (beyond the horizon scale at the matter-radiation equality) which, in turn, predict the large-scale peculiar velocities resulting from gravity produced by this matter distribution. In the course of our preceding work we have uncovered unexpected evidence for a peculiar velocity field with large amplitude and coherence possibly extending across the entire cosmological horizon. This motion hinges on the assumption of the CMB (as measured by its dipole moment) representing a rest-frame in the Universe. The flow -- dubbed the ``dark flow" by us in the original paper (Kashlinsky et al 2008, hereafter KABKE1) -- was proposed by us to originate in the pre-inflationary space-time landscape and may thus provide unique information on the global structure of the world and our Universe's place in it. This review is  devoted to the current state of measurements of large-scale peculiar flows, their robustness and consistency with the dark flow, and the implications of these results.

The measurements of the ``Dark Flow" (DF) take on additional importance due to the fact that the DF may be the most direct observational link thus far to the physics of the Quantum Cosmology era. Indeed, the DF may shed light on the following questions:
\begin{itemize}
\item Is our Universe part of much larger set, called the Multiverse?
\item What is the {\it pre}-inflationary structure of space-time (related to the initial configuration of the inflaton field)?
\item Is the $\Lambda$CDM concordance model (Dark Energy and Cold Dark Matter normalized to reproduce the Cosmic Microwave Background observations) broken in ways that reflect this structure?
\item Do we need to revise our understanding of gravity?
\end{itemize}

In this review we summarize the current status of the DF measurements and its cosmological implications. Our main motivation for this review is twofold: first, with more data and further studies confirming our initial findings, it now appears reasonably likely that the detected flow is real. This follows from a statistically significant CMB dipole remaining at the pixels associated exclusively with galaxy clusters, the signal being correlated to the clusters' X-ray luminosity and detected within apertures containing vanishing contributions from the CMB monopole. We believe that the only viable explanation of the detected dipole is the large-scale motion of galaxy clusters, and as of this writing no other explanation has appeared in the peer-reviewed literature. The time has thus come to summarize the available results, compare their consistency with other measurements, and discuss their implications for cosmology and fundamental physics alike. Second, several major issues related to the calibration of the detected CMB signal in terms of the velocity of the motion remain unsatisfactory -- or unresolved -- for now. While these do not affect the existence of the detected large-scale coherent motion, their resolution is important in terms of identifying with high accuracy the amplitude of the motion, its directional calibration, and its properties. Third, we use this opportunity to present -- in greater depth -- the details and the methodology of these measurements. This is important in no small part by the fact that the signal, while being highly statistically significant, is measured with a signal-to-noise ratio of only $S/N \sim 4$ and can be rendered insignificant if measurement flaws reduce it by only a factor of $\sim 2$. This review thus presents an opportunity to share the status of this investigation in the hope of facilitating further input and progress in this important area of research.

This review is organized as follows: Sec. \ref{sec:definitions} provides a summary of  frequently used definitions, quantities, and terms, followed by Sec. \ref{sec:standardmodel} which defines the standard cosmological model. After briefly outlining the standard formalism for the evolution of the homogeneous Universe within the Big Bang model, we discuss there the range of scales within which this homogeneity is expected to hold in the inflationary paradigm. We then introduce the density (and CMB temperature) field of the standard cosmological model which applies within the scale of the homogeneity. The standard gravitational-instability paradigm and its predictions for an inflationary class of models are described in Sec. \ref{sec:gravinstability}, along with the various  quantities necessary to characterize the velocity field.  Earlier results of peculiar-velocity studies are summarized in Sec. \ref{sec:earlyresults}. There we discuss the CMB dipole and its use as a reference frame, early results based on galaxies as distance indicators, measurements of the various dipoles in the distribution of cosmic structures as a probe of the local peculiar gravity (scales $\lsim 100h^{-1}$ Mpc or so), and address their mutual compatibility. Sec. \ref{sec:sz} discusses the Sunyaev-Zeldovich (SZ) effect and its usefulness for measuring peculiar velocities of X-ray luminous galaxy clusters. We review the method proposed  by Kashlinsky \& Atrio-Barandela (2000 -- hereafter KA-B) adopted by us as instrumental for this measurement. Sec. \ref{sec:instruments} describes briefly the X-ray and CMB instruments used to obtain the data used in this measurement. In Sec. \ref{sec:xraycat} we describe the construction of the cluster database upon which we relied, and in Sec. \ref{sec:cluster_tsz} we discuss how we establish the properties and profiles of the cluster gas distribution derived from this database which is a prerequisite to our peculiar velocity determinations. This is followed by Sec. \ref{sec:darkflow} which discusses the large-scale cluster bulk flow measurements which resulted in the unexpected "dark flow" detection. There we discuss how we assembled the cluster catalog(s) used for the DF measurement, the filtering schemes applied to the CMB maps, and the (analytical and numerical) basis for understanding the errors of our measurement. We then discuss at length the issue of ``systematics", addressed already in our previous studies, and show that it enters at an insignificant level and cannot reproduce the dark flow measurements. We further address the (currently still deficient) calibration which converts the measured CMB signal (in $\mu$K) into bulk velocity of the cluster sample (in km s$^{-1}$). In conclusion of that section we address the criticism that have so far appeared in the refereed literature and in unrefereed postings. Sec. \ref{sec:tests} discusses the various alternative tests of the DF which have been proposed so far. In Sec. \ref{sec:implications} we discuss the cosmological implications of the DF and its possible connection to either  ``new physics" or a new understanding of the global world structure. Sec. \ref{sec:future} discusses prospects of future measurements and the currently conducted SCOUT ({\bf S}unyaev-Zeldovich {\bf C}luster {\bf O}bservations as probes of the {\bf U}niverse's {\bf T}ilt) experiment designed by us. Our conclusions are presented in Sec. \ref{sec:conclusions}.

\section{Miscellaneous: definitions, units, abbreviations etc}
\label{sec:definitions}

Table \ref{tab:acronyms} lists the acronyms and abbreviations frequently used in this review. Unless otherwise specified, velocities will be measured in km/sec, X-ray luminosities in erg/sec, CMB temperature in $\mu$K, distances in Mpc (or at times in $h^{-1}$Mpc). $h$ will refer to dimensionless Hubble constant, $h_P$ is the Planck constant, $G$ is gravitational constant, $c$ is the speed of light, $m_e$ is the electron mass, $\sigma_T$ the Thomson cross section, $k_B$ the Boltzman constant, etc.

\begin{deluxetable}{c c}
\tabletypesize{\scriptsize}
\tablecaption{Alphanumerical summary of used acronyms and
abbreviations.}
 \startdata
\hline
\hline
2MASS & 2-Micron All Sky Survey\\
2MRS & 2-Micron Redshift Survey\\
6dFGRS & 6 Degree Field Galaxy Redshift Survey \\
ACT & Atacama Cosmology Telescope \\
AKEKE & Atrio-Barandela, Kashlinsky, Ebeling, Kocevski \& Edge (2010)\\
AKKE & Atrio-Barandela, Kashlinsky, Kocevski \& Ebeling (2008) \\
BOOMERanG & (Balloon Observations Of Millimetric Extragalactic Radiation and Geophysics \\
CDM & Cold Dark Matter \\
CIB & Cosmic Infrared Background \\
CMB & Cosmic Microwave Background \\
COBE & COsmic Background Explorer \\
CXB & Cosmic X-ray Background \\
DA & Differential Assembly \\
DCB & Diffuse Cosmic Background \\
DF & Dark Flow \\
DIRBE & Diffuse Infrared Background Experiment (on board COBE) \\
DMR & Differential Microwave Radiometers (on board COBE) \\
eROSITA & extended ROentgen Survey with an Imaging Telescope Array \\
FIRAS & Far-InfraRed Absolute Spectrometer (on board COBE) \\
FRW & Friedman-Robertson-Walker (metric) \\
FT & Fourier Transform \\
GA & Great Attractor \\
GC & Galactic Center \\
GR & General Relativity (theory) \\
GZ & Grischuk-Zeldovich (effect) \\
GUT & Grand Unification Theory \\
HZ & Harrison-Zeldovich (power spectrum) \\
HFI & ({\it Planck}) High Frequency Instrument\\
IRAS & InfraRed Astronomy Satellite \\
KA-B & Kashlinsky \& Atrio-Barandela (2000)\\
KABKE1 & Kashlinsky, Atrio-Barandela, Kocevski \& Ebeling (2008)\\
KABKE2 & Kashlinsky, Atrio-Barandela, Kocevski \& Ebeling (2009)\\
KAEEK & Kashlinsky, Atrio-Barandela, Ebeling, Edge \& Kocevski (2010)\\
KAE & Kashlinsky, Atrio-Barandela \& Ebeling (2011)\\
KSZ & Kinematic Sunyaev-Zeldovich (effect) \\
LFI & ({\it Planck}) Low Frequency Instrument\\
LG & Local Group \\
LOS & Line of Sight\\
LSR & Local System of Rest \\
MACS & The MAssive Cluster Survey \\
MAXIMA & Millimeter Anisotropy eXperiment IMaging Array \\
MDE & Matter-Dominated Era \\
PSCz & (IRAS) Point Source Catalog (with $z$)\\
PIXIE & Primordial Inflation Explorer\\
RASS & ROSAT All Sky Survey \\
RDE & Radiation-Dominated Era \\
ROSAT & R\"ont\-gen\-satellit (German)\\
SCOUT & Sunyaev-Zeldovich Cluster Observations as probes of the Universe's Tilt\\
SDSS & Sloan Digital Sky Survey \\
SFI++ & Spiral Field I-band ++ (survey)\\
S/N & Signal-to-Noise (ratio) \\
SN & Super Novae \\
SPT & South Pole Telescope \\
SW & Sachs-Wolfe (effect) \\
SZ & Sunyaev-Zeldovich (effect)\\
TSZ & Thermal Sunyaev-Zeldovich (effect) \\
WHIM & Warm Hot Intergalactic Medium\\
WMAP & {\it Wilkinson Microwave Anisotropy Probe}\\
ZoA & Zone of Avoidance \\
\enddata
\label{tab:acronyms}
 \end{deluxetable}

Throughout this review the
following notations are used, unless otherwise specified: $\rho$ denotes densities, $\Omega$
is the density parameter in units of the closure density
$3H_0^2/8\pi G$; $H_0\equiv100h$ km s$^{-1}$ Mpc$^{-1}$ is the present-day
Hubble constant; $L_X$ denotes the X-ray luminosity;
$\Lambda$ is the cosmological constant; $\delta$ is the
dimensionless density or CMB temperature contrast; $P(k)$ denotes
the 3-D power spectrum; $C_\ell$ is the angular power spectrum of the CMB at the $\ell$-multipole, $z$ stands for redshift. Other frequently used
symbols include $\tau$, the projected optical depth due to Compton
scattering;  $n_e$ and $T_X$, the electron density and temperature of the
X-ray emitting intra-cluster gas, respectively; $T_{\rm CMB}$,
the present-day CMB temperature; and $T_{\rm
e,ann}$, the electron annihilation temperature defined via $k_{\rm B}T_{\rm
e,ann}{=}511$ keV.

We present the results in Galactic coordinates. In this coordinate
system, $X$ points in the direction of the center of the Galaxy, $Z$
points in the direction of the Galactic North pole and $Y$ is
perpendicular to $X$ and $Z$. $(l,b)$ denotes the longitude and latitude of the Galactic coordinates respectively.

Table \ref{tab:pars} shows the cosmological parameters adopted throughout corresponding to the concordance cosmological model.
\begin{table}[h!]
\caption{Cosmological parameters used throughout.}
\begin{tabular}{l c p{3.0in} }
\hline
\tabletypesize{\scriptsize}
Parameter & Value & Source/Comments \\
\hline
\hline
$\Omega_{\rm m}$ & 0.26 & Komatsu et al (2011) \\
$\Omega_\Lambda$ & 0.74 & Komatsu et al (2011) \\
$\Omega_{\rm total}$ & 1 & Assumed within the directly observable Universe; inflation however requires that at very large scale deviations from flatness become noticeable\\
$h$ & 0.71 & Komatsu et al (2011) \\
$T_{\rm CMB}$ & 2.726K & Fixsen (2009) \\
\end{tabular}
\label{tab:pars}
\end{table}

\section{Standard cosmological model}
\label{sec:standardmodel}

\subsection{Background cosmology}
\label{sec:background}

We adopt below the standard cosmological model and the corresponding framework. The expanding and approximately homogeneous Universe is assumed to originate with a Big Bang, undergo a brief inflationary expansion, producing also density fluctuations with the standard $\Lambda$CDM power spectrum, with the present-day structures resulting from the later gravitational growth and evolution of these density perturbations. The cosmological redshift is defined through the ratio of observed to emitted wavelengths as $\lambda_{\rm obs}/\lambda_{\rm em} = (1+z)$, the expansion factor is $a=(1+z)^{-1}$ normalized to unity at present time ($z=0$).

Following observations we adopt flat cosmology described by the Friedman-Robertson-Walker (FRW) metric:
\begin{equation}
ds^2=c^2dt^2 - (1+z)^{-2}[dx^2+x^2d\omega]
\label{eq:metric}
\end{equation}
Geodesics are defined by $ds^2=0$ such that the comoving coordinate distances are given by $dx=(1+z) cdt$. The proper (physical) coordinates are given by $dx_{\rm phys}=dx/(1+z)$ Eq. \ref{eq:metric} would break down where space-time becomes significantly inhomogeneous; the scale where this happens within the framework of standard inflationary models is discussed in the following subsection.

The global evolution of the expansion factor, $a$, follows from the Einstein equations with the FRW metric and the dust energy-momentum tensor with pressure $p$:
\begin{equation}
\ddot{a} = -\frac{4\pi G}{3}(\rho + 3p/c^2) + \frac{2}{3}\Lambda a
\label{eq:einstein}
\end{equation}
In standard cosmology the vacuum energy-density, $\Lambda$, is assumed to be zero and the strong energy condition applies, $(\rho + 3p/c^2)>0$. Hence, one can show that the comoving Hubble scale-factor, $(aH)^{-1}$ decreases with increasing $z$ toward arbitrarily early times. Thus the standard cosmological paradigm is unable to explain - via natural causal processes - the observed homogeneity and isotropy of the Universe.

The global evolution of the Universe is then described by the Friedman equation:
\begin{equation}
\dot{a}^2 \;\; (\;+\;{\cal K}\;)\; =\frac{8\pi G}{3}\rho_{\rm tot}(z) a^2 + \frac{\Lambda}{3} a^2 = H_0^2(\Omega_{\rm m}a^{-1} + \Omega_\Lambda a^2)
\label{eq:friedman}
\end{equation}
coupled with the general-relativistic equivalent of the second law of thermodynamics (conservation of energy):
\begin{equation}
d(\rho a^3) = - \frac{p}{c^2} d(a^3)
\label{eq:pdv}
\end{equation}
Here ${\cal K}$ is the curvature parameter assumed zero in eq. \ref{eq:metric} and throughout most of the review. The total density $\rho_{\rm tot}$ is at present dominated by matter with $\rho_{\rm m}(z)=\frac{3H_0^2}{8\pi G}\Omega_{\rm m} (1+z)^3$, and $\Lambda = 3H_0^2\Omega_\Lambda$.

With ${\cal K}=0$ this specifies the cosmic time-redshift relation during the matter-dominated era (MDE) as:
\begin{equation}
\frac{dt}{dz} = \frac{H_0^{-1}}{(1+z)\sqrt{\Omega_{\rm m}(1+z)^3+\Omega_\Lambda}}
\label{eq:time}
\end{equation}
With the cosmological parameters from Table \ref{tab:pars} the age of the Universe is $t=\int_0^\infty \frac{dt}{dz}dz\simeq H_0^{-1}=14$ Gyr. With the radiation-matter decoupling occuring with recombination at $z\simeq 1,090$ the last scattering surface corresponds to the cosmic time of $\sim 500,000$ years. The comoving distance to the cosmological horizon is then given by:
\begin{equation}
d_{\rm hor}(z) = c\int_z^\infty (1+z) \frac{dt}{dz}dz = R_H \int_z^\infty \frac{dz}{\sqrt{\Omega_{\rm m} (1+z)^3+\Omega_\Lambda}}
\label{eq:horizon}
\end{equation}
Its value today ($z=0$) is $d_{\rm hor}(0)\simeq 3 cH_0^{-1}\simeq 14$Gpc.

\subsection{Inflation and the scale of homogeneity}
\label{sec:bubble}

Conventional inflationary paradigm has been so appealing precisely because it does not require fine-tuning in the initial conditions to explain the observed homogeneity and flatness of the Universe. Rather it posits that the original space-time state could have been arbitrarily inhomogeneous on scales larger than the particle horizon at those early epochs. The observed Universe then represents part of a homogeneous inflated region embedded in an inhomogeneous space-time. On much larger scales the initial inhomogeneous energy state of the inflation driving inflaton field would be preserved. In this picture the scale of inhomogeneity is merely stretched by inflationary expansion and so the points that have been separated by distances larger than the scale of the inflated homogeneous patch have never been in causal contact (Turner 1991, Kashlinsky et al 1994). One thus very generally expects to find large preinflationary fluctuations, $\delta_L\equiv (\delta\rho/\rho)_L \gsim 1$ at scales $L > L_0$, the scale of the homogeneity of the Universe.

So how far does this homogeneity extend or, in other words, what is the scale of homogeneity of our Universe, $L_0$, beyond which the FRW approximation breaks down? We discuss this here following closely the arguments presented in Kashlinsky, Tkachev \& Frieman (1994).

One can relate the size of the inflated patch, $L_0$, to the present day cosmological parameters by writing the Friedman equation as $-{\cal K} = [1-\Omega(t)]H^2(t)a^2(t)$. If we denote with subscript $s$ the values of parameters at the start of inflationary expansion, then $L_0 = L_S/a_s$. Thus the Friedman equation leads to:
\begin{equation}
[1-\Omega_s]H_s^2L_s^2 = [1-\Omega_{\rm total}]H_0L_0^2
\end{equation}
On the other hand, by the onset of inflation one can expect causal process to have smoothed out initial inhomogeneities on scales up to, at most, the horizon size at the time, so that $L_s\lsim c H_s^{-1}$. Thus the present size of the inflated patch is {\it at most} comparable to the present-day curvature radius of the Universe\footnote{It is assumed that the patch has slightly open geometry with $1-\Omega_s<1$ as favored in inflationary cosmologies.}:
\begin{equation}
L_0 \lsim R_{\rm curv} = \frac{c H_0^{-1}}{\sqrt{1-\Omega_{\rm m} - \Omega_\Lambda}}
\label{eq:ktf1}
\end{equation}
While large compared to the present-day horizon, the curvature scale is expected to be finite within the inflationary paradigm.

The {\it lower} limit on $L_0$ comes from the following argument: if the next inhomogeneity is too close to the present-day horizon it will induce significant CMB anisotropies via the Grischuk-Zeldovich (1978) (GZ) effect (discussed at more length in Sec. \ref{sec:tests}. Namely the (tidal) gradient in the resultant metric generated by superhorizon-scale non-linear fluctuations ($\delta_L\gsim 1$) with length-scale $L$ will induce quadrupole anisotropy in the CMB of order $Q_L\simeq \delta_L (R_H/L)^2$. Observationally the magnitude of the induced quadrupole  should not exceed the observed CMB quadrupole anisotropy, $Q_{\rm obs} \sim 2\times 10^{-6}$ in dimensionless units. This leads to the size of inflated patch being at least several hundred Hubble radii:
\begin{equation}
L_0 \gsim Q_{\rm obs}^{-1/2} cH_0^{-1} \sim 500 R_H
\label{eq:ktf2}
\end{equation}
Detailed calculations suggest that this scale is several times larger than the simple estimate above for the present-day cosmological parameters (Kashlinsky et al 1994, Castro et al 2003). This scale can be directly related to the number of e-foldings of the inflationary expansion if one assumes the energy scale when inflation happened. For the GUT energy scales, the minimal number of e-foldings required to solve the horizon problems are $N_{e, min} \sim 50-55$ with logarithmically larger amount for larger values of $L_0$ (e.g. Turner 1991).

These arguments suggest that within conventional inflationary models, the scale of homogeneity of the Universe must be at least several hundred Hubble radii and on the upper end should not exceed its curvature radius.

Discussions of this kind do not specify how far the next inhomogeneity can lie. In principle, this can lead to arbitrarily small GZ corrections. On the other hand, models of the kind proposed by Holman, Mersini-Houghton \& Takahashi (2008), where the observed Universe forms out of the landscape of string theory, $L_0$ and $\delta_L$ are not arbitrary but are related to the underlying properties of that landscape and the energy scale that drives inflation.

In principle, one could construct inflationary models with small GZ anistropy as discussed by Kashlinsky, Tkachev \& Frieman (1994). These would appeal to as yet uncertain quantum gravity physics and one such scenario can be based on Gott's (1982) suggestion for the overall space-time geometry. There the (open) Universe appears as a result of tunneling through a quantum barrier in the de Sitter space starting with an exponential expansion which turns into a standard big bang solution at late times. That model has no singularity and the observed isotropy is explained because the different regions we observe have all been in causal contact {\it prior} to tunneling.

To summarize, the inflationary paradigm, in its conventional form, generally predicts the edge of the currently observed homogeneity and isotropy of space time. While the scale of our patch, whose geometry is described with the underlying FRW metric, extends to scales well beyond the current cosmological horizon, it is nonetheless finite. One can, and should, then inquire about potential consequences, and testability, of this proposition.

\subsection{Standard cosmological model and matter/CMB fluctuations}
\label{sec:lcdm}

As the inflaton field rolls down its potential driving inflationary expansion of the homogeneous patch of $L_0$, the quantum uncertainty principle ensures the generation of (quantum) energy fluctuations within it (Guth \& Pi 1982). In their simplest non-contrived form, conventional inflationary models make specific predictions about the properties of these energy fluctuations: to high accuracy they should be Gaussian and their spatial distribution is scale-invariant with the Harrison-Zeldovich spectrum.

The density field of a stochastic Gaussian variable can be described with the 3-D Fourier decomposition with random phases:
\begin{equation}
\frac{\delta \rho}{\rho} (\vec{x}) = \frac{1}{(2\pi)^3}\int \delta_{\vec{k}} \exp(-i \vec{k}\cdot \vec{x}) d^3\vec{x} \; \; ; \; \; \delta_{\vec{k}} = \int \frac{\delta \rho}{\rho} (\vec{x}) \exp(i \vec{k}\cdot \vec{x}) d^3\vec{k}
\label{eq:ft_delta}
\end{equation}
We will denote in much of what follows $\delta\equiv \delta\rho/\rho$. The power spectrum is given by $P(k)=\langle |\delta_{\vec{k}}|^2\rangle$ after averaging over the phases of $\vec{k}$. Given this definition of Fourier transform the typical rms density contrast over a sphere of radius corresponding to the wavelength/scale $r=\pi/k$ is
\begin{equation}
\Delta(r\!\simeq\!\frac{\pi}{k}) = \sqrt{\frac{k^3P(k)}{2\pi^2}}
\label{eq:del_mass}
\end{equation}
The Harrison-Zeldovich (HZ) spectrum is given by $P(k) \propto k^n$ with $n=1$ and arises naturally during the inflationary rollover. In the HZ regime the rms density contrast decreases with scale as $\Delta \propto r^{-2}$.

The evolution of matter fluctuations in the post-inflationary period is by now well understood and can be described as follows: during radiation-dominated era, at $z\gsim 3200$ for the standard cosmological parameters, density fluctuations inside the horizon are frozen because the Universe expands faster (on timescale $t_{\rm exp}\sim (G\rho_{\rm rad})^{-1/2}$) than the gravitational growth rate of matter fluctuations (timescale $t_{\rm growth}\sim (G\rho_{\rm matter})^{-1/2}$). Outside the horizon all fluctuations grow self-similarly with the gravitational potential. Once the Universe becomes matter dominated all scales grow self-similarly at rate $\delta\rho/\rho \propto (1+z)^{-1}$ until $1+z\simeq \Omega_\Lambda^{-1/3}$ when the growth by and large seizes.

\begin{figure}[h!]
\includegraphics[width=4in]{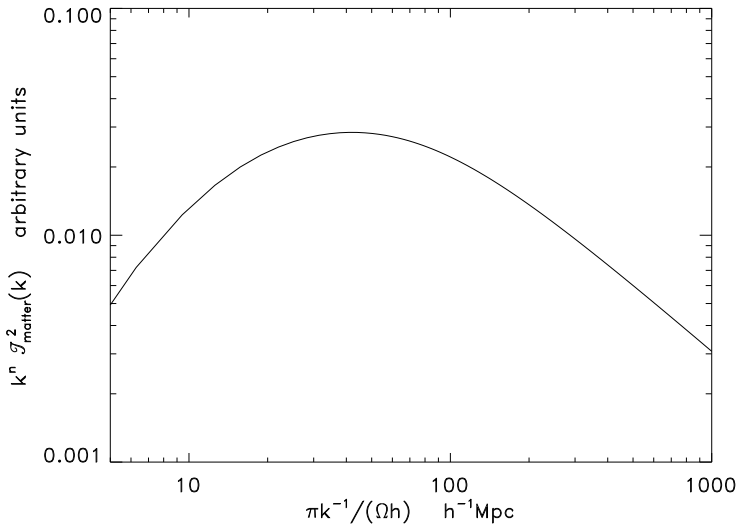}
 \caption[]{The shape of the matter power spectrum in inflationary models with HZ power spectrum ($n=1$). The turnover occurs at the scale corresponding to the horizon scale at matter-radiation equality which is proportional to $\propto (\Omega_m h)^{-1} h^{-1}$Mpc.}
\label{fig:del_cdm}
\end{figure}

Thus the shape of the matter fluctuations power spectrum is modified uniquely from its original HZ shape with the transition occurring at the scale corresponding to the horizon scale at the matter radiation equality and whose size if proportional to $(\Omega_{\rm m}h)^{-1} h^{-1}$Mpc. This transition is described by the transfer function, ${\cal T}$, so that the power spectrum of matter fluctuations leading to the present-day structure formation is $P(k) \propto k^n {\cal T}^2(k)$. Various accurate analytical approximations exist for the resultant power transfer (e.g. Bond \& Efstathiou 1985, Bardeen et al 1986, Sugiyama 1994). We use here the approximation from Bardeen et al (1986) as modified by Sugiyama (1994):
\begin{eqnarray}
{\cal T}_{\rm matter}(k) = \frac{\ln (1+2.34 q)}{2.34q}\; [1+3.89 q + (16.1q)^2+(5.46q)^3+(6.71q)^4]^{-1/4}\nonumber\\
q\equiv \frac{k}{\Omega_0h}\;(T_{\rm CMB}/2.7)^2\exp(\Omega_{\rm bar}+\sqrt{h/0.5}\Omega_{\rm bar}/\Omega_0)
\label{eq:transfer_cdm}
\end{eqnarray}
where $k$ is measured in $h$Mpc$^{-1}$. Numerically, with the WMAP parameters, the value of $\Omega_{\rm bar}+\sqrt{h/0.5}\Omega_{\rm bar}/\Omega_0\simeq 0.29$ in the above eq. \ref{eq:transfer_cdm} and varies little within the uncertainties of the cosmological parameters. Fig. \ref{fig:del_cdm} shows the resultant power spectrum under the assumption of the initially HZ spectrum ($n=1$).
The turnover occurs approximately at the horizon scale of matter-radiation equality. Its position is approximately independent of cosmology if plotted vs the scale $\Omega_mh$. At larger scales greater than horizon at matter-radiation equality the power spectrum remains in the HZ regime, so in concordance cosmology with parameters given in Table \ref{tab:pars} $P(k)\propto k$ at comoving scales $\gsim 100$ Mpc.

After the density fluctuations grow non-linear, the matter within them collapses and the power spectrum evolves by non-linear gravitational effects (e.g. Peebles 1980). Numerous semi-analytical approximations exist to extrapolate the power spectrum into such regime (e.g. Hamilton 1991, Peacock \& Dodds 1996). Observationally, one often measures the 2-point correlation of galaxies, $\xi(r)=\langle \delta(\vec{x}) \cdot \delta(\vec{x}+\vec{r})\rangle$ which strictly speaking traces the distribution of light. If the latter follows that of mass, as it happens to do on linear scales at least, then
\begin{equation}
\xi(r) = \frac{1}{2\pi^2} \int P(k) j_0(kr) k^2dk
\label{eq:xi2p}
\end{equation}
where $j_0$ is the spherical Bessel function of zero order. Non-linear evolution leads to the galaxy (light) two-point correlation function on small scales given by $\xi(r)=(r/r_0)^{-\gamma}$ with $\gamma\simeq 1.6$ and $r_0=5 h^{-1}$Mpc (Maddox et al 1990). The fluctuations in galaxy counts, $(\delta N/N)_{\rm rms}(r) = \frac{3}{r^3} \int_0^r \xi(x) x^2dx$   are then non-linear on scale $r_8=(\frac{3}{3-\gamma})^{\frac{1}{\gamma}}r_0 \simeq 8 h^{-1}$ Mpc, which is much less than the horizon at matter-radiation equality. The matter density fluctuations at $r_8$ are then normalized through WMAP7 CMB as $\sigma_8\simeq 0.8-0.9$ (Komatsu et al 2011). This provides a normalization point for the density fluctuation spectrum, and on significantly larger scales than $r_8$ the density field would remain in linear regime with the power spectrum given by eq. \ref{fig:del_cdm}.

As the Universe expands and cools, the CMB photons decouple from matter at $z\simeq 1,100$
when there are not enough photons in the in the Wien part of the CMB to maintain ionization
and the primordial hydrogen plasma recombines. Physical processes at that narrow {\it last scattering surface} lead to unique temperature fluctuations in the CMB via the Sachs-Wolfe effect (Sachs \& Wolfe 1967, hereafter SW).

The temperature field is generated at the last scattering surface over a very narrow redshift range. The formalism for describing the 2-dimensional field of these random fluctuations differs somewhat from eq. \ref{eq:ft_delta} suitable for 3-D variables. Since all the action happens at approximately the same project distance, one can expand the FT kernel of eq. \ref{eq:ft_delta} in terms of the spherical harmonics, $Y_{\ell m}$:
\begin{equation}
\exp(-i\vec{k}\cdot \vec{x})=4\pi \sum_{\ell,m} (-i)^\ell j_\ell(kx) Y_{\ell m}(\omega_k) Y^*_{\ell m}(\omega_x)
\label{eq:3d2d}
\end{equation}
where $j_\ell$ are spherical Bessel functions and $\omega$ is the solid angle in the specified direction.

The width of the last scattering surface is narrow, so primary CMB anisotropies can be approximated to lie at the same distance from the present-day observer. Their distribution can be described by the 2-D decomposition on a sphere into spherical harmonics $Y_{\ell,m}(\theta,\phi)$:
\begin{equation}
\delta T(\theta,\phi) = \sum_{\ell, m} a_{\ell m}Y_{\ell m}(\theta,\phi) \; \; ; \;\; a_{\ell m} = \frac{\int \delta T(\theta,\phi) Y_{\ell m}^*(\theta,\phi) d\omega}{\int Y_{\ell m}(\theta,\phi) Y_{\ell m}^*(\theta,\phi) d\omega}
\label{eq:dt2alm}
\end{equation}
Here $\omega$ is the solid angle covered by the temperature field measurements, $m=[-\ell,\ldots,0,\ldots,\ell]$ is the magnetic number analogous to the 3-D phase and the power spectrum is
\begin{equation}
C_\ell \equiv \frac{1}{2\ell+1}\sum |a_{\ell m}|^2
\label{eq:cl_power}
\end{equation}
If the sky is cut, $Y_{\ell m}$'s are no longer orthogonal and there is a cross-talk between different $\ell$'s in the second expression of eq. \ref{eq:dt2alm}.

The correlation function of the CMB field is given  by:
\begin{equation}
C(\theta)=\langle T(\vec{x}+\vec{\theta})\cdot T(\vec{x})\rangle = \frac{1}{4\pi} \sum (2\ell+1) C_\ell P_\ell(\cos\theta)
\label{eq:cmbcor}
\end{equation}
where $P_\ell$ denote Legendre polynomials. The variance of the CMB maps is given by $C(0)=\frac{1}{4\pi} \sum (2\ell+1) C_\ell$ where the sum extends to $\ell_{\rm max}$ specified b y the map pixel size. With this definition the rms temperature fluctuation over a radius $\pi/\ell$ radian can be approximated as:
\begin{equation}
(\delta T)_{\rm rms} \simeq [\frac{\ell^2 C_\ell}{2\pi}]^{1/2}
\label{eq:cl2dt}
\end{equation}

The cosmological density fluctuations can be generally represented as sum of two independent (initially) components: curvature and isocurvature. The latter are essentially perturbations of the equation of state which leave the overall metric, or total density, uniform. For this component $\delta\rho_{\rm total}=0$, so that at early times $\delta \rho_{\rm rad}/\rho_{\rm rad}=-(\delta \rho_{\rm m}/\rho_{\rm m})\times (\rho_{\rm m}/\rho_{\rm rad})$. Thus during the RDE times, $\rho_{\rm m}/\rho_{\rm rad}\ll 1$, these fluctuations are approximately isothermal.
In its conventional form, inflation, coupled with CDM models, predicts that the curvature fluctuations generated during the inflationary expansion up to $L_0$ were adiabatic with uniform entropy across the Universe, so that $(\delta T/T)_{\rm intrinsic} = -\frac{1}{3}\delta \rho_{\rm m}/\rho_{\rm m}$. The general evolution of these modes within the CDM models is reviewed by e.g. Efstathiou (1990) and Liddle \& Lyth (1993).

For curvature fluctuations there are three main contributions to CMB temperature fluctuations from the SW effect: 1) the intrinsic fluctuation $\delta T$, which depends on the type of the type of perturbations, 2) the component arising from Doppler effect due to peculiar velocities of the electrons at last scattering, 3) and the dominant on large scale fluctuations from the peculiar gravitational potential as photons climb of of potential wells existing at the last scattering surface. The general CMB temperature field produced by the SW effect at last scattering is given by eq. \ref{eq:sw_effect} and discussed later in the paper (Sec. \ref{sec:cmb_anisotropies}) after the suitable relevant quantities are defined.

 \begin{figure}[h!]
\includegraphics[width=6in]{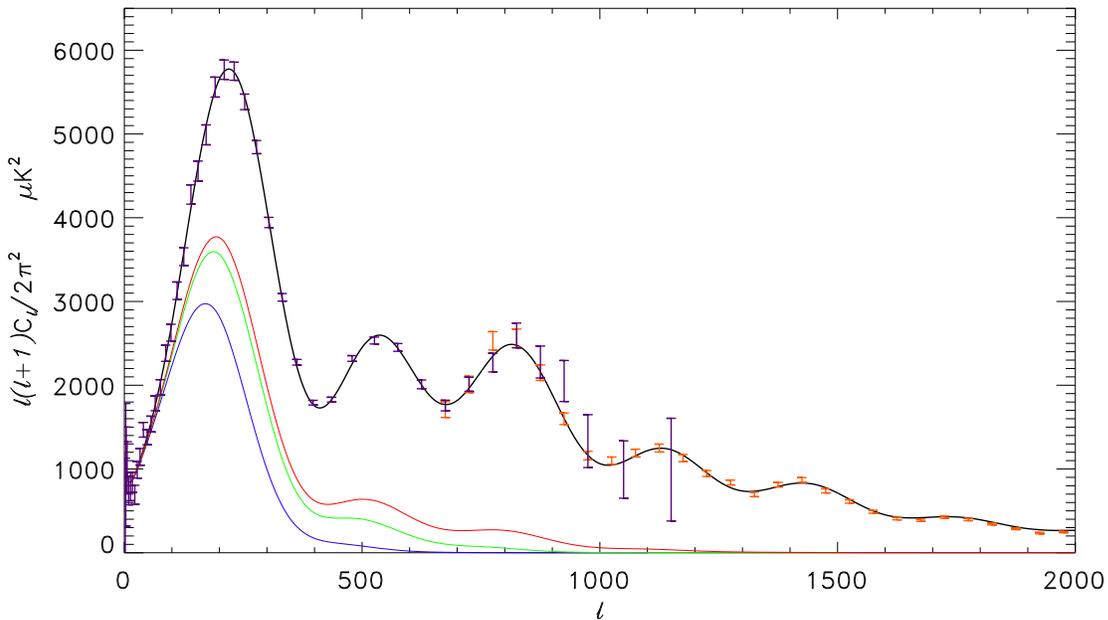}
 \caption[]{The spectrum of the mean squared CMB temperature fluctuation for the concordance $\Lambda$CDM model, which is used in the CMB filtering described in Sec. \ref{sec:filtering}. Blue, green and red lines show the spectrum when convolved with the WMAP $Q, V, W$ beams respectively. Violet error bars show the results from the 7-yr WMAP results (Jarosik et al 2011). Orange error bars show the results of the SPT measurements by Keisler et al (2011).}
\label{fig:cmb_lcdm}
\end{figure}

Fig. \ref{fig:cmb_lcdm} show the observed CMB temperature fluctuations measured with 7-yr WMAP maps (Jarosik et al 2011) and SPT (Keisler et al 2011). These are in excellent agreement with theoretical prediction of the concordance $\Lambda$CDM model on all scales probed to $\ell\sim 3,000$ (or angles down to $\sim 1^\prime$). This fact of the observationally established concordance cosmological model will prove central in what follows.

\section{Peculiar velocities and gravitational instability paradigm}
\label{sec:gravinstability}

\subsection{Characterizing (peculiar) velocity field}
\label{sec:vfield}

The velocity is a 3-D vector and so its properties must be characterized with more quantities than those of a random scalar field such as density or temperature fluctuations. Thus the bulk velocity, $\vec{V}_{\rm bulk}$, is the mean velocity over a given scale $r$:
\begin{equation}
\vec{V}_{\rm bulk}(r) = \frac{\int_{\vec{r}} \vec{v}(\vec{x}) \phi(\vec{x}) d^3\vec{x}}{\int_{\vec{r}}  d^3\vec{x}} = \frac{\sum w_i \vec{v}_i}{N}
\label{eq:bulk}
\end{equation}
where $\phi$ is a suitably normalized selection function of the survey describing the positions of test particles/sources probing the motion; in practice it is the comoving number density of objects in the survey normalized to $\int_{\rm survey\; volume} \phi d^3\vec{x}=1$. The last equality is valid when individual measurements of a 3-D velocity vector are available for a system of $N$ galaxies (or clusters) with $w_i$ describing the weights related to measurement errors.

Another important characteristic of the flow is the shear defined as:
\begin{equation}
{\cal S}_{ij} = \frac{1}{2} (\partial v_i/\partial x_j + \partial v_i/\partial x_j) - \frac{\delta_{ij}}{3} \vec{\nabla}\cdot \vec{v}
\label{eq:shear}
\end{equation}
where $\delta_{ij}$ is the usual Kroneker delta-symbol.

Several velocity correlation functions can be defined for a 3-D quantity such as velocity: the most general is $\Psi_{ij} = \langle v_i(\vec{x})\cdot v_j(\vec{x}+\vec{r}) \rangle$, which one usually divides into parallel and transverse correlations (e.g. Gorski 1988). However, if the flow is irrotational, the parallel and transverse correlations are not independent and the velocity correlations can be described with one {\it total} (sometimes called the ``dot") velocity correlation defined as:
\begin{equation}
\vartheta(r) = \langle \vec{v}(\vec{x})\cdot \vec{v}(\vec{x}+\vec{r}) \rangle
\label{eq:v-correlation}
\end{equation}
The rms velocity, or velocity dispersion, over a sphere of radius $r$ is then defined as:
\begin{equation}
v_{\rm rms}^2(r)=\frac{1}{2\pi^2} \int P_v(k)k^2 W_v(kr) dk
\label{eq:v_rms}
\end{equation}
where $P_v=\langle |\vec{v}_{\vec{k}}|^2\rangle$ is the velocity power spectrum and $W_v$ is the window function of the velocity survey. Some authors quote their results in terms of a {\it one-dimensional} velocity which is $v_{\rm rms}/\sqrt{3}$ assuming a statistically isotropic velocity field.

More generally let us consider peculiar motions within a sphere of radius $r$. The general velocity field at point $\vec{r}$ within that radius can be expanded as:
\begin{equation}
v_i(\vec{r}) = V_i + {\cal H} r_i + {\cal S}_{ij} r_j + \ldots
\label{eq:genvel}
\end{equation}
Here $V_i$ is the $i$-th component of the (mean) bulk flow across scale $r$ and the net (perturbed) Hubble ${\cal H}=H_0+\delta H$ constant within this volume defines the deviations of the 1-D radial velocity. The shear  contributes to the quadrupole of the flow across the scale $r$. The higher order tensorial expansion, neglected in eq. \ref{eq:genvel}, contains information about the octupole and higher moments. If the bulk flow term, $V$, dominates, the motion would be dipolar with small higher moments. The bulk flow (dipole moment) is sensitive to motion on scales larger than that of the system, while higher moments probe progressively smaller scales (see discussion in e.g. Feldman et al 2010).

The velocity field can be described as a superposition of two independent components:
\begin{equation}
\vec{v} = \vec{v}_g + \vec{v}_\bot
\label{eq:v_both}
\end{equation}
where $v_g$ is the peculiar velocity generated by peculiar gravity $g$ and the second term represents any primeval velocity, $\vec{v}_\bot$, which is statistically orthogonal to $\vec{g}$.

We note that in practice one determines from the distance indicator (and most other) measurements only the radial component of individual velocities. The latter is then used to constrain the full bulk flow as discussed in Sec. \ref{sec:vfromd}. In principle, the transversal component can be probed for clusters of galaxies via polarization measurements (Sunyaev \& Zeldovich 1980a, Audit \& Simmons 1999) or the Birkinshaw-Gull effect (Birkinshaw \& Gull 1983, Gurvits \& Mitrofanov 1986) generated by the gravitational lensing aberration of CMB photons by a moving cluster. In practice, however, the magnitude of the effect is too small to be of use in current state of measurements.

We now discuss the evolution of the two components in eq. \ref{eq:v_both} as the Universe expands.

\subsection{Relating peculiar velocity to gravity}
\label{sec:vfromg}

The first component, $v_g$, can be estimated fairly accurately from the following argument (e.g. Kashlinsky \& Jones 1991): Within the gravitational instability paradigm, peculiar velocities are
caused by gravitational instability, i.e.\ peculiar velocities are produced by local ("peculiar")
gravity, tracing the mass distribution on the corresponding scales. A simple and robust estimate of the velocity
amplitude $v_g$ can be obtained by considering that a particle that has
experienced a peculiar acceleration $g_p$ over a time period $t$ will have
acquired a velocity $v_g\simeq g_pt$. Writing $g_p= G\delta
M/r^2=4\pi G\delta_m \rho_m r/3 = \frac{1}{2} \Omega_m H_0^2 r \delta_m$ leads to
\begin{equation}
v_g(r)\simeq \frac{1}{3} f(\Omega_m) H_0r \delta_m
\label{eq:v-oom}
\end{equation}
where $f(\Omega) \equiv \frac{3}{2}\,H_0\Omega t\simeq \Omega^{0.6}$. In the HZ regime $\delta_m\propto r^{-2}$, which leads to peculiar velocity field decreasing as $v_g(r)\propto r^{-1}$ on these scales.

Essentially the same result is obtained through proper calculation. The formalism for this is well known and we do not report it here other than what is minimally necessary. Everything here can be done in Newtonian approximation which then describes the evolution of peculiar velocity $v_g$ resulting from peculiar gravitational acceleration $g_p$ via the following system of equations (e.g. Weinberg 1972):
\begin{equation}
\left\{\begin{array}{c}
\partial \rho/\partial t +\vec{\nabla}\cdot(\rho \vec{v})=0\nonumber\\
\partial \vec{v}/\partial t+(\vec{v}\cdot\vec{\nabla}) \vec{v} = -\vec{g}\nonumber\\
\vec{\nabla}\cdot \vec{g} = -4\pi G \rho_m\\
\vec{\nabla}\times \vec{g}=0\nonumber\\
\end{array}
\right.
\label{eq:velfrommotion}
\end{equation}
The first of these is the continuity equation which describes the conservation of mass, the second is the Euler formulation of Newton's law of motion under the influence of peculiar gravity $g_p$, the penultimate expression is the Poisson equation for Newton's gravity and the last equation says that gravity is a conservative force describable via a gradient of its scalar potential.

The next step is to perturb all the quantities around their FRW values, e.g. $\rho_m=\frac{3H_0^2\Omega_m}{8\pi G} (1+\delta_m),  v= H_0r + v_g$, etc and linearize the resultant system of equations ignoring terms of higher than linear order, i.e. $\delta_m^2\ll \delta_m$ etc.  Linearizing then the continuity equation leads to $\dot{\delta}=-\vec{\nabla}\cdot \vec{v_g}$, and noting that gravity is a conservative force that must lead to an irrotational flow, i.e. the Fourier component of the velocity, $v_{\vec{k}}$ is aligned with $\vec{k}$. Then the $k$-th component of the velocity field is given by:
\begin{equation}
\vec{v}_{\vec{k}} = -i H_0 f(\Omega) \frac{\hat{k}}{k} \delta_{\vec{k}}
\label{eq:v_k}
\end{equation}
The $k$-th Fourier component of shear is given by:
\begin{equation}
{\cal S}_{ij, \vec{k}} = -H_0 f(\Omega_m) (\hat{k}_i\hat{k}_j - \delta_{ij}) \delta_{\vec{k}}
\label{eq:shear_ft}
\end{equation}
The rms amplitude of the three-dimensional velocity $v_g$ is then related to the power spectrum of the underlying density field via
\begin{equation}
v_{g, {\rm rms}}^2(r) = \frac{\Omega^{1.2} H_0^2}{2\pi^2} \int P(k)W(kr) dk
\label{eq:vg_rms}
\end{equation}
where $W$ is the window function of the velocity survey. Some studies measure the one-dimensional velocity dispersion component of the statistically isotropic random velocity field which is related to the above expression via $\sigma_{g,1}^{1D}=v_{g, {\rm rms}}/\sqrt{3}$.

Eq. \ref{eq:vg_rms} gives the rms value of the predicted random velocity field: more precisely in an isotropic model it is the dispersion around the zero mean of the bulk flow. Since the flow is measured only in our local realization of the density field of the $\Lambda$CDM model these predictions are subject to (scale-dependent) cosmic variance. In order to compare with the actual measurements of the bulk flow amplitudes it is therefore imperative to evaluate how likely the different realizations are. This can be done analytically via a simple prescription outlined in e.g. Gorski (1991). For a Gaussian density field the peculiar velocity distribution on
linear scales is Maxwellian, with the probability density of
measuring a 1-D bulk velocity being:
\begin{equation}
p(V) dV \propto V^2\exp\left(-\frac{V^2}{2v_{g, {\rm rms}}^2}\right) dV
\label{eq:velprob}
\end{equation}
The probability of finding a region with $V <V_0$ is then $P(V_0)$=$\Gamma(\frac{3}{2},\frac{3V_0^2}{2\sigma_V^2})$ where $\Gamma$ is the incomplete gamma-function normalized to $\Gamma(n,\infty)$=1. The (68\%, 95\%) c.l. require $V_0=(1.08,1.6)\sigma_V$. The distribution of shear is discussed in Jaffe \& Kaiser (1995).

Within the gravitational instability picture one can relate the velocity correlation function to the {\it observed} 2-point galaxy correlation function as follows (Kashlinsky 1992, Juszkiewicz \& Yahil 1989):
\begin{equation}
\nabla^2\vartheta_g(r) = -H_0^2 f^2(\Omega) \xi(r)
\label{eq:k92}
\end{equation}
This equation assumes that light (galaxies) traces the peculiar mass accurately, an assumption which appears to hold at least on linear scales, whereas on small scales the above can be corrected downward by the bias parameter $b$ which measures the ratio of light-to-mass fluctuations. This formulation is advantageous as it connects two directly observed/measured quantities and, in principle allows to isolate the velocity component from gravitational instability.

Because this component of the net velocity field is coupled with the underlying correlation function of the matter/light distribution, it generates the  Kaiser (sometimes called ``rocket") effect (Kaiser 1987) which leads to a predictable anisotropic distortion in the transverse and parallel (to the $z$-direction) measurements of the 3-dimensional $\xi$. Thus one can use observational limits on the distortions of the galaxy correlation function in order to obtain information on the gravitational instability component of the observed velocities. This would be important in identifying the nature of the measured flows.

\subsection{Inflation and large-scale velocity field}
\label{sec:vfrominflation}

The inflation-based theories, such as the concordance $\Lambda$CDM model, make specific - and practically model-independent - predictions on the magnitude and distribution of peculiar velocities due to peculiar gravity. Indeed in these models on scales that were outside the horizon during the radiation-dominated era, the peculiar density field remains in the Harrison-Zel`dovich regime (i.e., $P(k)\propto k$ or $\delta_{\rm rms} \propto r^{-2}$) set during the inflationary
epoch and verified in CMB data.  On these scales, the peculiar bulk velocity
caused by gravitational instability should decrease as $V_{\rm rms} \propto r^{-1}$
and be quite small. Since the horizon scale of matter-radiation equality is $\lsim$100
Mpc, the concordance cosmology predicts a component due to gravitational
instability of $V_{\rm rms} \simeq 250 (r/100h^{-1}{\rm Mpc})^{-1}$km s$^{-1}$ at
$r\ga100h^{-1}$Mpc. Note that this expression specifies both the typical amplitude
and the fall-off with $r$ of the peculiar velocity field generated by
gravitational instability. The one-dimensional component of the velocity field would be $1/\sqrt{3}$ factor smaller. {\it Any deviation from this prediction would indicate a break-down of the standard cosmological model and may represent evidence of "new physics"}.

 \begin{figure}[h!]
\includegraphics[width=4in]{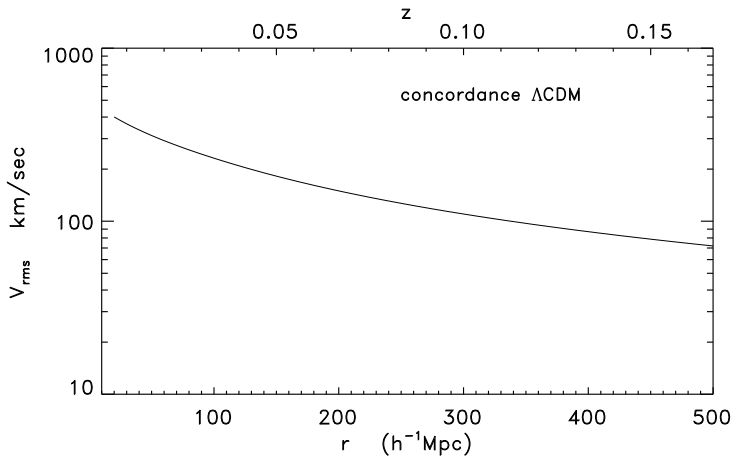}
 \caption[]{The rms peculiar velocity for the concordance $\Lambda$CDM model. At $100h^{-1}$Mpc the rms velocity field field is about $\simeq 235$ km/sec dropping to $\simeq 72$ km/sec at $500h^{-1}$Mpc. }
\label{fig:vel_lcdm}
\end{figure}

Fig. \ref{fig:vel_lcdm} shows the rms velocity field expected in the standard cosmological model.
Thus in the concordance $\Lambda$CDM model with
$v_{\rm rms}=(150,109)$ km/sec at $(200,300)h^{-1}$Mpc, the 95\% of
cosmic observers should measure bulk flow velocities less than
(240,180) km/sec at these scales.

\subsection{Evolution of ``primeval" velocity field}
\label{sec:vprimeval}

As long as Newton's laws hold,  this velocity component must be present -- but {\it any additional components would point to an incompleteness of the standard cosmological model}. This incompleteness can either direct to a new understanding of the global world structure, or require new physics. Of particular interest in this discussion is the evolution of any ``primeval" velocity component, $v_\bot$.

This component is statistically orthogonal to peculiar gravity and its evolution can be described using the conservation of angular momentum, preserved here because of absence of tangential forces. The angular momentum of unit proper (physical) volume is then $\rho [\vec{x}_{\rm phys}\times\vec{v}_\bot] d^3\vec{x}_{\rm phys}$ whose magnitude is $ \propto \rho v_\bot a^4=$ const, which leads to the following evolution (e.g. Collins \& Hawking 1973):
\begin{equation}
v_\bot \propto \;\; \left\{ \begin{array}{cc}
{\rm constant} & ;\;\; {\rm radiation}\\
(1+z)^{-1} & ;\;\; {\rm matter}
\end{array}
\right.
\label{eq:v_primeval}
\end{equation}
This leads to an interesting situation: if at some early epoch, $z\gg1$, both matter and radiation were coupled and moving with respect to the uniform expansion frame, then after the matter-radiation decoupling at $z\sim 10^3$ the peculiar velocity of matter will have decayed to a negligible magnitude, and that that of the CMB would then be shifted with respect to matter because the radiation should have preserved any initial velocity and will now appear moving with respect to the matter and uniform expansion frames (Matzner 1980).

This solenoidal component - if present - is unrelated to the physics determined by the cosmological conditions which arise as a result of the slow rollover of the inflation-driving field. They therefore do not probe, confirm or rule out the standard cosmological model ($\Lambda$CDM) which is so successful in accounting for various aspects of structure formation and evolution in the Universe and is confirmed by the CMB fluctuation measurements. Rather, this possible component of the velocity field would probe - within the inflationary paradigm - the primeval state of the inflaton field and the overall preinflationary structure of space-time.

Importantly, this component does not generate Kaiser effect as it is unrelated to the observed galaxy correlations, and its existence across the cosmological horizon is equivalent to a primeval CMB dipole. It also is unlikely to have shear.

\section{Results from earlier measurements}
\label{sec:earlyresults}

In this section we review briefly the earlier measurements of peculiar velocities starting with the CMB dipole anisotropy.

There are two ways of getting at the peculiar velocities: 1) by direct measurements of deviations from the Hubble expansion using various phenomenologically established distance indicators, and 2) measuring the peculiar gravity from the distributions of selected sources (galaxies or clusters) and determining the expected direction and amplitude of the infall flows induced by this gravitational potential. The two methods should agree in the absence of the solenoidal component. The prior results from them are reviewed briefly below following the CMB dipole subsection.

In a nutshell, the problem was - and remains - that the two are not entirely consistent in the sense that they give, at statistically significant levels, different directions, coherence length and even the amplitude of the peculiar flow. It can, in principle, be argued that the light-traces mass assumption is not strictly valid in methods 2) or that some massive structures loom behind the Zone-of-Avoidance (ZoA) obscured from view by the Milky Way galaxy. Nevertheless, this unsatisfactory - at face value - situation has led Gunn to state already in 1988 that ``Most of the problem ... would disappear if the [CMB] did not, in fact, provide a rest frame" since the basic cornerstone of these analyses is that the entire CMB dipole arises from the Doppler effect due to local peculiar motion.

Excellent reviews on the subject exist in the literature, such as early discussions by Davis \& Peebles (1983) on the Virgo infall, Gunn (1988) on the Great Attractor era measurements, followed by the later overviews of the subject by Strauss \& Willick (1995) and Willick (2000) among many others. We refer the reader to these reviews for further details.

\subsection{CMB dipole and the local motions}
\label{sec:cmbdipole}

The dipole anisotropy of the CMB has by now been well established from the {\it COBE}
FIRAS (Fixsen \etal 1994a) and DMR (Kogut et al 1993) measurements. With respect to the Sun restframe, the CMB has a dipole amplitude of $3.346\pm
0.017$~mK in the direction of $(l,b)_{\rm CMB} =  (263.85\pm 0.1,
48.25 \pm 0.04)^\circ$ (Hinshaw \etal\ 2009).

For a black-body background, such as the CMB, the motion at local velocity $v_{\rm loc}\equiv \beta c$ generates a CMB dipole in the direction of the motion of:
\begin{equation}
T(\theta)= \frac{T_{\rm CMB}}{\sqrt{1-\beta^2} (1-\beta \cos\theta)} \simeq T_{\rm CMB}[1 + \beta \cos\theta - \beta^2  \cos^2\theta + O(\beta^3)\times {\rm octupole}]
\label{eq:doppler}
\end{equation}
where the expansion assumed $\beta \ll 1$ with the dominant fluctuation generated in the (second) dipole term, followed by small contributions to the quadrupole and negligible contributions to octupole and higher moments (Peebles \& Wilkinson 1968).

If the entire CMB
dipole is of kinematic origin, its observed amplitude in the Sun-centric system corresponds to velocity
of $V=370$ km/sec in that direction. At least a substantial part of this motion must originate from the local motions of the Sun and the Galaxy,
so the conventional paradigm has been that {\it all} of the CMB dipole
can be accounted for by motions within the nearby $30-100$~Mpc
neighborhood.
This then allows to translate the measured CMB dipole into the corresponding velocity under the assumption that its origin is purely kinematic. However, in order to isolate truly cosmological motions, one should subtract from the CMB dipole measured in the Sun-centric system the motions of the Sun with respect to the Galaxy and the motion of the latter with respect to the Local Group. Table \ref{tab:localmotion} adopts the values from Kogut et al (1993, Table 3) in this conversion to the motion of the Local Group with respect to the CMB.

\begin{deluxetable}{c c c c}
\tabletypesize{\scriptsize}
\tablecaption{Summary of local motion data reproduced from Kogut et al (1993).}
 \startdata
  & $V$ (km/sec) &$(l_{\rm Gal}, b_{\rm Gal})^\circ$ & Refs\\
\hline
\hline
Sun-CMB & $369.5\pm 3.0$ & $(264.44\pm 0.3, 48.4 \pm 0.5)$ & Kogut et al (1993) \\
(COBE/DMR-based) & & & \\
\hline
Sun-LSR & $20.0\pm 1.4$ & $(57\pm 4, 23\pm 4)$ & Kerr \& Lynden-Bell (1986) \\
LSR-GC & $222\pm 5$ & $(91.1\pm 0.4, 0)$ & Fich et al (1989) \\
GC - CMB & $552.2 \pm 5.5$ & $(266.5 \pm 0.3, 29.1\pm 0.4)$ & Kogut et al (1993) \\
\hline
Sun - LG & $308\pm 23$ & $(105\pm 5, -7 \pm 4)$ & Yahil et al (1977) \\
\hline
\hline
 & & & \\
LG-CMB & $627\pm 22$ & $(276\pm 3, 30\pm 3)$ & Kogut et al (1993) \\
\enddata
\label{tab:localmotion}
 \end{deluxetable}

CMB dipole is not equal to the Local Group velocity and is in the direction $\sim 45^\circ$ away from Virgo. The motion of the Local Group with respect to the CMB appears to be at $\simeq 630$ km/sec in the direction  given in the last entry of Table \ref{tab:localmotion}. The difference was used by Lilje et al (1986) to predict bulk motion of the Local Supercluster at $\sim 600$  km/sec.

The big question in this context is at what scale is the matter frame, as defined by the uniform and isotropic expansion of galaxies and clusters, at rest with respect to the CMB as defined by its dipole. We next move to discussing the early (mostly late 20th century) important attempts at answering this question. The progress toward this goal was advanced in two independent directions: 1) reconstructing the peculiar velocity field directly with measurements of deviations from the Hubble flow using various galaxy distance indicators, and 2) reconstructing peculiar gravitational field, using galaxy and cluster surveys, which presumably drives this flow. In the gravitational instability paradigm the direction of the latter should coincide with that of the CMB dipole and the velocity field, which in turn should vanish at some distance.

Before we move to discussing the reconstructions of peculiar velocity field and peculiar gravity, we note the basis limitation of addressing the the issue of misalignments with the CMB dipole direction: in any velocity measurement with the signal-to-noise $S/N\sim$ a few, the directional uncertainty is large at $\Delta \theta \simeq \sqrt{2}(S/N)^{-1}$ radian. Whereas, the CMB dipole has by now been measured at $S/N> 10^2$ and so its direction is determined remarkably accurately with $\Delta \theta_{\rm CMB}\ll1^\circ$, the peculiar velocity measurements are at best measured with $S/N\sim 3-5$ and hence any alignment (or misalignment) is established only down to the limits of at best a few (or even few tens) degrees.

\subsection{Distance indicator based measurements}
\label{sec:vfromd}

Most determinations of the peculiar velocities are based on
surveys of individual galaxies using empirical correlations
for distance indicators, e.g. the Tully-Fisher (TF - Tully \& Fisher 1977) relation for
spirals, the Fundamental Plane (FP) relations for elliptical
galaxies (Djorgovski \& Davis 1987, Dressler et al 1987) and using SNIa as standard candles. The methodology is very generally as follows: given the spectroscopic determination of galaxy's redshift {\it and} determination of its distance via an empirical relation leading to its absolute luminosity, the line-of-sight (direction $\hat{r}$) peculiar velocity is determined as:
\begin{equation}
\vec{v}_p\cdot \hat{r} = cz - H_0 {\cal D}
\label{eq:vp-los}
\end{equation}
where the distance ${\cal D} \propto l^{-1/2}$, the square-root of the apparent luminosity. Importantly, {\it here the redshift $z$ is corrected to the coordinate frame with zero CMB dipole anisotropy}.

Given the measured quantities given by eq. \ref{eq:vp-los} one determines the bulk velocity of a given volume spanned by the survey by maximizing the likelihood, or minimizing the variance or $\chi^2$. As an example, if one were to minimize the $\chi^2$ or the likelihood, ${\cal L}$:
 \begin{equation}
\chi^2=\sum_{i=1}^N w_i (\vec{v}_{p,i}\cdot \hat{r}_i -\vec{V}_{\rm bulk}\cdot \hat{r}_i)^2\;\; ;\;\; \ln{\cal L}\propto -\sum\ln\sigma_i \ -\chi^2
 \label{eq:mle}
 \end{equation}
with the sum taken over $N$ galaxies, the solution for the $\alpha$-direction component of the bulk flow would be for $\chi^2=\min$:
\begin{equation}
V_{\rm bulk, \alpha}=\frac{\sum_i w_i \hat{r}_{i, \alpha} (\vec{v}_{p,i}\cdot \hat{r}_i)}{\sum_i w_i \hat{r}_{i, \alpha}^2}
\label{eq:vfromchi}
\end{equation}
where the weights $w_i=1/\sigma_i^2$ are related to the errors in each measured galaxy. One should also be careful as the effects of geometry may be important (cf. Kaiser 1988) since the above expression assumes that over the survey volume the cross terms vanish: $\sum_i \hat{x}_i\hat{y}_i =0$ with constant weights etc. If the weights are constant and barring geometrical irregularities, the denominator reduces to $\sum_i \hat{r}_{i, \alpha}^2=1/3$. This simplified expression has been revised by Kaiser (1988) who gives a more rigorous formalism of deducing the 3-dimensional bulk flow from realistic data.

SNIa give the most accurate distance indicators individually ($\sim 8\%$ per source) but provide small samples, TF relation is accurate to $\sim 15\%$ and probes late-type galaxies located predominantly in the field, whereas FP is valid for early-type galaxies populating clusters and is accurate to $\sim25\%$ per individual galaxy.  Important corrections have to be made throughout in order to arrive at peculiar velocity component with respect to the CMB rest frame: for reddening, Malmquist bias, subtracting the CMB dipole after assuming it originates entirely from the Doppler effect, etc.

Early measurements by Rubin et al (1973, 1976) found
peculiar flows with respect to the CMB of $\sim 600-700$ km/sec but were widely dismissed at the time. Interestingly, the direction of the motion of Rubin et al pointed toward Galactic coordinates $\sim(l, b) \sim (290^\circ, 30^\circ)$, the direction which - within the error bars - will appear again later.

On the other hand, Aaronson et al (1986) probed motions of late-type spirals in clusters, measured with Arecibo, in a direction {\it perpendicular} to that of the CMB dipole and find that the clusters in {\it that direction} are at rest with respect to the CMB dipole frame to within the observational errors of $\sim 200$ km/sec.

A major next advance was made using the FP relation for a sample of several hundred galaxies in a direction roughly orthogonal to that probed by Aaronson et al (1986). The the implication of that study was that elliptical
galaxies within $\sim 60h^{-1}$Mpc were streaming at $\sim 600$
km/sec with respect to the rest frame defined by the CMB (Dressler et al 1987). Lynden-Bell et al (1987) suggested that elliptical galaxies were streaming at $\sim 600$ km/sec into a massive concentration of about $\sim 10^{16}M_\odot$ centered around Hydra-Centaurus clusters about $\sim 35h^{-1}$Mpc away and dubbed it the ``Great Attractor".

Mathewson et al (1992a) used the TF relation of a large sample of almost $\sim 1,500$ galaxies (Mathewson et al 1992b) and found
that this flow of amplitude 600 km/sec does not converge until
scales much larger than $\sim 60 h^{-1}$ Mpc and claimed no back-side infall into the Great Attractor suggesting that ``there is bulk flow in the supergalactic plane over very large scales greater than 130/h Mpc". These results were in agreement with those of Willick (1990) using TF relation for over 300 galaxies in the Perseus-Pisces Supercluster (see also Willick 1999).

A sample of 24 SNIa by Riess et al
showed no evidence of significant bulk flows out to $\sim 100
h^{-1}$ Mpc, and a similar conclusion was reached with a TF-based
study of spiral galaxies of Courteau et al (2000).

Lauer \& Postman (1994) measured the velocity of the Local Group with
respect to an inertial frame defined by the 119 Abell and ACO (Abell, Corwin,
\& Olowin 1989) clusters contained within $150h^{-1}Mpc$ and found
the motion of the LG with respect to the Abell sample to be inconsistent
with its motion inferred from the CMB dipole anisotropy. Using brightest cluster galaxies as
distance indicators they find motion toward
$(l,b)=(220,-28)^\circ$ with uncertainty of $\pm27^\circ$. However, a re-analysis of these data by Hudson \& Ebeling (1997) taking
into account the correlation between the luminosities of
brightest-cluster galaxies and that of their host cluster found a
reduced bulk flow pointing in a different direction.

Using the FP relation for early type
galaxies in 56 clusters Hudson et al (1999) find a bulk flow of a
similarly large amplitude of $\sim 630$ km/sec to
Lauer \& Postman on a comparable scale, but in a different
direction.

On the other hand, a survey of spiral galaxies by Courteau et al (2000) used TF relation and claimed negligible velocities ($\lsim 300$ km/sec) with respect to the CMB at distances $\gsim 50h^{-1}$Mpc. These results are also in agreement with the Giovanelli et al (1998a,b) extensive compilation of the TF-based distances.

While important, such measurements are affected by
complex systematics (e.g. Malmquist bias), sparse (and inhomogeneous) sampling, do not measure velocities directly in the CMB frame,  and at times yielded discrepant results for the
amplitude and direction of peculiar flows.
More critically in the context of this review, they also fail to probe scales $\gsim 100$ Mpc, a particularly
important range for testing the gravitational-instability paradigm
for the origin of the flows. It is thus paramount to develop
alternative methods that enable accurate measurements of
large-scale flows.

For a more comprehensive review of these findings see the extensive review by Strauss \& Willick (1995).

\subsection{Dipoles of mass tracers and peculiar gravitational field}
\label{sec:vfromdips}

Since in the gravitational instability picture the velocities must
trace peculiar gravitational potential one can and should compare
the two independent measures. Within gravitational instability paradigm the velocity is given by:
\begin{equation}
\vec{v}(\vec{r}) = \frac{2f(\Omega_{\rm m})}{3H_0\Omega_{\rm m}}\;\;\vec{g}(\vec{r})
\label{eq:vfromg}
\end{equation}
where the peculiar gravitational acceleration is can be derived from the observed (galaxy or cluster) density distribution provided the sources trace the overall mass:
\begin{equation}
\vec{g}(\vec{r}) = G \int_r^\infty \delta\rho_{\rm m}(\vec{r}^{\; \prime}) \frac{\vec{r}^{\; \prime}-\vec{r}}{|\vec{r}^{\; \prime}-\vec{r}|^3}\; d^3\vec{r}^{\; \prime}
\label{eq:g_pec}
\end{equation}
It is important to emphasize that the direction of the peculiar gravitational acceleration remains fixed in linear regime, which is described by eqs. \ref{eq:velfrommotion}. Thus the peculiar velocity component produced by gravitational instability should point in the same direction as at present. Because both the fluxes and gravity decrease with distance as $r^{-2}$, then assuming that the gravitational
potential is traced by galaxies or clusters one can evaluate the
amplitude and convergence of the various dipoles measured in the
corresponding surveys. Thus a complementary technique aimed at constraining bulk motions reconstructs directly the peculiar gravity of the observed galaxy distribution and uses
measurements of the dipole in the distributions of light and matter.

Very generally, existence of a statistically significant misalignment between the reconstructed gravity dipole and the observed all-sky CMB dipole would argue for a part of the latter being intrinsic/primordial.

\subsubsection{Peculiar acceleration using galaxies}




The observations of large scale streaming motions described in Sec.~\ref{sec:vfromd}
were difficult to reconcile with the biased Cold Dark Matter cosmology
which was the standard model during that period.
By comparing the CMB velocity vector with the acceleration vector,
it was possible to investigate the cause of the LG motion
and its cosmological implications.
The LG acceleration can be estimated using galaxy surveys that
trace the matter distribution.
This technique was first applied by Yahil, Sandage \& Tammann (1980) using
the revised Shapley–Ames catalogue and later by Davis \& Huchra (1982)
using the Centre for Astrophysics (CfA) catalogue.

Analyses using optically selected galaxies were
limited due to limited sky coverage, especially at low Galactic latitude.
These studies improved significantly when full-sky galaxy samples became available.
The IRAS Point Source Catalog allowed to construct full-sky galaxy catalogs
due to the large (96\%) sky coverage of the satellite and the negligible
Galactic extintion in the infrared.  Yahil, Walker \& Rowan-Robinson (1986),
Meiksin \& Davis (1986), Harmon, Lahav \& Meurs (1987),
Villumsen \& Strauss (1987) and Lahav, Rowan-Robinson \& Lynden-Bell (1988)
used only the position and fluxes of the sources to obtain the LG dipole.
The dipole vectors derived by these authors are in agreement with each other
and the CMB dipole vector to within $10^\circ-30^\circ$.
Yahil et al (1986) and Meiksin \& Davis (1986)
found that the dipoles from the optical and IRAS
galaxies are nearly aligned with each other, but are misaligned
from the CMB dipole by $\sim 30^\circ$
Furthermore, both showed that the peculiar velocities identified by
7S were essentially absent in the IRAS
accelerations (see discussion in Gunn 1988).

Using galaxy catalogs with galaxy redhsifts allowed to estimate the
scale of convergence, i.e.,  the distance at which most of
the peculiar velocity of the LG is generated. The earlier work by
Rowan-Robinson et al (1990) was extended by
Strauss et al (1992) who used a redshift survey of 5288 IRAS galaxies, covering
87.6\% of the sky. Their computed acceleration
of the LG pointed $18^\circ-28^\circ$ away from the direction of the LG peculiar
velocity vector, the differences depending on the model used for the velocity
field and the window through which the acceleration was measured. Strauss et al (1992)
argue misalignments of this amplitude were to be expected due to
shot noise, finite window and nonlinear effects. They concluded that
the data were consistent with the acceleration being mostly
due to galaxies within $40h^{-1}$Mpc. However, since
the peculiar acceleration
of the LG was calculated using redshifts instead of real distances
it differs from the actual acceleration of the local group
due to redshift distortions (Kaiser 1987; Kaiser \& Lahav 1989)
Strauss et al (1992) noted that the convergence of the
acceleration vector depended on the corrections made for the motion of the LG.
Later Webster, Lahav \& Fisher (1997) using IRAS galaxies,
Lynden-Bell, Lahav \& Burstein (1989) with optically selected galaxies
and da Costa et al.  (2000) with a sample of early-type galaxies concluded that
the LG acceleration is mostly due to galaxies within $\sim 50h^{-1}$Mpc.
These latter results are in contrast with those derived from cluster samples
that indicate that there is a significant contribution up to $200 h^{-1}$Mpc
(see Sec.  \ref{sec:clusterdipole}).

Juskiewicz, Vittorio \& Wyse (1990) discuss a related
discriminant which would be the misalignment
angle between the peculiar velocity $\vec{v}_R$ induced on the LG by the
matter within a sphere of radius $R$ and the apex of the CMB dipole anisotropy. In order to test the cosmological models popular at the time, they confronted the model predictions with the data, provided by IRAS survey of galaxies, but found that the data sets were not deep enough
to discriminate the predictions although they did discuss the variation of the misalignment
angle between the direction of the LG velocity and the
apex of the dipole anisotropy of the CMB within cosmological models.

More recently Erdogdu et al (2006)
estimated the acceleration on the LG from  a sample of 23,200 galaxies from
the 2MASS Redshift Survey (2MRS). They computed the flux-weighted dipole of the sample arguing that this is a robust statistic that closely approximates the mass-weighted dipole, and so it
is not affected by redshift distortions and requires no preferred reference frame.
The redshift information enabled them to determine the variation and convergence of the dipole
with distance. They found a statistically significant misalignment between the LG and the CMB dipole: the misalignment angle in their compilation is $12^\circ\pm 7^\circ$ at $\sim 50 h^{-1}$Mpc, and {\it increases} to $21^\circ\pm 8^\circ$ at $130h^{-1}$Mpc although they argue that is within
$1\sigma$ of the dipole probability distribution in a $\Lambda$CDM model
with $\Omega_m=0.3$.


\subsubsection{Peculiar acceleration using clusters}
\label{sec:clusterdipole}


Since the space density of galaxy clusters is much lower than that of galaxies, clusters are less well suited to map the distribution of matter (or the peculiar-velocity field) on small scales. Marking the deepest gravitational potential wells they are, however, more reliable tracers of the mass distribution -- and thus the distribution of gravitational attractors -- on large scales.

Cluster-based dipole studies suggested early on that the motion responsible for the CMB dipole does not converge out to very large distances. Scaramella et al.\ (1991) used the distribution of Abell clusters to argue that the acceleration of the LG converges at about 180 $h^{-1}$ Mpc, with roughly one third of the acceleration being generated at distances greater than 60 $h^{-1}$ Mpc. The suspected source of at least part of the large-scale component of the acceleration, the Shapley Concentration (SC) located at approximately 140 $h^{-1}$ Mpc and well aligned with the direction of the CMB dipole, was independently identified by Plionis \& Valdarnini (1991) by measuring the dipole anisotropy in the distribution of the X-ray brightest Abell-type clusters (XBACs; Ebeling et al.\ 1996). Complementing the XBACs sample with clusters at very low Galactic latitude from the CIZA sample (Ebeling et al.\ 2002; Kocevski et al.\ 2007), Kocevski et al.\ (2004) estimated the contribution of the Shapley Concentration to the LG's motion to be as high as 50\%. Whether the motion really converges at that distance remains unclear though: Raychaudhury (1989) questioned whether the SC is in fact massive enough to account for the contribution to the local velocity field attributed to it. The suspicion that the Shapley Concentration is not the sole large-scale contributor to the LG's acceleration is strengthened by the fact that back infall into the SC has yet to be observed. Indeed, Kocevski \& Ebeling (2006) find the cluster dipole amplitude to only flatten, but not decrease at distances beyond 180 $h^{-1}$ Mpc, in contrast to the signature of (temporary) back infall observed at 50 $h^{-1}$ Mpc, the distance of the Great Attractor (Fig.~9 of Kocevski \& Ebeling 2006).

Probing the mass-density or the peculiar-velocity field at distances approaching or even exceeding 200 $h^{-1}$ Mpc is currently possible only with cluster-based studies. However, even the largest statistically complete all-sky cluster samples become increasingly sparse at these distances, preventing a robust direct determination of the amplitude and direction of the cluster dipole. Fortunately, clusters offer another path to measuring large-scale flows, namely by using them as tracers of the velocity field rather than as beacons of the mass distribution. Details of this alternative approach are provided in the following section.

\section{Sunyaev-Zeldovich effect and peculiar velocity measurements}
\label{sec:sz}

An alternative method of measuring peculiar velocities is by utilizing the Sunyaev-Zeldovich effect produced by Compton scattering of the CMB photons by the hot X-ray emitting gas in clusters of galaxies. In this section we provide a brief overview of the SZ effect relevant
for this discussion. Excellent review of the physics of the SZ effect
is given in Birkinshaw (1999) to which the reader is referred for additional details.

\subsection{Compton/Thomson scattering and SZ effect}
\label{sec:thomson}

We begin by discussing the change in photon frequency from a single
electron scattering.
When a photon is scattered by an electron, both particles exchange
energy and momentum.  In the rest frame of the electron, a photon with
frequency $\nu_e$ moving in the direction $\hat{x}_e$ will change its
frequency to $\nu_e^\prime$ and direction to $\hat{x}_e^\prime$
according to the standard Compton scattering formula:
\begin{equation}
\frac{\nu^\prime}{\nu}=\left[1+\frac{h_P\nu}{m_ec^2}
	(1-\hat{x}_e\cdot \hat{x}_e^\prime)\right]^{-1}
\label{eq:compton}
\end{equation}
Given that for clusters
one reasonably expects $v/c\ll 1$ and CMB observations are
conducted in the microwave range of the electromagnetic spectrum,
the scattering is almost elastic ($\nu\simeq\nu^\prime$) and causes
a considerable simplification in the physics.
If the electron is moving with velocity $\vec{v}$ in the reference
frame of the observer, standard Lorentz transformations give:
\begin{equation}
\frac{\nu^\prime}{\nu} = \frac{1-\vec{v}\cdot \hat{x}/c}{1-\vec{v}\cdot
\hat{x}^\prime/c +\frac{h_p\nu}{\gamma m_ec^2}(1-\hat{x}\cdot \hat{x}^\prime)}
\label{eq:nu2nuprime}
\end{equation}
where we have dropped the subscript $e$ to refer to quantities in
the observer reference frame (see e.g. Appendix of
Sunyaev \& Zeldovich 1980). For the physical conditions
present in clusters of galaxies and for the range of frequencies probed
by WMAP ($\nu\lsim 100$ GHZ) and Planck ($\nu \lsim 1$THz),
where $h_P\nu/\gamma m_ec^2\simeq 8.2\times 10^{-9} (\nu/1{\rm THz})$
and $v_e\ll c$, the denominator in
eq.~(\ref{eq:nu2nuprime}) can be expanded in Taylor series.
On average, one ``typical" scattering leads to the following fractional
frequency shift:
\begin{equation}
\frac{\Delta \nu}{\nu}
\simeq -\frac{\vec{v}}{c}\cdot(\hat{x}-\hat{x}^\prime)
- (\frac{\vec{v}}{c}\cdot\hat{x})(\frac{\vec{v}}{c}\cdot\hat{x}^\prime)
-(\frac{\vec{v}}{c}\cdot \hat{x}^\prime)^2
\label{eq:dnu_scattering}
\end{equation}
In this expression, the direction of the incoming photon is random, but
the outgoing photon must be in the direction of observation $\hat{x}_{obs}=-
\hat{x}^\prime$. When eq.~(\ref{eq:dnu_scattering}) is
averaged over all possible incoming
photon directions then all terms linear in $\hat{x}$ are zero.

To obtain the net energy-momentum transfer between the CMB radiation and
the hot X-ray gas one needs to average over the entire electron population.
On average CMB photons undergo $\tau$ scatterings, where $\tau$ is
the projected cluster optical depth due to Thomson scattering.
If $n_e(r)$ denotes the number density of electrons in the cluster measured
from the center of the cluster, $\tau$ is given by
\begin{equation}
\tau = \sigma_T \int n_e(r) dl \sim 6\times 10^{-3} \;\;
(\frac{n_e}{10^{-3}{\rm cm}^{-3}})\; (\frac{R_{\rm cluster}}{3\; {\rm Mpc}})
\label{eq:opticaldepth}
\end{equation}
where the integration is taken along the line of sight. The averaged
frequency shift induced on a photon of frequency $\nu$
by the cluster electron population is
\begin{equation}
\langle\tau\frac{\Delta\nu}{\nu}\rangle=
-\langle\tau\frac{\vec{v}}{c}\rangle\hat{x}_{obs}
-\langle\tau\left(\frac{\vec{v}}{c}\cdot\hat{x}_{obs}\right)^2\rangle ,
\label{eq:dnu_dtt}
\end{equation}
with averages taken over the cluster distribution.
The two terms on the RHS above are different in nature. The first of these
corresponds to the temperature shift due to the motion of the
cluster as a whole; it is termed the Kinematic (sometimes called kinetic) SZ (KSZ) effect.
The second term is due to the thermal motions of electrons in the
cluster potential well. It is usually expressed in terms of the electron
temperature, i.e. $\langle (\vec{v}\hat{x}_{obs})^2\rangle \propto k_BT_e/m_ec^2$ and
is commonly known as the Thermal SZ (TSZ) effect.

The KSZ effect has a clear physical interpretation.
If we neglect the random motion of the electrons in the potential well of the
cluster, and consider only their motion due to the peculiar
velocity $\vec{v}_{cl}$ of the cluster, then $\langle \vec{v}\rangle =
\bar{\tau}\vec{v}_{cl}$. If in the observer rest frame the CMB photons
are isotropic,
in the cluster rest frame the radiation field will have a dipole pattern.
At the velocities and frequencies involved, Compton scattering is
elastic, and there is no net transfer of energy. Scattering produces only
a change in the direction of the photons.
The effect on the observer reference frame is different, however. For example,
if the cluster is moving away along the direction of observation $\hat{x}_{obs}$,
photons coming towards the observer are blue-shifted in the reference
frame of the cluster as they are scattered off their trajectory
while other photons, less energetic,
are scattered back towards the observer. The net effect is
to change the radiation temperature in the direction of the cluster.
This change will appear as a decrement (negative temperature) for
clusters moving away (receding) from the observer.

This effect also occurs if cluster and observer
are at rest with respect to each other and the CMB dipole is intrinsic, or
primordial. We can define two different reference frames: the matter
rest frame (MRF) and the isotropic CMB frame (ICF). If the CMB dipole
is intrinsic, these two sets of observers do not coincide and both
assign a peculiar velocity $v_{CMB}$ to each other.
In this case, in the cluster rest frame the photon
distribution will not be homogeneous but will have a distribution
pattern $\nu[1+(v_{cmb}/c)\cos\theta]$ where $\theta$ is the angle
with respect to the direction of observation $\hat{x}_{obs}$.
Then
\begin{equation}
\frac{\nu^\prime[1-\vec{v}_{cmb}\cdot\hat{x}^\prime/c]}{\nu
[1-\vec{v}_{cmb}\cdot\hat{x}/c]} = \left(1-\frac{h_p\nu}{\gamma m_ec^2}
(1-\hat{x}\cdot \hat{x}^\prime)\right)^{-1}\approx 1
\label{eq:nu2nuprime2}
\end{equation}
and the average frequency shift adding the contributions
of photons incoming from all directions is:
\begin{equation}
\int d\omega_{\hat{x}}\left(\frac{\Delta\nu}{\nu}\right)=
\frac{\vec{v}_{cmb}}{c}\hat{x}^\prime=
-\frac{\vec{v}_{cmb}}{c}\hat{x}_{obs}
\label{eq:dnu_scattering2}
\end{equation}
i.e., identical result to  eq.~(\ref{eq:dnu_scattering}).
In both cases, the radiation field has a dipole pattern in
the cluster rest frame and Compton scattering removes
photons from its initial trajectory and adds them from other directions,
altering the initial dipole pattern in the frames of the cluster and
of the observer.

\subsection{Spectral dependence of the TSZ and KSZ effects.}
\label{sec:tsz}

The undistorted CMB blackbody spectrum is given by the Planck function.
Its spectral intensity in terms of the dimensionless frequency $x=h\nu/kT_{CMB}$
is given by
\begin{equation}
I_{CMB}(x)=\frac{2(k_BT_{CMB})^3}{(h_Pc)^2}\frac{x^3}{e^x-1}=I_o i(x) ,
\label{eq:intensity_cmb}
\end{equation}
where $T_{CMB}$ is the CMB temperature. In the last
expression, $I_o\equiv 2(k_BT_{CMB})^3/(h_Pc)^2$ and
$i(x)=x^3/(e^x-1)$ gives the spectral frequency dependence of the CMB.
This spectrum is distorted when CMB photons are scattered
by the free electrons residing in the potential wells of
clusters of galaxies (Sunyaev \& Zeldovich 1972, 1980a, 1980b)

Commonly, CMB measurements are presented as temperature anisotropy maps.
If we denote by $x^\prime=h\nu/(k_BT^\prime)$ with
$T^\prime=T_{CMB}+\Delta T$, $\Delta T$ being a temperature fluctuation, then the
spectral distortion produced by a small change in temperature
can be derived by Taylor expansion around $x=h\nu/(k_BT_{CMB})$:
\begin{equation}
I_{CMB}(x^\prime)=I_{CMB}(x)+\left(\frac{dI_{CMB}}{dx}(x^\prime-x)\right)=
I_o\left(i(x)-H(x)\frac{\Delta T}{T_{CMB}}\right)
\label{eq:cmbH}
\end{equation}
with $H(x)=x^4e^x(e^x-1)^{-2}$. In the cluster direction
the temperature anisotropy has two additional contributions: the
KSZ and TSZ components (see eq.~\ref{eq:dnu_dtt}).

Since the cluster peculiar velocity
is uncorrelated with cluster properites such as mass, temperature, electron
distribution, etc, then the KSZ contribution of eq.~(\ref{eq:dnu_dtt}) is
\begin{equation}
\langle\tau(\vec{v}/c)\rangle\hat{x}_{obs}=\langle\tau\rangle
\langle \frac{\vec{v}}{c}\rangle\hat{x}_{obs}=\bar{\tau}(v_{cl,r}/c)
\label{eq:tau_uncorrelated_v}
\end{equation}
where $v_{cl,r}$ is the component of the peculiar velocity along the line of sight.
The sign of the KSZ effect depends on the direction of the peculiar velocity
and is chosen such that, as indicated in the previous section, the effect would
give rise to a temperature decrement for receding clusters
of galaxies. In non-relativistic limit, the KSZ effect induces
a thermodynamic temperature shift on the incoming CMB photons that
is independent of the photon
frequency. Therefore, for the incoming CMB spectrum, the frequency
shift will be common to all scattered photons and the black-body temperature
will be shifted by
\begin{equation}
\left(\frac{\Delta T}{T_{CMB}}\right)_{KSZ}\equiv-
\langle\tau \frac{\vec{v}\cdot\hat{x}}{c}\rangle=-\bar{\tau}\frac{v_{cl,r}}{c} .
\label{eq:ksz_def}
\end{equation}

The contribution of the TSZ component is not as simple. The average frequency
shift will, in this case, depend on the incoming photon frequency. The scattering of
an isotropic and unpolarized radiation with a thermal distribution of
electrons changes the photon number density $n=c^2I(\nu)/2h_P\nu^3$
in a rather complex way. In the limit of the temperatures
being negligible with respect to the electron rest mass, the evolution
of the photon occupation number can be described
by the Kompaneets equation
\begin{equation}
\frac{\partial n}{\partial
t}=\frac{\sigma_T N_e h_P}{m_e c} \frac{1}{\nu^2}\frac{\partial}{\partial \nu}[\nu^4
{\large{(}}\frac{k_B T_e}{h_P}\frac{\partial n}{\partial \nu}+n+n^2{\large{)}}]
\label{eq:kompaneets}
\end{equation}
One can define the Comptonization parameter
\begin{equation}
y=\frac{\sigma_T k_B}{m_ec^2}\int dl T_e(r)n_e(r) \approx \frac{k_BT_e}{m_ec^2}\bar{\tau} ,
\label{eq:y_compton}
\end{equation}
where $r$ is the direction to the center of the cluster. The y-parameter
provides a measure of the electron pressure integrated along the line of sight.
The last approximation in eq~(\ref{eq:y_compton}) is correct if the intergalactic gas in the cluster
is isothermal. In the limit of $y\ll 1$  and for the initially
black-body radiation, $n=1/[\exp(x)-1]$, eq.~\ref{eq:kompaneets} specifies
the change in the photon spectrum as (e.g. Stebbins 1997):
\begin{equation}
\Delta n \simeq y
\frac{x\exp(x)}{[\exp(x)-1]^2}[x {\rm coth}\frac{x}{2}-4]
\end{equation}
Then, the change induced in the CMB black-body spectrum is:
\begin{equation}
I_{CMB}(x^\prime)-I_{CMB}(x)= \Delta I_{CMB}(x)=\frac{\Delta n}{n}I_{CMB}
=I_oH(x)[x {\rm coth}\frac{x}{2}-4]y
\label{eq:tsz1}
\end{equation}
and comparison with eq.~\ref{eq:cmbH} yields the temparature anisotropy generated
by the TSZ effect
\begin{equation}
\left(\frac{\Delta T}{T_{CMB}}\right)_{TSZ}=yg(\nu)
\label{eq:tsz_def}
\end{equation}
where $g(x)=[x {\rm coth}(x/2)-4]$. This expression leads to $g \simeq -2$ in the
Rayleigh-Jeans part ($x<1$ or $\nu \simeq 60$ GHz), crosses zero at $\sim$
220 GHz, but formally diverges as $x^2$ at higher frequencies. The (formal)
divergence arises because of the paucity of CMB photons at the Wien
part of the spectrum and their repopulation by Thomson scattering from the
CMB photons at initially lower frequencies.

Note that, while the KSZ effect corresponds to a frequency-invariant thermodynamic
temperature change leaving its energy spectrum identical to that of primary CMB anisotropies, the TSZ effect changes the amplitude and
sign in CMB maps of different frequencies as $g(x)$
(in brightness as $G(x)=g(x)H(x)$). This behavior is different from that
of any known foreground and makes the TSZ a very useful tool to detect
clusters in CMB data by comparing its brightness at different frequencies.
Also, being a distortion of the CMB spectrum, once imprinted, it will
not be thermalized and, due to adiabatic expansion of the ambient space-time, its $\Delta T$ will decay with redshift in the same way as the CMB temperature, making the SZ signature
independent of redshift.

\begin{figure}[h!]
\includegraphics[width=4in]{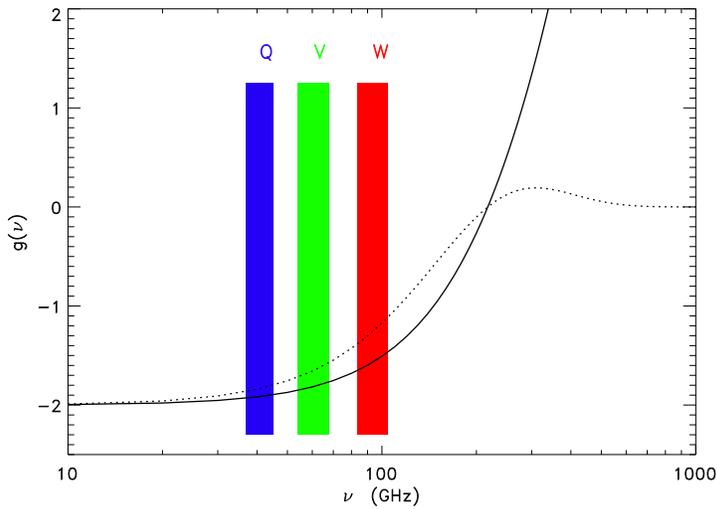}
\caption[]{Frequency dependence of temperature anisotropies,
$g$, due to TSZ effect is shown with solid line. Dotted line corresponds to the spectral dependence if measured by the effective antenna temperature, which is still occasionally used in CMB studies. WMAP frequency bands are also shown.}
\label{fig:tsz_sed}
\end{figure}

In the Rayleigh-Jeans
regime $g(\nu)$  is close to $-2$, vanishes near 217 GHz
and becomes positive at higher frequencies.
For the WMAP Q, V, W bands one gets $g(x) \simeq
-1.84, -1.65, -1.25$. Additionally, there may be non-thermal
components and relativistic corrections (Birkinshaw 1999).
The spectral dependence of the TSZ effect is shown in Fig. \ref{fig:tsz_sed}.

It is illustrative now to provide, from eqs.~(\ref{eq:ksz_def}),
(\ref{eq:y_compton}) and (\ref{eq:tsz_def}), an estimate of the
relative amplitudes of the KSZ and TSZ effects over a given
aperture $\theta_{\rm ap}$:
\begin{equation}
\frac{{\rm KSZ}}{{\rm TSZ}} \sim \left(
\frac{V_{\rm bulk}}{1,100\; {\rm km/sec}}\right) \;\;
\left(\frac{k_B\langle T_X(\theta)\rangle|_{\theta_{\rm ap}}}{1\;
{\rm KeV}}\right)
\label{eq:tsz2ksz}
\end{equation}
This estimate shows that, while KSZ is small compared to TSZ in
central regions of X-ray luminous cluster where $T_X\sim 10$KeV, the
two contributions may become comparable as soon the X-ray temperature
drops to the value expected in poor and/or outer parts of luminous clusters.

\subsection{Relativistic Effects.}
\label{sec:6.3}

The original Sunyaev-Zeldovich formula of the TSZ effect (eq.~[\ref{eq:tsz_def}])
was derived using the Kompaneets equation, a non-relativistic diffusion
approximation to the full kinetic equation describing the change of the photon
distribution due to scattering. This approximation may be insufficient
for massive clusters due to the high electron velocities and
the low scattering probability of photons in the intracluster medium.
The first relativistic treatment of the scattering
process was given by Wright (1979). Rephaeli (1995) showed the
effect to be more significant in the Wien part of the spectrum,
where deviations of the spectral shape of the intensity $I(x)$ from
eq. \ref{eq:tsz1} could reach a factor of 20\% at $300-400$GHz for
a cluster with $T_X\simeq 15$KeV at the same time also increasing the cross over
frequency to $223$GHz from 217GHz in the non-relativistic limit.
Nozawa et al (1996) presented analytic fitting formulae which
are accurate for $T_X\le 25$KeV. Itoh et al. (1998) adopted a
relativistically covariant formalism to describe the Compton
scattering process obtaining higher-order corrections in the form
of the Fokker-Planck expansion showing excellent agreement with
numerical results for clusters with $T_e\le 15$KeV.

\begin{figure}[h!]
\includegraphics[width=6in]{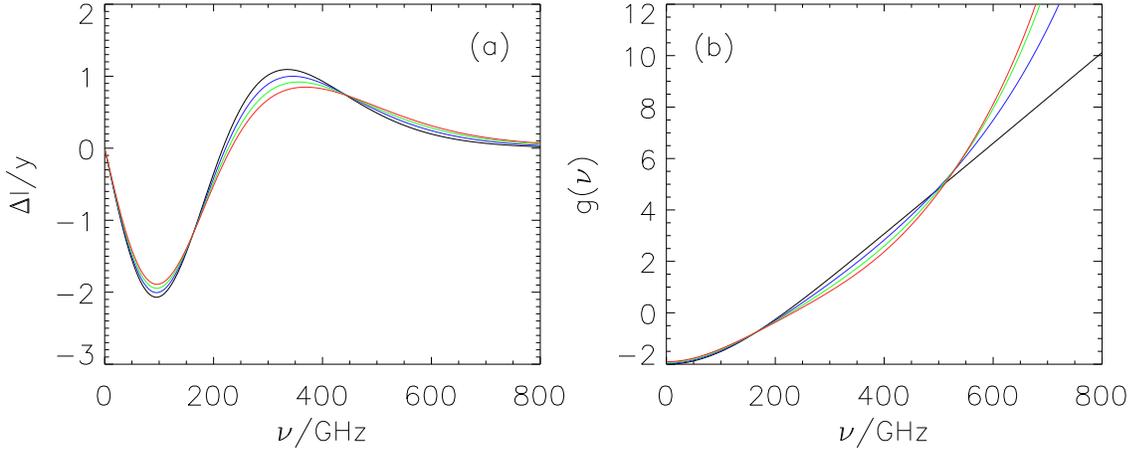}
\vspace*{-3cm}
\caption[]{(a) Spectral intensity distortion in units of the Comptonization
parameter as a function of frequency for 4 cluster temperatures
$T_x=0,5,10$ and $15$KeV, represented by the black, blue, green and red
curves, respectively. (b) Frequency dependence $g(\nu)$ of the TSZ spectrum
as a function of frequency for the same cluster temperatures and the
same line conventions as in (a).
}
\label{fig:itoh_formula}
\end{figure}

In Fig~\ref{fig:itoh_formula} we show the relativistic corrections
obtained using Itoh et al (1998) formula up to 3rd order in
the cluster temperature expressed in units of the electron mass
$\theta_e=k_BT_x/m_ec^2$. In Fig~\ref{fig:itoh_formula}a we
represent the corrections
to the spectral intensity distortion of the CMB spectrum in units
of the Comptonization parameter defined in eq.~(\ref{eq:y_compton}).
In Fig~\ref{fig:itoh_formula}b, the corrections to the frequency
dependence of the temperature anisotropies $g(\nu)$ (see
eq.~\ref{eq:tsz_def}). The black solid line
corresponds to the SZ non-relativistic formula, while the
blue, green and red solid lines represents the corrections
for clusters with X-ray temperature $T_X=5,10,15$KeV, respectively.
The Itoh (1998) series expansion converges very quickly in the
Rayleigh regime but shows a much slower convergence in the
Wien part of the spectrum. The approximation is valid for all
temperatures and for frequencies up to $\nu\simeq 450$GHz. At WMAP frequencies,
the corrections are 7-8\% for clusters with temperature $T_x=15$KeV.
At  217GHz, one of the {\it Planck}/HFI channels, the corrections are relatively very large
since the relativistic corrections shifts the zero cross frequency.
In the Wien part of the spectrum the relativistic corrections
are more important, about 20\% at 353Gz, also one of PLANCK
frequencies of operation, for a cluster of $T_X=15$KeV.

The KSZ effect is also affected by relativistic corrections.
Sazonov \& Sunyaev (1998) and also Nozawa, Itoh and Kohyama (1998)
computed the corrections which make the KSZ depart from the
frequency scaling of the primary CMB temperature anisotropies,
but the differences were smaller than 8\%
for a cluster of $T_X=10$KeV moving at $1000$km/s, with the correction
being largest at the cross over frequency. The relativistic correction introduces
cross terms between the kinematic and thermal effects. In the present
context, the most interesting term is the first order correction to the
dipole given by  Nozawa et al (1998)
\begin{equation}
\frac{\Delta T}{T_0}=\tau\frac{\vec{v}\cdot\hat{x}}{c}[1+C_1\theta_e]
\label{eq:ksz_rel_correction}
\end{equation}
where $\theta_e\equiv\frac{k_BT_X}{m_ec^2}$, $C_1=-10+(47/2)\tilde{X}-(7/5)\tilde{X}^2+(7/10)\tilde{S}^2$
with $\tilde{X}=x\cosh(x/2)$, $\tilde{S}=x/\sinh(x/2)$ and
$x$ is the frequency of observation in units of the CMB temperature,
$x\equiv h_{\rm P}\nu/k_{\rm B}T_{\rm CMB}$.

\begin{figure}[h!]
\includegraphics[width=4in]{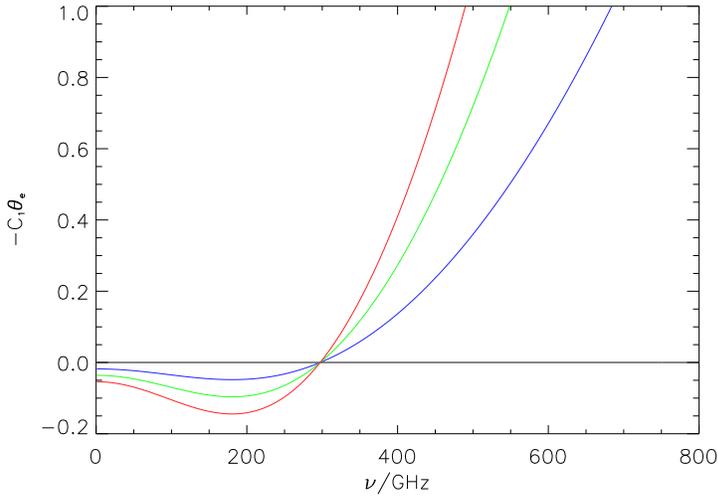}
\vspace*{2cm}
\caption[]{
Frequency dependence of the KSZ-TSZ cross term induced by relativistic
corrections for 4 different cluster temperatures. Line colors follow the
same conventions as in Fig~\ref{fig:itoh_formula}.
}
\label{fig:ksz_tsz_cross}
\end{figure}

In Fig~\ref{fig:ksz_tsz_cross} we show the relativistic corrections
to the dipole induced by the TSZ-KSZ cross term of eq.~\ref{eq:ksz_rel_correction}.
Lines represent different cluster temperatures, with
$T_x=0,5,10$ and $15$KeV represented by the black (no correction),
blue, green and red curves, respectively.
At WMAP frequencies the correction is small, at the few percent level for the hottest
clusters. In the Wien part of the spectrum, eq.~\ref{eq:ksz_rel_correction}
is no longer sufficiently accurate and higher order terms need to be
taken into account.


\subsection{SZ observations and prospective measurements.}
\label{sec:6.4}

Early measurements of the TSZ effect carried out since 1970's
concentrated on targeting known X-ray clusters.
Existing all-purpose radio telescopes were used, but few detections
were obtained despite considerable observational effort (Birkinshaw 1999).
Capability to observe the SZ effect has improved in recent years
thanks to improvements both in low-noise detection systems and
in observing techniques (Carlstrom et al 2002). At present,
dedicated instruments provide high resolution images of individual clusters
(Kitayama et al. 2004, Halverson et al. 2009, Nord et al. 2009).
Very recently, large scale
blind surveys of clusters of galaxies using dedicated telescopes
like the South Pole Telescope (SPT, Lueker et al 2010), the Atacama Cosmology
Telescope (ACT, Dunkley et al 2010) from the ground and the PLANCK
satellite from space have started to produce the first discoveries of
clusters through their SZ signature (Plagge et al 2009, Staniszewski et al. 2009,
Ade et al 2011). The TSZ effect,
being a spectral distortion of the CMB, does not suffer from cosmological
surface-brightness dimming as has been recently demonstrated by the
PLANCK mission which detected a cluster at redshift $z=0.94$ (Aghanim et al. 2011).

Prior to the Planck, ACT or SPT instruments which enabled a direct detection of
SZ clusters, WMAP allowed statistical determination of
the SZ effect by cross-correlating WMAP data with the location of known
clusters of galaxies (Fosalba et al. 2003, Rubi\~no Mart{\'\i}n \& Hern\'andez-Monteagudo
2004, Hern\'andez-Monteagudo et al 2004, Myers et al 2004, Afshordi et al 2007,
AKKE). Hern\'andez-Monteagudo et al (2004) cross-correlated a template
of projected galaxy density constructed from the 2MASS
Extended Source Catalog with the first year data from
WMAP and found a temperature decrement of $-35\pm 7\mu$K over
26 deg$^2$, this measurement having
the highest S/N at the time. The decrement was associated
with the clusters of galaxies within the 2MASS galaxy catalog.
Atrio-Barandela et al (2009, AKKE) carried out a similar analysis but used the positions of
identified X-ray clusters to measure the SZ signal using WMAP 3 year data. The measured
temperature decrement in the central cluster regions was $-28.5\pm 2.3\mu$K
for all clusters with redshift $z\le 0.2$, in agreement with the
Hern\'andez-Monteagudo et al (2004) result, and $-72.2\pm 4.8\mu$K
for clusters with X-ray luminosity $L_X(0.1-2.4 keV)\ge 3\times 10^{44}$
erg s$^{-1}$. The high statistical significance of these measurements,
close to $15\sigma$, were obtained in the central cluster regions enclosing 99\% of the X-ray emitting flux. The SZ emission was
found to be significant out to $\simeq 3-4$ times this identified X-ray extent, with no statistically significant TSZ signal detected at larger apertures.
Due to the high level of significance of the AKKE measurement,
we were able to measure the average pressure profile
of the cluster population. The implications of these results and
their connection with the DF measurement are deferred
to Sec. \ref{sec:cluster_tsz}.

The CMB distortions generated by the unresolved cluster population
give rise to an extra anisotropy component whose power spectrum has
been determined analytically (Cole \& Kaiser 1988; Bartlett \& Silk 1994)
and from simulations (Refregier et al. 2000). The TSZ power spectrum is
challenging to model
accurately because it includes significant contributions from galaxy clusters
spanning a wide range of masses and redshifts, and encompassing systems in various dynamical states.
The earlier calculations showed it to be  sensitive to the gas distribution
(Atrio-Barandela \& M\"ucket 1999; Hern\'andez-Monteagudo et al. 2000;
Molnar \& Birkinshaw 2000)
and $\sigma_8$, the amplitude of the matter density fluctuations over the $8 h^{-1}$Mpc radius (Komatsu \& Kitayama 1999; Komatsu \& Seljak 2002;
Bond et al 2005).
Numerical simulations have been carried out to make more accurate predictions
but at present the differences in treatment of the cluster gas physics
result in predictions differing in amplitude up to 50\% for a given
cosmology (Shaw et al 2010; Trac et al 2011; Battaglia et al 2010).
Komatsu et al (2011) carry out systematic analysis of the SZ signal
measured by WMAP 7yr data. By stacking the data over 29 nearby clusters,
they found that the universal cluster profile of Arnaud et al (2010)
and the Komatsu \& Seljak (2002) profile agree in general with the
profile derived from the data, although the difference was up to $\sim 30\%$
for the Arnaud et al (2010) profile.
By fitting the TSZ power spectrum templates to the data, they found
that both profiles over-predicted the unresolved SZ contribution to
WMAP data by $\sim 30\%-50\%$.

Recently, direct determination of the SZ power spectrum has been
carried out by  Lueker et al (2010) using the
SPT instrument. Improved measurements over an area of $200 deg^2$
by Shirokoff et al (2011) indicated the amplitude of the SZ power
spectrum was $4.5\pm 1.0\mu$K$^2$ at multipole $\ell=3000$ and frequency 152 GHz.
This result is consistent with the measurement obtained with the ACT
at the same multipole which measured the power of $6.8\pm 2.9\mu$K$^2$
at 148 GHz (Dunkley et al 2011), both being consistent with the earlier WMAP results of Komatsu et al (2011).
This discrepancy between the measured spectrum and the predictions
derived using universal electron pressure profiles or obtained
from numerical simulations indicate that systematically these calculations
over-predict the TSZ power spectrum or $\sigma_8$ is significantly smaller
than the preferred value of $\sim 0.8$ derived from WMAP 7yr data (Larson
et al 2011).
Recently, Efstathiou \& Migliaccio
(2011) developed a model for the unresolved contribution of clusters
of galaxies, which included parameters to describe departures from self-similar
evolution, and was compatible with the low TSZ amplitude inferred from those
observations.

Cluster atmospheres are not the only objects to contain electron populations
with significant energy content to produce CMB
spectrum distortions. The ionized gas content of the Universe as a whole and the
hot gas within our Galaxy are two other important sources of SZ effect.
Hydrodynamical simulations predict that a large fraction of all
baryons resides in mildly nonlinear structures that are partly
shock-confined gas filaments heated up to temperatures of
$10^5-10^7$K, called the warm-hot intergalactic medium (WHIM).
Atrio-Barandela \& M\"ucket (2006) and Atrio-Barandela, M\"ucket
\& G\'enova-Santos (2008) discussed the possible contribution
of the TSZ and KSZ effects due to the WHIM to CMB temperature
anisotropies. G\'enova-Santos et al (2009) deduced that this component
could represent a 3\% of the total power measured by the WMAP 5-yr data.
Hern\'andez-Monteagudo et al. (2004) searched for this diffuse component
by cross-correlating templates of projected galaxy density with the WMAP 1-yr data, but found no evidence of any contribution outside known galaxy clusters.

The KSZ effect observations are still in their infancy.
While recent experiments have been able to offer
high resolution images of the TSZ distortion induced by clusters,
no detection has been reported on the measurement of individual
clusters. This is no surprising since,
as indicated by eq.~(\ref{eq:tsz2ksz}) the amplitude of the KSZ
effect is much weaker than the TSZ in the central
regions of highly luminous (and hot) clusters, which are the easiest targets to observe.
As remarked in Sec.~\ref{sec:tsz}, the KSZ effect has
the same spectral dependence as the primary CMB temperature
anisotropies and so frequency information is not useful to disentangle
this component from the cosmological contribution.
Haenhelt \& Tegmark (1996) suggested to use the statistical
properties of the intrinsic CMB signal to design filters that
could remove the cosmological contribution
at cluster locations and discussed the performance
of different filters. Filtering out the intrinsic CMB contribution
is the basis of the KA-B method which we discuss
in Sec \ref{sec:ka-b}.

\subsection{KA-B method and CMB filtering}
\label{sec:ka-b}

\subsubsection{Method and filtering}

If a cluster at angular position $\vec{y}$ has the line-of-sight
velocity $v$ with respect to the CMB, the SZ CMB fluctuation at
frequency $\nu$ at this position will be $\delta_\nu(\vec
y)=\delta_{\rm TSZ}(\vec y)G(\nu)+ \delta_{\rm KSZ}(\vec
y)H(\nu)$, with $ \delta_{\rm TSZ}$=$\tau T_{\rm X}/T_{\rm e,ann}$
and $\delta_{\rm KSZ}$=$\tau v/c$. Here $G(\nu)\simeq-1.85$ to
$-1.35$ and $H(\nu)= 1$ over the range of frequencies probed
by the WMAP data, $\tau$ is the projected optical depth due to
Compton scattering, $T_{\rm X}$ is the cluster electron
temperature and $k_{\rm B}T_{\rm e,ann}$=511 KeV.

For a typical individual cluster the KSZ-induced fluctuation is of order
\begin{equation}
\delta T \sim 10 \; \left(\frac{\tau}{10^{-3}}\right) \left(\frac{V}{1,000\; {\rm km/sec}}\right)\;\; \mu{\rm K}
\label{eq:ksz_oom}
\end{equation}
which is small compared to the typical microwave signal from primary CMB, foregrounds and/or instrument noise. However, if averaged over
many isotropically distributed clusters moving at a significant
bulk flow with respect to the CMB, the kinematic term may dominate
enabling a measurement of $V_{\rm bulk}$. Thus KA-B suggested
measuring the dipole component of $\delta_\nu(\vec y)$.

When computed from the total of $N_{\rm cl}$ positions the
dipole moment of $\delta_\nu(\vec y)$ also will have positive contributions from 1) the
instrument noise, 2) the cosmological/primary CMB fluctuation component arising from the
last-scattering surface, 3) the thermal SZ (TSZ) component, and 4) the various foreground components at the WMAP frequency range. The latter contribution can be
significant at the two lowest frequency WMAP channels (K \& Ka)
and, hence, we restricted our analysis to the WMAP Channels Q, V \&
W which have negligible foreground contributions.

For $N_{\rm cl}\gg1$ the dipole of the observed $\delta_\nu$
becomes:
\begin{equation}
a_{1m} \simeq a_{1m}^{\rm KSZ} +a_{1m}^{\rm TSZ} + a_{1m}^{\rm
CMB} + \frac{\sigma_{\rm noise}}{\sqrt{N_{\rm pix}}}
\label{eq:dipole}
\end{equation}
Here $a_{1m}^{\rm CMB}$ is the residual dipole produced at the
cluster positions occupying $N_{\rm pix}$ pixels by the primordial CMB anisotropies. The amplitude
of the dipole power is $C_1= \sum_{m=-1}^{m=1} |a_{1m}|^2$. We
use the notation for $C_{1,{\rm kin}}$ normalized so that a
coherent motion at velocity $V_{\rm bulk}$ would lead to
$C_{1,{\rm kin}}= T_{\rm CMB}^2 \langle \tau \rangle^2 V_{\rm
bulk}^2/c^2$, where $T_{\rm CMB} =2.725$K is the present-day CMB
temperature.

While the noise term integrates down as $1/\sqrt{N_{\rm pix}}$ with the increasing cluster sample, the situation with primary CMB anisotropies is more complex because they are significantly correlated on sub-degree scales due to the Doppler peaks. However, the power spectrum of the primary CMB is now  accurately known and KA-B proposed to reduce the primary contribution to the dipole by utilizing this information. One has to be careful with filtering though, because the goal of filtering is to reduce the contribution of primary CMB to the dipole while not affecting as much the KSZ signal which is proportional to the (filtered) projected optical depth. {\it Not every filter can achieve this}. The filter designed for this purpose in KABKE1,2 proved to work remarkably well and was shown in AKEKE to remove the primary CMB down to the fundamental limit of cosmic variance while preserving much of the KSZ amplitude.

TSZ component presents another obstacle to isolating the KSZ dipole term in eq. \ref{eq:dipole}. In our analysis it was reduced to negligible levels by measuring the final dipole over an aperture containing vanishing monopole. This presents an important cornerstone in applying the KA-B method. The decrease in the TSZ terms then occurs because of the X-ray temperature decreasing toward cluster outskirts as was demonstrated by AKKE phenomenologically to occur in our clusters. This too is discussed at length below and in KABKE1,2 and later DF papers.

Additional contributions to eq. \ref{eq:dipole} come from
non-linear evolution/collapse of clusters (Rees \& Sciama 1968),
gravitational lensing by clusters (Kashlinsky 1988), unresolved
strong radio sources (present, for instance, in WMAP 5 year data,
Nolta et al 2008) and the Integrated Sachs-Wolfe effect from the
cluster pixels. All these effects have a dipole signal only when
clusters are inhomogenously distributed on the sky and is in turn
bounded from above by the amplitude of the monopole. The magnitude
of these contributions is at most $\sim 10\mu$K$^2$ in power (see
Aghanim, Majumdar \& Silk 2008 for a review on secondary
anisotropies) a factor of 10 smaller than the TSZ monopole amplitude. Moreover, as we discuss
below, we find a dipole signal when the monopole vanishes, so our
measurements can not be significantly affected by all these
effects.

In the following sections we detail out the process that enabled
us to isolate the KSZ term in eq. \ref{eq:dipole}.

\subsubsection{Statistics of the measured quantities}

In order to properly interpret the measurement one has to understand the underlying statistical distribution of the determined quantities and how - or whether that distribution differs from similar quantities determined with alternative methods. This is briefly addressed below inasmuch as may be required later.

It is important to emphasize in this context that the method determines cluster velocities directly with respect to the CMB rest frame defined by its dipole component. In the KA-B method one requires a cluster catalog with a (quasi-)isotropic coverage on the sky. The determined quantities are the three dipole components of the CMB temperature field evaluated at cluster positions and for a quasi-isotropic spherical distribution of clusters these components are statistically independent, i.e. $\langle a_{1m} a_{1n}\rangle \simeq 0$ for $m\neq n$.  Thus the overall probability of measuring the three dipole components is the product of the individual probabilities.

It is important to emphasize for comparison to other measurements that the KA-B method measures the full 3-D velocity, $V_{\rm bulk}$, whose rms value in the gravitational instability picture is given by eq. \ref{eq:v_rms} and whose individual components are statistically independent to good accuracy. At the same, the measured signal is the CMB dipole and additional calibration steps coupled with good understanding of the cluster properties are required to convert the measured CMB signal into an equivalent velocity.

The measured dipole is little affected by relativistic corrections. To
be of any significance, massive clusters need to be aligned with the
direction of the flow to induce a correction of $\sim 5\%$. Since clusters
are randomly distributed on the sky, the effect of the cross-term
of eq.~\ref{eq:ksz_rel_correction} can be absorbed into the monopole
term and does not affect the residual dipole at cluster
locations and measured by us at zero monopole.

\subsubsection{Higher moments}
\label{sec:kab-highmoments}

Can the KA-B method be used to constrain the shear of the flow? Here the situation is less clear for the following reason: the KSZ induced temperature fluctuation induced at each cluster position is $\Delta T_{\rm KSZ} \propto \tau v$. Since each cluster's optical depths vary, we expand $\tau = \langle \tau\rangle + \delta \tau$. The velocity can be expanded as in eq. \ref{eq:genvel} as $v=V_{\rm bulk}\cos\theta + ({\rm shear \; terms})+...$. Thus the multipoles resulting from the KSZ contributions at cluster positions would be:
\begin{equation}
a_{\ell m}^{\rm KSZ} \propto \int (\langle \tau\rangle + \delta \tau)[V_{\rm bulk}\cos\theta + ({\rm shear \; terms})+...] Y^*_{\ell m} d\omega
\label{eq:kab-shear}
\end{equation}
This expression shows that the dipole terms are cleanest in the sense that the measured signal is directly proportional to $\langle \tau\rangle V_{\rm bulk}$. On the other hand, if one wants to measure the shear of the flow directly from the quadrupole (or higher) moments, one needs to know the cluster catalog properties very accurately in order to subtract the contribution from the cross-terms $\int  \delta \tau V_{\rm bulk}\cos\theta Y^*_{\ell m} d\omega$.

\subsection{ Other KSZ methods.}

Measuring peculiar velocities for individual clusters of galaxies is
difficult since primary CMB anisotropies have the same spectral dependence
as the KSZ effect. Several other methods have been proposed for this measurement in addition to KA-B.

Haenhelt \& Tegmark (1996) suggested to filter the
intrinsic CMB temperature anisotropies using statistical properties
of the CMB temperature field. If clusters are approximately Poisson distributed
on the celestial sphere, while temperature anisotropies are correlated on
angular scales of $\theta\sim 1^\circ$, a Wiener-type filter can be constructed
to remove the intrinsic CMB in the neighborhood of each cluster. The
residual CMB and noise would allow to measure peculiar velocities of
individual clusters with an estimated
error bar of $1-2\times 10^3$km/s depending on cluster mass.

This was used by Aghanim, Gorski \& Puget (2001) to suggest that by directly
averaging the measured CMB signal over a cluster sample could allow to
determine the bulk velocity of the cluster population.  Since clusters can
be detected up to $z\sim 1$ using the redshift-independent SZ effect, this allowed the possibility
of measuring bulk flows up to that redshift.

Atrio-Barandela, Kashlinsky \& M\"ucket (2004) suggested to cross-correlate
CMB temperature anisotropies with cluster redshift information to obtain
peculiar velocity of shells of clusters and to determine the Mach number
of the velocity field at different redshifts.

A direct attempt at the measurement of peculiar velocity
of a cluster sample was tried by Benson et al (2004).
They averaged the peculiar velocity of six clusters between $z\simeq 0.22$ and $z\simeq 0.55$
observed with the SuZie telescope and
set a 95\% confidence limit of $1410$kms$^{-1}$ to the bulk flow in the direction of the CMB dipole on a poorly determined scale.

In addition to statistical uncertainties, astrophysical contaminants
are also a source of error on SZ measurements in such measurements.
Aghanim, Hansen \& Lagache (2005) analyzed the effect of residual contamination
by astrophysical sources radiating at SZ frequencies such as dust emission,
infrared galaxies and radio sources. They showed that for PLANCK and ACT type
of experiments the systematic errors due to contamination by foreground emissions
can be significantly larger than systematic errors, making the measurement of peculiar velocities exceedingly difficult. This shows the limitations of straight averages
of individual cluster peculiar velocities. In contrast, the KA-B method
computes the CMB dipole at the cluster locations on the sky.
While this limits the method to determine the immediate bulk flow of a region
surrounding the Local Group, it strongly reduces the effect of astrophysical
foregrounds since dust or infrared galaxies do not correlate with cluster locations
and their contribution to the error bar can be estimating by computing
the dipole at the position of clusters placed randomly on
the sky but outside the location of known clusters. Radio galaxies within
clusters can reduce the Compton parameter of individual clusters, but
their contribution to the dipole is bounded from above by the amplitude of its monopole.
Therefore, while astrophysical foregrounds can be a severe problem
for the Aghanim et al (2001) method, it is not nearly as significant for the
KA-B method, as we discuss in Sec. \ref{sec:darkflow}.


\section{Instruments used}
\label{sec:instruments}

The KA-B method requires an all-sky catalog of X-ray clusters as well as all-sky CMB maps. The cluster catalog used in the DF studies was compiled from ROSAT All-Sky Survey  detections (RASS; Voges et al.\ 1999) with additional cluster characteristics derived from deeper Chandra (and XMM?) observations. The CMB maps used thus far were collected by the {\it Wilkinson Microwave Anisotropy Probe} (WMAP); future applications of our method proposed and planned by us will use CMB data currently being gathered by {\it Planck}. In this section we describe the surveys and instruments used to obtain these datasets.

\subsection{X-ray missions relevant to the DF project}
\label{sec:xray-instruments}

In the following we provide technical information on the two X-ray missions that have, so far, been most important for the measurements discussed in this review. The first one, ROSAT, provided the wide areal coverage that allowed the compilation of the required all-sky cluster sample. The second one, Chandra, performed a large number of deep follow-up observations of individual clusters from this sample, thereby enabling precise measurements of the physical characteristics of the X-ray emission.

\subsubsection{ROSAT}\label{sec:rosat}

Carrying two imaging Wolter-type X-ray telescopes, ROSAT (from the German ``R\"ont\-gen\-satellit"; Fig.~\ref{fig:rosat}) covered both the EUV and X-ray passband. In the soft X-ray regime ROSAT provided a choice of two detectors in the focal plane: the High Resolution Imager (HRI), which provided high angular resolution (5\arcsec) but no spectral information, and the Position Sensitive Proportional Counter (PSPC), which provided modest spectral resolution and 20\arcsec\ angular resolution in pointing mode. With the PSPC at the telescope focus, ROSAT offered a circular field of view of 1\,deg radius, a peak effective area of 240 cm$^2$ at 1\,keV,  and, in scan mode, an effective angular resolution of roughly 45\arcsec. The latter characterizes the data obtained in 1990 during the ROSAT All-Sky Survey (RASS), conducted in great circles through the ecliptic poles and providing a median exposure of 390 seconds in the soft X-ray band (0.1--2.4 keV). Although this median exposure time can be taken as representative of the depth of the RASS on the whole, significant deviations from the median do occur. Specifically, 12\% of the sky (mainly in the areas of the South-Atlantic Anomaly and the Magellanic Clouds) were observed for less than 200\,s, while the accumulated exposure time exceeded 1,000\,s for a few percent of the sky around the ecliptic poles.  The RASS data remained proprietary and access restricted to collaborations approved by MPE (Max Planck Institute for Extraterrrestial Physics) until their public release in 1999. The catalogue of over 150,000 X-ray sources resulting from the RASS  enabled detailed statistical studies of nearly all classes of astrophysical objects, from comets through neutron stars to galaxy clusters. As of 2012, i.e., over 20 years after its completion, the RASS remains the only all-sky survey conducted with an imaging X-ray telescope.

 \begin{figure}[h!]
\begin{center}
\end{center}
 \caption[]{THE FILE WITH THE FIGURE IS AVAILABLE FROM \url{http://www.kashlinsky.info/bulkflows/physicsreport}. -- The ROSAT satellite. The X-ray telescope, surrounded by the satellite electronics, dominates the central body; the Wide-Field Camera, a much smaller extreme UV telescope, rides on top of the array in this representation.}
\label{fig:rosat}
\end{figure}

After the end of the 6-month all-sky survey, ROSAT continued to operate  as an observatory, available to researchers from the entire astronomical community, until the end of 1998. The much deeper, pointed observations  conducted in this mode were almost equally divided between the PSPC and the HRI, with clusters of galaxies constituting about 10\% of the over 6,000 unique targets. The resulting data were instrumental for more quantitative studies of the X-ray properties of clusters (such as the temperature and density of the intra-cluster gas) and also proved a valuable resource for the compilation of samples of serendipitously detected clusters. A more detailed discussion of the resulting cluster samples is provided in Section~\ref{sec:xraycat}.

ROSAT re-entered the atmosphere on Oct 23, 2011, over 10 years after the end of the satellite's science mission.

 \subsubsection{Chandra}

 \begin{figure}[h!]
\hspace*{2cm}
\includegraphics[width=0.8\textwidth]{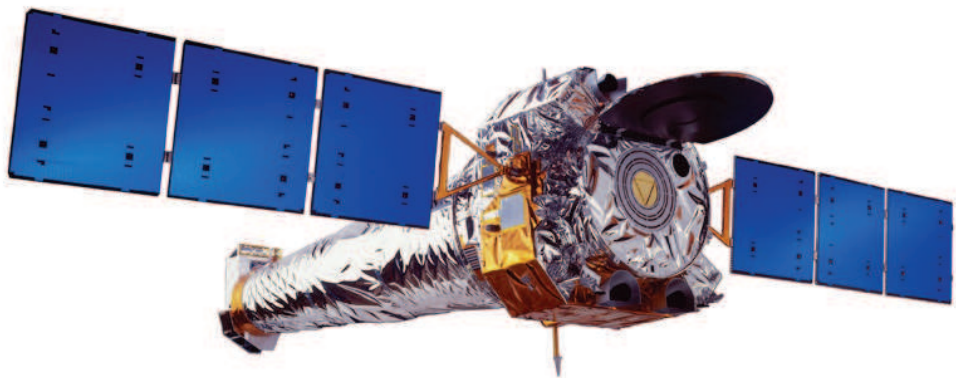}
 \caption[]{The Chandra X-ray Observatory. A focal length of 10m and extremely smooth iridium-coated mirrors provide high angular resolution up to 10 keV. Chandra's two detectors, the High Resolution Camera (HRC) and the Advanced CCD Imaging Spectrometer (ACIS) are mounted at the far end of the spacecraft.}
\label{fig:rosat}
\end{figure}

The Chandra X-ray Observatory, launched into a highly elliptical orbit from the Space Shuttle Columbia in 1999, opened a new era in imaging X-ray astronomy. A nested set of four Iridium-coated Wolter Type-I  (paraboloid / hyperboloid combination) mirrors provide an unprecedented (and presumably for several more decades unequalled) on-axis resolution of better than 0.5\arcsec at the focal plane.

In addition to two transmission-grating spectrometers, Chandra carries two imaging detectors, one of which (the Advanced CCD Imaging Spectrometer, ACIS) also provides modest energy resolution. With ACIS in the focal plane, Chandra offers an effective area of almost 500 cm$^2$ at 1 keV (roughly twice that or the ROSAT PSPC) and still more than 200 cm$^2$ at 6 keV. The most commonly used energy band for Chandra data is 0.5 to 6 keV, although the telescope offers some effective area up to 10 keV. ACIS consists of two arrays of CCD detectors each of which covers an area of $8\times 8$ arcmin$^2$ on the sky. The first one, ACIS-S, combines six CCDs in a linear arrangement. Although mainly used in combination with the transmission gratings, ACIS-S is occasionally also selected for non-grating observations. Only the back-illuminated S3 CCD of the ACIS-S array is used in this case, because of its higher sensitivity at low energies and better energy resolution compared to the front-illuminated devices making up ACIS-I. The latter is a square array of four CCDs covering $16.9 \times 16.9$ arcmin$^2$ and Chandra's primary imaging spectrometer. Although less sensitive to low-energy photons than the back-illuminated S3 chip of ACIS-S, the front-illuminated CCDs of ACIS-I are usually preferred for observations of extended sources (such as galaxy clusters) because they feature only about half the quiescent background rate.

\begin{figure}[h!]
\includegraphics[width=0.32\textwidth]{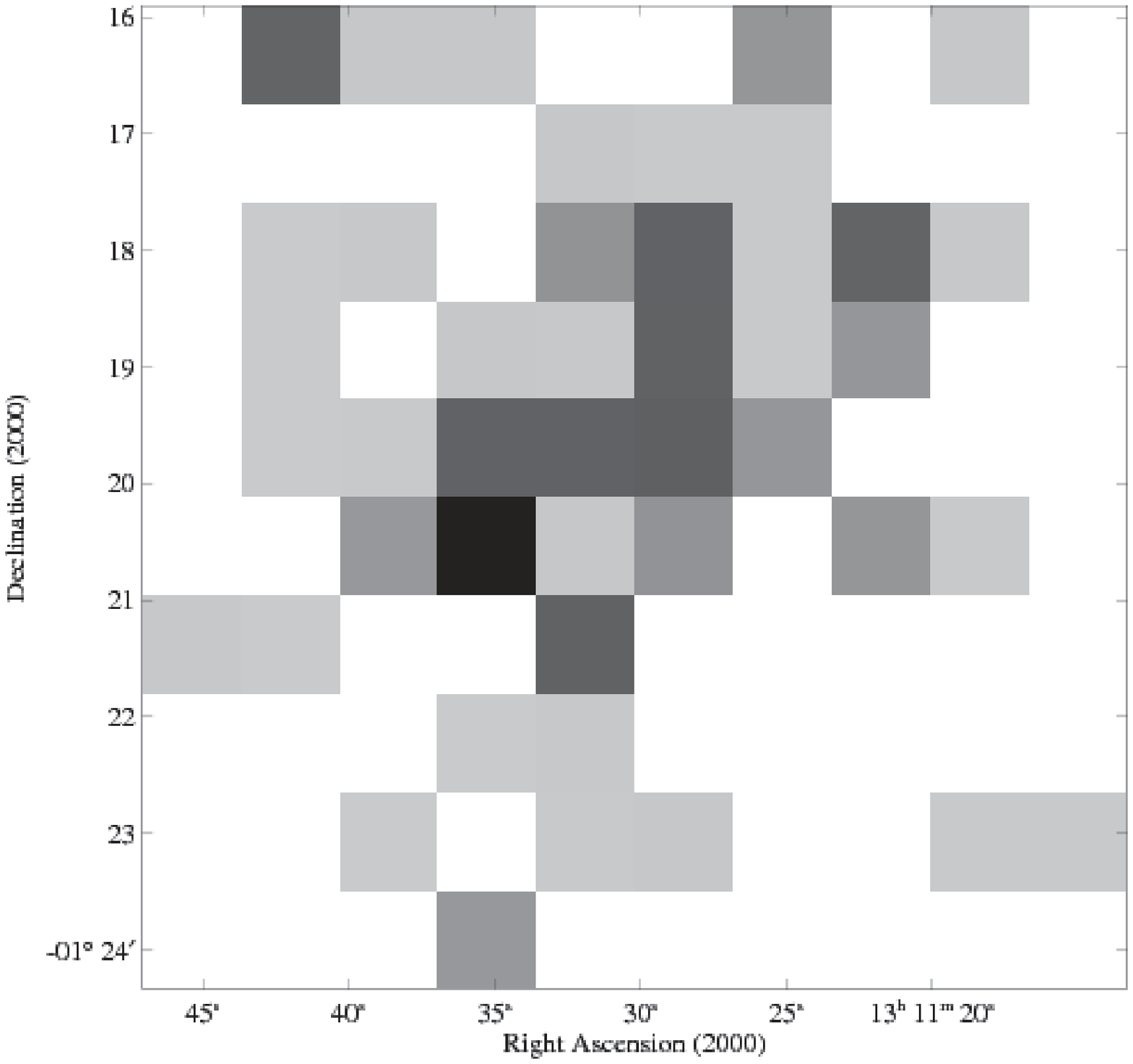}
\includegraphics[width=0.32\textwidth]{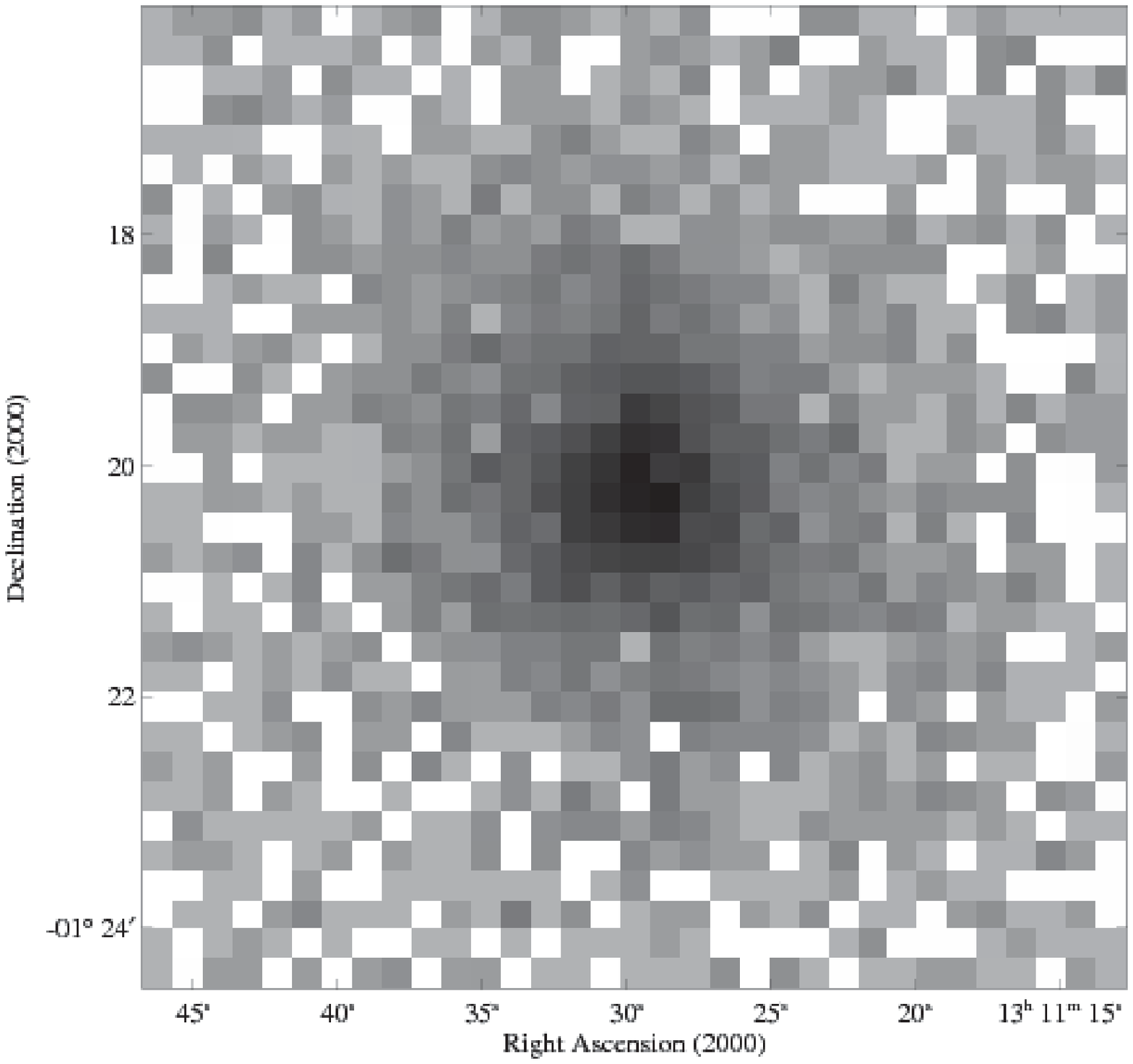}
\includegraphics[width=0.32\textwidth]{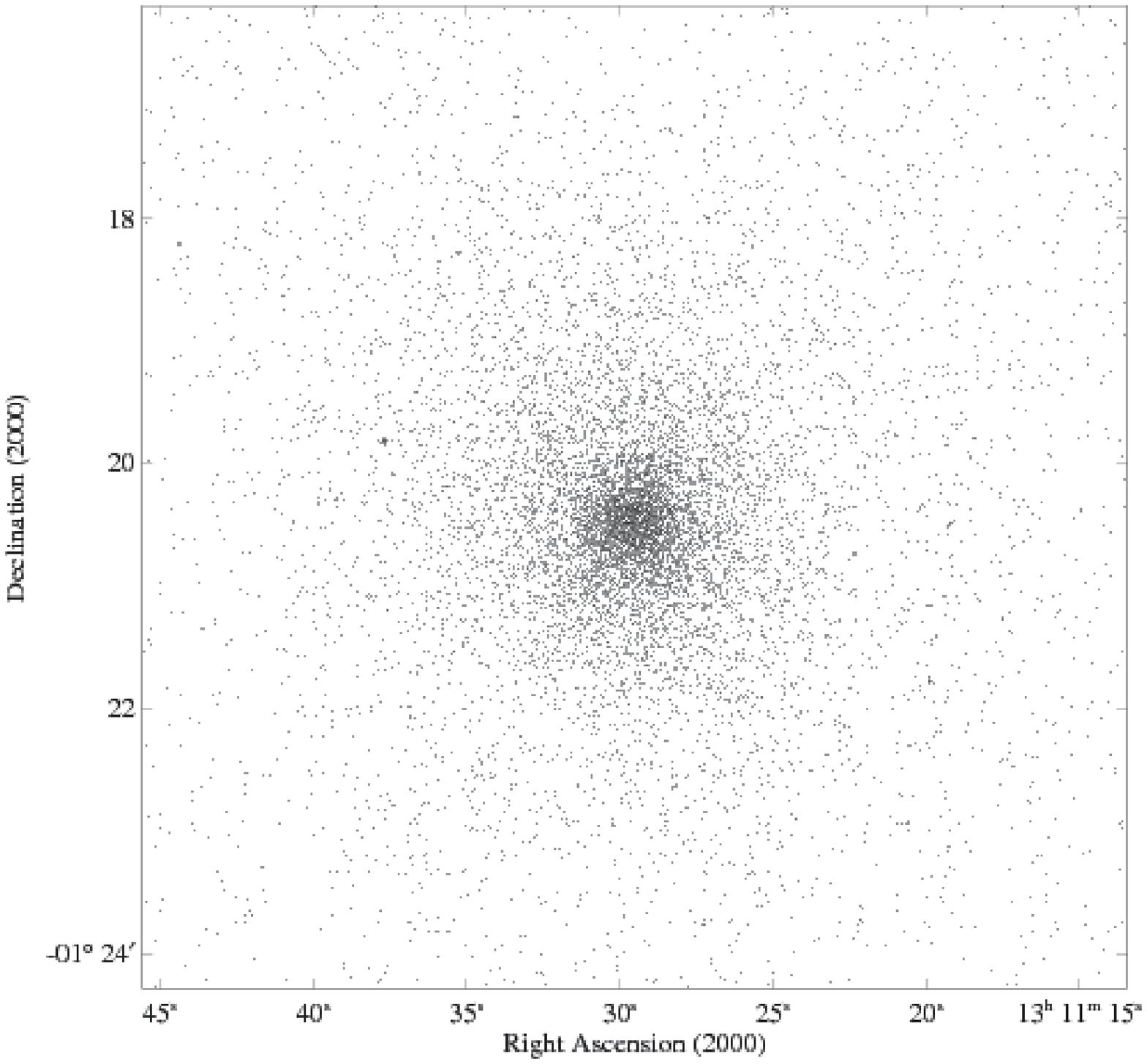}
\caption[]{The massive cluster A\,1689 ($z=0.183$) as observed in the RASS (left; 0.1--2.4 keV, average exposure time 280\,s, pixel size 45\arcsec), in a pointed observation with the ROSAT PSPC (center; 0.1--2.4 keV, on-axis exposure time 14\,ks, pixel size 15\arcsec), and in a pointed observation with Chandra / ACIS-I (right; 0.3--10 keV, on-axis exposure time 20\,ks, pixel size 0.5\arcsec). The shown images cover $1.1\,h^{-1}$ Mpc on the side at the cluster redshift. Their angular size of $8.5\times 8.5$ arcmin$^2$ exceeds only slightly the area of a single WMAP pixel in CMB maps with HEALPix $N_{\rm side}=512$, rendering A\,1689 barely resolvable with WMAP instruments.}
\label{fig:ximcomp}
\end{figure}

Figure~\ref{fig:ximcomp} illustrates the difference in angular resolution and throughput by comparing images of A\,1689, a very X-ray luminous cluster at $z=0.18$, as obtained during the RASS, with the ROSAT PSPC in pointed mode, and with Chandra / ACIS.

\subsection{CMB}
\label{sec:cmb-instruments}

The Wilkinson Microwave Anisotropy Probe (WMAP) was launched by NASA in Summer 2001 with the goal to map the all-sky CMB anisotropies and polarization with sub-degree angular resolution. The satellite observations have been stopped in Aug 2010 and the 9-yr CMB maps should be made publicly available in 2012. As of this review writing, the 3-yr, 5-yr and 7-yr integration maps have been released and are used below.

\begin{deluxetable}{c c c c c}
\tabletypesize{\scriptsize}
\tablecaption{WMAP instrumental characteristics.}
 \startdata
Band & Frequency (GHz) & \# of DA's & Beam (arcmin) & $\sigma_{\rm noise}$(WMAP7) ($\mu$K$\cdot\sqrt{\rm yr}$) \\
\hline
\hline
Q & 41 & 2 & 30 & $\sim170$ \\
V & 61 & 2 & 20 & $\sim 210$\\
W & 94 & 4 & 13 & $\sim 360$ \\
\enddata
\label{tab:wmap}
 \end{deluxetable}
WMAP uses HEMT amplifiers to make differential observations in 5 frequency channels, K, Ka, Q, V, W. The two shortest frequency channels, K and Ka, are used to isolate and subtract the foreground emissions at the Q, V, W channels and are not used in our studies. The three longest frequency channels have two differential assemblies (DAs) at Q and V channels and four DAs at the W channel with the best angular resolution. Table \ref{tab:wmap} gives the WMAP instrumental characteristics relevant to this study. More information on the WMAP instrumentation is given the mission website \url{http://wmap.gsfc.nasa.gov/}.

 \begin{figure}[h!]
\includegraphics[width=4.5in]{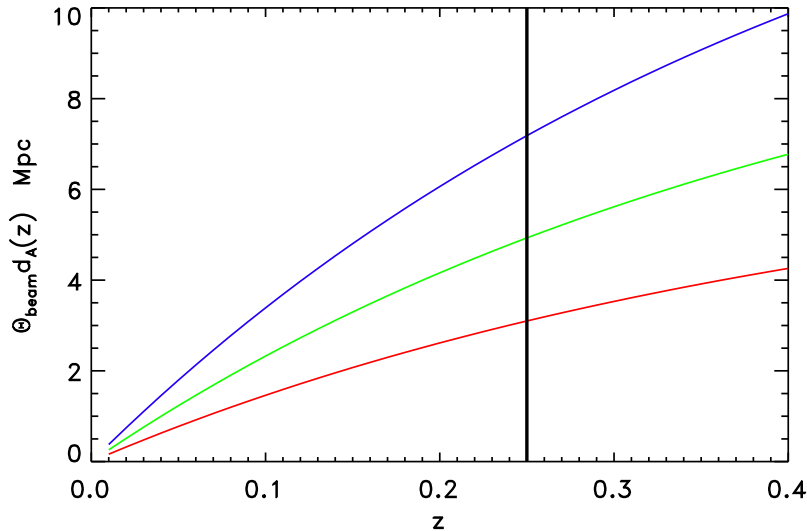}
 \caption[]{Comoving scales subtended by the WMAP beams. Blue, green, red lines correspond to Q, V, W bands respectively. Black vertical line shows the current (pre-2011) redshift limit of the DF studies. }
\label{fig:kae3}
\end{figure}
The limited angular resolution of the WMAP coupled with the higher instrument noise for the four W-band, where the angular resolution is finest at $\sim 12^\prime$, makes it necessary to stack galaxy clusters for any meaningful SZ cluster studies (cf. Komatsu et al 2011). Fig. \ref{fig:kae3} shows the comoving linear scale subtended at a given $z$ by the WMAP beams for each frequency channel. The figure clearly shows that individual clusters are unresolved and mostly undetected (because of noise) by WMAP at the cosmological distances of interest.

We can now illustrate with selected - for this purpose - well-studied hot and reasonably relaxed X-ray clusters the limitations of the WMAP instruments for this measurement. The clusters we select for this are listed in Table \ref{tab:wmap-clusters} and cover the redshift range in Fig. \ref{fig:kae3}. The table lists the mean CMB temperature decrement within a given radius aperture of the clusters after averaging it over the four W-channel DA's, which were selected because of their finest angular resolution. The approximate estimate of the instrument noise contribution is shown in the last row of the table. The TSZ component is identifiable by a negative CMB temperature decrement over a given aperture around the cluster centers.
\begin{deluxetable}{c | c c c | c c c}
\tabletypesize{\scriptsize}
\tablecaption{Selected X-ray luminous relaxed clusters as seen by WMAP7 - W-band}
 \startdata
 & $z$ & $kT_X$ & $(l,b)$ & $\langle \delta T\rangle$ (W-band) & $\langle \delta T\rangle$ (W-band) & $\langle \delta T\rangle$ (W-band) \\
 & & KeV & deg & central pixel & aperture radius $15^\prime$ & aperture radius $30^\prime$ \\
\hline
\hline
Coma & 0.023 & 8.2 & $(58.08, 87.96)$ & $-235\;\mu$K & $-268\;\mu$K & $-206\;\mu$K \\
A2029 & 0.078 & 8.2 & $(6.47, 61.13)$ & $6\;\mu$K & $-34\;\mu$K & $-9\;\mu$K\\
A478 & 0.088 & 8.0 & $(182.43, -28.29)$ & $-9\;\mu$K & $-78\;\mu$K & $-35\;\mu$K \\
A1689 & 0.183 & 9.2 & $(313.36, 61.13)$ & $-57\;\mu$K & $97 \;\mu$K & $103\;\mu$K \\
\hline
$\sigma_{\rm noise} N_{\rm pixels}^{-\half}$ & & & & $\sim 65\; \mu$K & $\sim (16-17)\;\mu$K & $\sim (8.3-8.6)\;\mu$K  \\
\enddata
\label{tab:wmap-clusters}
 \end{deluxetable}

It is now obvious from the numbers in Table \ref{tab:wmap-clusters} that:
 \begin{enumerate}
 \item One needs to stack many clusters in order to identify robustly any possible, but necessarily faint, KSZ contribution.
 \item Whereas nearby bright clusters, such as Coma, are resolved well by the WMAP W-channel, by $z\sim 0.2$ even clusters as bright as A1689 are undetected individually with the WMAP best resolution channel. At the intermediate redshifts the extended clusters, such as A478 at $z\simeq 0.09$, are only partially detectable. Thus the larger apertures are critical for detection. (Note that the inner parts of A478 lie within the KP0 CMB mask although its TSZ component is clearly detectable). This clearly defines the limits beyond which adding more (unresolved by WMAP) clusters will not add to the S/N of any KSZ measurement with WMAP.
 \item The primary CMB is highly correlated on sub-degree scales which is seen explicitly by the CMB temperature remaining positive out to the $\sim 0.5^\circ$ aperture at the location of A1689. This correlated component with dispersion around $\sim 80\mu$K creates significant noise in the KSZ measurements of the type discussed in KA-B. It is thus critical to be able to filter it out {\it while leaving the KSZ component intact}. This can be achieved with certain, {\it but not any}, filters as we discuss below.
 \item The numbers in the table demonstrate that it is critical to increase the cluster aperture in order to decrease the noise and encompass the entire X-ray gas extent of the clusters.
 \item At the same time, the clusters are clearly diluted by the beam and so even clusters as bright in X-rays as A1689 at $z\sim 0.2$ become unresolved with their entire hot gas content concentrated in just a few WMAP pixels. Thus adding more clusters at $z\gsim 0.2-0.3$, for the WMAP instrument resolution and noise, may not increase dramatically the S/N of the measurement.
 \item We also note a relative robustness to aggressive CMB masking (the central pixel of the A478 lies within the KP0 mask, yet the negative temperature decrement, as required by the TSZ effect is apparent).
 \end{enumerate}


\section{X-ray cluster catalog(s)}
\label{sec:xraycat}

\subsection{X-ray detections of clusters: a brief history}\label{sec:xhistory}
First studies of celestial X-ray sources were performed in the early 1960s using rocket-borne detectors. They led to the discovery of the diffuse X-ray background as well as of Sco X-1, a low-mass X-ray binary that is the brightest source in the X-ray night sky (Giacconi et al.\ 1962). Subsequent balloon and rocket experiments employing proportional as well as scintillation counters added supernova remnants, black-hole binaries, and the Galactic center to a rapidly growing list of discrete cosmic X-ray sources (Clark 1965; Bowyer et al.\ 1965). The first detections of extragalactic sources, M87 and Cygnus A, followed soon after (Byram et al.\ 1966).

The first X-ray detection of a cluster of galaxies was claimed by Boldt et al.\ (1966) for the Coma cluster. Although the detection was soon shown to be spurious (Friedman \& Byram 1967) it prompted Felten et al.\ (1966) to propose that the origin of such emission would have to be bremsstrahlung from a diffuse plasma at roughly $10^8$ K. Extended X-ray emission from the intra-cluster gas in Coma was indeed detected shortly thereafter, and independently, by Meekins et al.\ (1971) and Gursky et al.\ (1971), causing astronomers to realize that  clusters in general should be extended, highly luminous X-ray sources (Cavaliere et al.\ 1971; Gursky et al.\ 1972).

The first all-sky survey in hard X-rays (2--10 keV), conducted by the Uhuru satellite from 1970 to 1973, confirmed this suspicion when 45 of the 339 X-ray sources detected were identified as galaxy clusters (Forman et al.\ 1978). More quantitative studies of the X-ray emission from clusters became possible with the advent of imaging X-ray telescopes, an era that began in 1978 with the launch of the Einstein Observatory (Giacconi et al.\ 1979). Like all imaging X-ray telescopes since, the Einstein Observatory's telescope used Wolter-type nested mirrors that combine parabolic and hyperbolic profiles to focus X-rays at grazing incidence (Wolter 1952). With the Imaging Proportional Counter (IPC) in the focal plane, the Einstein Observatory's telescope offered an effective area of 100 cm$^2$ in the soft X-ray regime (0.4--4 keV),  a field of view of $75\arcmin \times 75\arcmin$, and a spatial resolution of about 1\arcmin, enough to allow a classification of the X-ray morphology and modeling of the gas density profiles of nearby clusters (Jones et al.\ 1979; Jones \& Forman 1984). Although groundbreaking in many ways, the mission was devoted entirely to targeted observations and thus did not  improve dramatically upon the census of X-ray sources from the Uhuru and Ariel-V large-area surveys.

Prior to the 1990s, X-ray satellite missions had thus opened a new window to the sky by identifying a wide range of celestial X-ray sources and by characterizing the origin and nature of their emission through spectroscopy and resolved imaging of selected targets. The next milestone, the first all-sky survey with an imaging X-ray telescope, was achieved by ROSAT, described in more detail in Section~\ref{sec:rosat}. Owing to the huge solid angle covered, the ROSAT All-Sky Survey (RASS)  allowed, for the first time, the compilation of sizable samples of even the rarest, most massive clusters of galaxies (e.g., Ebeling, Edge \& Henry et al.\ 2001) and the study of correlation functions on very large scales. The two-decade reign of the RASS is destined to come to an end with the launch of eROSITA, another German satellite, currently scheduled to commence a vastly deeper all-sky survey in 2013.

\subsection{The X-ray advantage}

Until X-ray astronomy had reached a sufficient level of maturity, visual searches for overdensities of galaxies on optical plates were the most effective way to compile large cluster catalogs, the most widely used ones being the compilations of Zwicky, Herzog \& Wild (1961-1966), Abell (1958), and Abell, Corwin \& Olowin (1989). While large, these samples of optically selected clusters had the disadvantage of suffering from significant projection effects (van Haarlem, Frenk \& White 1997; Hicks et al.\ 2008).

A nearly unbiased way of selecting statistical cluster samples is through X-ray surveys, as the X-ray emission, which originates from the diffuse intra-cluster gas trapped in the clusters' gravitational potential well and heated to virial temperatures of typically $10^{7-8}$ K, represents direct proof of the existence of a three-dimensionally bound system (see Section~\ref{sec:xhistory}). Also, the X-ray emission is much more peaked at the cluster center than is the projected galaxy distribution, making projection effects in X-ray selected cluster samples highly improbable.

\begin{figure}[h!]
\baselineskip0.5mm
\parbox{0.49\textwidth}{
\includegraphics[width=0.49\textwidth]{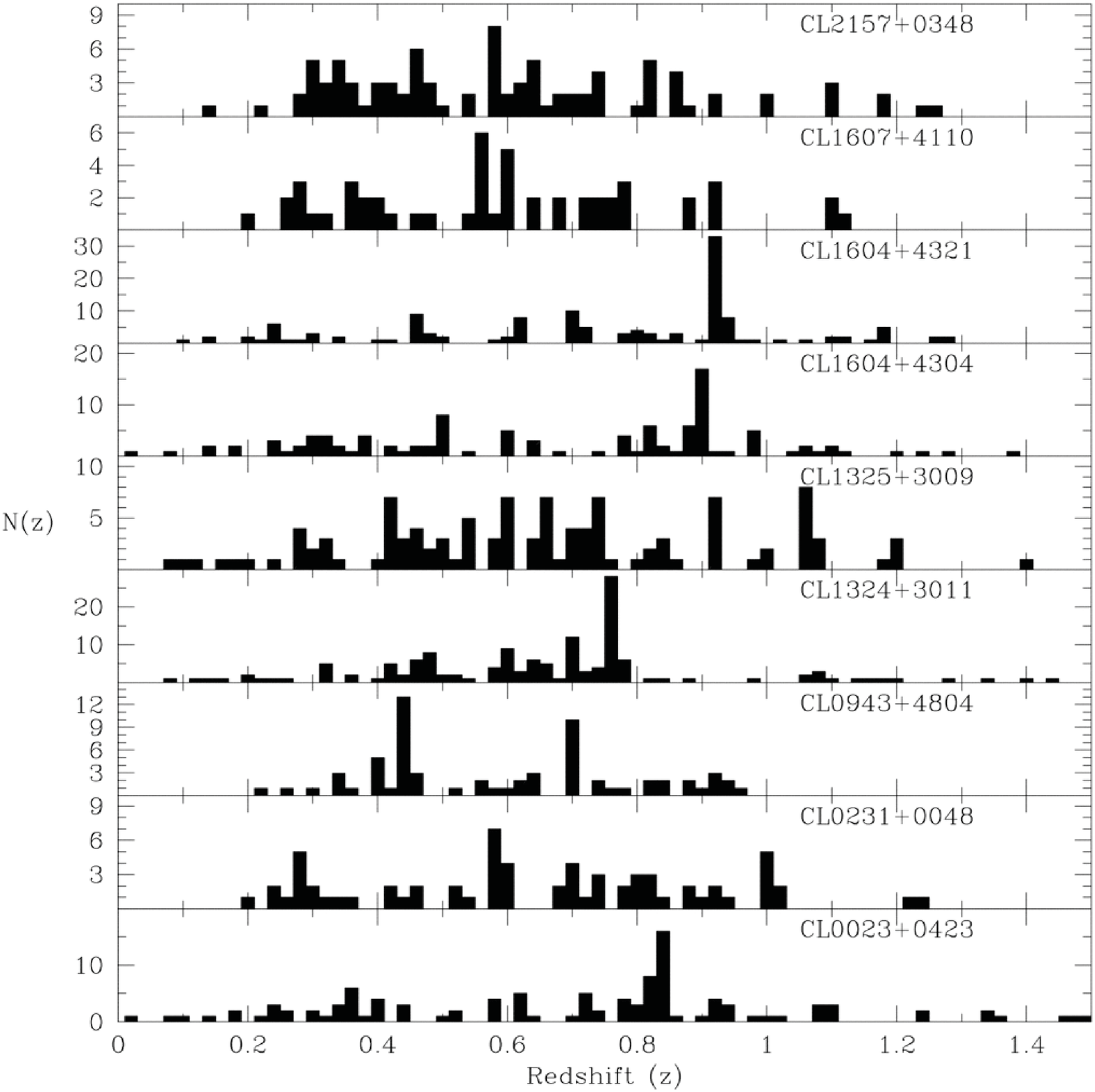}}
\parbox{0.5\textwidth}{
\includegraphics[width=0.51\textwidth]{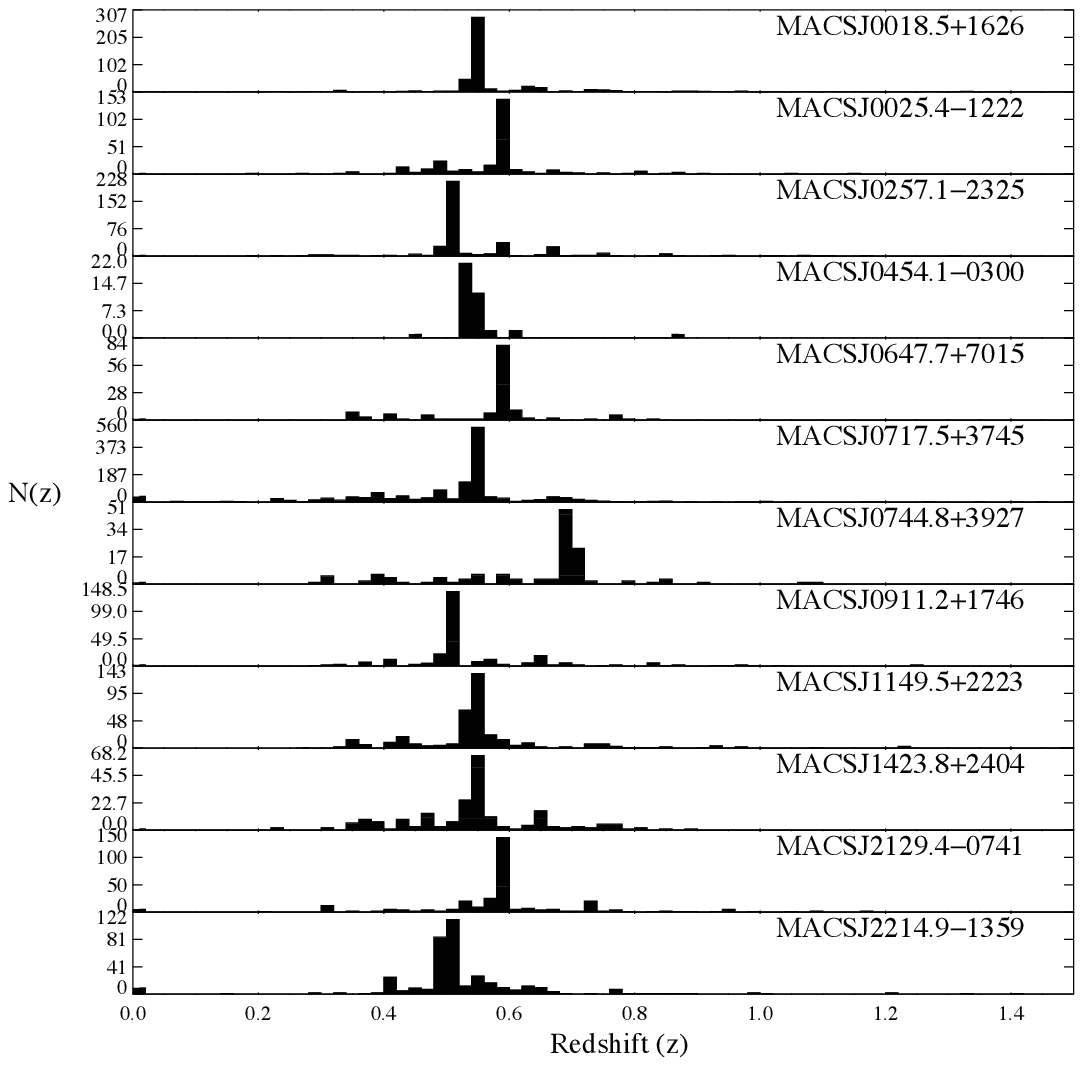}}
\hfill
\caption[]{Histograms of
  galaxy redshifts in the fields of the nine optically selected
  systems of the PDCS (left; Oke et al.\ 1998) and the twelve most
  distant X-ray selected MACS clusters (right; Ebeling et al.\
  2007). For ease of comparison the MACS data are shown over the same
  redshift range and with the same binning as used in the published
  PDCS figure. Note that both surveys used similar criteria to select
  galaxies for spectroscopic follow-up observation. (Reproduced from Ebeling et al.\ 2010.)}\label{fig:xoptclusters}
\end{figure}

The advantage of X-ray cluster surveys over optical surveys is illustrated in Fig.~\ref{fig:xoptclusters} which compares the redshift histograms of the nine optically selected systems of the Palomar Distant Cluster Survey (PDCS; Oke, Postman \& Lubin, 1998) with those of the twelve most distant X-ray selected  clusters from the Massive Cluster Survey (MACS; Ebeling et al.\ 2007). The severe contamination by fore- and background structures seen in projection in the PDCS is endemic in optically selected cluster samples. Pure projection effects like, e.g., CL0231+0048 (Fig.~\ref{fig:xoptclusters}, left) can be largely eliminated by including information on galaxy colors or redshifts (photometric or spectroscopic) in the original cluster detection phase. However, even the latest, state-of-the-art optical cluster samples remain biased, as they are prone to select intrinsically poor systems whose apparently compact cluster core, high optical richness, and high velocity dispersion are inflated by line-of-sight alignment and infall (``orientation bias''; Hicks et al.\ 2008; Horesh et al.\ 2010).  By contrast, X-ray selected cluster samples such as those compiled by the BCS or MACS are almost entirely free of projection effects since they, by virtue of the X-ray selection criteria, comprise exclusively intrinsically massive, gravitationally collapsed systems.


\subsection{All-sky versus serendipitous cluster surveys}

\begin{figure}[h!]
\hspace*{8mm}
\includegraphics[width=0.8\textwidth]{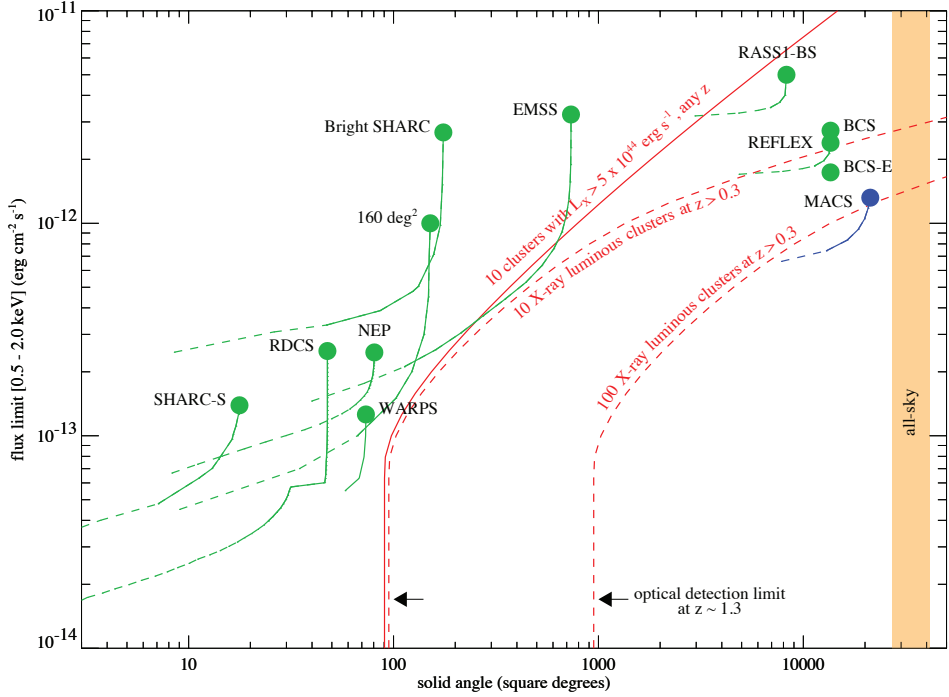}
\caption[]{Selection functions of all major X-ray cluster surveys of the 1990s. The red lines indicate the solid angle required at a given flux limit to (statistically) detect a given number of X-ray luminous clusters within different redshift regimes. Note how large-area surveys are uniquely suited to finding the most X-ray luminous systems, while serendipitous surveys such as the 160-deg$^2$ survey or WARPS  sample the cluster population primarily at low to intermediate luminosity. (Reproduced from Ebeling et al.\ 2001).   \label{fig:xsurveys}}
\end{figure}

Already prior to the advent of ROSAT and the RASS in the 1990s, samples of X-ray luminous clusters were being compiled in two complementary ways. Although limited in depth, the enormous solid angle covered by wide-angle (ideally: all-sky) surveys allowed the detection of the rarest most massive clusters at low to moderate redshift. By contrast, pointed observations of individual targets (most of them not galaxy clusters) probed deeply over small areas, enabling the serendipitous detection of clusters of low to average mass out to much higher redshift. The dichotomy of the resulting samples is clearly visible in Fig.~\ref{fig:xsurveys} which illustrates the effect of depth and breadth of an X-ray cluster survey on the mass range of the clusters discovered.

\begin{figure}[h!]
\hspace*{8mm}
\includegraphics[width=0.8\textwidth]{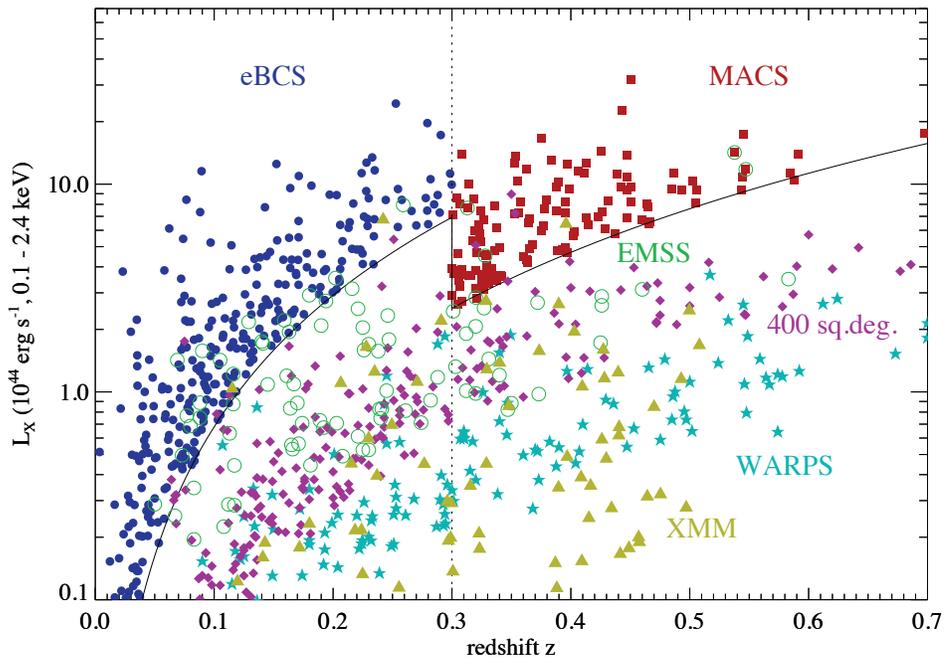}
\caption[]{$L_{\rm X}$-$z$ distribution of clusters from various X-ray selected samples. Note how large-area cluster surveys, such as the eBCS and MACS, contain a significant number of systems that are about 10 times more X-ray
  luminous, and thus much more massive, than the most extreme clusters found in deeper serendipitous cluster surveys such as the EMSS, WARPS, 2XMMi/SDSS, or the 400 sq.-deg.\ project.   \label{fig:lx-z}}
\end{figure}

The availability of large, representative, X-ray selected samples compiled from RASS data (Ebeling et al.\ 1996, 1998, 2000; De Grandi et al.\ 1999; Ebe\-ling, Mullis \& Tully 2002; Cruddace et al.\ 2002; B\"ohringer et al.\ 2004; Kocevski et al.\ 2007) has allowed greatly improved, unbiased measurements of the properties of clusters in the local universe ($z\le 0.3$). Especially the ROSAT Brightest Cluster Sample (BCS, Ebeling et al.\ 1998, 2000) and the REFLEX sample (B\"ohringer et al.\ 2004) have been used extensively for studies of the local cluster population from groups to extremely massive systems. More recently, the Massive Cluster Survey (MACS; Ebeling et al.\ 2001), also compiled from RASS data, extended the redshift baseline for evolutionary studies of very X-ray luminous systems to $z\sim0.6$ (Ebeling et al.\ 2007, 2010). The $L_{\rm X}-z$ distribution of RASS-based cluster samples is shown in Fig.~\ref{fig:lx-z} and compared to that of the EMSS, WARPS, 400-sq.-deg., and 2XMMi/SDSS cluster samples (Gioia \& Luppino 1994; Perlman et al.\ 2002; Burenin et al.\ 2007; Horner et al.\ 2008; Takey et al.\ 2011), where all of the latter are serendipitous cluster surveys. Note the striking difference in the X-ray luminosity ranges of these samples: while serendipitous cluster surveys excel at finding clusters of low to intermediate X-ray luminosity out to significant redshifts, their small solid angle (typically a few 100 square degrees, comparable to the largest deep optical cluster surveys; see Fig.~\ref{fig:xsurveys}) unavoidably prevents them from finding the rare extreme systems.



\subsection{The DF cluster catalogs}

Any experiment attempting to measure a large-scale flow across distances of, potentially, hundreds of Mpc will benefit from the availability of a set of probes that is as uniformly distributed as possible over the largest possible volume. Ideally this means a large, deep (in redshift space), and homogeneously selected all-sky sample of tracers. For applications of the K-AB method, these tracers are clusters of galaxies, and the databases of choice for their selection are the source lists of all-sky X-ray or SZ surveys.

The DF cluster catalog was compiled with two main considerations in mind: a) the cluster sample had to be as homogeneous and isotropic as possible and, ideally, cover the entire sky, and b) it should contain as many truly massive clusters as possible, as they will produce the strongest SZ signal in the CMB maps. With all-sky SZ-selected cluster catalogs unavailable at the time\footnote{In 2007, large-area SZ surveys were still in their infancy, and even in 2012 they are only beginning to provide cluster samples of competitive size. }, these criteria were best met by cluster compilations based on RASS data (cf.\ Figs.~\ref{fig:xsurveys} and \ref{fig:lx-z}).

The original DF cluster catalog used by KABKE was thus created by combining the three largest RASS-based cluster samples published at the time: the eBCS covering the northern extragalactic sky (Ebeling et al.\ 1998, 2000), the REFLEX sample covering the southern extragalactic sky (B\"ohringer et al.\ 2004), and the CIZA catalog covering the Galactic plane (Ebeling et al.\ 2002; Kocevski et al.\ 2007). Although all of these three samples are X-ray selected, they differ in the way cluster fluxes are measured and also adopt different X-ray flux limits. In order to create a homogenous combined catalog, fluxes were recomputed from the RASS raw data as detailed in Kocevski et al.\ (2006). Removal of duplicate entries (the solid angles of the three samples overlap slightly) and application of a global flux limit of $3\times 10^{-12}$ erg s$^{-1}$ (0.1--2.4 keV) then resulted in an all-sky sample of 782 clusters\footnote{The KABKE sample thus constructed is slightly larger than that of Kocevski et al.\ (2006) because it includes clusters from the unpublished third release of the CIZA sample (Kocevski et al.\ 2012).}.

\begin{figure}[h!]
\hspace*{8mm}
\includegraphics[width=0.5\textwidth]{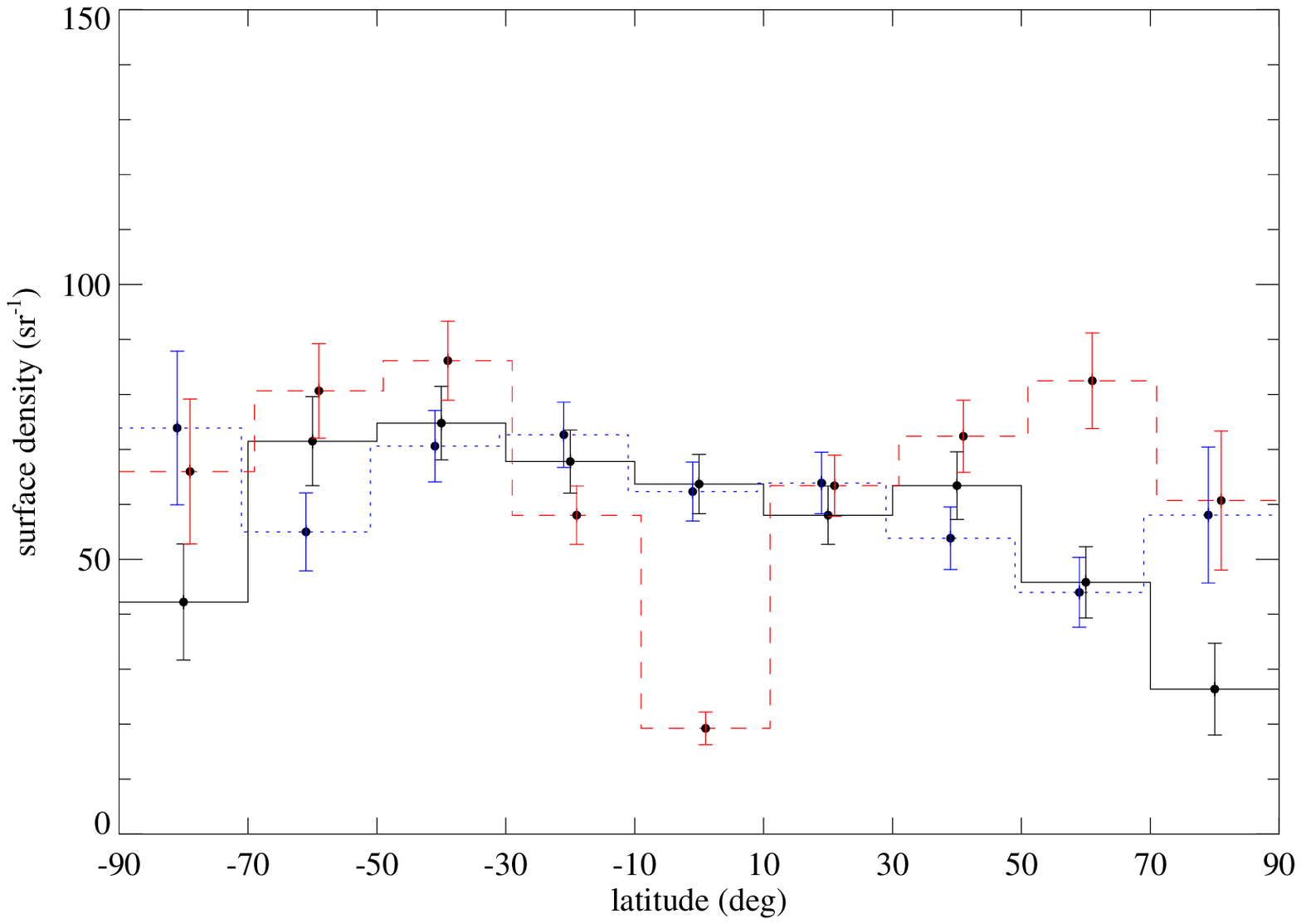}
\includegraphics[width=0.5\textwidth]{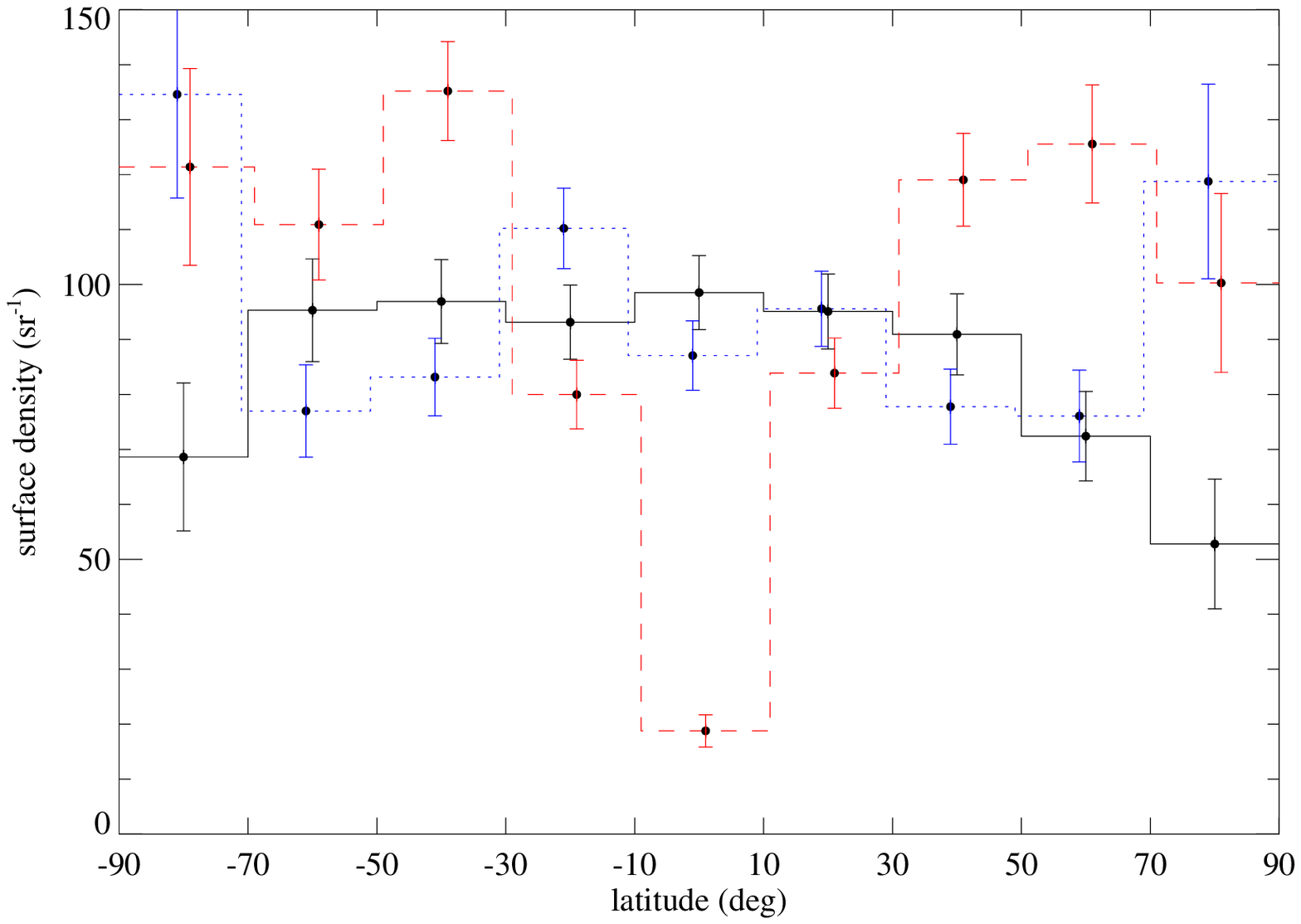}
\caption[]{The projected surface density of clusters in the DF catalogs as a function of celestial (black, solid histogram), ecliptic (blue, dotted histrogram), and Galactic (red, dashed histrogram) latitude for the samples used by KABKE (left) and KAEEK (right). Both catalogs are significantly incomplete at $|b|<10$ deg, reflecting the challenges of securely detecting extragalactic X-ray sources behind the plane of the Galaxy and of obtaining robust optical identifications, including spectroscopic redshifts. Mild incompleteness is observed near the celestial poles, most likely because very high declinations are hard to reach from optical observatories in either hemisphere. Finally, the longer accumulated exposure times of the RASS at high ecliptic latitudes results in an increased detection efficiency near the poles; this effect is beginning to become significant for the larger cluster sample used by KAEEK (right panel). \label{fig:catdensity}}
\end{figure}

Figure~\ref{fig:catdensity} (left) shows the projected surface density of clusters in the KABKE sample as a function of celestial, ecliptic, and Galactic latitude. No significant variations are observed in the celestial and ecliptic coordinate frames. However, absorption of soft X-rays by neutral hydrogen in the Galactic plane, compounded by the challenge of optical identification of X-ray selected cluster candidates in regions of very high stellar densities, causes  the KABKE sample to be significantly incomplete at low Galactic latitude. As the areas of most severe incompleteness also feature strong foreground emission in the CMB maps (the unmasked area at $|b|<10$ deg amounts to less than 15\% of the sky), they are largely masked out, greatly reducing the impact of this incompleteness on the DF analysis.

In order to test the importance of homogeneously computed fluxes for the design of the DF cluster catalog, KAE repeated the KABKE analysis on a cluster catalog obtained simply by merging the eBCS, REFLEX, and CIZA samples as published. Combined with 7-year WMAP data, this cluster sample is found to exhibit a large-scale flow that is fully consistent with the earlier DF measurements in both amplitude and direction, although the correlation of the SZ amplitude with cluster X-ray luminosity (an important diagnostic) is, unsurprisingly, less clear than for the more carefully compiled KABKE sample (see Section 10.3.2 for details). The latitude distribution of the clusters in the KAE sample is, within the errors, consistent with that used by KABKE (Fig.~\ref{fig:catdensity}, left). The anisotropies of the cluster samples on the robustness of the DF results is addressed via different error calculation methods in Section \ref{sec:darkflow}.

The SCOUT project, finally, aims to extend the aforementioned catalogs significantly, both in terms of X-ray flux limit and redshift range covered. SCOUT uses and expands the list of identifications obtained during the compilation of the MACS sample (Ebeling et al.\ 2001) without adopting {\it a priori}\/ constraints regarding redshift or X-ray flux for sources detected in the RASS. Complementing the MACS identifications, which are limited to the extragalactic sky ($|b| > 20$ deg) and $\delta < 80$ deg, SCOUT also draws from the source list established by the CIZA team at low Galactic latitude. A first version of the SCOUT cluster catalog was used by KAEEK. Almost doubling the median redshift of the KABKE sample, the new catalog, in conjunction with 5-year WMAP data, enabled KAEEK to trace the DF to at least 800 Mpc $h_{\rm 70}^{-1}$, although its significance decreases after peaking at 500 Mpc $h_{\rm 70}^{-1}$. The completeness of the KAEEK catalog as a function of celestial, ecliptic, and Galactic latitude can be assessed from Figure~\ref{fig:catdensity} (right), which shows the respective distributions for all 1144 RASS-selected clusters in the catalog that meet the requirements of $f_{\rm X}\geq 1\times 10^{-12}$ erg s$^{-1}$ cm$^{-2}$, $L_{\rm X}\geq 5\times 10^{43}$ erg s$^{-1}$, and $z\le 0.25$ (985 of these fall within the WMAP CMB map). The incompleteness at low Galactic latitude remains; its impact on the DF study is, however, mitigated again by the large overlap between the areas of incompleteness and those eliminated by the CMB mask because of strong foreground emission. As before for the KABKE sample a slight deficiency of clusters is noted near the celestial poles, made more prominent in the northern hemisphere by the limitation of the MACS database to $\delta < 80$ deg. The improved statistics also increase the significance of the excess cluster detections near the ecliptic poles, facilitated by the greater effective exposure times accumulated there during the RASS.

\section{Establishing cluster SZ properties}
\label{sec:cluster_tsz}

Clusters are extended objects and models based on observations or
N-body simulations have been used to parametrize their electron density and pressure
profiles. These are the two basic quantities relevant for our study since
determine the amplitude of the KSZ and the TSZ effects, respectively.
In Section~\ref{sec:6.4} we briefly discussed the observations of
the TSZ and KSZ effect. Following this we now expand the previous discussion,
and in connection with our most relevant finding needed for the DF measurement:
clusters of galaxies are not isothermal and their temperature drops with
increasing distance from the center.

\subsection{SZ cluster radial profiles.}
\label{sec:7.3}

Much of the work on TSZ has used radial profiles modeled on their X-ray properties.
Most commonly, clusters were assumed to be isothermal
$\beta$ models (Cavaliere \& Fusco-Femiano 1976, 1978) where the
electron density is described as
\begin{equation}
n_e(r)=\frac{n_c}{[1+(r/r_c)^2]^{3\beta/2}} .
\label{eq:betamodel}
\end{equation}
In this expression, $n_c$ is the electron central density, $r_c$ is the core radius
of the cluster and $\beta$ is a free parameter chosen to fit the cluster
profile. For most clusters, the observed value falls in the range
$0.5\le\beta\le 0.7$ (Markevitch et al, 1998).
The functional form of this model can be derived from a parametrization
of the density assuming clusters to be isothermal. Strictly speaking,
isothermality and hydrostatic equilibrium are incompatible for $\beta$
models, but they are compatible for a large range of radius when
$\beta\approx 2/3$ (Lancaster et al 2005). For this reason,
and also for mathematical convenience, this value has been widely used.
However, isothermality is a poor assumption for many clusters, in particular
for merging and cooling core systems. Measurements of the X-ray temperature
profiles of 15 nearby clusters, carried out by Pratt et al. (2007)
using XMM-Newton data and numerical simulations showed significant
departures from the $\beta$ model outside the cluster core
(Vikhlinin et al. 2005; Hallman et al. 2007). Different generalizations
of the original $\beta$-model have been considered: a double $\beta$ model with a common value
of $\beta$ and a cusped $\beta$-model (Pratt \& Arnaud 2002; Lewis et al. 2003)
designed to account for the sharply peaked surface brightness
in the centers of relaxed X-ray systems.
LaRoque et al. (2006) examined two extensions of the $\beta$-model using a
sample of 38 clusters with both X-ray and SZ data. They found that
a nonisothermal model fit the gas distribution out to $r_{2500}$,
the radius where the mean density of the cluster is 2,500 times the
critical density. Nevertheless,
all these generalizations do not provide a good description in
the outskirts of clusters, regions where departures from the X-ray
measured profiles are expected to be most pronounced.

A different set of profiles were derived assuming the gas to
be in hydrostatic equilibrium within the potential wells of Dark Matter
(DM) haloes (Komatsu \& Seljak 2001).
Numerical simulations suggest that the DM
distribution in galaxy clusters is described by a universal
density profile (the Navarro-Frenk-White [NFW] profile; Navarro et al. 1997)
\begin{equation}
\rho_{dm}(x)=\frac{\rho_s}{x(1+x)^2} ,
\label{nfw}
\end{equation}
where $x=r/r_s$ and
$r_s$ and $\rho_s$ are the characteristic scale radius and density, respectively.
Usually, the scale radius $r_s$ is expressed in terms of the concentration parameter
$c=r_{vir}/r_s$, where $r_{vir}$ is the halo virial radius. This parameter depends
only weakly on mass, with less massive systems being more concentrated, and
thus having larger $c$. While the electron density for the $\beta=2/3$
model scales as $r^{-2}$ at large radii, the NFW
model is much steeper, scaling as $r^{-3}$.

The Komatsu-Seljak (KS) model assumes a
constant polytropic equation of state for the gas, $P_{gas}=A\rho_{gas}^\gamma$.
Their model contains two parameters, the polytropic index $\gamma$ and
the normalization constant $A$. Further, they fix the index $\gamma$  by taking
the same slope around the virial radius for the gas and DM density profiles.
The KS gas pressure profile is given by (see
Komatsu \& Seljak 2002 and Komatsu et al. 2011 for details):
\begin{equation}
P_{gas}(x)=P_{gas}(0)y^\gamma(x) ,
\label{eq:p_gas}
\end{equation}
where distances are measured in units of the scale radius $r_s$: $x=r/r_s$.
The function $y(x)$ is defined by
\begin{equation}
y_{gas}(x)=\left[1-B(1-x^{-1}\ln(1+x))\right]^{1/(\gamma-1)}
\end{equation}
with
\begin{eqnarray}
B&=&3\eta^{-1}(0)(\gamma-1)\gamma^{-1}[c^{-1}\ln(1+c)-(1+c)^{-1}]^{-1} ,\\
\gamma&=&1.137+0.0894\ln(c/5)-3.68\times 10^{-3}(c-5) ,\\
\eta(0)&=&2.235+0.202(c-5)-1.16\times 10^{-3}(c-5)^2 ,
\end{eqnarray}
where $c$ is the concentration parameter defined above.
For a hydrogen mass fraction of $X=0.76$, the electron pressure profile is
given by $P_e(x)=0.518 P_{gas}(x)$ and
\begin{equation}
\rho_{gas}(x)=\rho_{gas}(0)y(x),\qquad
T_{gas}(x)=T_{gas}(0)y_{gas}^{\gamma-1}(x)
\end{equation}
The central gas density $\rho_{gas}(0)$
is fixed by assuming that the gas density at the virial radius is equal to
the cosmic mean baryon fraction, $f=\Omega_b/\Omega_m$, times the dark mater
density at the same radius. Then
\begin{equation}
\rho_{gas}(0)=7.96\times 10^{13}\,
\left(\frac{\Omega_b}{\Omega_m}\right)\,
\frac{M_{vir}/(10^{15}M_\odot/h)}{[r_{vir}/(1 Mpc/h)]^3}\,
\frac{c^2(1+c)^{-1}y^{-1}_{gas}(c)}{(1+c)\ln(1+c)-c} \,
{\rm \left[\frac{M_\odot/h}{(Mpc/h)^3}\right]}\,,
\end{equation}
where $M_{\rm vir}$ is the virial mass of the cluster. The central gas temperature
and density are respectively given by
\begin{eqnarray}
k_BT_{\rm gas}(0)&=&8.8\,\eta(0)\,
	\frac{M_{vir}/(10^{15}M_\odot/h)}{r_{\rm vir}/(1 h^{-1}{\rm Mpc})}\, {\rm [KeV]},\\
P_{gas}(0) &=& 55.0\, \frac{\rho_{\rm gas}(0)}{10^{14}h^2M_\odot{\rm Mpc}^{-3}}\,
\frac{k_BT_{gas}(0)}{8KeV}\, {\rm \left[\frac{eV}{cm^3}\right]} .
\end{eqnarray}
Finally, the concentration parameter can also be expressed in terms of the
virial mass. Recently, Duffy et al (2008) derived the following fitting
formula from N-body simulations
\begin{equation}
c=\frac{5.72}{(1+z)^{0.71}}\left(\frac{M_{vir}}{10^{14}M_\odot/h}\right)^{-0.081}
\end{equation}
This formula predicts that clusters of galaxies with masses $M\le 10^{14}M_\odot$
are more concentrated than in Komatsu \& Seljak (2002).

\begin{figure}[h!]
\includegraphics[width=7in]{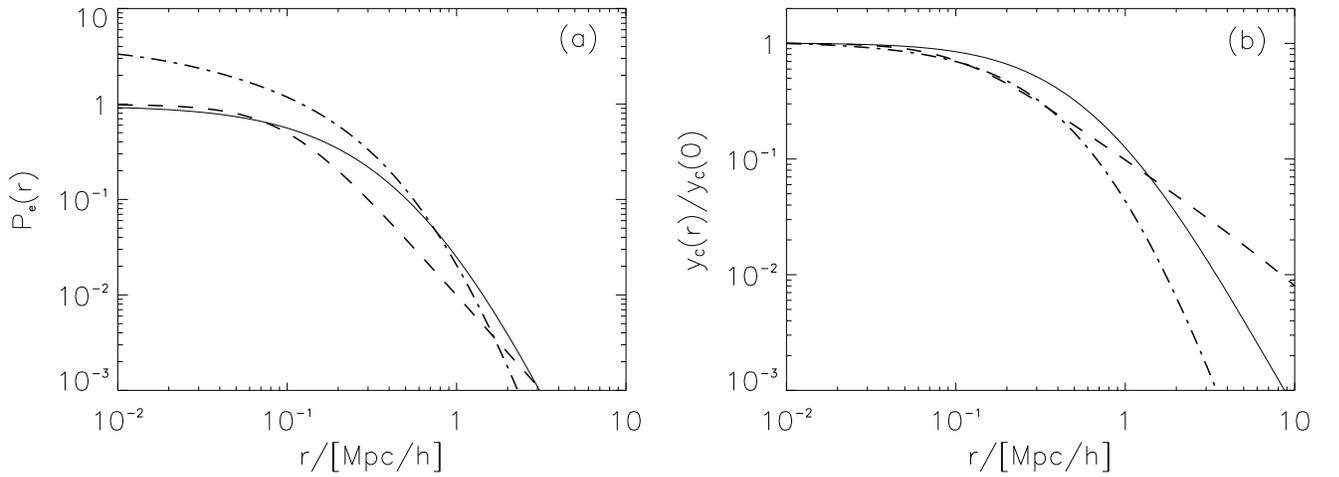}
\caption[]{Compasion of
Komatsu \& Seljak (2002) (solid lines) and Arnaud et al (2010) 3-D profiles which are seen to be rather different in their properties. In this plot, concentration is: c=[5,10,15] corresponding
to the solid black, blue and green lines and the Arnaud
models correspond to: $(\alpha,\beta,\gamma,c_{500})=$
([1.05,5.45,0.31,1.18] -black-; [1.0,5.5,0.5,1.0] -blue-; and
[1.05,5.0,0.4,1.3] -green-) dashed lines, corresponding to the data
derived Arnaud et al (2010), Plagge et al (2010) and Nagai et al (2007),
respectively. For simplicity we took $r_{vir}=2r_{500}$.
}
\label{fig:arnaud-ks}
\end{figure}

Nagai et al. (2007) proposed a generalization of the NFW model that
is expected to describe the cluster pressure out to a significant
fraction of the virial radius. The model has been investigated by Mroczkowski
et al. (2009). Arnaud et al. (2010) rescaled by
mass and redshift the measured pressure profile of 33 clusters
observed with XMM-Newton and produced a ``universal pressure profile'' after
taken the median of the scaled profiles. This phenomenological electron
pressure profile is given by
\begin{equation}
P_e(r)=3.37h^2E^{8/3}(z)\left[\frac{M_{500}}{2.1\times 10^{14}M_\odot/h}
\right]^{2/3+\alpha_p} p(r/r_{500})\,
{\rm \left[\frac{eV}{cm^3}\right]} \, .
\end{equation}
In this expression $E(z)=H(z)/H_0$, $r_{500}$ is the radius at which the mean
overdensity of the cluster is 500 times the critical density of the Universe
and $M_{500}$ is the mass enclosed within $r_{500}$: $M_{500}=(4\pi/3)
[500\rho_c(z)]r_{500}^3$ and $\rho_c(z)=2.775\times 10^{11}E^2(z)h^2M_\odot$Mpc$^{3}$.
The slope is $\alpha_P=0.12$ and shows a small decrease
with radius (see Arnaud et al. 2010, for details).
The function $p(x)$ is defined by (Nagai et al 2007)
\begin{equation}
p(x)=\frac{4.92h^{3/2}}{(c_{500}x)^\gamma[1+(c_{500}x)^\alpha]^{(\beta-\gamma)/\alpha}} ,
\label{nagai}
\end{equation}
where the parameters $(\alpha,\beta,\gamma,c_{500})=(1.05,5.49,0.31,1.18)$
were derived from the X-ray observations. The slopes $\alpha,\beta,\gamma$
in eq.~\ref{nagai} are not independent but are strongly correlated with
$r_s=r_{500}/c_{500}$. Of those three parameters, $\beta$ is unconstrained
by data in the central parts of the cluster $(r\le r_{500})$. Then, using
only X-ray data will result in large uncertainties in $\beta$ and,
consequently, in the profile beyond $r_{500}$ and in the integrated SZ signal.
To overcome this difficulty, Arnaud et al (2010) determined $\beta$
by using the results of numerical simulations in the region $(1-4)r_{500}$.

In Figure~\ref{fig:arnaud-ks} we compare the profiles of Komatsu\& Seljak (2002)
with those of Arnaud et al (2010). The electron pressure profile have been normalized
to unity at the center of the cluster. This normalization makes more evident
the difference in slope of the two profiles. Since the KS profile is given
in terms of the virial radius $R_{\rm vir}$, for the purpose of
comparison  we have taken $R_{vir}=2r_{500}$. Without a thorough statistical
analysis, Komatsu et al (2011) used WMAP 7yr data to compare both profiles with
the data and found them to be consistent.

\begin{figure}[h!]
\includegraphics[width=6in]{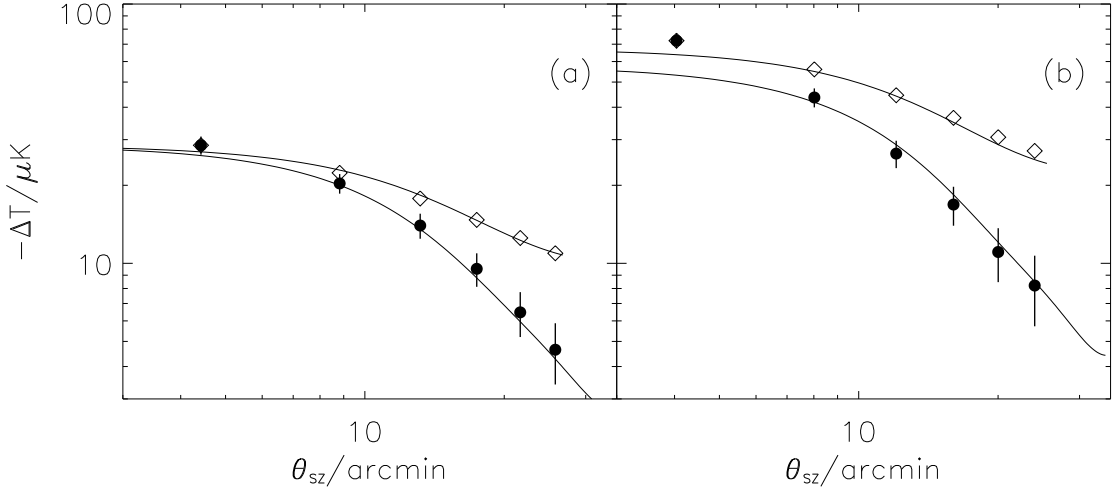}
\caption[]{
Measured and predicted electron-pressure profile vs. angular clustercentric
radius. The filled circles and open diamonds correspond to the measured
profile and the profile predicted for the isothermal $\beta$ model,
respectively. The solid lines correspond to the $\beta$ model (upper solid line)
and the NFW model (lower solid line), assuming a single cluster at $z\sim 0.12$.
We show results for two cluster subsamples: (a) $z\le 0.2$ and (b) $
L_X[0.1-2.4keV]\ge 3\times 10^{44}$ergs/s.
In (a),the best fit corresponds to $c=8$ and $r_s= 350$kpc/h, and in (b),
the to $c=15$ and $r_s=250$kpc/h.
}
\label{fig:tsz_profile}
\end{figure}

\subsection{Observed Pressure Profiles}

The first determination of the TSZ profile of clusters of galaxies
was carried by us and presented in AKKE, using WMAP 3yr data.
We derived the average TSZ profile of a set of 661 clusters by
stacking the radial temperature anisotropy of WMAP data at the locations of
the X-ray clusters cataloged for this project.
The catalog used in AKKE listed the position, flux,
luminosity, X-ray temperature, redshift, angular and physical extent
of the region containing the measured X-ray flux.
Positions, fluxes and luminosities were
measured directly from the ROSSAT All Sky Survey (RASS). Electron
temperature derived from the $L_x-T_e$ scaling relation of White et al (1997).
Redshifts were measured spectroscopically, while the angular extent
was determined directly from the RASS imaging data using a growth-curve
analysis. For each cluster, central density and core radii were determined
by fitting to the data a $\beta$-model profile (Cavaliere \& Fusco-Femiano 1976)
convolved with the RASS point-spread function. To avoid the uncertainties
associated with the correlation between $r_c$ and $\beta$, the latter
was held at the canonical value $\beta=2/3$ (Jones \& Forman 1984).

The stacking of the cluster signal in WMAP 3yr data was carried out in
units of the X-ray extent $\theta_X$, the angular size of the cluster emitting
99\% of the total X-ray flux in RASS data. The TSZ amplitude was detected at the
the 15$\sigma$ level, the highest statistical significance reported at the time.
Thanks to the large number of clusters, we measured the TSZ profile out to large
cluster-centric radii and found them to be in good agreement with
the KS profile. The region over which the TSZ signal was detected was, on
average, $\sim 4$ times larger in radius than the X-ray emitting region,
extending to $2-3h^{-1}$Mpc.
Figure~\ref{fig:tsz_profile} shows the measured cluster profile, and
the KS prediction compared with an expectation for an isothermal $\beta=2/3$ model,
in units of $\theta_X$.
Error bars were computed by generating 1000 cluster templates with
clusters randomly placed outside the measured region.
While more massive clusters are less concentrated,
our results indicated that the concentration parameter was larger for the more
rather than the less massive clusters, but the data for subsets of
clusters was not good enough to quantify the statistical significance of
this result.

Since AKKE, deep observations of SZ effect in 11 clusters selected from the
highest luminosity REFLEX clusters in the range of elevation angles accessible
to the SPT allowed Plagge et al. (2010) to make a model independent
estimate of the radial profile for each individual cluster and place
constraints on the pressure profile out to large radius.
The $\beta$-model and universal pressure profile fit the SPT data
comparably well. The $\beta$-model parameters that were consistent with
the average SZ profile out to the virial radius
were $\beta=0.86$, $r_{core}=0.2r_{500}$, corresponding an electron
density scaling as $n_e\sim r^{-2.6}$, steeper than $r^{-2}$ of the X-ray
data and closer to $r^{-3}$ of AKKE. For the universal pressure profile,
the parameters were $[\alpha,\beta,\gamma,c_{500}]=(1.0,5.5,0.5,1.0)$,
consistent with the values found by Nagai et al (2007) and Arnaud et al (2010).

\subsection{Cluster temperature radial decline}

\begin{figure}[h!]
\includegraphics[width=0.5\textwidth,height=0.435\textheight]{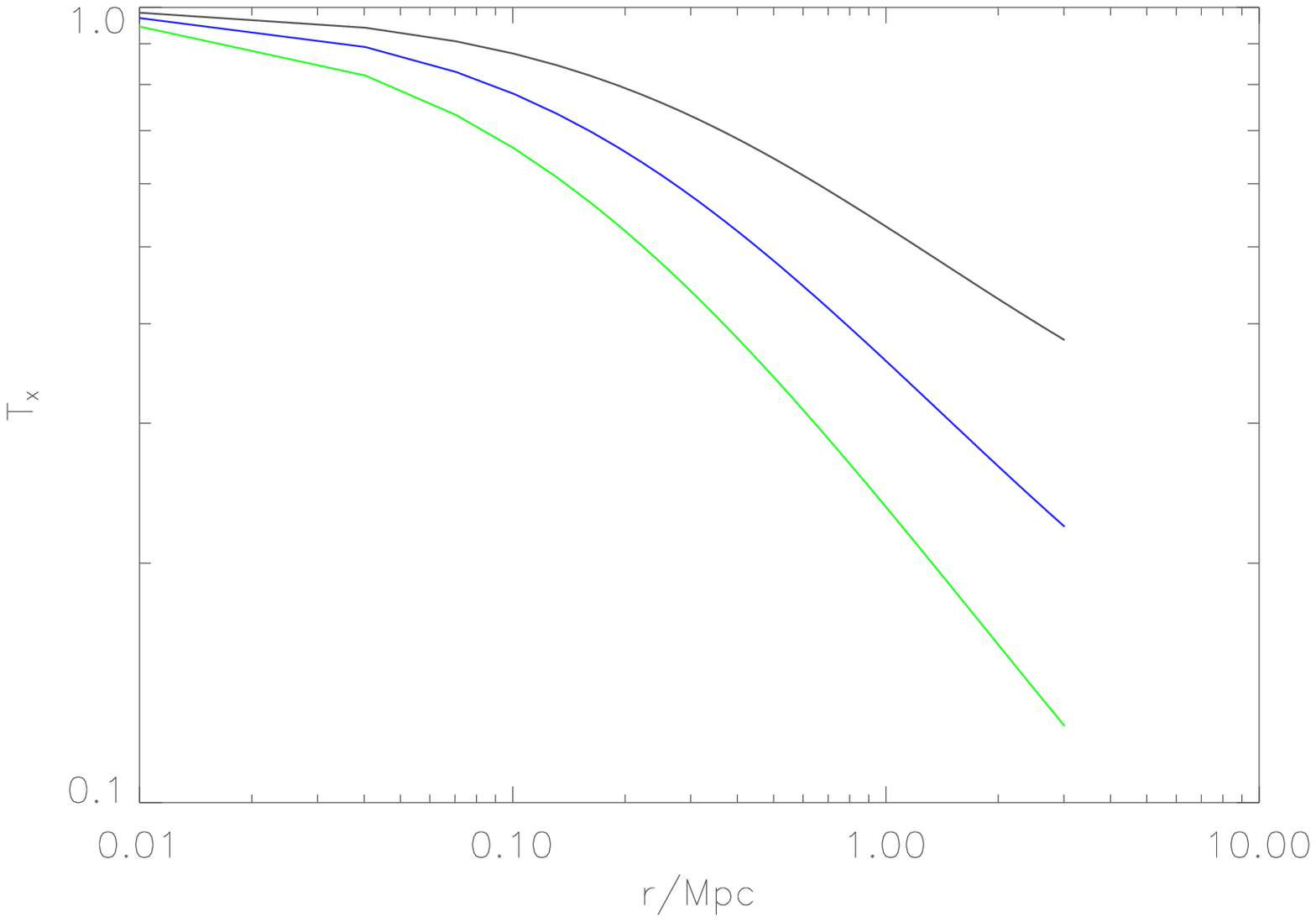}
\includegraphics[width=0.5\textwidth,height=0.4\textheight]{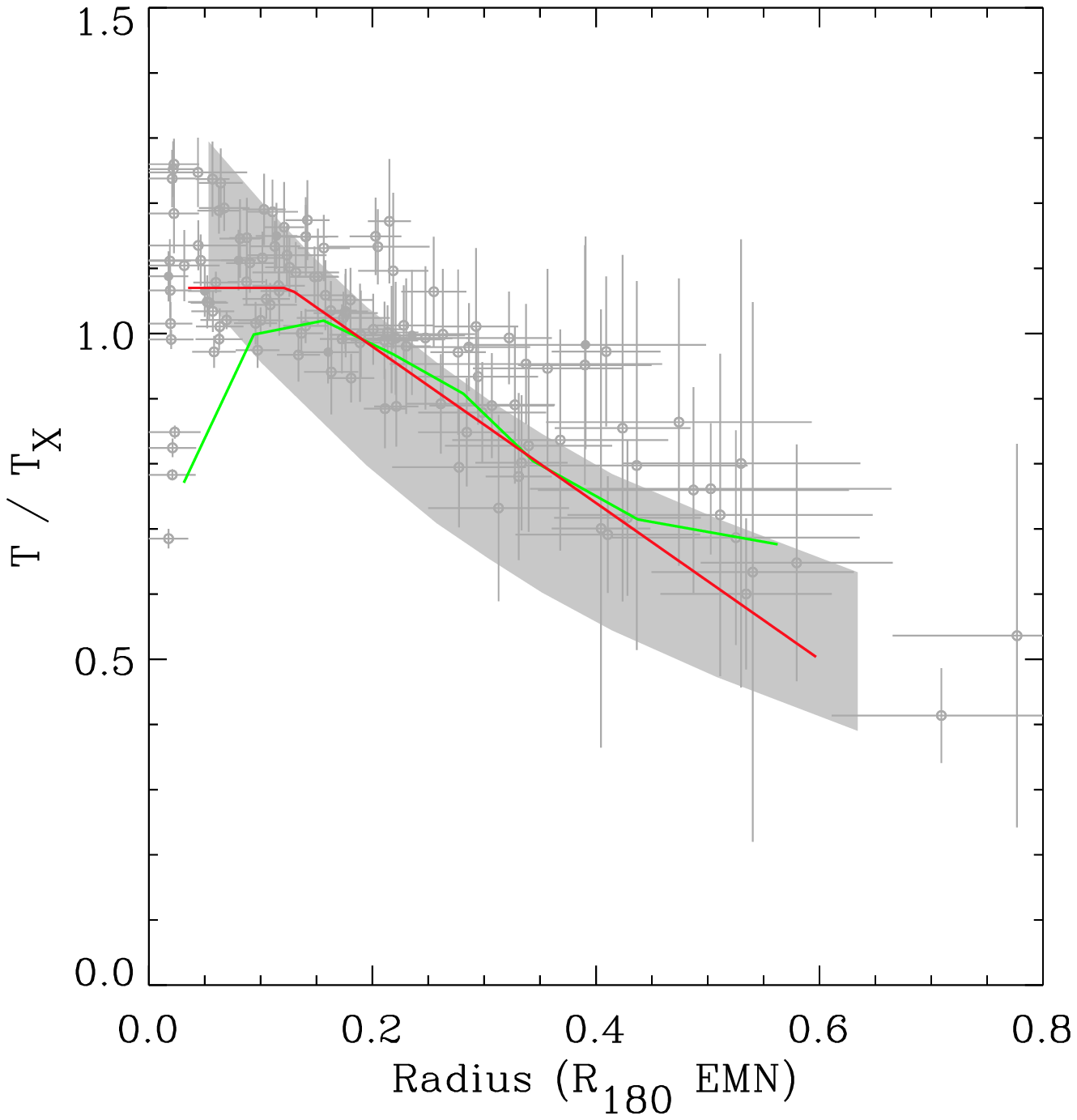}
\caption[]{{\bf Left}: Radial variation of the electron temperature with radius
in arbitrary units. From top to bottom, black, blue and green solid lines correspond
to concentration parameter $c=4,7,12$. {\bf Right}: Compilation of scale-projected
temperature profiles of 15 well measured X-ray clusters from Pratt et al (2007).
Green/red lines correspond to BeppoSAX/Chandra observations of cooling core systems
as discussed in the original reference. Reproduced from Fig. 5 from Pratt et al (2007)
with permission.
}
\label{fig:temp_profile}
\end{figure}

Komatsu \& Seljak (2001) discussed that when the gas density follows the NFW
distribution (eq.~\ref{nfw}), hydrostatic equilibrium was inconsistent
with the gas being isothermal in the outer parts of clusters.
The universal gas temperature profiles of Komatsu \& Seljak (2001) was
found to fit WMAP 3yr data in AKKE.
Consequently, the central temperature must decrease by a significant factor
at the virial radius, this decrease being steeper for the more
concentrated (less massive) clusters. This result is in agreement
with an earlier analysis of X-ray temperature profiles of 15
nearby clusters, carried out by Pratt et al (2007) using
XMM-Newton data. They measured temperature profiles declining by a
factor $\sim 2$ at half the virial radius, in good agreement with
numerical simulations outside the core region.
In Fig~\ref{fig:temp_profile} we plot the variation of the
electron temperature as a function of radius for three concentration
parameters $c=(4,7,12)$ corresponding to black, blue and green lines,
respectively and also reproduce the results from the compilation
by Pratt et al (2007, Fig. 5 there).

%

The decline of the electron temperature with distance from the cluster center
is very relevant in the context of the DF measurement.
In Section \ref{sec:darkflow} we discuss that the DF appears as a dipole at
cluster location {\it measured at zero monopole}. That would be impossible
if clusters were isothermal since then the TSZ and the KSZ effects would be
proportional to one another and a TSZ dipole, generated from the inhomogeneity
of the cluster distribution in the sky, could not be separated from a dipole generated
by the bulk motion of the cluster population. That is, while the
monopole is dominated by the TSZ component the dipole has contributions
from the KSZ and the TSZ. By increasing the cluster aperture we  demonstrate
that the monopole vanishes (within the noise) isolating the KSZ-produced dipole
since the monopole amplitude is an upper bound to the TSZ-produced dipole.
In other words, the CMB dipole at cluster locations
computed at zero monopole is free from the TSZ contamination.
This fundamentally important fact was first established by us
in KABKE1,2 and was key to isolate the dipole due to the KSZ effect
from any other contributions.

\section{Measuring peculiar motion of X-ray clusters}
\label{sec:darkflow}

A procedure has to be constructed such that all other contributors to the dipole at cluster locations be reduced more than the KSZ component. This section discusses the processing that we have developed and applied to the WMAP 3, 5 and 7-yr data in order to achieve the measurement of the KSZ contribution with a statistically significant S/N (in our case S/N$\simeq 3.5-4$). Briefly, the steps are summed up below before we delve into the technical details. The potential obstacles to isolating the KSZ contribution to the dipole measurement proposed by KAB and the way to reduce them are:
\begin{itemize}
\item Primary CMB - has been removed by filtering which is demonstrated to work down to the fundamental limit imposed  by cosmic variance; the remainder is then integrated down over many clusters as $N_{\rm cl}^{-\half}$.
\item Foreground emissions - have been removed by the WMAP processing at the frequencies used by us; additionally our signal will be restricted to clusters, whereas foreground should produce dipoles across the entire sky.
\item Instrument noise - is reduced by stacking many pixels as $N_{\rm pix}^{-\half}$.
\item Cluster TSZ contributions - are removed by increasing the cluster aperture to produce zero monopole at cluster positions.
\end{itemize}

\subsection{Procedural steps to measurement}

The process that enabled us to isolate the kSZ term is described
in detail in KABKE1,2. Briefly:
\begin{enumerate}
\item An all-sky catalog of X-ray selected galaxy clusters was
constructed using available X-ray data extending to $z\sim 0.3$.

\item As indicated, we only used WMAP Q, V and W bands, where the
foreground contamination is smallest.  We applied the 3- and 5-yr
version of the Kp0 mask to remove those pixels where galactic or
point source contributions dominate. Next, to prevent any power
leakage from the dipole generated by our peculiar velocity, it was
removed from the pixels that survived the mask. Furthermore, KAEEK
explicitly removed dipole {\it and} quadrupole from the original
maps and demonstrated that the quadrupole did not contribute to
the results. This removes $v_{\rm local}$ down to $O[(v_{\rm
local}/c)^3]$ contribution to the octupole.

\item The cosmological CMB component is removed from the WMAP data
using the Wiener-type filter, eq. \ref{eq:filter_kabke}, constructed using the $\Lambda$CDM
model that best fit the data.  It is designed in order to
minimize the difference $\langle (\delta T - {\rm
noise}))^2\rangle$. Next, filtered maps were constructed using all
multipoles with $\ell\ge 4$ and keeping the same phases as in the
original maps. Modes with $\ell\le 3$  were not included to avoid
any possible contributions that could be introduced by the
alignment of those low order multipoles and also because those
modes would potentially be the most affected by any hypothetical
power leakage.

\item The Wiener filter is constructed (and is
different) for each DA channel because the beam and the noise are
different. This prevents inconsistencies and systematic errors
that could have been generated if a common filter was applied to
the eight channels of different noise and resolution.

\item In the filtered maps, the monopole and dipole are computed
exclusively at the cluster positions, using Healpix {\bf
remove\_dipole} routine ascribing to each cluster a given circular
aperture. Due to the variations of the Galactic absorbing column
density and ROSAT observing strategy, cluster selection function
and X-ray properties may vary across the sky introducing possible
systematics. In KABKE1,2 we used the measured X-ray extent of each
cluster, $\theta_X$ and computed the dipole for different
apertures, in multiples of $\theta_X$ and, to avoid being
dominated by a few very extended nearby clusters like Coma, we
introduced a cut so the final extent of any cluster was always
smaller than 30$^\prime$. There we computed core radii directly from the
data and from an $L_X-r_c$ relation. Analyses using both sets gave
consistent results, consistent with the X-ray systematic effects
not affecting our results significantly. More important,
variations in the final aperture were already tiny in the KABKE1,2
analysis and KAEEK used altogether a fixed aperture were the mean
monopole vanishes. The KAEEK results are consistent with the
previous (KABKE1,2) measurements. Fixing the same aperture for all
clusters simplifies the statistical analysis and this is the
approach taken in our post-KABKE measurements.

\item We compute the monopole and dipole for different angular
apertures. At small apertures ($\sim 10^\prime$), clusters show a
clear TSZ decrement, but the amplitude of the signal falls off
with increasing angular aperture. The final dipole is computed at
the aperture where the mean monopole of the clusters vanishes.
This ensures that the TSZ contribution to the measured dipole is
negligible and does not confuse the KSZ component.

\item Our final result is a dipole measured in units of
thermodynamic CMB temperature. To translate the three measured
dipoles into three velocity components, we need to determine the
average cluster optical depth to the CMB photons,
$\langle\tau\rangle$, on the filtered maps. Since filtering
reduces the intrinsic CMB contribution, it also modifies its
optical depth, $\tau$. In KABKE1,2 we introduced a calibration
factor $C_{1,100}$ that gave the kSZ dipole in $\mu$K of a bulk
motion of amplitude $V_{\rm bulk}=100km/s$. The calibration factor
depends both on the filter and on the cluster profile. In KABKE1,2
and KAEEK it was estimated using a $\beta$ model and the angular
X-ray extent of the cluster. This procedure is still deficient in the sense that our current cluster modeling is i) inadequate for high-precision calibration (it is accurate to within 20-30\%) and ii) more importantly, a change of sign in the KSZ term can occur because we measure the dipole from the filtered maps, and the convolution of the intrinsic KSZ signal with a filter with wide side-lobes (as in KABKE) {\it can change the sign} of the KSZ signal for NFW clusters. While there is indeed evidence for this sign change in the KSZ signal which affects the direction of the KAEEK-measured dipole, we emphasize again that a definitive answer will have to await a more complete, expanded and recalibrated SCOUT catalog.
\end{enumerate}

We now move on to technical discussion of the procedure along with the estimates of the residual systematics and then present the results and compare them to other data.

\subsection{CMB datasets, (required) filtering and data processing}
\label{sec:filtering}

\clearpage
\begin{figure}[h]
\includegraphics[width=2in, angle=+90]{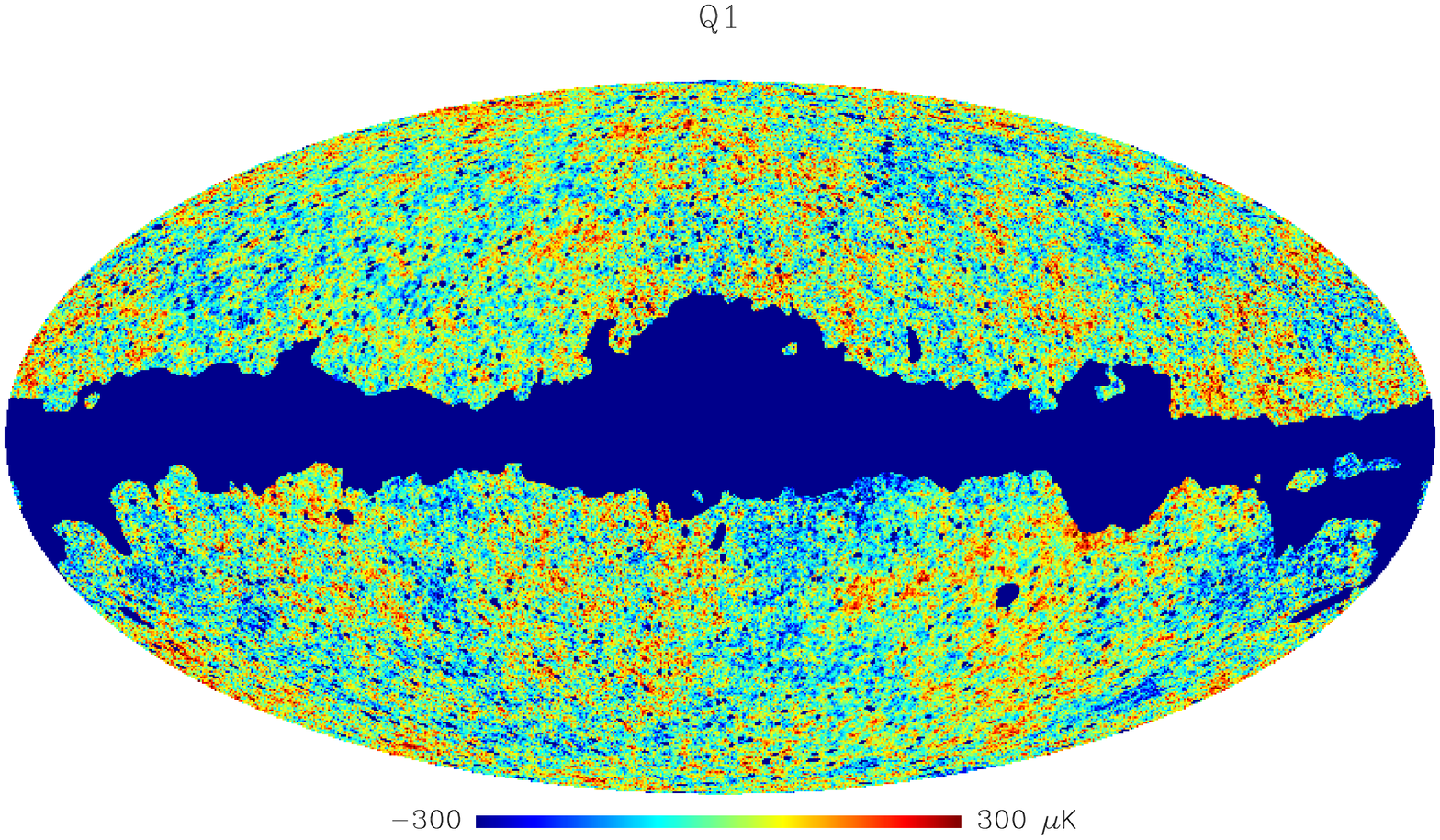}
\includegraphics[width=2in, angle=+90]{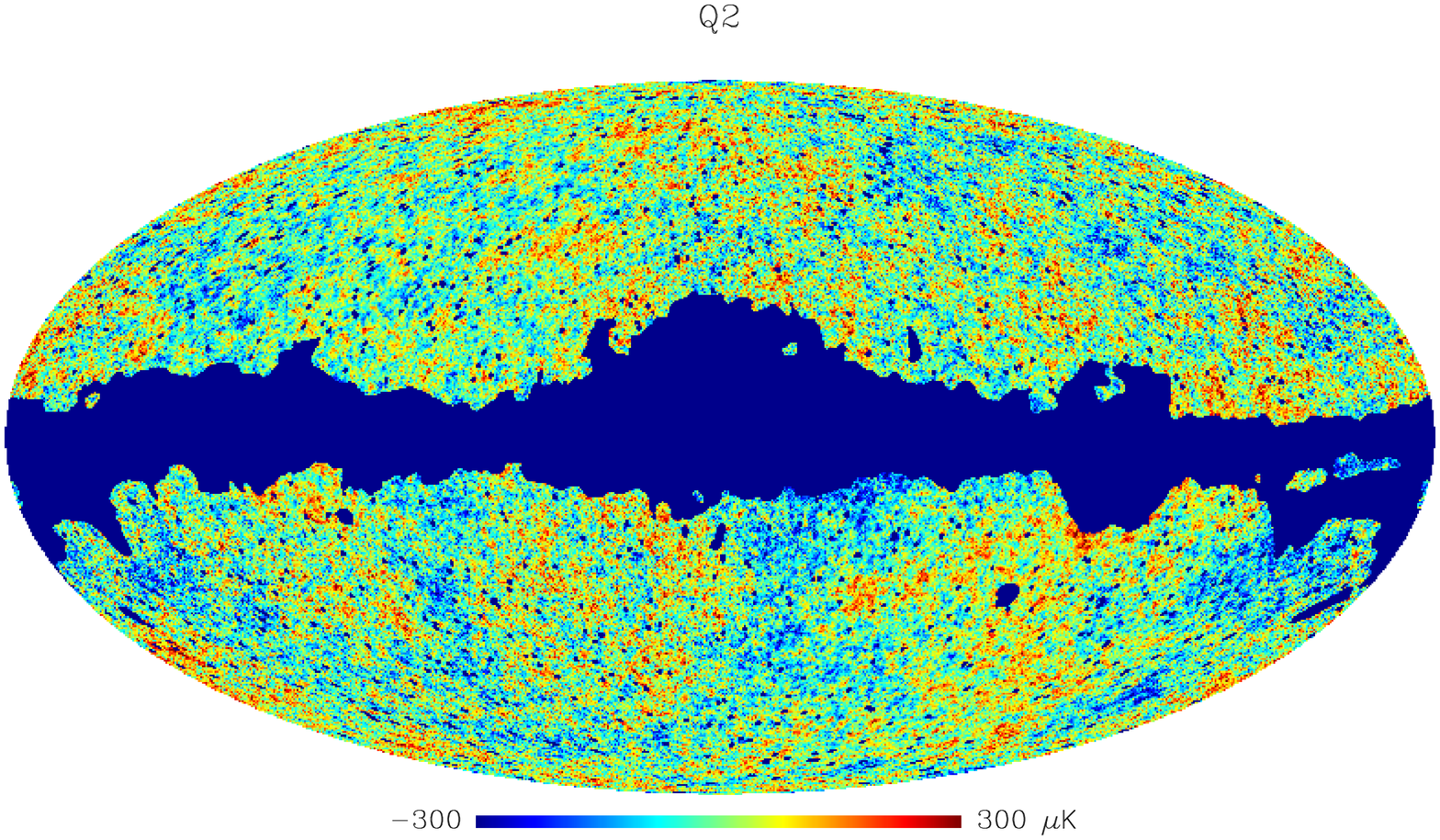}\\
\includegraphics[width=2in, angle=+90]{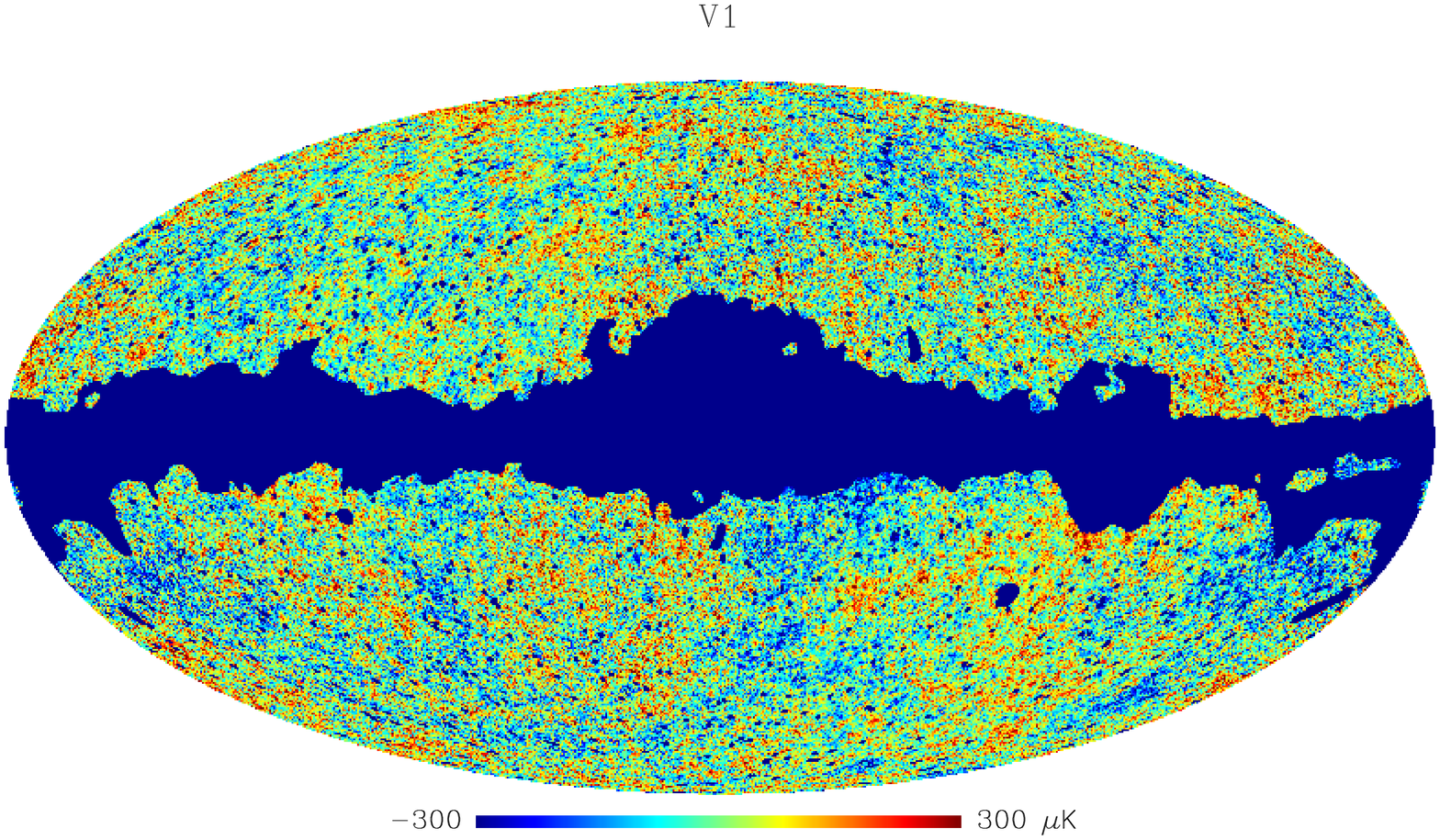}
\includegraphics[width=2in, angle=+90]{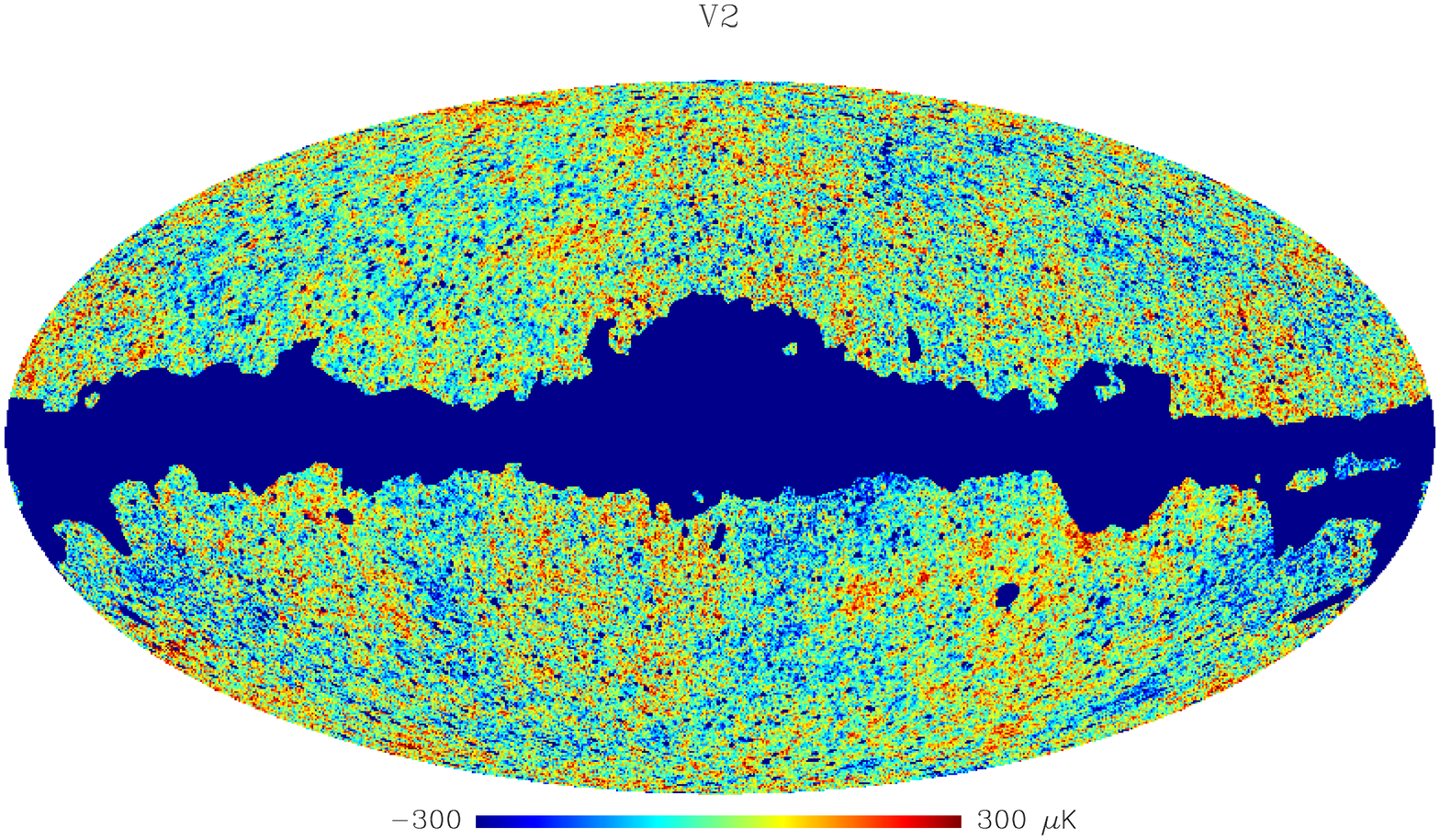}\\
\includegraphics[width=2in, angle=+90]{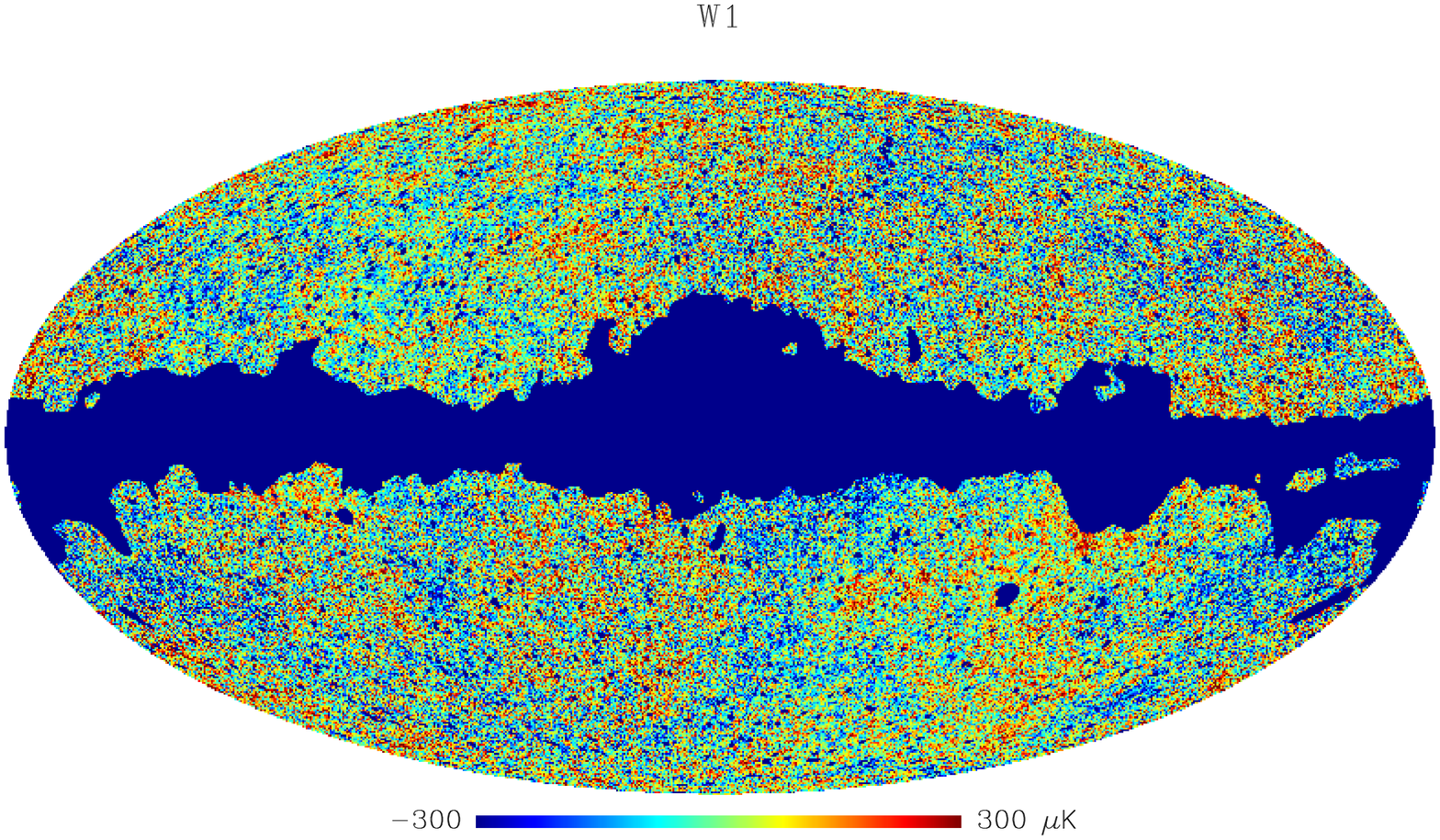}
\includegraphics[width=2in, angle=+90]{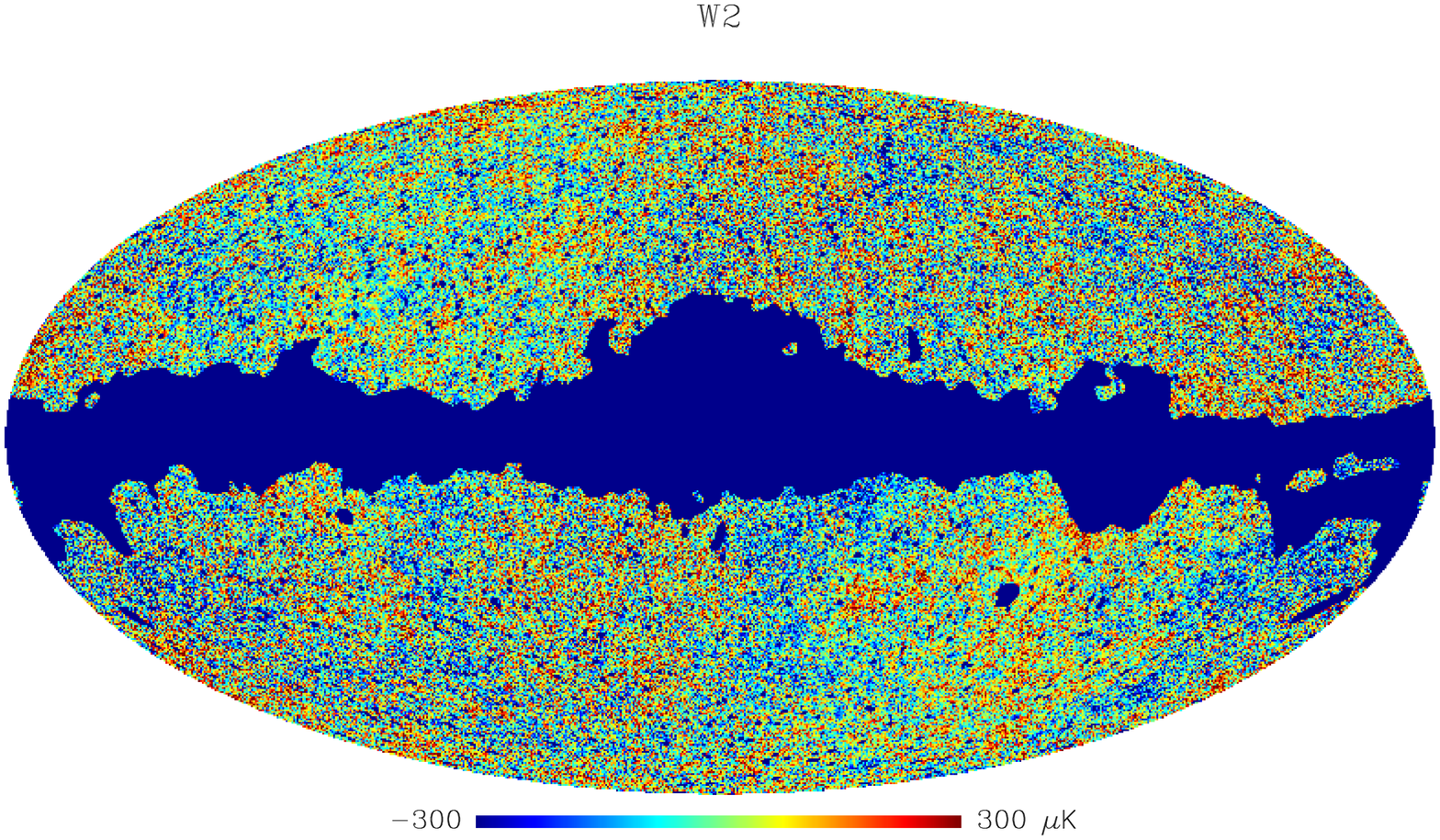}\\
\includegraphics[width=2in, angle=+90]{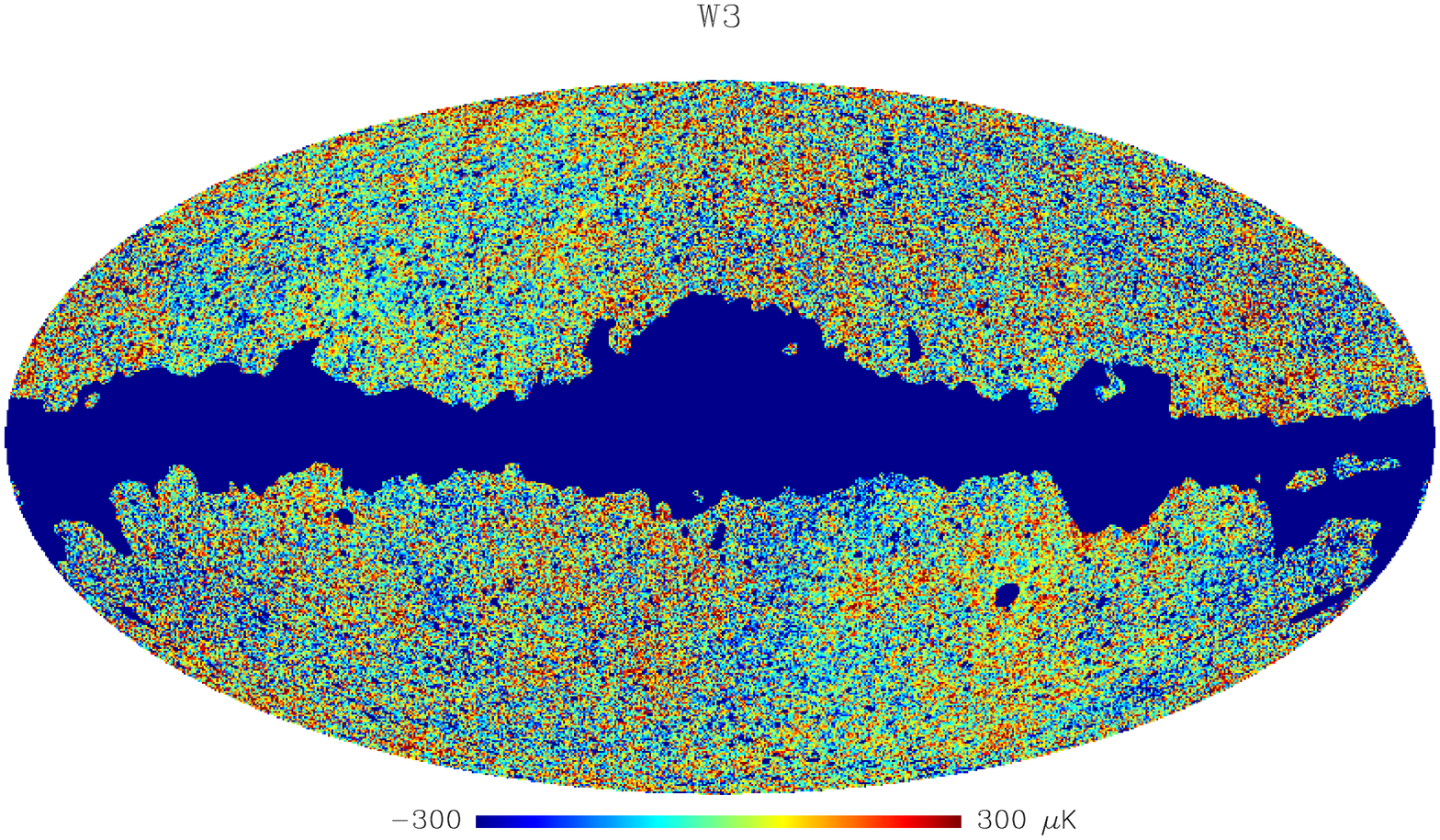}
\includegraphics[width=2in, angle=+90]{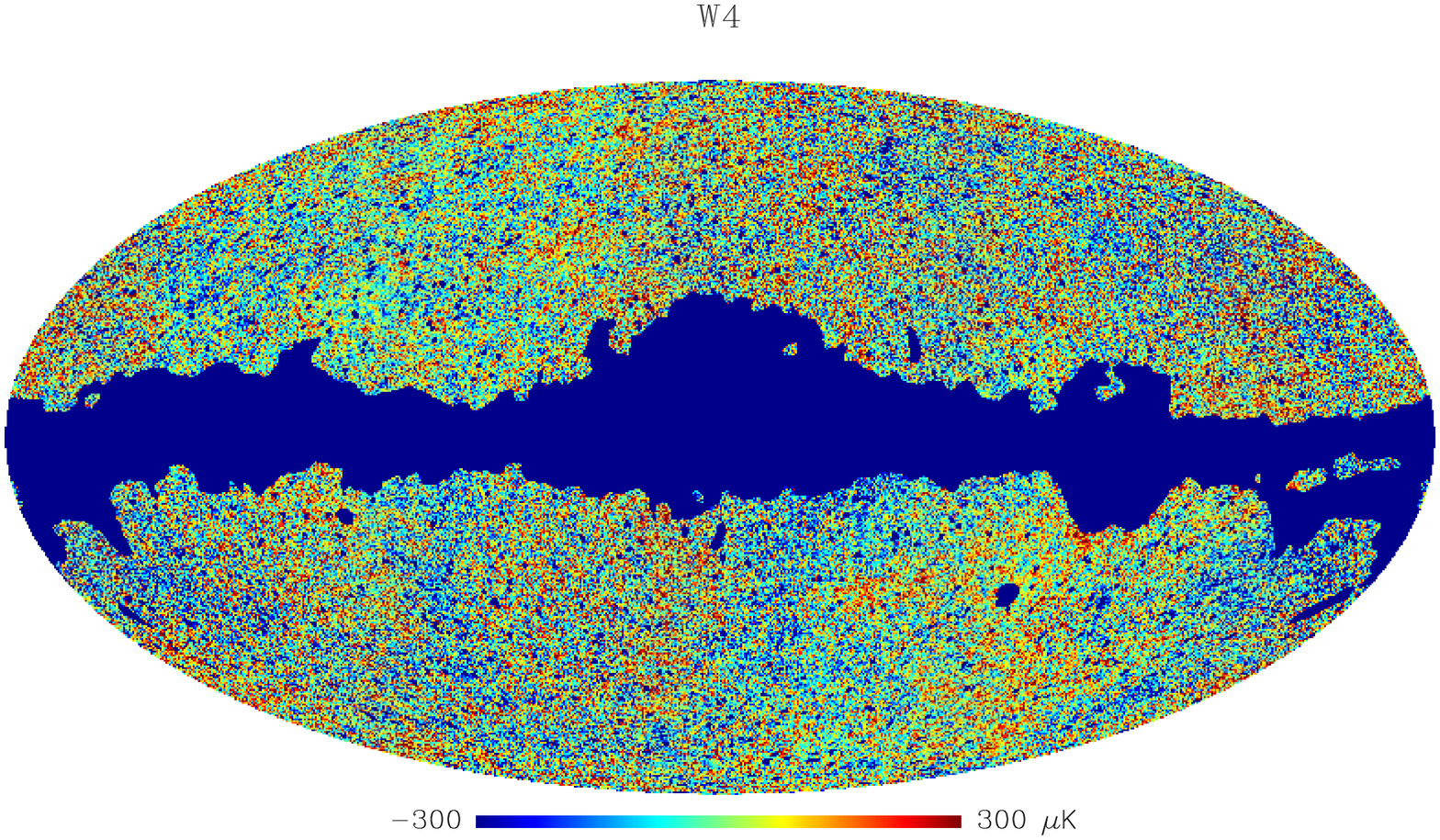}\\
 \caption[]{Original masked (KP0) WMAP7 CMB maps. }
\label{fig:maps_orig}
\end{figure}
\clearpage
\begin{figure}[h]
\includegraphics[width=2in, angle=+90]{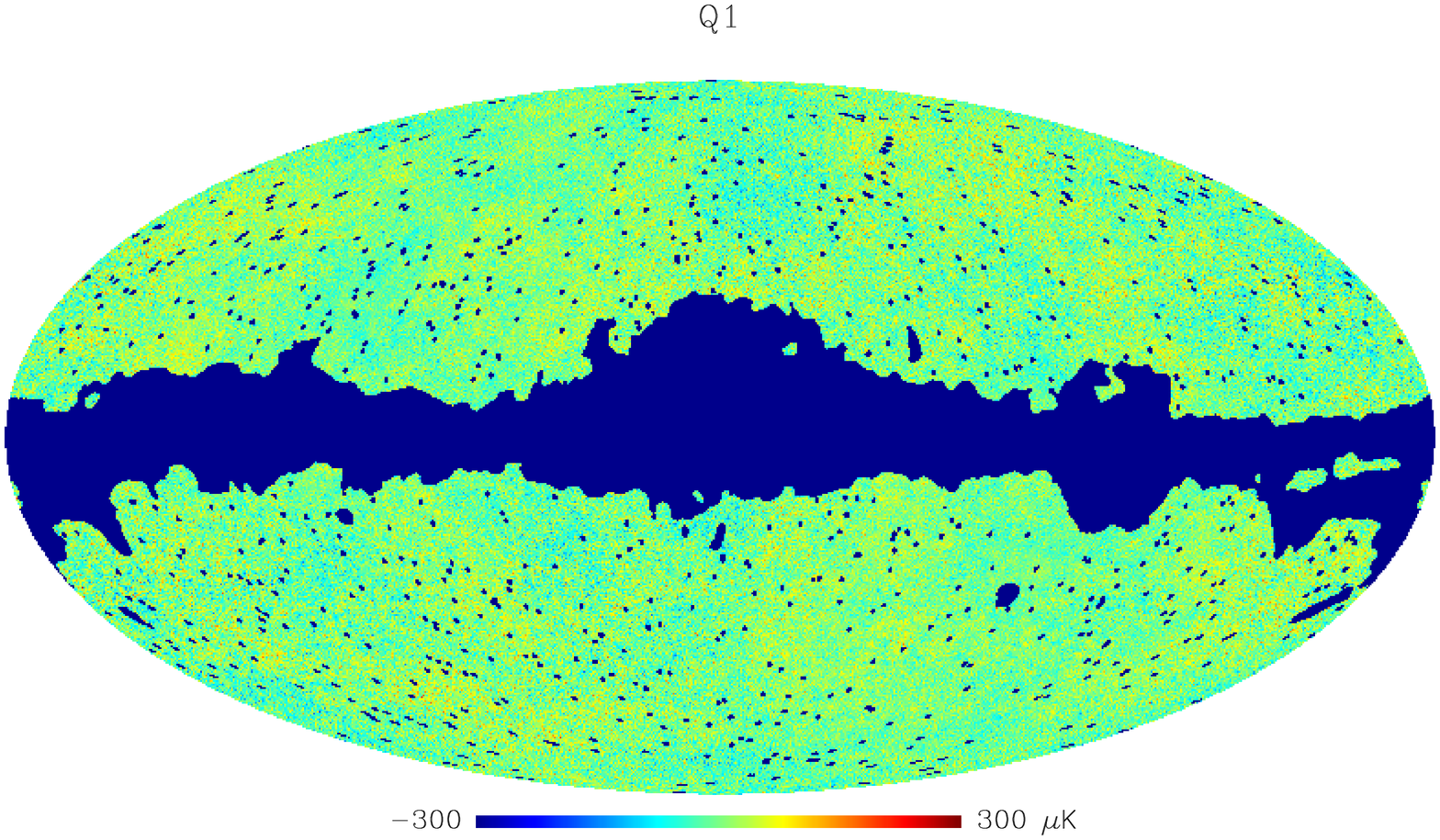}
\includegraphics[width=2in, angle=+90]{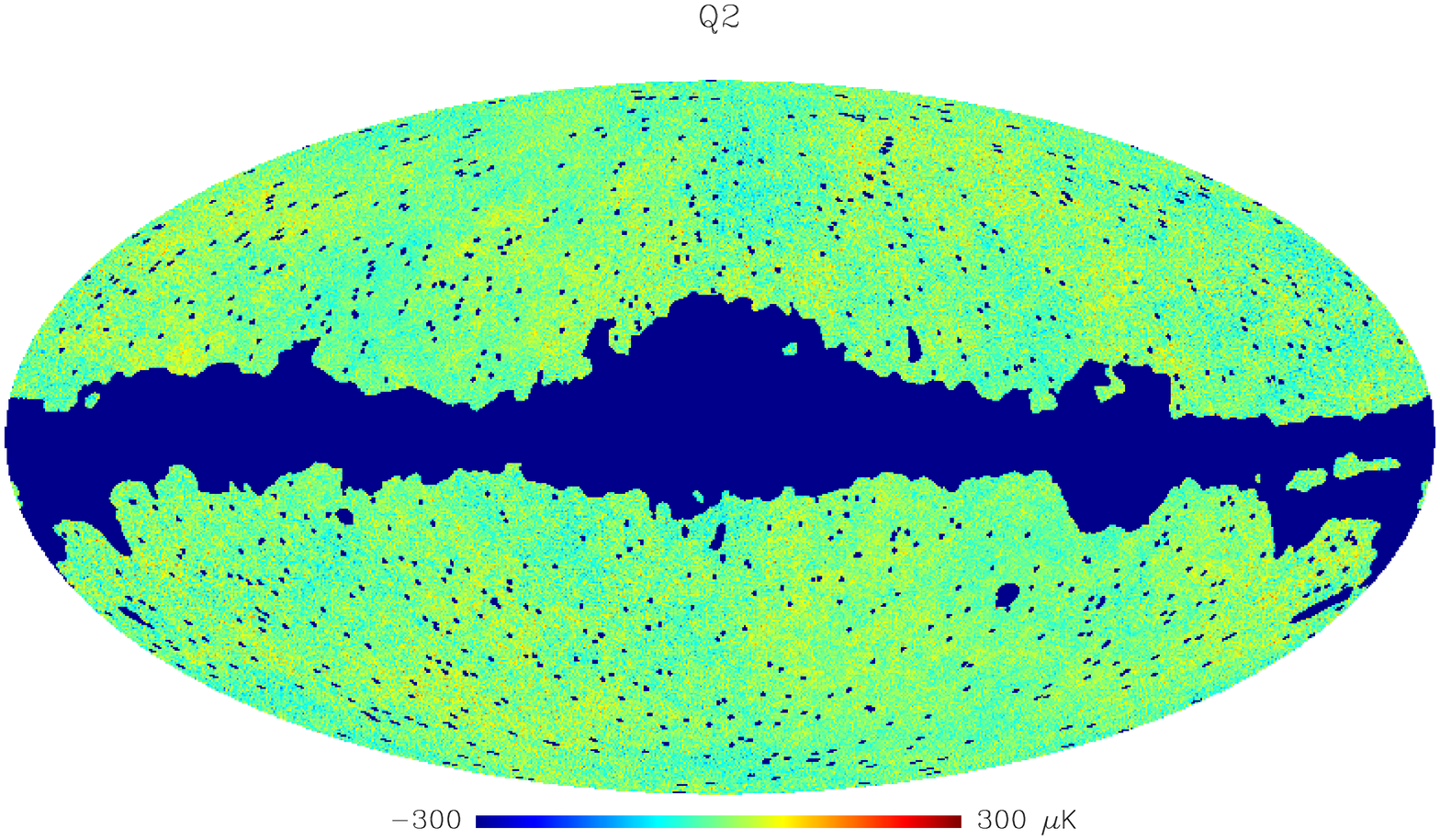}\\
\includegraphics[width=2in, angle=+90]{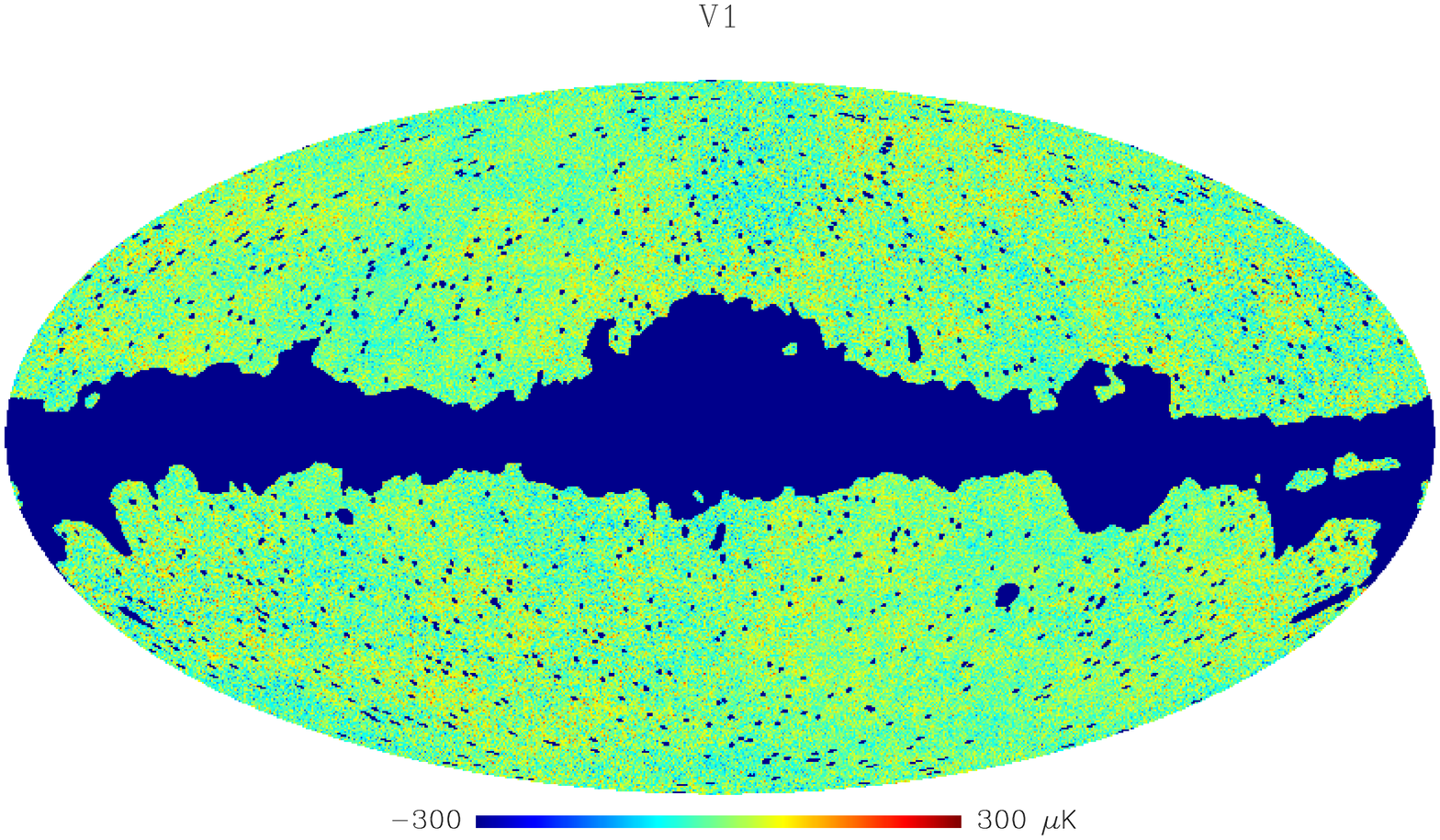}
\includegraphics[width=2in, angle=+90]{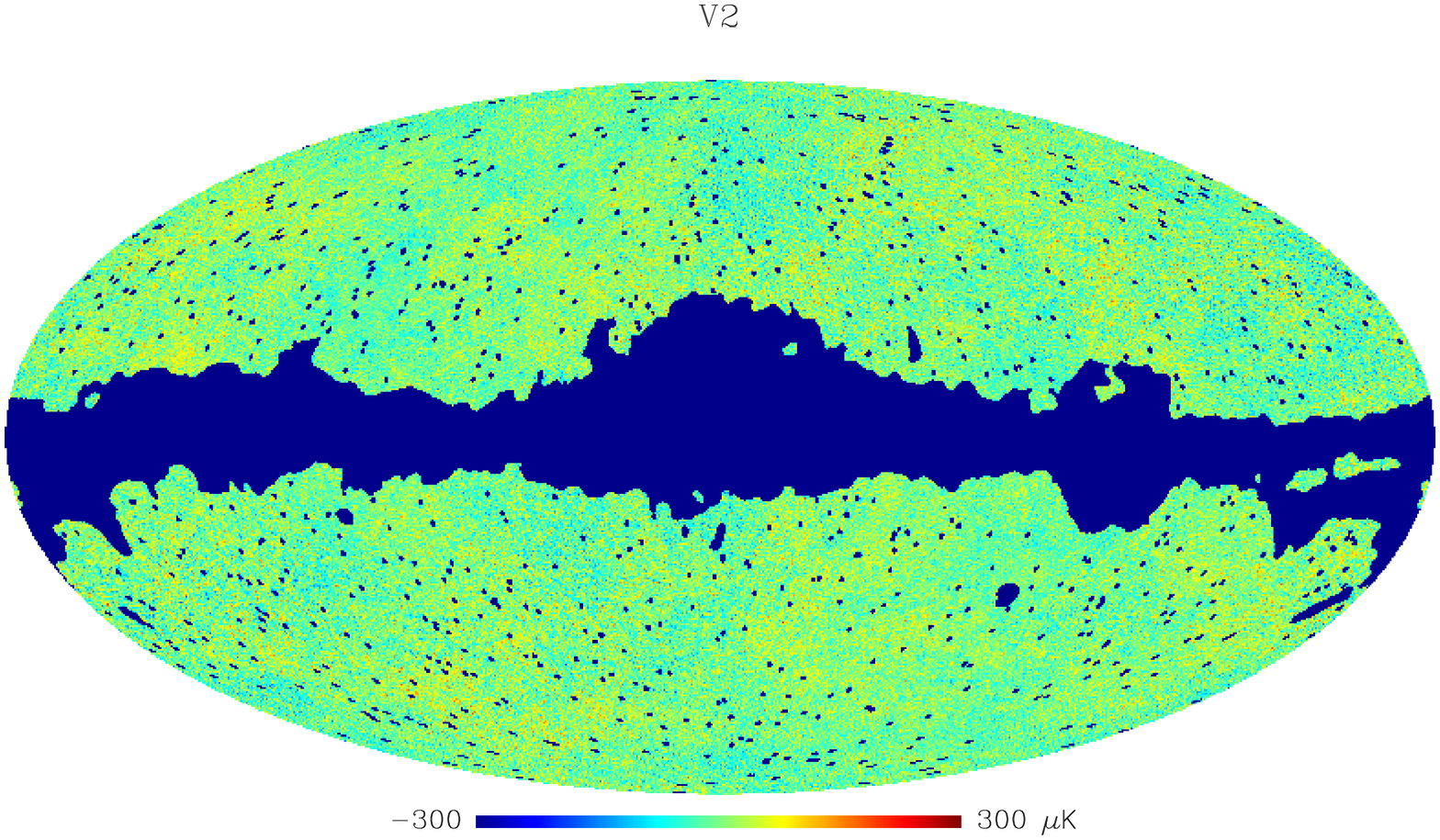}\\
\includegraphics[width=2in, angle=+90]{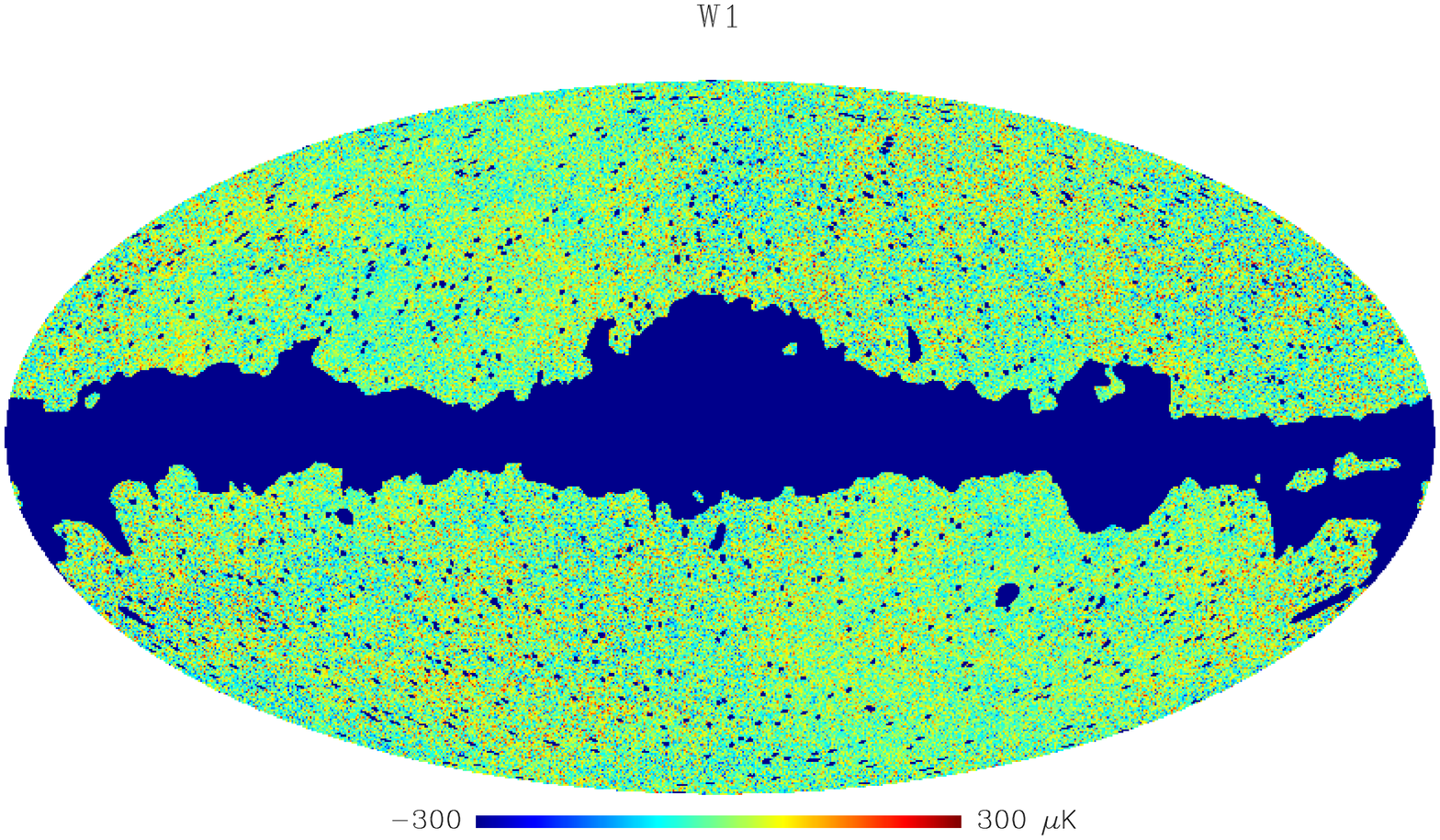}
\includegraphics[width=2in, angle=+90]{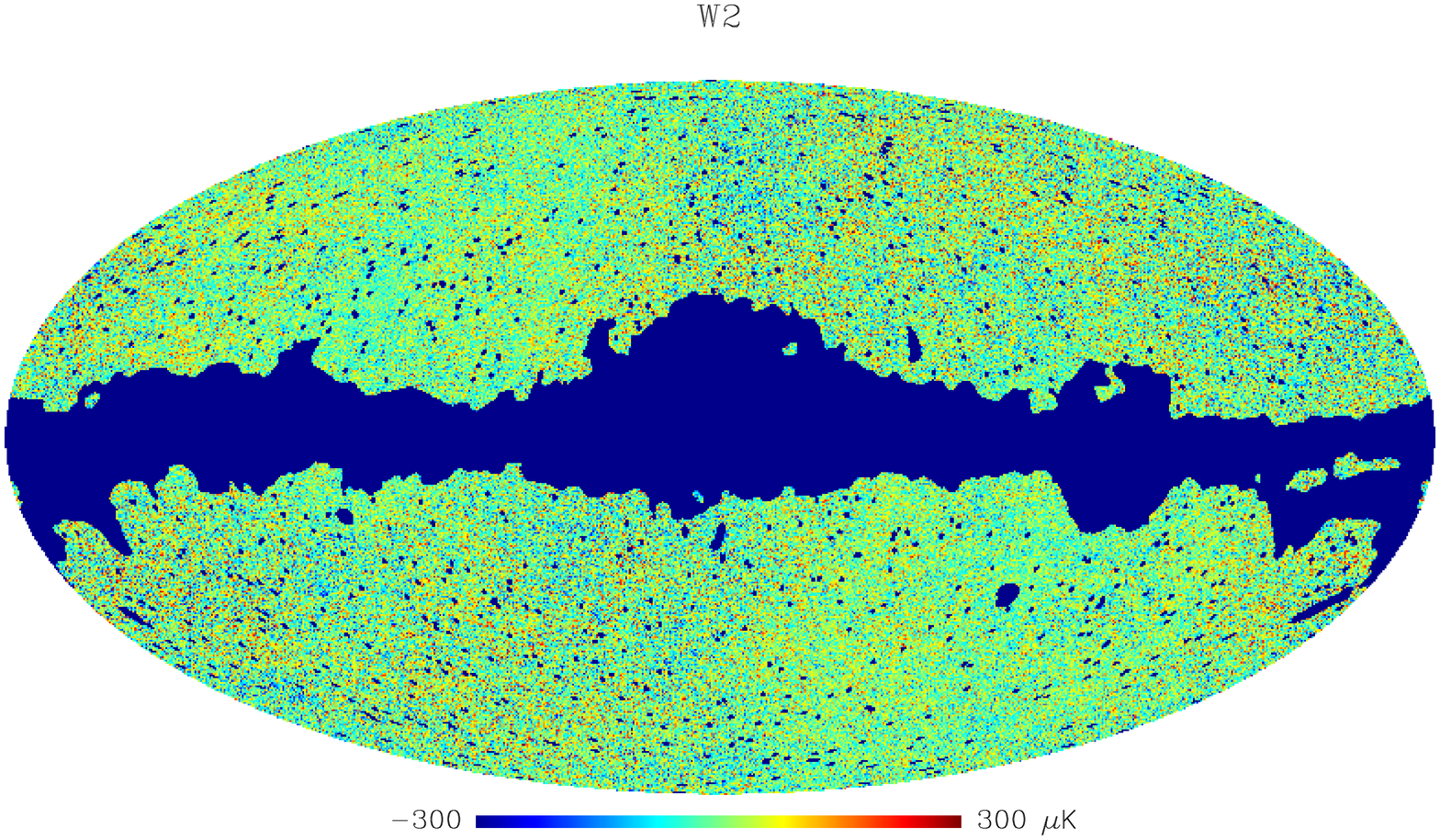}\\
\includegraphics[width=2in, angle=+90]{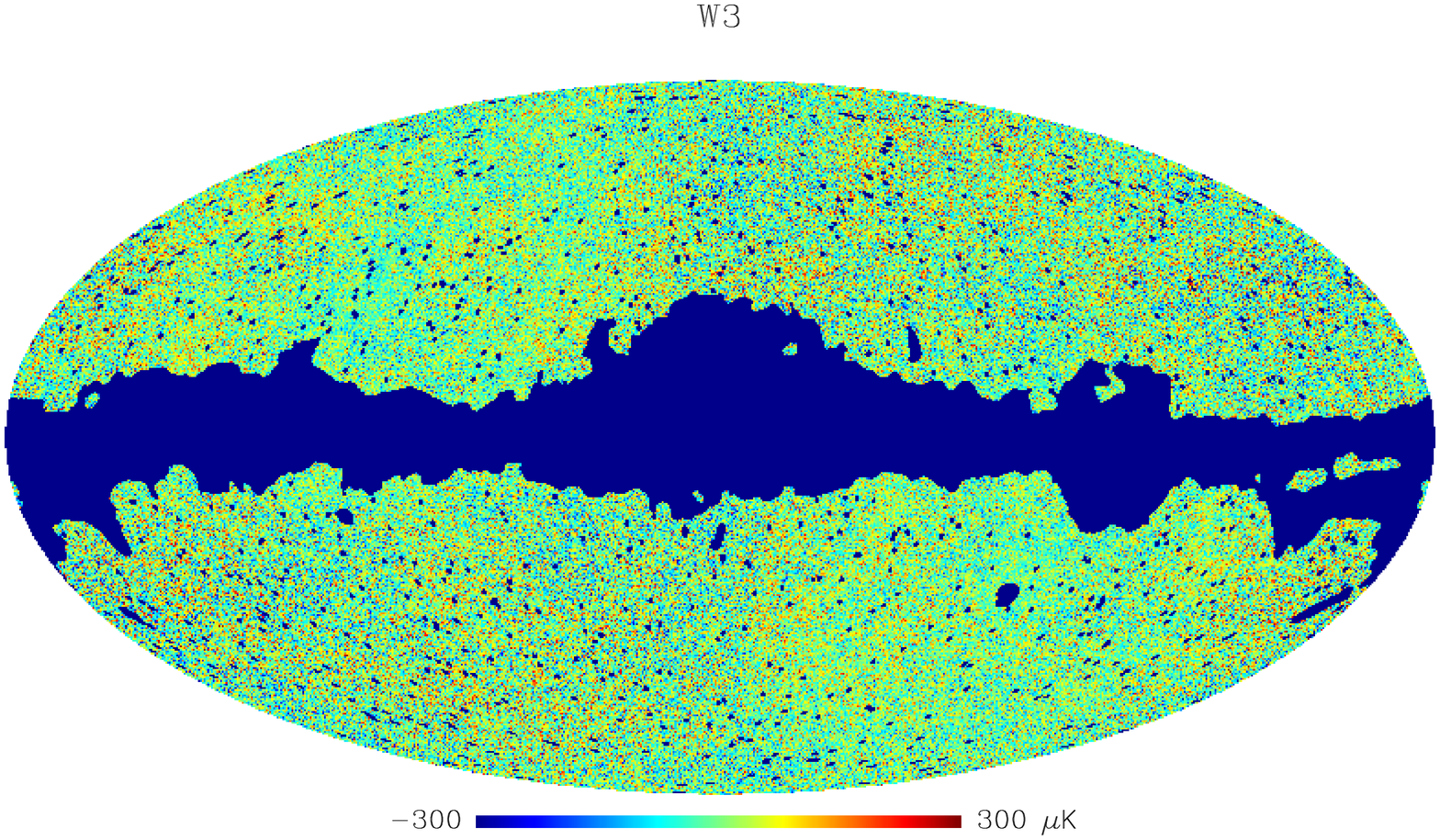}
\includegraphics[width=2in, angle=+90]{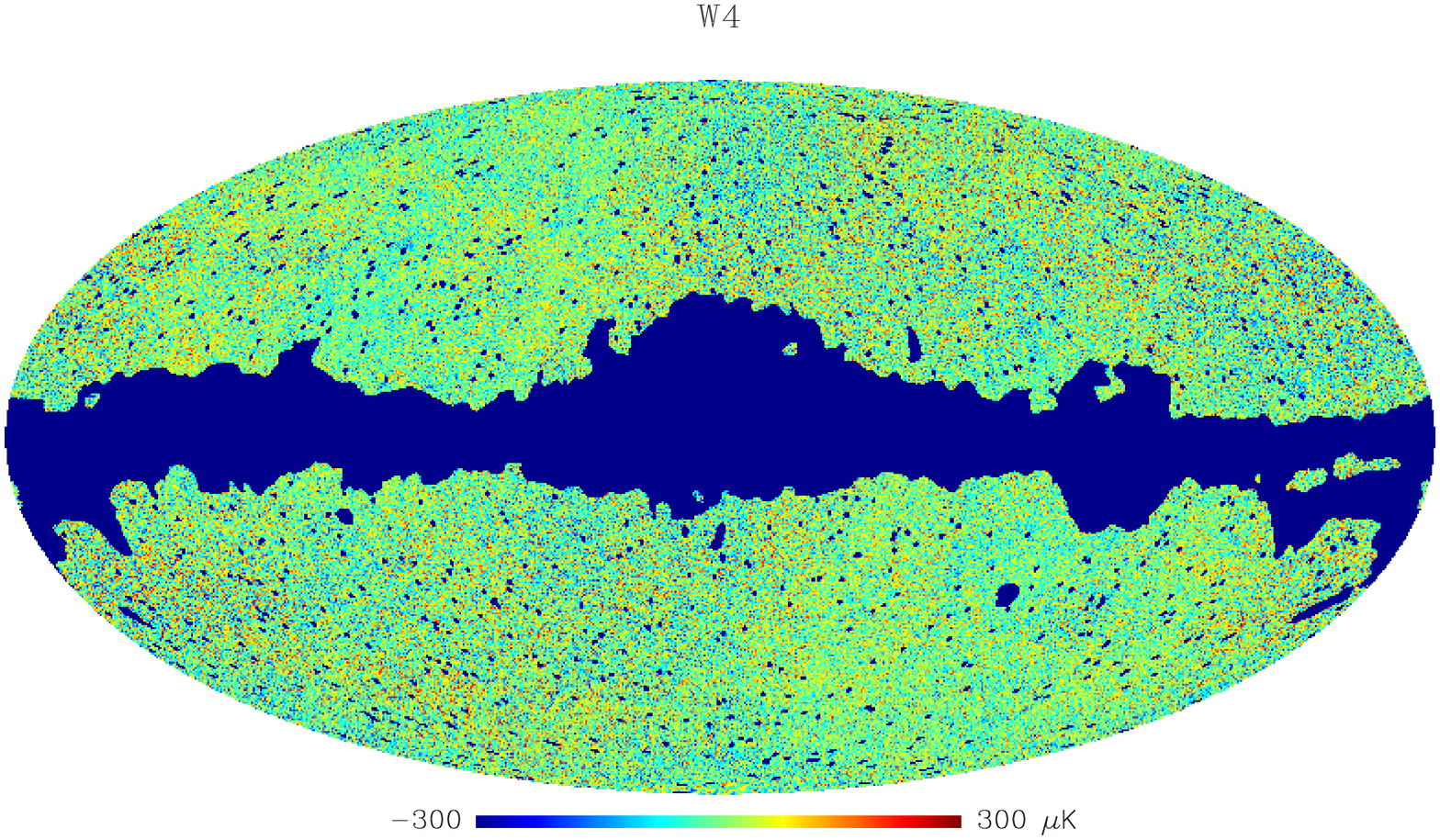}\\
 \caption[]{Filtered maps using the KABKE1,2 filter. }
\label{fig:maps}
\end{figure}
\clearpage
\begin{figure}[h]
\includegraphics[width=2in, angle=+90]{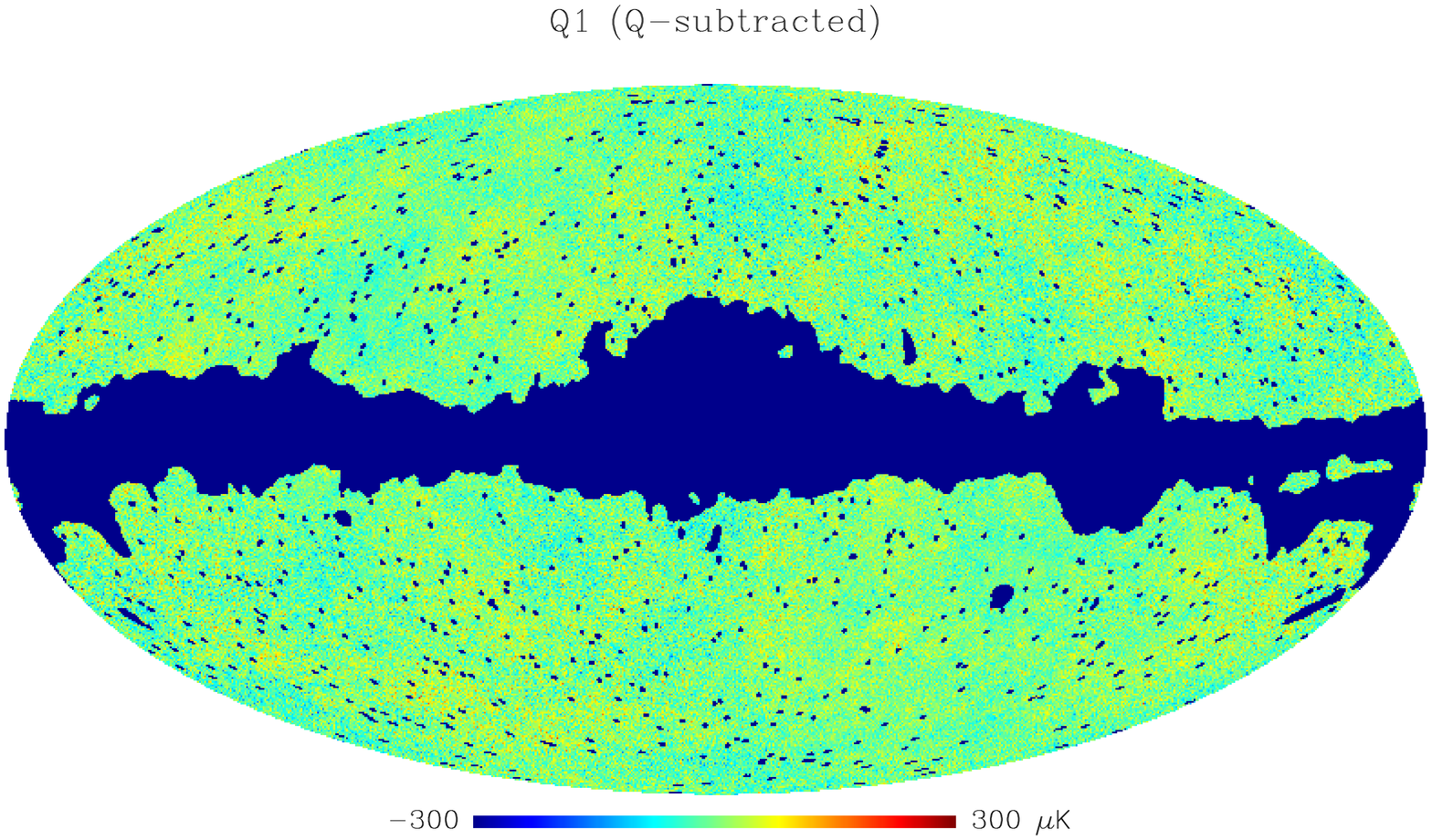}
\includegraphics[width=2in, angle=+90]{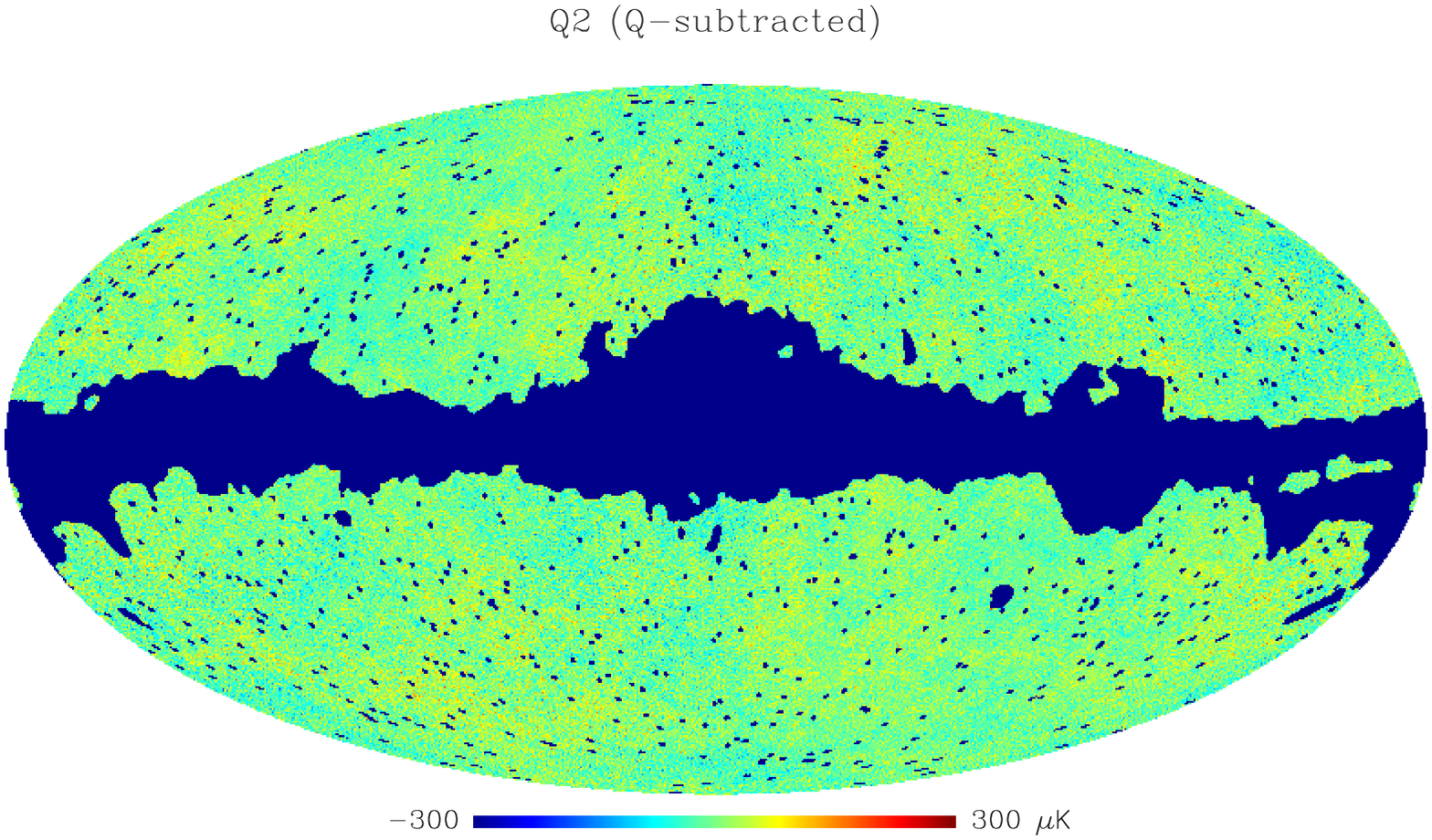}\\
\includegraphics[width=2in, angle=+90]{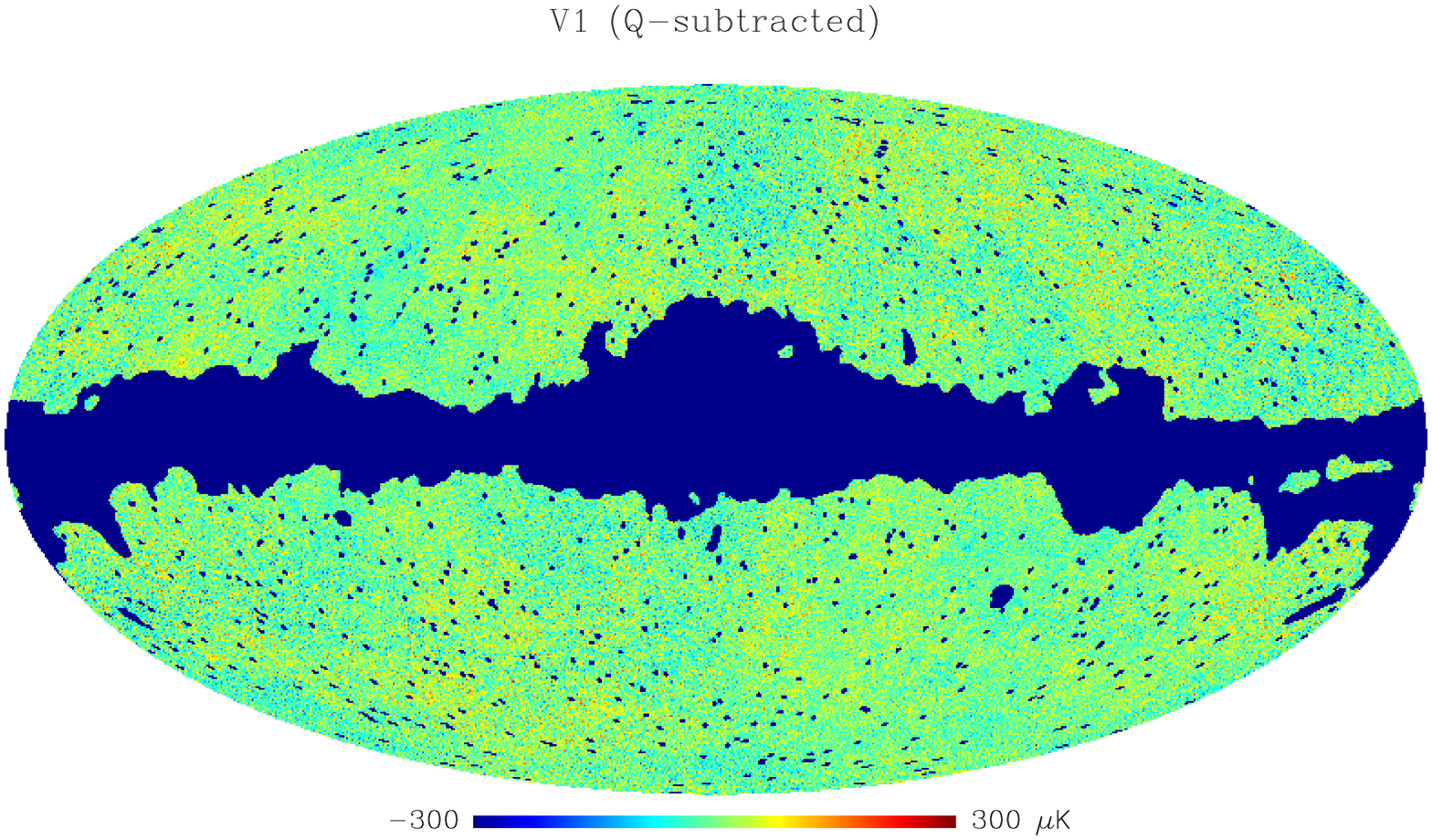}
\includegraphics[width=2in, angle=+90]{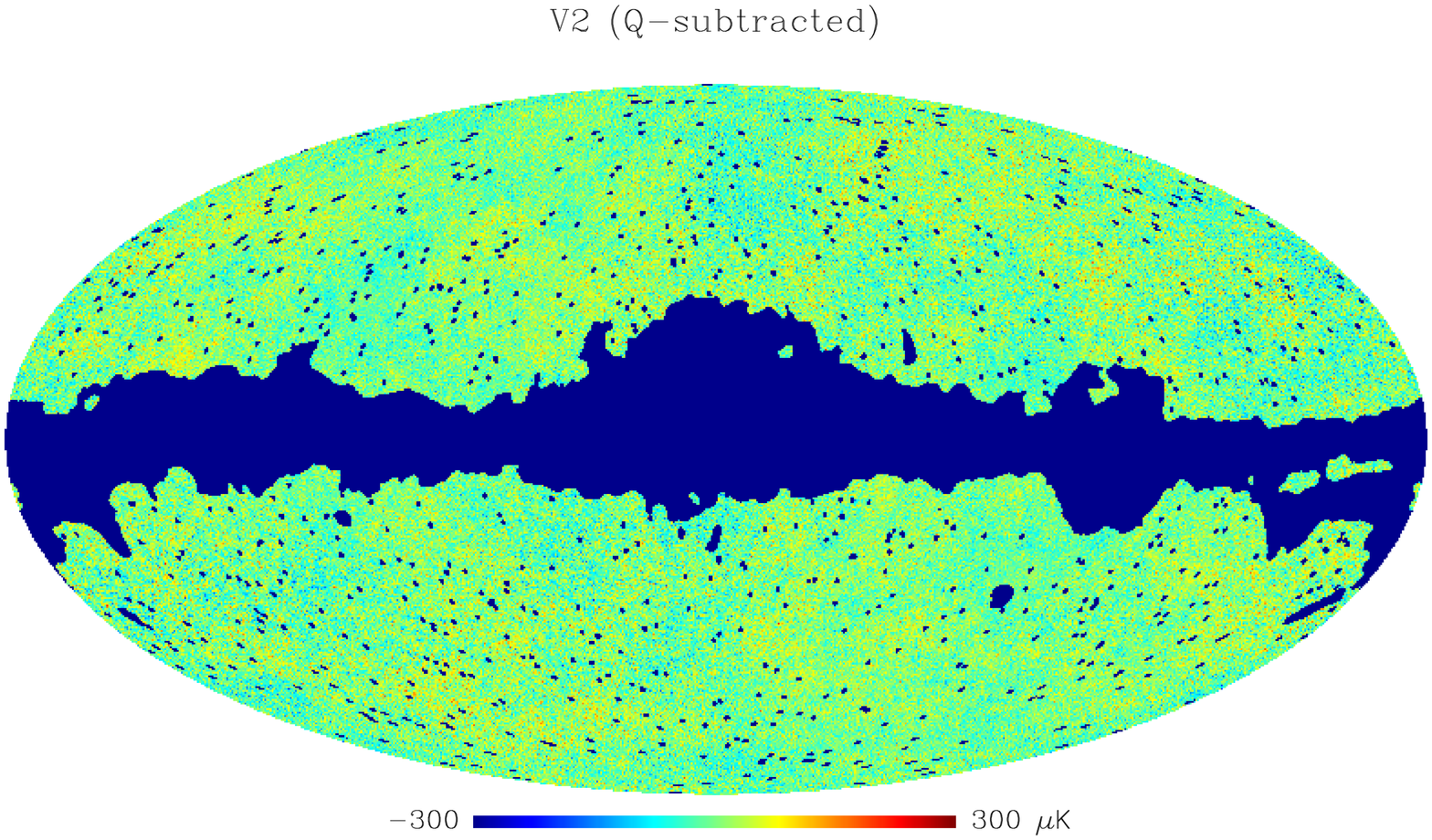}\\
\includegraphics[width=2in, angle=+90]{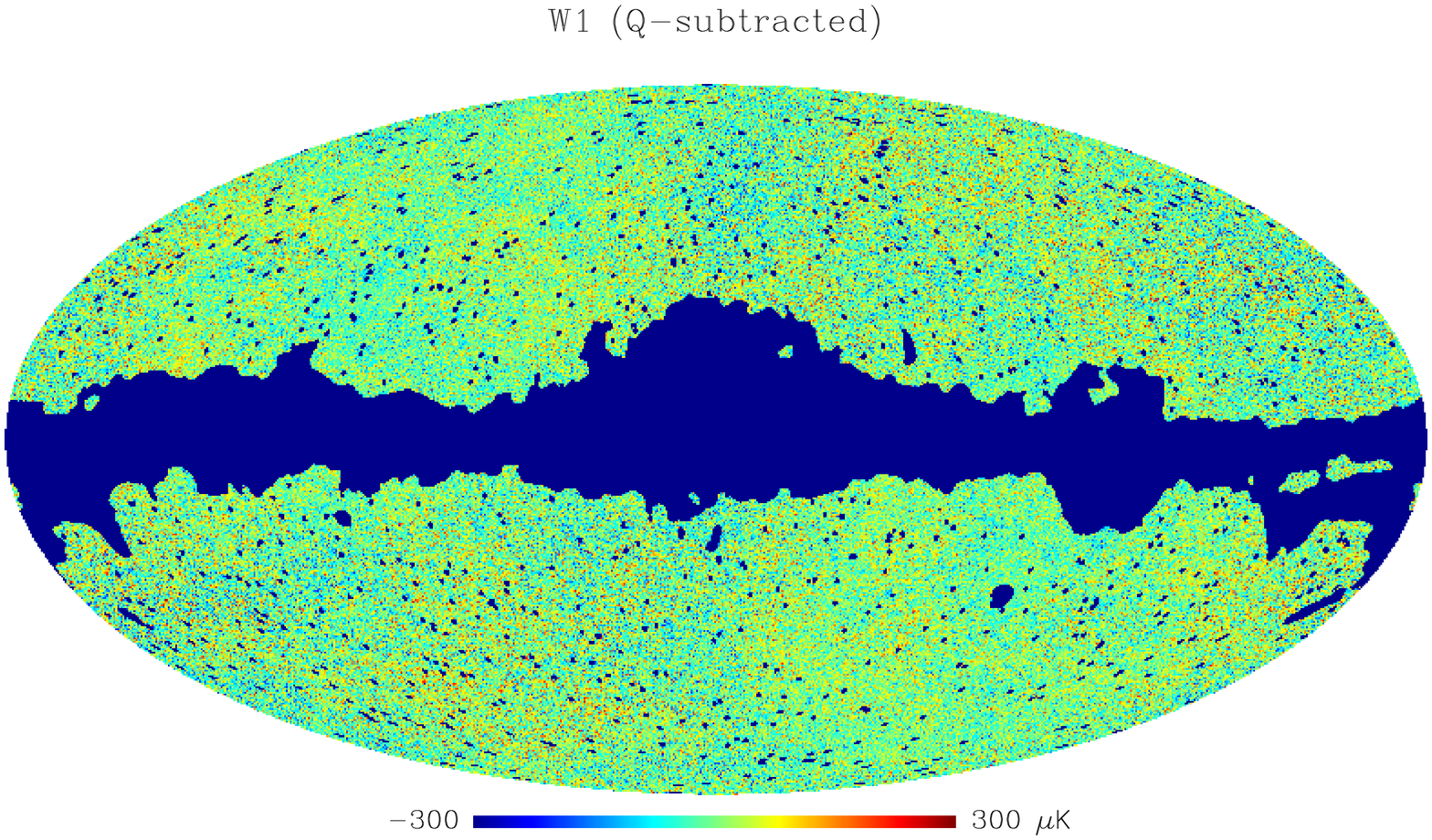}
\includegraphics[width=2in, angle=+90]{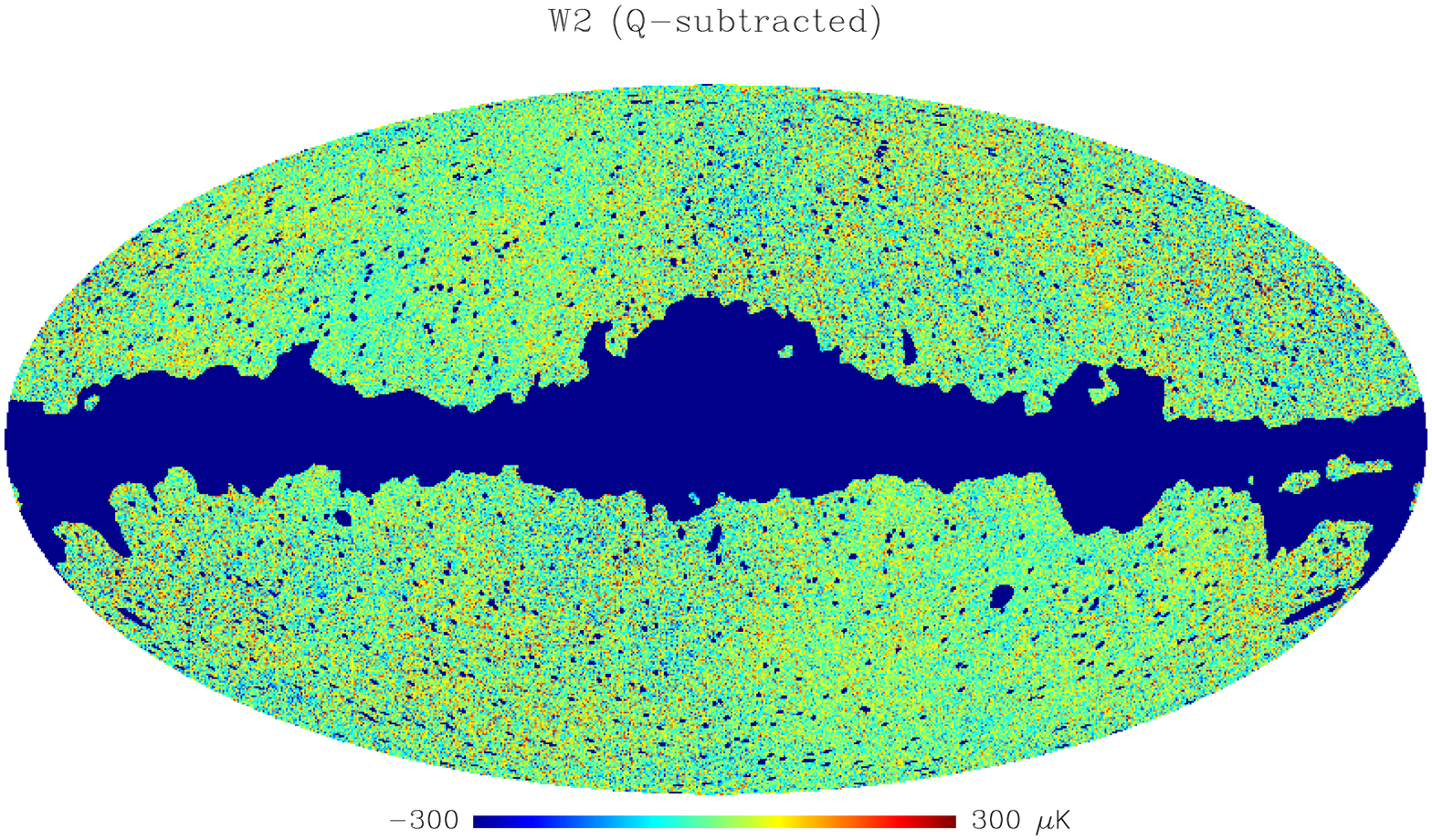}\\
\includegraphics[width=2in, angle=+90]{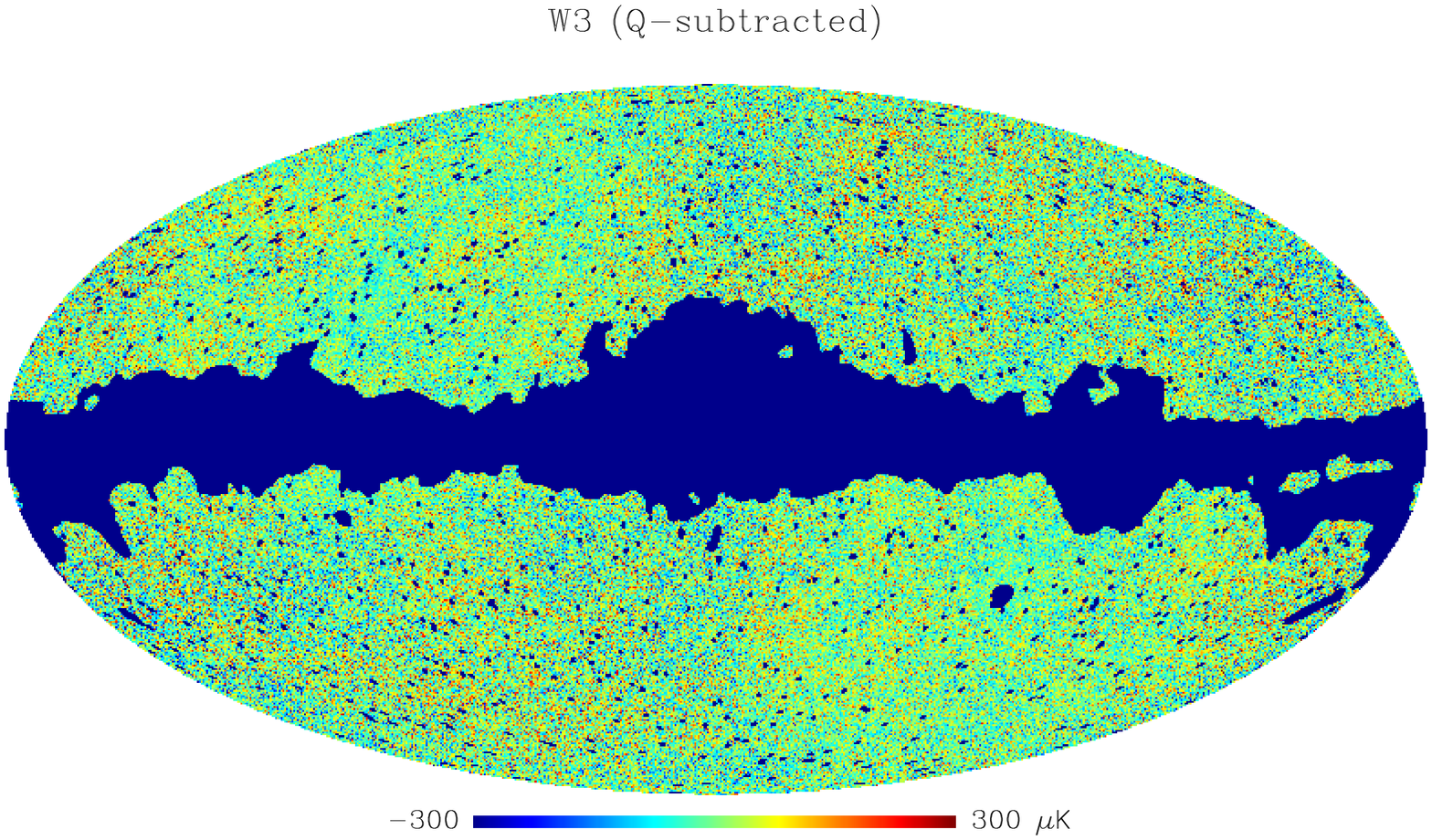}
\includegraphics[width=2in, angle=+90]{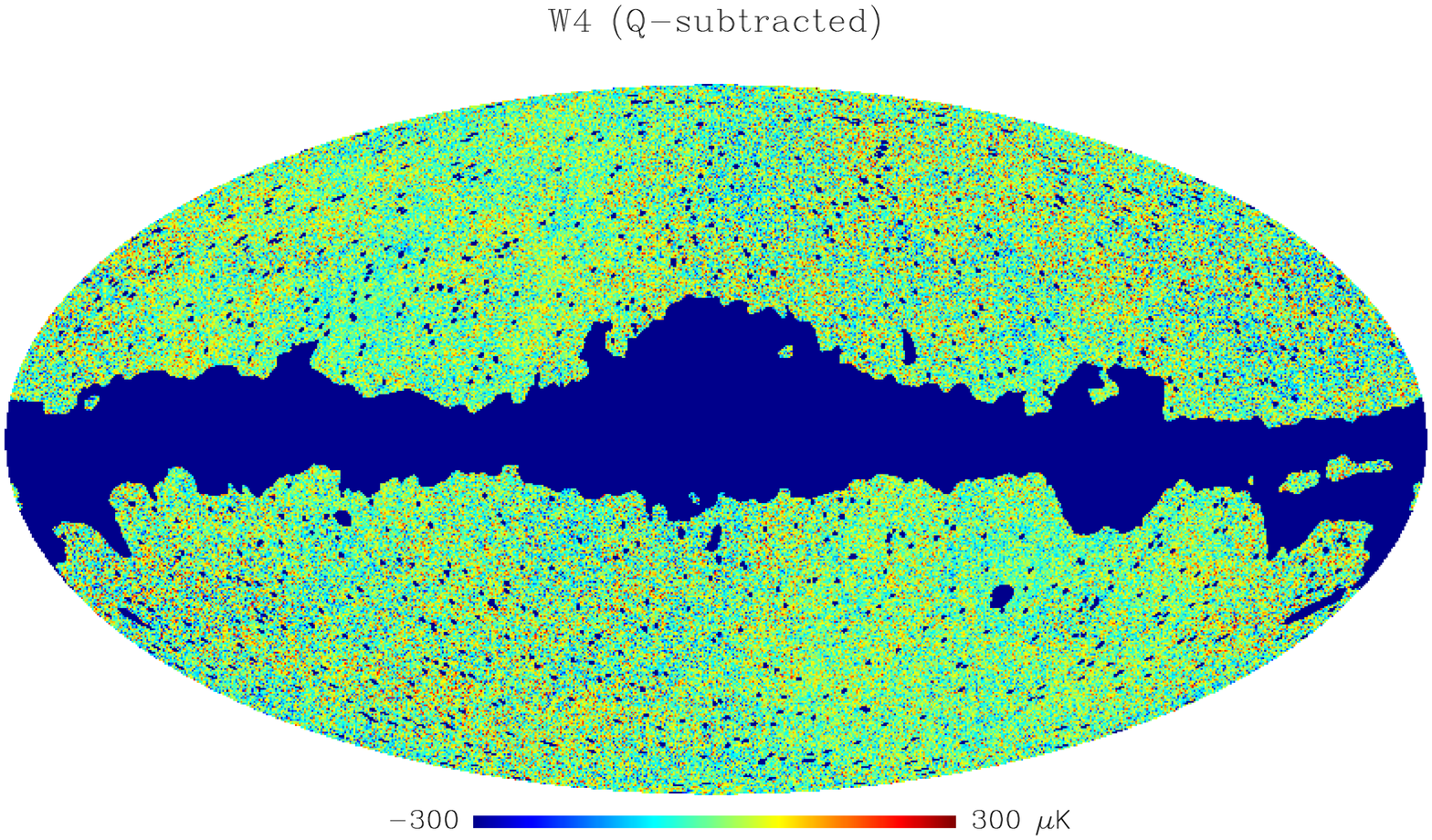}\\
 \caption[]{Same as in Fig. \ref{fig:maps} but the original maps had their quadrupole subtracted outside the mask prior to filtering. }
\label{fig:maps_nomdq}
\end{figure}
\clearpage
\begin{figure}[h!]
\includegraphics[width=6in]{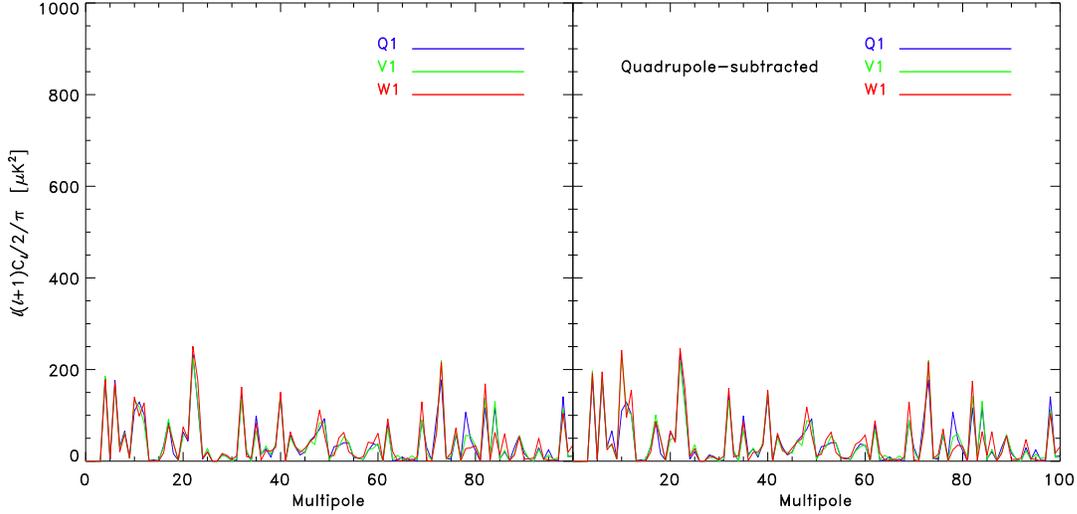}
 \caption[]{Spatial spectrum of the KABKE-filtered 7-year WMAP maps.  {\bf Left}: monopole and dipole subtracted maps. {\bf Right} Monopole, dipole {\it and} quadrupole-subtracted maps. Adapted from KAE.}
\label{fig:kae1}
\end{figure}

Because the primary CMB is spatially highly correlated, a filter needs to be designed
that removes this component without significantly
attenuating $A_{\rm KSZ}$; clearly {\it not every filter will achieve this}.

KABKE1,2 defined a filter, described in detail in KABKE2 and AKEKE,
which belongs to a Wiener variety and removes the primary CMB fluctuations from the concordance $\Lambda$CDM model by minimizing the mean squared deviation of the CMB measurements
from noise, $\langle (\delta_{\rm CMB}-n)^2\rangle$. In multipole $\ell$-space it is given by:
\begin{equation}
F_\ell = \frac{(C_\ell - C_\ell^{\rm \Lambda CDM})}{C_\ell}
\label{eq:filter_kabke}
\end{equation}
where $C_\ell, C_\ell^{\rm \Lambda CDM}$ are the power spectra of the CMB map and the theoretical model convolved with the beam, respectively.

AKEKE develop a formalism to quantify the errors in the resultant dipole determination that can be applied to {\it any} filtering scheme and show that the KABKE filter removes the primary CMB fluctuations down to the fundamental limit of the cosmic variance. This is discussed with extra details below.

\subsection{Statistical uncertainties}
\label{sec:errors}

We now turn to estimating - and understanding - the statistical uncertainties of the dipole measurement in the KA-B method by analytical and numerical means and verify the developed formalism empirically from the observed CMB data. This subsection is thus structured as follows: first, we present an analytical discussion of estimating the errors in the measurement, which can be applied to {\it any} filtering scheme. Following this, we show - analytically and empirically - that the filter employed by us in KABKE1,2 removes the primary CMB down to the fundamental limit imposed by cosmic variance. Then we present results of results of Monte-Carlo type simulations employing several (four, in fact) methods to estimate the errors. We show that they all agree to within a few percent, demonstrate the diagonality of the dipole covariance matrix and finish this subsection by presenting an analytical approximation for the errors expected in the post-5yr WMAP CMB integrations.

\subsubsection{Analytical evaluation and empirical confirmation}

 Our dipole is measured at cluster pixels, so the error on this measurement is determined by the distribution of the random dipoles in the CMB maps away from the actual clusters. Outside of the cluster pixels the diffused microwave emissions in WMAP maps are composed of two dominant terms: primary CMB and instrumental noise. In what follows we will ignore contribution from foregrounds of various kinds since it will be shown that there remains little trace of them after accounting for primary CMB and noise. The instrument noise in unfiltered CMB WMAP7 maps is given by (Jarosik 2011):
\begin{equation}
\sigma_{\rm noise} = \; \left[\frac{1}{4\pi} \sum (2\ell+1)  {\cal N}_\ell \right]^\half \;
= \left\{
\begin{array}{c}
        65 \; \mu {\rm K} \;\;\; Q-{\rm band}\\
        80 \; \mu {\rm K} \;\;\; V-{\rm band}\\
       130 \; \mu {\rm K} \;\;\; W-{\rm band}
\end{array}
\right.  \label{eq:wmapnoise}
\end{equation}
There is no convolution with the beam in the above sum. The noise is independent from pixel-to-pixel and between the different DA's. Thus the dipole contribution supplied by this term decreases as $N_{\rm pix}^{-\half}N_{\rm DA}^{-\half}$ when evaluated over $N_{\rm pix}$ independent pixels and $N_{\rm DA}$ DA's.

The contribution from primary CMB behaves differently: its multipoles are convolved with the beam and the temperature field correlates between adjacent pixels and different DA's. This term presents a major obstacle in measuring the KSZ dipole in the KA-B method and has to be filtered out. The filtering has to preserve (at least much of) the KSZ terms intact or at least attenuate the KSZ term in a way that does not decrease the signal-to-noise of the bulk flow measurement.

AKEKE have developed a precise analytical and numerical formalism to understand the errors and applied it to the KABKE filtering scheme which we review below. Their formalism can be applied to {\it any} filtering. We revisit the AKEKE discussion as applied to the KABKE filter, eq. \ref{eq:filter_kabke}.

The standard deviation of {\it any} filtered map is given by eq. 3 of AKEKE:
\begin{equation}
\sigma_{\rm map}^2 = \frac{1}{4\pi} \sum (2\ell+1) F_\ell^2 C_\ell
\label{eq:sigma_fil_map}
\end{equation}
Here $C_\ell = C_\ell^{\Lambda CDM} + {\cal N}_\ell$ is the power spectrum of the original map containing the primary $\Lambda$CDM signal (convolved with the beam) and instrument noise, ${\cal N}_\ell$. The uncertainty in measuring the monopole and three dipole terms from such maps is then $\simeq \sigma_{\rm map} \sqrt{1/N_{\rm cl}}$ and  $\sigma_{\rm map} \sqrt{3/N_{\rm cl}}$ respectively. Equation  \ref{eq:sigma_fil_map} can be applied to {\it any} filtering scheme and is the key to estimating the uncertainties of the eventual measurement using the KA-B methodology.

Because of the cosmic variance, the power spectrum of the CMB sky at the particular realization in our Universe, $C_\ell^{LOC}$, differs from the theoretical model
$C_{\ell}^{th}$ and so a residual CMB signal from primary
anisotropies is left in the filtered maps. To estimate the
contribution of noise and the CMB residual to the total power in
these maps, we expand the filtered temperature field as $\delta T(\hat{n}) = \sum F_\ell a_{\ell m}Y_{\ell
m}(\hat{n})$. The variance of any filtered map is given by the middle term below with the RHS valid for our filter:
\begin{equation}
\sigma_{fil}^2=\frac{1}{4\pi}\sum (2\ell+1)F_\ell^2 C_\ell=
\frac{1}{4\pi}\sum
(2\ell+1)\frac{(C_\ell^2-C_\ell^{th}B_\ell^2)^2}{C_\ell^2} .
\label{eq:sigmafil}
\end{equation}
As indicated, $\delta T(\hat{n})$ contains the cosmological CMB
signal and noise, $C_\ell=C_\ell^{LOC}B_\ell^2+{\cal N}_\ell$. The
power spectrum of our C<B realization differs from the underlying power
spectrum by a random variable of zero mean and (cosmic) variance
$\Delta_\ell=(\ell+\frac{1}{2})C^{th}_\ell/f_{sky}$, where
$f_{sky}$ is the fraction of the sky covered by the data (Abbot \&
Wise 1984). Then, the cosmic variance leads to
$C_\ell^{LOC}=C_{\ell}^{th}\pm\Delta_\ell^{1/2}$. The above limits
on $C_\ell$ bound the range of $\sigma_{fil}$,
eq.~(\ref{eq:sigmafil}), to:
\begin{equation}
\sigma_{fil}^2=\frac{1}{4\pi}\sum (2\ell+1)\left[\frac{\Delta_\ell^2}
{C_\ell^{th}+\Delta_\ell+{\cal N}_\ell}+\frac{{\cal N}_\ell^2}{C_\ell^{th}+\Delta_\ell+{\cal N}_\ell}\right]=
\sigma_{CV,fil}^2+\sigma_{{\cal N},fil}^2(t_{\rm obs})
\label{eq:sigmafil2}
\end{equation}
In this last expression, the variance of the filtered map depends
on two components: 1) the residual CMB left due to cosmic variance
$\sigma_{CV,fil}$ and 2) the noise $\sigma_{N,fil}$, that is
not removed by the filter. The latter component integrates down with
increasing observing time $t_{\rm obs}$ as
$t_{\rm obs}^{-1/2}$ {\it and} $N_{\rm DA}^{-\half}$ when several DA's are used. It becomes progressively less important in
WMAP data with longer integration time. The first component, however, does not decrease with longer integration time and remains present at all FA's. It then presents a fundamental floor which limits the S/N of the measurement which can be lowered only by increasing the number of clusters used in the measurement.

We can now study the two terms in the RHS of eq. \ref{eq:sigmafil2} separately. Since all the terms in the denominators are positive, the upper limits on them become:
\begin{equation}
\sigma_{CV,fil}^2\equiv\frac{1}{4\pi}\sum (2\ell+1)\left[\frac{\Delta_\ell^2}{C_\ell^{\rm th}+\Delta_\ell+{\cal N}_\ell}\right] < 261 \mu K^2
\label{eq:sigma_cv}
\end{equation}
\begin{equation}
\sigma_{{\cal N},fil}^2\equiv\frac{1}{4\pi}\sum (2\ell+1)\left[\frac{{\cal N}_\ell^2}
{C_\ell^{th}+\Delta_\ell+{\cal N}_\ell}\right] < \sigma_{\rm noise}^2
\label{eq:sigma_noise}
\end{equation}
The upper limit on the first of these terms was evaluated assuming concordance $\Lambda$CDM model for WMAP7 data and zero instrument noise (${\cal N}_\ell=0$). Thus  the cosmic variance from the CMB field realized in {\it our}  Universe (in which the KSZ dipole is measured) adds up to $\sigma_{CV,fil}\simeq 15-16\mu$K. Its contribution to the dipole/monopole integrates down as $1/N_{\rm cl}$ independent of the integration time and the aperture size.

\begin{figure}[h!]
\includegraphics[width=4in]{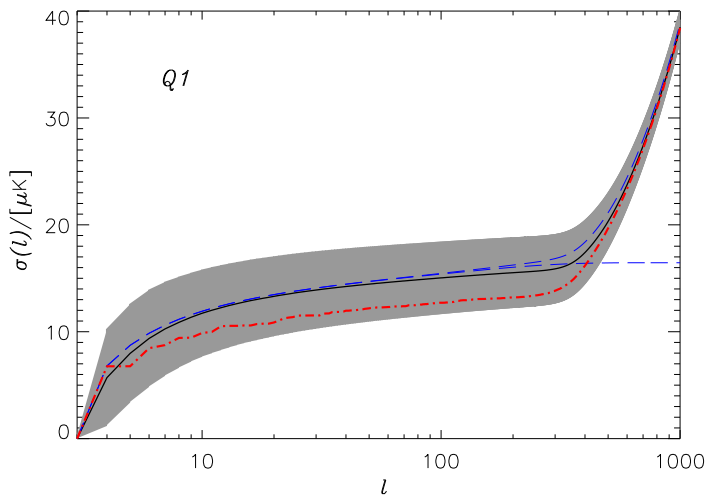}
 \caption[]{Cumulative rms deviation as
a function of multipole. Solid line and shaded area show mean and
rms of 4,000 simulated Q1 filtered maps. Dashed lines represent
the residual CMB component of the filtered maps due to cosmic
variance, computed using eq.~(\ref{eq:sigmafil2}) and the residual
CMB plus the noise components. The dot-dashed line corresponds to
the actual Q1 band of WMAP 5-yr data. Adapted from AKEKE. }
\label{fig:akeke1}
\end{figure}

Consider now the contribution to the total variance in the filtered maps from noise, $\sigma_{{\cal N},fil}$. The uncertainty in measuring the monopole and three dipole terms from such maps will then integrate down with the number of pixels (fixed by the aperture size), $N_{\rm pix}$, and integration time, $t_{\rm obs}$, as $N_{\rm pix}^{-1/2} t_{\rm obs}^{-1/2}$. We use the WMAP data pixelized with $\simeq 7^\prime$ pixels, so the number of pixels subtended by a given aperture of radius $\theta_A$ is $N_{\rm pix} = \pi \theta_A^2/(7^\prime)^2 N_{\rm cl} \simeq 58  N_{\rm cl} (\theta_A/30^\prime)^2$ where everything is normalized close to the final aperture used. For W-band there are $N_{\rm DA}=4$ DA's and Q, V bands have two DA's each. Thus for the parameters in table \ref{tab:wmap} the combined 7-yr CMB data in each band will lead to:
\begin{equation}
\sigma_{{\cal N},fil} < \left(\frac{30^\prime}{\theta_{\rm app}}\right) \times \left\{
\begin{array}{c}
6 \; \mu{\rm K \; \; Q-band}\\
7 \; \mu{\rm K \; \; V-band}\\
9 \; \mu{\rm K \; \; W-band}\\
\end{array} \right.
\label{eq:sig_noisefil}
\end{equation}
When added in quadrature to $\sigma_{\rm CV, fil}$ this term - for post 7-year WMAP data and the final cluster apertures - gives a negligible contribution ($\lsim 15\%$) to the overall error budget in each band of $\sigma_{fil}$ given by eq. \ref{eq:sigmafil2}. If analysis is performed in each DA, the noise may still contribute even in the final 9-yr WMAP dataset.

Thus in post-WMAP5 CMB maps, the cosmic variance contribution, eq. \ref{eq:sigma_cv} dominates the error budget. To better understand this term's role we follow AKEKE and denote by $\sigma_q^2\equiv\frac{1}{4\pi}(2q+1)
(\Delta^2_q+N^2_q)(C_q^{th}+\Delta_q+N_q)^{-1}$ and let
$\sigma^2(\ell)=\sum_{q=4}^\ell\sigma_q^2$ be the cumulative
variance of the residual map. With these definitions, the total
variance of the filtered map is $\sigma_{fil}^2=
\sigma_{fil}^2(\ell_{max})$. For Healpix maps with $N_{side}=512$
the maximal multipole is $\ell_{max}=1024$ (Gorski et al 2005). In
Figure~\ref{fig:akeke1} we plot this cumulative contribution of each
multipole $\ell$, $\sigma_{fil}(\ell)$, to the total rms of the
map. The solid lines represent the mean and rms
$\sigma_{fil}(\ell)$ of filtered maps of 4,000 realizations of the
Q1 DA; the shaded area represents the dispersion of those
realizations, the dot-dashed line is the same quantity but for the
filtered Q1 WMAP 5-year data. The lower dashed lines represent
$\sigma_{CV,fil}$, the residual CMB component, and upper dashed
line, the total variance of the map [eq~(\ref{eq:sigmafil2})]. The
dot-dashed line also contains any contributions from foreground
emissions; the fact that it lies so close to the to the region
expected from simulating CMB sky implies that foreground emission
contributions to $\sigma_{fil}$ are small. Figure~\ref{fig:akeke1}
clearly shows that for multipoles below $\ell\sim 200$ the
cumulative variance of the 5-year WMAP maps $\sigma^2(\ell)$ is
dominated by the residual primary CMB signal from the cosmic
variance, even though the total variance of the filtered maps is
dominated by noise. For the Q1 WMAP channel, the mean variance of
our simulations was $\sigma_{fil}^2\sim 2000 (\mu$K)$^2$ out of
which $\sim 200 (\mu$K)$^2$ come from the residual primary CMB
signal.

An important self-consistency check, which also serves as a diagnostic of the accuracy of any pipeline, is to understand a connection between the monopole and dipole uncertainties.  The filtered maps have no {\it intrinsic monopole or dipole} by construction. Since we measure
these two moments from a small fraction of the sky, our limited
sampling generates an error due to (random) distribution of these
quantities around their mean zero value. The sampling variances of
$\langle a_0\rangle$ and $\langle a_{1i}\rangle$ are $Var(\langle
a_0\rangle)=\langle a_0^2\rangle/N$, $Var(\langle
\sigma_{i}\rangle)=\langle a_{1i}^2\rangle/N$, where $N$ is the
number of independent data points. Direct computation shows that:
\begin{equation}
\sigma_0^2\equiv \langle a_0^2\rangle = \langle (\Delta
T)^2\rangle \qquad \sigma_i^2\equiv \langle a_i^2\rangle =
\frac{\langle (\Delta T)^2\rangle}{\langle \hat{n}_i^2\rangle}, \quad
i=(x,y,z)
 \label{eq:sqrt3}
\end{equation}
In this expression, $\hat{n}_i$ are the direction cosines of clusters.
If clusters were homogeneously distributed on the sky then
$\langle \hat{n}_i^2\rangle=1/3$ and one should recover the dipole
errors of
\begin{equation}
\sigma_i=\sqrt{3}\sigma_0
\label{eq:sigma_dip2mon}
\end{equation}
Thus the error on the
monopole serves as a consistency check in any such computation. If
one does not recover the above scaling between the two, then there
are either problems with the catalog used or errors in the
analysis (we discuss such cases in Sec. \ref{sec:rebuttals}).

Finally, Figure~\ref{fig:akeke1} indicates that {\it our filter removes the
intrinsic CMB down to the fundamental limit imposed by cosmic
variance}. In this sense the filter is close to {\it optimal}, since it
minimizes the errors contributed to our measurements by primary
CMB. In principle, one can define a more aggressive filter that,
together with the intrinsic CMB, also removes the noise leaving
only the SZ signal. But filtering is not a unitary operation and
does not preserve power. Such a filter would then remove an
important fraction of the SZ component and would probably reduce
the overall S/N. In general, a different filter would give
different dipole (measured in units of temperature) and would
require a different calibration. Discussion of filtering schemes
that maximize the S/N ratio and minimize the systematic error on
the calibration will be given elsewhere (Atrio-Barandela et al 2012,  in preparation).

\subsubsection{Directly evaluated uncertainties}

We have used four methods to evaluate the errors directly. We emphasize that for correct error analysis it is important to estimate the errors in the CMB realization of {\it our} Universe, the only one where we determine the dipole at the cluster locations. With some 99\% of the maps pixels left outside the clusters this is straightforward and doable. To account for the correlations between the different DA's (Keisler 2009, KAEEK, AKEKE) all the results presented below are shown after selecting the same cluster positions in {\it every} DA and then averaging over all the eight DA's to produced the final distributions. The four methods are (Method II is subdivided into IIa and IIb):
\begin{itemize}
\item {\bf Method I}: we estimate error bars by placing random
clusters in the real filtered maps outside clusters in the catalog and outside the CMB mask. Errors estimated using Method I include any contribution
originated by foreground residuals. This method also accounts for the mask effects. The distribution of the dipole components for various values of $N_{\rm cl}$ is shown in Fig. \ref{fig:kaeek1b}. The distribution is Gaussian with the width shown in the rightmost panel of the Figure. The width of that distribution scales as $N_{\rm cl}^{-\half}$ to a very good accuracy. As expected and as explained in this section the error is largest for the X-component and smallest for the Z-component, but the differences are small at the levels below $\sim 10\%$.

 \begin{figure}[h!]
 \hspace{-1in}
\includegraphics[width=7.5in]{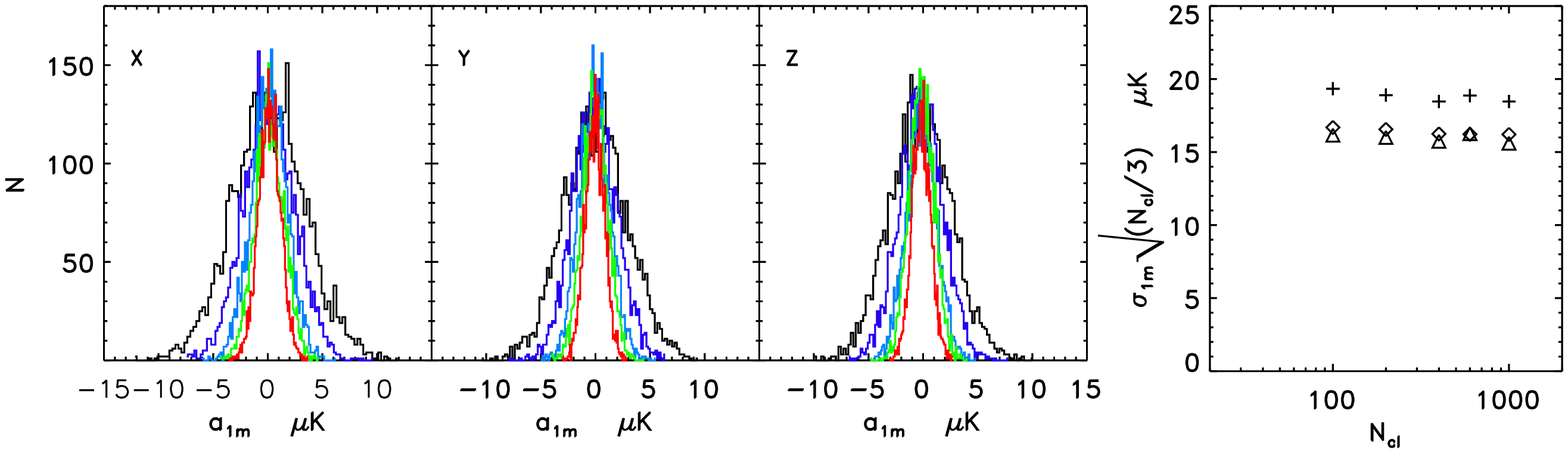}
 \caption[]{The errors on the dipole components using Method I. Left panels show histograms for the distribution of randomly place pseudo-clusters with different colors corresponding to $N_{\rm cl}$ shown in the rightmost panel (adapted from KAEEK). The histograms are to a high accuracy Gaussian. The rightmost panel shows that the dispersion of the distribution, which is the error on our measured dipole, scales to a good accuracy as $N_{\rm cl}^{-\half}$. Pluses, diamonds and triangles correspond to $X, Y, Z$ components respectively.}
\label{fig:kaeek1b}
\end{figure}

\item {\bf Method II}: Here the cluster template is fixed and the sky is
simulated; the spectrum are gaussian realizations of the measured
power of the filtered maps. The method is then applied to:

{\bf IIa} - the CMB sky is simulated from random realizations of the CMB power spectrum given by the concordance $\Lambda$CDM model, and

{\bf IIb} - the CMB sky is simulated  from random realizations of the CMB power spectrum given by the {\it observed} CMB anisotropies in {\it our} Universe.

This method accounts for the geometric irregularities of the cluster catalog. Method IIa has an {\it upward} bias since Fig. \ref{fig:akeke1} shows that in {\it our} Universe the CMB realization happens to be on the low side of the cosmic variance. Thus Method IIb is a more correct approximation for the error budget using simulated maps. Fig. \ref{fig:akeke4} shows the quantities relevant here and illustrates the errors recovered in applying this method.

 \begin{figure}[h!]
\includegraphics[width=6in]{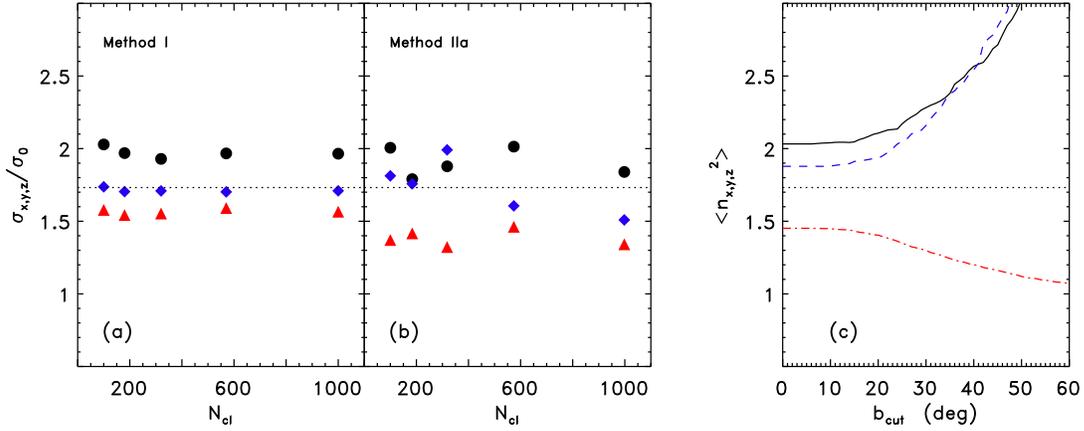}
 \caption[]{Dipole to monopole error
bar ratio. (a) (Black) circles, (blue) diamonds and (red) circles
correspond the the ratio of the (x,y,z) component of the dipole to
the monopole, respectively. Monopole and dipole were computed
using Method I. (b) Same as (a) but monopole and dipoles are
computed using Method IIa. (c) Ratio of the dipoles to monopole
error bars for our cluster catalog. The horizontal axis, $b_{cut}$
indicates that clusters with $|b|\le b_{cut}$ are excised from the
catalog. In all three plots, the dotted line represents the ratio
for a perfectly isotropic cluster catalog. Adapted from AKEKE. }
\label{fig:akeke4}
\end{figure}

\item {\bf Method III}: Here for each cluster configuration given by the $L_X$ and $z$ limits containing $N_{\rm clus}$ clusters in total, we select many smaller subgroups of clusters with $N_1<N_{\rm clus}$ chosen at random. We then study the distribution of the measured dipole for many such realizations as function of $N_1$. This method accounts for the masking, the average effects of geometries of the cluster samples, and - additionally - it enables to check whether various randomly chosen subgroups of clusters participate - on average - in the same motion. Fig. \ref{fig:method3} shows the distribution of the dipoles for clusters with $L_X\geq 2\times10^{44}$erg/sec at $z\leq 0.25$ corresponding to one configuration of KAEEK, Table 1 (see also Table \ref{tab:a1mfromkaeek} below, last row). That configuration was selected for display because it has sufficiently large overall numbers of clusters (322) enabling a reliable subselection with at least $>100$ cluster subsamples. For this configuration the dipole measurements are $(3.7\pm1.8, -4.1\pm1.5, 4.1\pm1.5)\mu$K in the $(X,Y,Z)$ directions. Figure \ref{fig:kaeek1b} shows that the distribution of the dipoles, evaluated from 1,000 random selections each for various $N_1$, is in good agreement with that derived in Methods I and II and also with the analytical insight developed above.

 \begin{figure}[h!]
\hspace{-1in}
\includegraphics[width=7.5in]{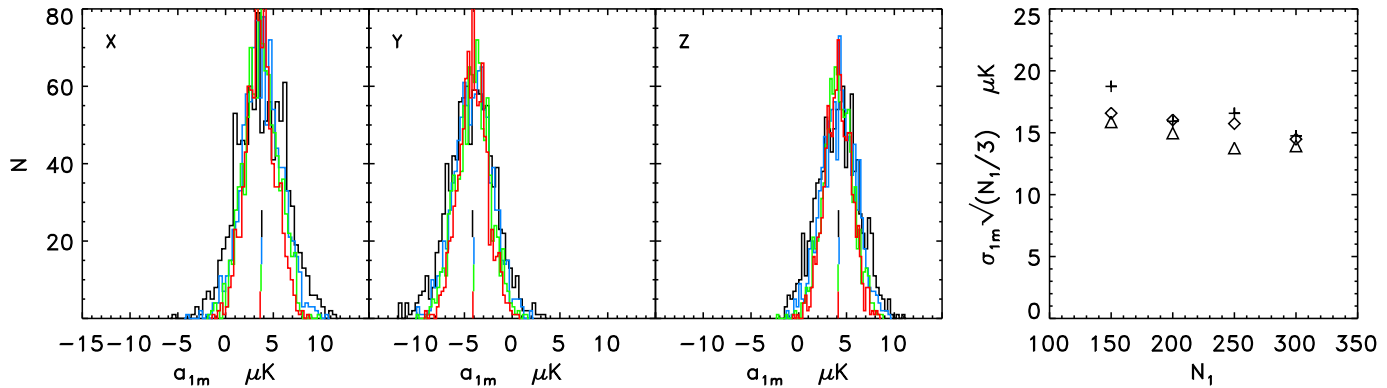}
 \caption[]{Method III for clusters at $z \leq 0.25, L_X \gt 2\times 10^{44}$erg/sec. Leftmost panels show the histograms for $N_1=150, 200, 250, 300$ clusters subselected at random out of the total 322 available in that configuration. The distribution of the dipoles is to a good accuracy Gaussian. The average of each histogram is shown with vertical bars of the corresponding color; the overall configuration leads to the dipole shown in Table \ref{tab:a1mfromkaeek} gives $(a_{1x}, a_{1y},a_{1z})= (3.7\pm1.8, -4.1\pm 1.5, 4.1\pm 1.5)\mu$K. The rightmost panel shows the scaling of the standard deviation determined in this way vs the $N_1$. }
\label{fig:method3}
\end{figure}

\end{itemize}
Since the final error is dominated by the cosmic variance terms in eq. \ref{eq:sigmafil2}, as realized independently by Keisler (2009), KAEEK and AKEKE, when computing the errors in the above prescriptions one needs to assign pseudo-clusters to the {\it same locations in each of the eight DA's} (KAEEK). All of the methods lead to only small variations in the errors of up to $\sim 10-15\%$ which testifies to the robustness of the measurement.

\subsubsection{Diagonality of the dipole error matrix}

One of the strengths of the KA-B method is that it determines directly (albeit within the current calibration uncertainties discussed below) the full 3-dimensional bulk velocity of the cluster sample with each component determined independently at least for the cluster catalogs and the filtering scheme used in our studies. The latter is an important advantage over the galaxy distance indicator determinations where the errors between the different component can have significant coupling (Kaiser 1988, Watkins et al 2009). This can be seen empirically from the fact that the cross-correlation matrix, ${\cal A}_{mn}=\langle a_{1m} a_{1n}\rangle$ evaluated over the CMB filtered with the KABKE filter is to good accuracy diagonal. The following equation shows that it is indeed diagonal using the 5-yr W-band filtered CMB maps:
\begin{equation}
\stackrel{N_{\rm clus}=200}{{\cal A}_{mn}=\left(\begin{array}{ccc} 5.99 & 0.36 & 0.49 \\0.36 & 4.62 & -0.31 \\
0.49 & -0.31 & 4.34\end{array}\right)\mu{\rm K}^2} \;\; ; \;\;
\stackrel{N_{\rm clus}=400}{{\cal A}_{mn}=\left(\begin{array}{ccc} 2.88 & 0.09 & 0.18 \\0.09 & 2.23 & -0.13 \\
0.18 & -0.13 & 2.21\end{array}\right)\mu{\rm K}^2}\label{eq:a1mn_cross}
\end{equation}
In evaluating the matrix we used each of the four W-channel maps separately placing the (pseudo)-clusters randomly outside the CMB mask and the cluster pixels from our catalog. 4,000 realizations of the cluster positions were used with the pseudo-clusters occupying the same pixels in all four W1-W4 DA's thereby accounting for the correlation between the DA's  from the residual primary CMB. The pseudo-clusters were all taken to have a fixed aperture of $30^\prime$ radius. The dipoles in each DA from the pixels associated with the pseudo-clusters in each realization were then averaged to give the combined W-band realization and the above matrix is evaluated over 4,000 such realizations. Note that 1) the non-diagonal terms do not exceed $(5-7)\%$ of the diagonal terms, 2) the terms scale to good accuracy as $1/N_{\rm cl}$, and 3) the diagonal ${\cal A}_{mn}$ terms are well described by the eq. \ref{eq:sigmas} below with the slight excess ($\lsim 10\%$) arising from the contribution due to the instrument noise in the 5-yr WMAP data (which would be negligible in the upcoming 9-yr WMAP release).

\subsubsection{Approximations for errors in post WMAP5 CMB data}

Finally we note that the following approximations are valid for the errors in post-5yr WMAP maps using eq. \ref{eq:sigmafil2}, and assuming 1) the concordance WMAP5 $\Lambda$CDM model, 2) an isotropic X-ray cluster catalog, and 3) a KP0 CMB mask:
\begin{equation}
\sigma_{\rm mon}\simeq \frac{15}{\sqrt{N_{\rm cl}}} \;\mu K \;\; ;\;\; \sigma_{\rm dip}^{x,y,z} \simeq (15-18)\sqrt{\frac{3}{N_{\rm cl}}} \;\mu K
\label{eq:sigmas}
\end{equation}
The factors for the dipole errors span the three components with the largest error describing $\sigma_x$ and the smallest corresponding to the $z$-component of the dipole; these approximations are good to about $\sim 10\%$ for the methods above. In addition, the ratio of the dipole-to-monopole errors serves as an important self-consistency check.

\subsection{Isolating/removing the TSZ contribution by aperture increase}
\label{sec:tsz_removal}

As discussed above, we demonstrated in AKKE that our cluster catalog applied to the
{\it unfiltered} CMB data indicates that the gas in X-ray clusters
is well described by the Navarro-Frenk-White (1996, NFW) density
profile theoretically expected from the non-linear evolution of
the concordance $\Lambda$CDM model. Because in the NFW-type models the hot gas  X-ray temperature must decrease with increasing cluster-centric radius in order to maintain hydrostatic equilibrium (Komatsu \& Seljak 2001), we can expect that the TSZ component produced by such clusters would decrease with increasing aperture. This was indeed found by us to be the case for clusters in our catalog and this property can be - and should be in any other attempt to reproduce the DF measurement at the frequency bands of the WMAP - used to reduce the TSZ contribution to the dipole and isolate the dipole produced by the KSZ from cluster bulk motions.

\begin{figure}[h!]
\includegraphics[width=6.5in]{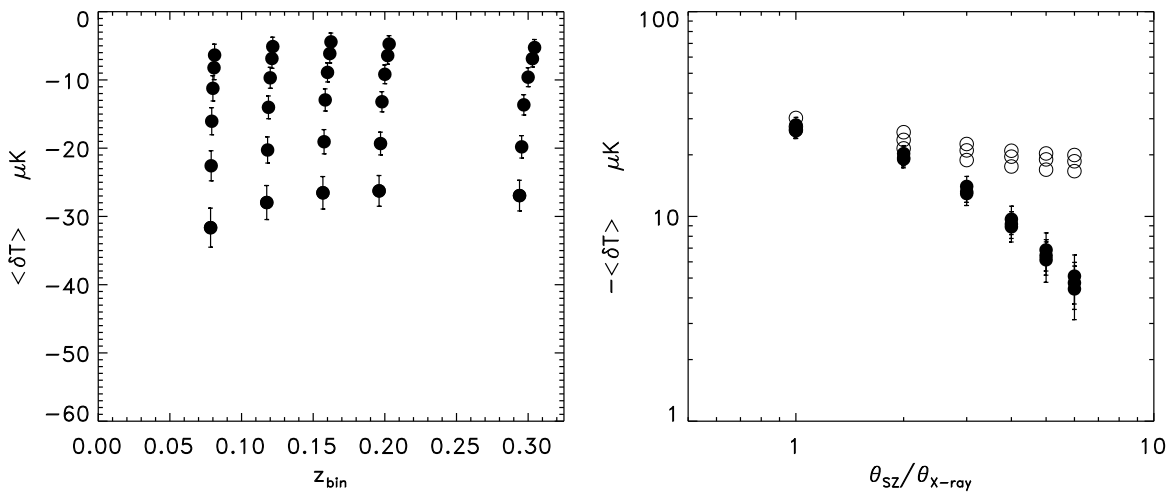}
 \caption[]{\small{{\bf Left}: The mean CMB temperature decrement
 averaged over the Q, V, W channels. The results are for unfiltered
maps with 0.5$^\circ$ cut in cluster extent/aperture shown for the outer
$z$-bins for progressively increasing $\alpha=\theta_{\rm
SZ}/\theta_{\rm X-ray}$. Filled circles from bottom to top
correspond to $\alpha=1,2,3,4,5,6$. {\bf Right}: Solid circles
show the mean TSZ decrement profile in the unfiltered CMB data vs
$\alpha$ for three farthest $z$-bins. Open circles correspond to
the isothermal $\beta=2/3$ model evaluated as described in Sec. 6. Adapted from KABKE2.
}}
\label{fig:kabke2_9}
\end{figure}

In the left panel of Fig. \ref{fig:kabke2_9} we show the mean TSZ decrement at the cluster
positions evaluated from the WMAP maps for the various total
cluster extent limits. The mean
temperature decrement from each of the eight DA's were
averaged with their corresponding uncertainties to give the
final $\langle \delta T \rangle$ shown in the figure. The strong
decrease in the mean TSZ decrement with the increasing angular
size is apparent from the figure.

The middle panel of the figure shows the mean CMB temperature
profile of the TSZ decrement in the unfiltered maps for three
outer redshift bins from KABKE2. The expectation
from the isothermal $\beta$-model for these bins is shown with the open circles. It fits
well the data at the cluster inner parts, but deviates strongly
from the measurements at larger radii. The fits from the NFW
profiles using a method similar to Komatsu \& Seljak (2001) are
shown with solid lines for two concentration parameters (see AKKE
for details). These profiles provide a good fit to the data.

We emphasize again in this context that the gas with the
NFW profile which is in hydrostatic equilibrium with the cluster
gravitational field must have the X-ray temperature decreasing
with radius (Komatsu \& Seljak 2001). This is confirmed by
numerical simulations of the cluster formation within the
$\Lambda$CDM model (Borgani et al 2004) as well as by the
available observations of a few nearby clusters (Pratt et al
2007). The latter cannot yet probe the $T_{\rm X}$ profile all the
way to the virial raidus, but do show a decrease by a factor of
$\sim 2$ out to about half of it (see e.g. Fig. 5 of Pratt et al
2007). In the NFW profile the gas density profile in the outer
parts goes as $n_e \propto r^{-3}$ with the polytropic index which
is approximately constant for all clusters at $\gamma \simeq 1.2$
(Komatsu \& Seljak 2001). Thus the X-ray temperature must drop at
least as $T_{\rm X} \propto r^{-0.6}$ at the outer parts and for
larger values of $\gamma$ the drop will be correspondingly more
rapid. The temperature profile implied by the NFW density profile
normalized to the data in the middle panel is shown in the right
panel of Fig. \ref{fig:kabke2_9}.

\subsection{CMB dipole}
\label{sec:cmbdip_df}

We now move to the measurements of the bulk flow using the KA-B method, having 1) discussed the processing of the data required for this analysis, 2) laid foundation for understanding the error budget in this measurement (which any filtering scheme must satisfy or else it has errors in the pipeline and/or implementation), and 3) addressed the filtering requirements and expectations.

The original study of KABKE, first released in 2008, used a flux-limited all-sky sample of $\sim 700$ X-ray clusters from Kocevski \&  Ebeling (2006). KABKE1,2 have developed the machinery for the application of the KA-B method to the cluster sample and, at the time the best available, 3-yr WMAP data. They detected, for the first time, the (unexpected) dipole signal located at high significance exclusively at cluster pixels, which remained at apertures containing zero monopole. At smaller apertures the CMB monopole was robustly negative a clear sign of the presence of hot X-ray gas there producing the SZ effect. The flow in that study extended to $\sim 300h^{-1}$Mpc and within the errors the dipole pointed in the same direction for all cluster $z$-bins. In this way, they identified the signal as arising from the KSZ effect produced by clusters moving coherently in the same direction. Within the bounds of their (admittedly imperfect) calibration, the measured velocity field was highly coherent and well above that expected on these scales from gravitational instability due to the established $\Lambda$CDM model. Thus KABKE1 proposed that the origin of the flow results from the tilt which extends across the entire cosmological horizon, dubbing it ``the dark flow".

A later reanalysis by KAEEK and AKEKE revised the statistical treatment of KABKE1,2, correcting for correlations between the signals in different DA's due to the primary CMB remaining from cosmic variance. AKEKE have also developed a precise formalism to understand, and predict, the errors of such measurements analytically. Keisler (2009) also has independently noted these correlations, but has overestimated the errors as pointed out in AKEKE and Sec. \ref{sec:rebuttals}. With the revised statistical treatment, the signal measured in KABKE1,2 was still statistically significant, albeit at a reduced level of $S/N \simeq 3$. (Clearly the effect of correlations between CMB in $N_{\rm DA}=8$ WMAP DA's used can at most increase the errors quoted in KABKE2 by $\sqrt{N_{\rm DA}}$; so in reality the errors should be multiplied by a constant factor of $\simeq 2.6$).

The KABKE analysis has been now superseded by an improved study of KAEEK which used 5-yr WMAP data with lower noise and a much larger cluster catalog, binning it by the cluster luminosity. The latter step allowed to identify the flow better and to larger scales, and, equally importantly, the correlation with the $L_X$-threshold identified in KAEEK shows that the signal very likely originates from the KSZ effect, rather than some putative systematics from primary CMB and/or noise.

We discuss below the results from the KAEEK analysis below; these results supersede KABKE, but at the same time are fully consistent with our original 2008 detection. We also discuss then the results from a recent study by KAE that showed how to uncover - and verify - the DF results with data which is publicly available at this time.

\subsubsection{Using early SCOUT X-ray cluster catalog}
\label{sec:kaeek}

To improve upon the all-sky cluster catalogue of Kocevski \&
Ebeling (2006) used by KABKE1,2, for KAEEK we have screened the ROSAT
Bright-Source Catalogue (Voges et al.\ 1999) using the same X-ray
selection criteria (including a nominal flux limit of $1\times 10^
{-12}$ erg/sec, 0.1--2.4 keV) as employed during the Massive
Cluster Survey (MACS, Ebeling et al.\ 2001), as well as the same
optical follow-up strategy. These tasks were performed by the DF collaborator, Harald Ebeling. As a result, the interim all-sky cluster catalogue used in KAEEK comprised in excess of 1,400 X-ray selected clusters, {\it all} of
them with spectroscopic redshifts. X-ray properties of all
clusters (most importantly total luminosities and central electron
densities) were computed as described in KAEEK. Within the same
$z$-range ($z\la0.25$), where clusters can still be (at least partially) resolved, that catalog
comprised 1,174 clusters outside the KP0 CMB mask. To eliminate
low-mass galaxy groups KAEEK required that clusters feature $L_X\geq 2
\times 10^{43}$ erg/sec; 985 systems met this criterion.

 \begin{figure}[h!]
\includegraphics[width=3in]{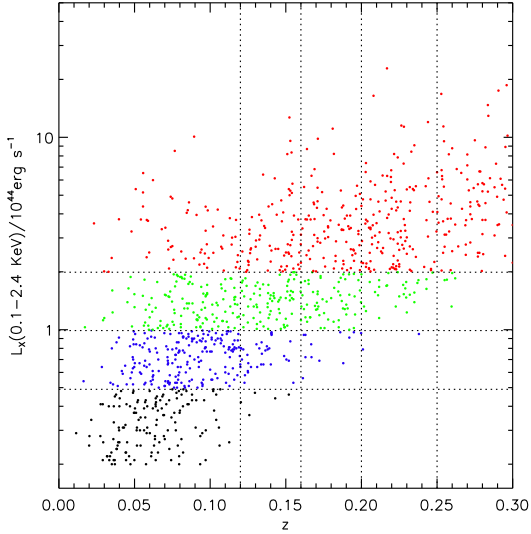}
 \caption[]{
Plotted is the $z$-distribution of clusters in the $L_X$-bins
used in the KAEEK analysis. Black, blue, green and red colored dots correspond to
$L_X = (0.2-0.5, 0.5-1, 1-2, >2) \times 10^{44}$ erg/sec respectively. The vertical dotted lines show the redshift bins in Table \ref{tab:a1mfromkaeek}. Adapted  from KAEEK.
}
\label{fig:kaeek1a}
\end{figure}

\begin{deluxetable}{l | c | c | c | c c c c }
\tablewidth{0pt} \tabletypesize{\scriptsize} 
\tablecaption{CMB DIPOLE RESULTS  from KAEEK } 
\startdata
 $z\leq$ & $L_X$-bin & $N_{\rm cl}$ &
$z_{\rm mean}/z_{\rm median}$  & $\bar{a}_{1,x}$ & $\bar{a}_{1,y}$ &
$\bar{a}_{1,z}$ & $\sqrt{C_1}$ \\
 & $10^{44}$ erg/s &  &  & $\mu$K & $\mu$K & $\mu$K & $\mu$K \\
 \hline
 \hline
0.12$^{*}$ & 0.2--0.5 & 142 & 0.061/0.060 & $-4.2\pm2.7$ & $-0.7\pm2.3$ & $0.5\pm2.3$  & $4.3\pm2.7$ \\
0.12 & 0.5--1 & 194 & 0.081/0.082 & $-2.7\pm2.3$ & $-2.3\pm2.0$ & $1.4\pm2.0$  & $3.9\pm2.2$ \\
0.12 & $>1$ & 180 & 0.083/0.086 & $4.9 \pm 2.4$ & $-4.5 \pm 2.1$ & $1.5\pm 2.0$ & $6.8\pm2.2$ \\
\hline
 0.16 & 0.5--1 & 226 & 0.089/0.087 & $-1.5\pm2.2$ & $-0.6 \pm1.9$ & $2.1 \pm 1.8$  & $2.7\pm1.9$ \\
0.16 & 1--2 & 191 & 0.106/0.107 & $1.9\pm2.3$ & $-2.8 \pm 2.0$ & $-0.5 \pm 2.0$  & $4.1\pm2.2$ \\
0.16 & $>2$ & 130 & 0.115/0.125 & $4.2\pm2.8$ & $-8.0\pm2.4$ & $4.9\pm2.4$  & $10.3\pm2.5$ \\
\hline
0.20 & 0.5--1 & 238 & 0.093/0.089 & $-2.5\pm2.1$ & $-1.3 \pm 1.8$ & $1.0 \pm 1.8$ & $3.0\pm2.0$ \\
0.20 & 1--2 & 248 & 0.122/0.123 & $0.1\pm2.0$ & $-1.8 \pm 1.8$ & $-0.3\pm 1.7$ & $1.8\pm1.8$ \\
0.20 & $>2$ & 208 & 0.140/0.151 & $3.6\pm2.2$ & $-5.8 \pm1.9$ & $4.5\pm1.9$ & $8.1\pm2.0$ \\
\hline
0.25 & 0.5--1 & 240 & 0.094/0.090 & $-2.3\pm2.1$ & $-1.1 \pm 1.8$ & $0.9 \pm 1.8$ & $2.7\pm2.0$ \\
0.25 & 1--2 & 276 & 0.133/0.133 & $-0.2 \pm 2.0$ & $-1.4\pm 1.7$ & $0.7 \pm 1.6$ & $1.6\pm1.7$  \\
0.25 & $>2$ & 322 & 0.169/0.176 & $3.7\pm1.8$ & $-4.1\pm1.5$ & $4.1\pm1.5$ & $6.9\pm1.6$ \\
\enddata
\label{tab:a1mfromkaeek}
\tablecomments{Adapted from KAEEK, Table 1. }
\end{deluxetable}
Fig. \ref{fig:kaeek1a} shows the distribution of the X-ray clusters used in KAEEK. The improved cluster catalog allowed KAEEK to extend our study to
higher $z$,  and to further test the impact of systematics. Thus KAEEK
created $L_{\rm X}$-limited subsamples which achieves two important
objectives. 1) As the $L_X$ threshold is raised, fewer clusters
remain and the statistical uncertainty of the dipole increases
($\propto 1/\sqrt{N_{\rm cl}(L>L_X)}$). If, however, all clusters
are part of a bulk flow of a given velocity $V_{\rm bulk}$, very
X-ray luminous clusters will produce a larger CMB dipole ($\propto
\tau V_{\rm bulk}$), an effect that might overcome the reduced
number statistics, giving a higher S/N in the measured dipole. 2)
Since, as outlined under (1), the dipole signal should increase
with cluster luminosity, whereas systematic effects can be
expected to be independent of $L_{\rm X}$, an actual observation
of such a correlation would lend strong support to the validity of
our measurement and the reality of the "dark flow".
Fig. \ref{fig:kaeek1a} shows that indeed the depth to which we probe the flow increases
dramatically as the $L_X$-threshold is raised.

Table \ref{tab:a1mfromkaeek} (reproduced from Table 1 of KAEEK) shows the
dipole results for each subsample. The dipole is shown at the aperture
where monopole vanishes. Of the three dipole components, the
$y$-component is best determined, its value always remaining
negative and its S/N increasing strongly with increasing $L_X$.

\begin{figure}[h!]
\includegraphics[width=6in]{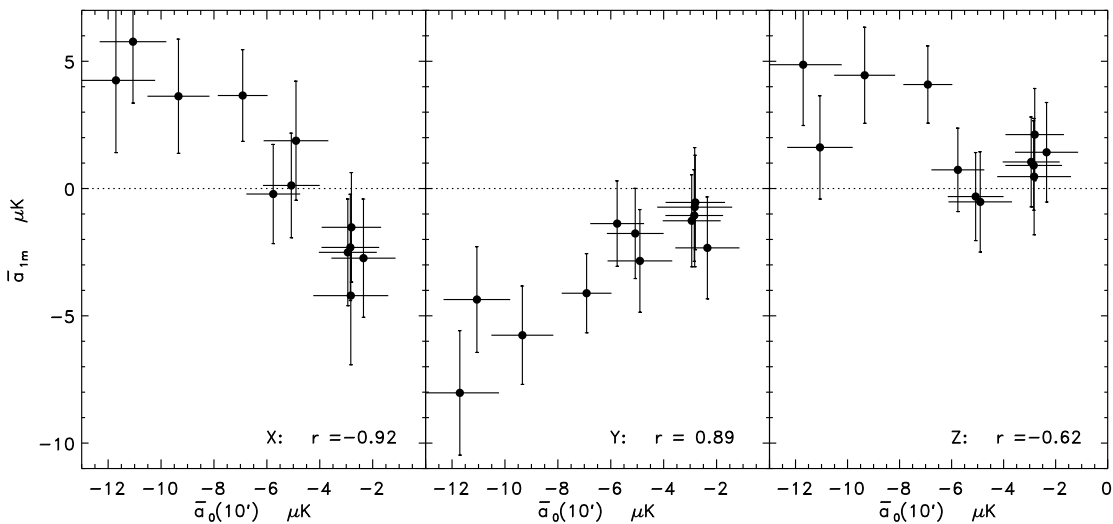}
 \caption[]{From KAEEK. Correlation between the
three dipole components and the central monopole from Table \ref{tab:a1mfromkaeek}. Both quantities increase in amplitude with $L_X$
because the optical depth and central temperature increase for
more massive clusters: KSZ dipole scales as $\tau$ and the
monopole as $\tau T_X$. The linear correlation coefficient for the
circles is shown in the lower left.}
\label{fig:kaeek1c}
\end{figure}

As shown in Fig. \ref{fig:kaeek1c} the amplitudes of all the dipole components and of the monopole
in the central parts are strongly, and approximately linearly,
correlated. {\it This correlation provides strong evidence against
unknown systematics causing our measurement}.

\begin{figure}[h!]
\includegraphics[width=6in]{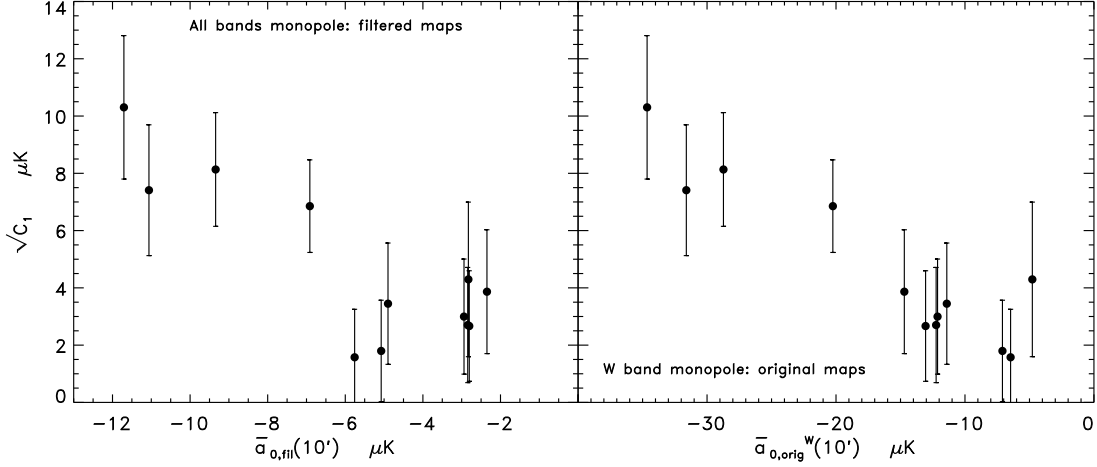}
 \caption[]{Correlation between the
total dipole ampltiude and the central monopole from Table \ref{tab:a1mfromkaeek}. Left panel shows the central monopole at cluster positions in the filtered maps after averaging over all eight DA's. Right panel shows the correlation with the central monopole at cluster positions in the original foreground subtracted WMAP CMB maps after averaging over four DA's of the W-band, which has the best angular resolution.}
\label{fig:kaeek_c1}
\end{figure}

Fig. \ref{fig:kaeek_c1} plots the values of the dipole amplitude from Table \ref{tab:a1mfromkaeek} vs the mean central monopole from all DA's measured in filtered maps (left) and vs the mean monopole measured in {\it unfiltered} maps in the four W-band DA's. The figure shows the correlation between the dipole from filtered maps evaluated at zero monopole and the central monopole in W-band from unfiltered maps. The latter is a more accurate measure of the TSZ component owing to the finer resolution of the W-band.

In addition Fig. \ref{fig:kaeek1b} demonstrates that on average and within the statistical uncertainties all subsets of clusters participate in the same flow.

\subsubsection{Using public X-ray cluster data}
\label{sec:public}

To facilitate independent tests of the intermediate results of our data-processing pipeline as well as of our final results concerning the presence and properties of the Dark Flow, we provided in KAE step-by-step instructions on how to compile a basic version of a cluster catalog from public data prior to completion of the SCOUT catalog. The data produced there have been made publicly available while the work on the SCOUT catalog continues at \url{http://www.kashlinsky.info/bulkflows/data\_public} encouraging the community to test our findings using the tools provided there. (They have been verified ``exactly" by numerous colleagues).

This verification can be achieved using three publicly available catalogues of X-ray selected clusters compiled from ROSAT All-Sky Survey data (RASS, Voges et al.\ 1999) which are 1) the extended BCS sample (Ebeling et al.\ 1998, 2000) in the northern equatorial hemisphere, 2) the REFLEX sample (B\"ohringer et al.\ 2004) in the southern equatorial hemisphere, and 3) the CIZA sample (Ebeling, Mullis \& Tully 2002; Kocevski et al.\ 2007) in the regions of low Galactic latitude ($|b|<20^\circ$) excluded from both of the first two samples. All data contained in these catalogues can be obtained in electronic form at:

\noindent
{\small \verb# http://vizier.cfa.harvard.edu/viz-bin/VizieR?-source=J/MNRAS/301/881/#}  (206 clusters)\\
{\small \verb# http://vizier.cfa.harvard.edu/viz-bin/VizieR?-source=J/MNRAS/318/333/#}  (99 clusters)\\
{\small \verb# http://vizier.cfa.harvard.edu/viz-bin/VizieR?-source=J/A+A/425/367/#} (447 clusters)\\
{\small \verb# http://vizier.cfa.harvard.edu/viz-bin/VizieR?-source=J/ApJ/580/774/#} (73 clusters)\\
{\small \verb# http://vizier.cfa.harvard.edu/viz-bin/VizieR?-source=J/ApJ/662/224#} (57 clusters)

\begin{figure}[h!]
\includegraphics[width=4in]{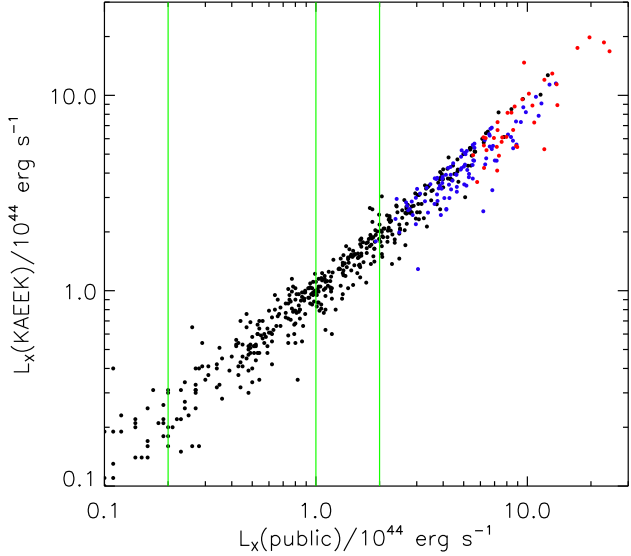}
 \caption{ \small{KAEEK vs the X-ray luminosities available from public compilations (such as in Osborne et al, 2011)for the broad
ROSAT band. Green vertical lines correspond to the $L_X$ bins used
in dipole computations. Black dots correspond to clusters at
$z\leq 0.16$, blue dots to $0.16 < z \leq 0.25$ and red dots to
$z>0.25$. At $z=0.25$ the radius subtended by the W-channel WMAP
beam (13$^{\prime\prime}$ radius) corresponds to 3 Mpc.  Adapted from KAE.}} \label{fig:kae2}
\end{figure}

More details are given in KAE and in Sec. \ref{sec:xraycat}. Fig. \ref{fig:kae2} illustrates the differences in cluster X-ray luminosity (scaled to the concordance $\Lambda$CDM model) between the KABKE sample and a simple cluster catalogue compiled from literature sources  as described. While the impact of the corrections applied for KABKE (and KAEEK) are obvious, the good overall agreement supports our notion that the existence of a statistically significant bulk flow can be successfully tested from immediately available public cluster data.

\begin{figure}[h!]
\includegraphics[width=6in]{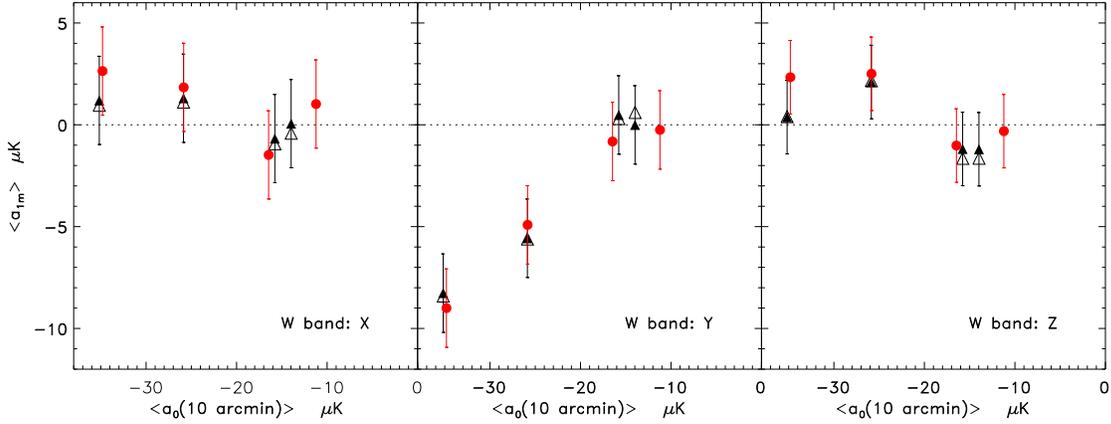}
 \caption[]{The values of the X,Y,Z dipole evaluated at
constant aperture of zero monopole for each sample marked with
green lines in Fig. \ref{fig:kae2}. The horizontal axis shows the
value of the central (10$^{\prime}$ radius aperture)
monopole in the unfiltered W-band maps which is a reflection of
the $L_X$-threshold imposed. WMAP5 data are used; our WMAP7
pipeline is not yet finalized, but the differences should be small
since primary CMB residuals dominate the errors. The last two
points at the largest negative $a_0$ correspond to $z\leq 0.16$
(142 clusters) and $z\leq 0.25$ (281 clusters). Red circles correspond to 5-yr WMAP maps. Black triangles are for 7-yr WMAP data: filled are for monopole and dipole subtracted maps; open triangles correspond to maps where quadrupole outside the mask was subtracted as well prior to filtering. Adapted from KAE.}
\label{fig:kae4}
\end{figure}

We binned the resulting cluster sample as shown by the green lines
in Fig. \ref{fig:kae2} and evaluated the dipole {\it at the constant
aperture corresponding to zero monopole} for each subsample. As discussed throughout the W-band present the best channel for resolving the clusters and isolating the KSZ component. Hence, in  KAE the results were computed for each W DA and averaged; errors are
computed as discussed above (accounting for residual
primary CMB correlations). They are shown in Fig. \ref{fig:kae4} taken from KAE
where, for each subsample, we plot the final dipole against the central monopole in
the unfiltered maps.

The results are clearly statistically significant and fully
consistent with those of KAEEK. In addition, there is a clear correlation
with the $L_X$ threshold, as expected if the signal is caused by the KSZ effect (the dipole is
computed at zero monopole; hence the TSZ contribution is small).
For clusters with $L_X\geq 2\times 10^{44}$ erg s$^{-1}$, the value of the $y$ component of the dipole obtained with this catalog in the W band is for 7-yr[5-yr] WMAP W-channel data:
\begin{eqnarray}
a_{1y}=-(8.3[9.0] \pm 2.6) \;\mu K \; ; \;  z\leq 0.16 \;; \; z_{\rm mean/median}=0.115/0.125 \; ; \; (l_0,b_0)=(278\pm 18, 2.5\pm 15)^\circ\\
a_{1y}=-(5.6[4.9] \pm 1.6) \;\mu K \; ; \;  z\leq 0.25 \;; \; z_{\rm mean/median}=0.169/0.176 \; ; \; (l_0,b_0)=(283\pm 19, 20\pm 15)^\circ\nonumber
\label{eq:dipole}
\end{eqnarray}
Here $(l_0, b_0)$ is the direction of the dipole in Galactic coordinates, and the results on the $y$-component represent 3- to 4-sigma detections using 142 and 281 clusters,
respectively. The errors are evaluated from eqs 4,6 of AKEKE. The decrease in amplitude between $z\leq 0.16$ and
$z\leq 0.25$ is consistent with the effects of beam
dilution decreasing the optical depth of the more distant
clusters. For comparison, for the same configuration KAEEK obtain with 5-yr WMAP data and the first version of the SCOUT catalog the $y$-component and the direction as $a_{1y}=(-8.0\pm 2.4)\;[(-4.1\pm 1.5)] \mu$K and $(l_0,b_0)=(292\pm 21, 27 \pm 15)^\circ\;[(296\pm 29, 39 \pm 15)^\circ]$ for $z\leq 0.16\;[0.25]$.

Comparison with Table 1 of KAEEK and Table \ref{tab:a1mfromkaeek} here and Figs. \ref{fig:kaeek1c},\ref{fig:kaeek_c1} shows that the publicly available cluster sample used here is adequate to verify the basic dark flow result. However, the same comparison also demonstrates a clear superiority already of the preliminary SCOUT cluster sample (used in KAEEK) for a more accurate measurement of the components and properties of the dark flow.

\subsection{Quadrupole and higher moments of the velocity field}
\label{sec:quadrupole}

Is the detected flow dipolar or is its shear significant? It is hard to answer this important question decisively within the framework of the KA-B method unless the cluster catalog properties are known to high accuracy as demonstrated by eq. \ref{eq:kab-shear}. Nevertheless, it is important to pursue this subject which we have done thus far by computing the quadrupole of the CMB field at cluster positions after binning by $L_X$ as in the dipole computation.

\begin{figure}[h!]
\plotone{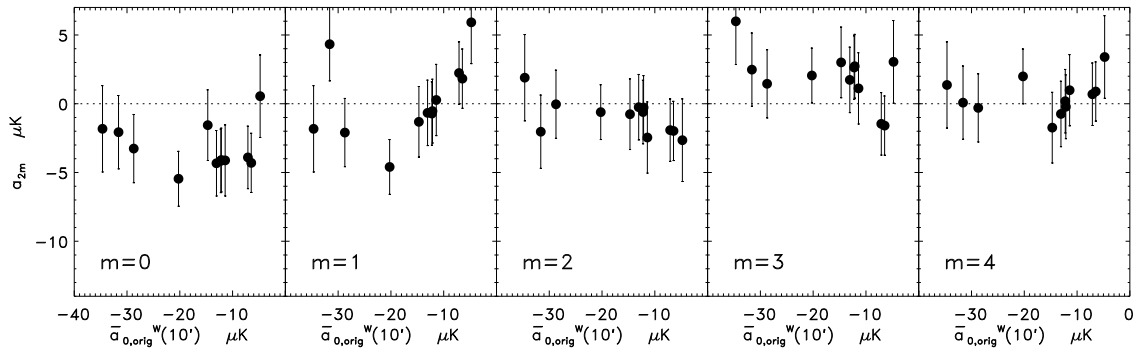} \caption{\small{Quadrupole in filtered WMAP5 data vs central values of the monopole in the original unfiltered maps. The data represent the W-band averages. Errors are estimated as $\sigma_{2m}\simeq 16 \sqrt{5/N_{\rm clus}}\mu$K.
 }} \label{fig:quadrupole}
\end{figure}

Fig. \ref{fig:quadrupole} shows the quadrupole computed over cluster apertures containing zero monopole ($\simeq 30^\prime$) vs the central monopole derived from the W-band DA's, which have the most suitable angular resolution. The figure shows the absence of any significant quadrupole and - unlike for the dipole measurement - the measured quadrupole amplitudes do not correlate with $L_X$. The absence of any measurable quadrupole here also shows that the cluster catalog assembled for this measurement is highly uniform resulting in no significant cross-terms between the dipoles of optical depth of the catalog prepared for KABKE and KAEEK analysis and the bulk flow (eq. \ref{eq:kab-shear}).

\label{sec:quad}

This coupled with the fact that all subsets of clusters appear to be moving - on average and within the uncertainties - in the same direction (as demonstrated in Figs. \ref{fig:kaeek1b},ref{fig:method3}) argues for a dipolar flow with no detectable shear across the scales probed (more detailed discussion will be given in Kashlinsky et al 2012, in preparation).

\subsection{Calibration: from $\mu$K to km/sec}
\label{sec:calibration}

The measured signal in the KA-B method is the dipole in temperature units of $\mu$K, whereas the velocity has to be deduced in km/sec. Hence, the need to calibrate our measurements, a procedure which is still in progress. While this problem is still not resolved satisfactorily leading to a systematic uncertainty in the measured velocity, we note that, as discussed in KABKE2, the calibration deficiencies do not not affect the fact of the significant and coherent large-scale motion out to the largest distance probed. Rather, within the limits of this systematic uncertainty ($\sim 20-30\%$, KABKE2), which arises due to our current modeling of clusters as isothermal $\beta$-models, this calibration uncertainty merely affects the amplitude of the velocity of the motion which is then determined (likely overestimated) within these limits. However, more importantly, the current cluster calibration also leads to ambiguity in the sign of the KSZ effect in the {\it filtered} maps.

\subsubsection{Current calibration: $C_{1,100}$}

KABKE2 and KAEEK presented the current estimates of the calibration of the measured dipole in terms of the equivalent velocity. This has been accomplished - with a systematic bias (upward) in the conversion to velocity - using the calibration factor $C_{1,100}$ introduced in KABKE2 and revisited below.

In order to translate the CMB dipole in $\mu$K into the amplitude
of $V_{\rm bulk}$ in km/sec, KABKE2 proceeded as follows. First, we
verified that our catalog reproduces accurately the measured TSZ
properties of the measured CMB parameters: Table 3 of KABKE2 compares the directly determined TSZ
contributions in the redshift bins with those determined from the
parameters in the catalog used in KABKE1,2. The latter was constructed
for each cluster from a TSZ map in each WMAP channel/DA using
the catalog values for the electron density, core radius, X-ray
temperature and total extent, and assuming an isothermal $\beta$-model with $\beta=2/3$. The agreement was demonstrated to be fairly good for the apertures corresponding to the X-ray cluster extent ($\theta_X$), which demonstrated the good accuracy of the assembled catalog.

Since the cluster properties in the catalog are determined
reasonably well (at least out to $\theta_X$), one can use the current catalog to estimate the translation factor between the CMB
dipole amplitude and the bulk flow velocity. To account for the
attenuation of the clusters' $\tau$ values by both the beam and
the filter, the gas profile of each cluster was evaluated using the current ($\beta$-model) catalog parameters and then convolved with the
beam and the filter over the WMAP pixels associated with each cluster. Each cluster was then given a bulk flow
motion of 100 km/sec in the direction of the
measured dipole, so that each pixel of the $i$-th cluster has $\delta
T=T_{\rm CMB}\tau_i(\theta) V_{\rm bulk}/c$, with $\theta$ being
the angular distance to the cluster center. We then computed the
CMB dipole of the resulting cluster map and averaged the results
over all channel maps. This allowed us to estimate the dipole amplitude,
$C_{1,100}$, contributed by each 100 km/sec of bulk-flow. In general, the $C_{1,100}$ would be a matrix coupling cross-directional terms, but in reality the cross-terms are small in the dipole determination, consistent with eq. \ref{eq:a1mn_cross} and Fig. \ref{fig:monleak}, and the diagonal terms vary less than the systematic uncertainty of this calibration procedure. For the overall current catalog, the calibration leads to $\sqrt{C_{1,100}}\simeq 0.3 mu$K pr each 100/km/sec.

\begin{deluxetable}{c c c c c}
\tablewidth{0pt} \tabletypesize{\scriptsize} 
\tablecaption{Calibration parameters derived from KAEEK.}
\tablehead{ \colhead{$z\leq$} & \colhead{$L_X$-bin} &
\colhead{$\langle\tau_0\rangle$} & \colhead{$\sqrt{C_{1,100}}$ ($\mu$K)} &  \colhead{$\sqrt{C_{1,100}}$ ($\mu$K)} \\
 & $\times 10^{44}$ erg/s & $\times 10^{-3}$ & at $5^\prime$ & at $\Theta_X$
}
 \startdata
0.12 & 0.2--0.5 &  2.8 & 0.2301 &  0.1942 \\
0.12 & 0.5--1 &  3.5 & 0.2989 & 0.2561 \\
0.12 & $>1$ & 5.4 & 0.4610 & 0.3496  \\
\hline\\
 0.16 & 0.5--1 & 3.5 & 0.2843 & 0.2363  \\
0.16 & 1--2 & 4.4 & 0.3480 & 0.2894 \\
0.16 & $>2$ & 6.8 & 0.4930 & 0.4238 \\
\hline\\
0.20 & 0.5--1 & 3.5 &  0.2828 & 0.2390 \\
0.20 & 1--2 & 4.4 & 0.3231 & 0.2835 \\
0.20 & $>2$ & 6.6 & 0.4644 & 0.4218 \\
\hline \\
0.25 & 0.5--1 & 3.5 & 0.2848 & 0.2444 \\
0.25 & 1--2 & 4.4 & 0.3162 & 0.2806 \\
0.25 & $>2$ & 6.6 & 0.4434 & 0.4160 \\
\enddata
\tablecomments{Calibration of the entire catalog in KABKE2 gives $\sqrt{C_{1,100}}\simeq 0.3\mu$K.}
\label{tab:calibration}
\end{deluxetable}

KAEEK have further modified the calibration procedure to account for the luminosity binning. There we generated CMB
temperatures from the KSZ effect for each cluster and estimated the
dipole amplitude, $C_{1,100}$, contributed by each 100 km $s^{-1}$
of bulk-flow in each $L_X,z$-bin. Since the $\beta$-model still
gives a fair approximation to cluster properties around
$\Theta_X$, we present in Table {tab:calibration} the final calibration
coefficients evaluated at apertures of $5^\prime$ and $\Theta_X$
in radius. When averaged over clusters of all X-ray luminosities
the mean calibration is $\sqrt{\langle C_{1,100}\rangle} \simeq
0.3 \mu$K in each of the $z$-bins as in KABKE2. Within the uncertainties, the
dependence of the calibration on $L_X$ is in good agreement with
the measured dipoles, particularly for the most accurately
measured $y$-component. Table \ref{tab:calibration} also shows the mean central
optical depth, $\langle\tau_0\rangle$, evaluated from the cluster
catalog. Its variation with $L_X$ is also in good agreement with
that of the measured dipole, which indicates that the clusters can
indeed be assumed to have similar profiles.

Both $\sqrt{C_{1,100}}$ and $\langle \tau_0\rangle$ scale approximately
linearly with the better measured dipole coefficient, $a_{1,y}$,
as they should in case of a coherent motion; the linear
correlation coefficients are $r=$0.92/0.93 for correlation of
$-a_{1,y}$ with columns 7/8. This then can allow for a translation of the measured dipole into an equivalent velocity.

\subsubsection{Sign of the filtered KSZ dipole}

A potentially more serious and still unresolved problem is as follows: the dipole is measured from the maps produced by DA's at different angular resolution (and noise) that, in addition, are individually filtered in order to reduce the primary CMB contribution to the dipole at cluster positions. The former determine whether, and with what $L_X$, clusters, located at different $z$, contribute as resolved or point sources. The latter affects not only the final cluster optical depth profile, but in filtered data, also the sign of it and the KSZ term as the different $L_X$-clusters contribute to the measured KSZ signal.

\begin{figure}[h!]
\includegraphics[width=6in]{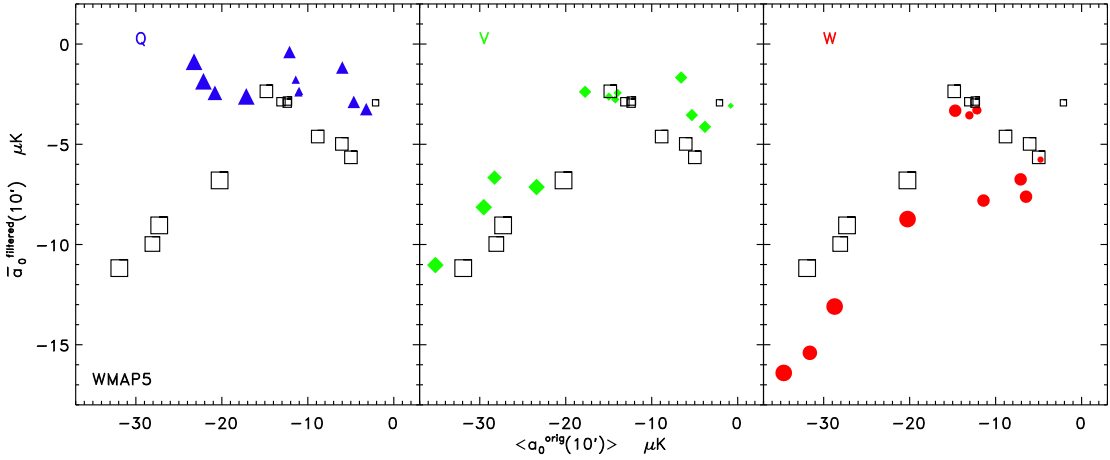}
 \caption[]{Central monopole in filtered maps is plotted vs that in the original unfiltered WMAP data. Blue triangles correspond to the average of two Q-band DA's, green diamonds to the average of two V-band DA's and red circles correspond to the average of four W-band DA's. Open squares show the mean of all WMAP bands. The size of the symbol is proportional to the $L_X$-bin from Table 1 of KAEEK. }
\label{fig:a0_origvsfil}
\end{figure}

Because of it  has the best angular resolution, the W channel four DA's are most suitable for calibrating the measured dipole from the filtered maps. Fig. \ref{fig:a0_origvsfil} shows the central monopole (with a 10$^\prime$ radius aperture) in filtered vs. unfiltered maps for the various $L_X$-bins (denoted by the different size symbols) considered in KAEEK. The figure demonstrates explicitly that in the Q-band DA's even the largest $L_X$-clusters remain unresolved. On the other hand, in the four DA's at W-band most of the clusters are at least partially resolved and this band presents the best dataset for calibrating the measured dipole signal.

\begin{figure}[h!]
\includegraphics[width=4in]{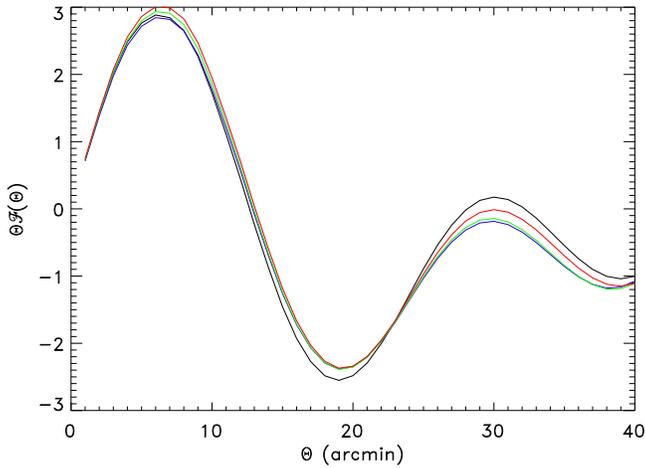}
\caption{ \small{${\cal F}(\Theta)\equiv \sum (2\ell+1) F_\ell B_\ell P_\ell(\cos\Theta)/\sum (2\ell+1) B_\ell$. Black, blue, green, red correspond to the W1, W2, W3, W4 DA channels respectively. Adapted from KAE.  }} \label{fig:kae5}
\end{figure}

The change of sign in the KSZ term can occur because we measure the dipole from the filtered maps, and the convolution of the intrinsic KSZ signal with a filter with wide side-lobes (as in KABKE) {\it can change the sign} of the KSZ signal for NFW clusters. The TSZ signal, which is more concentrated towards the inner cluster regions, will be less susceptible to this effect. The maps we use include SZ clusters and are convolved with the filter $F_\ell$ in $(\ell,m)$ space. This is equivalent to a convolution in the 2-D angular space $(\theta,\phi)$. After this convolution, the cluster properties clearly depend on the intrinsic profile of the clusters. As shown by AKKE, the latter are well described by an NFW model and are poorly matched by an isothermal $\beta$ model. Convolution will thus lead to different behavior of the SZ profiles, including the sign of the convolved SZ terms. In a measurement of the SZ signal from filtered maps, the intrinsic properties of clusters are first convolved with the beam ($B_\ell$ in $\ell$-space) and then with the filter. Fig. \ref{fig:kae5} (from KAE) shows this filtering function, ${\cal F}(\theta) \equiv \sum_\ell (2\ell+1) F_\ell B_\ell P_\ell(\cos\theta)/\sum _\ell (2\ell+1) B_\ell$, where $P_\ell$ are Legendre polynomials. Because the convolution is performed in two dimensions, ${\cal F}(\theta)$ is multiplied by $\theta$ in the figure.  The significant side-lobes of the filter can affect the sign of the KSZ term in the outer parts differently than the more concentrated (for NWF profiles) TSZ terms. Because of the particular form of the KABKE filter, the sign of the KSZ dipole measured from the inner parts ($<10^\prime-15^\prime$ in angular radius where ${\cal F}$ remains positive) would be the opposite of the one of the signal we measure from the final apertures ($\simeq 30^\prime$). Clearly, the sign of the KSZ dipole term would then depend on the cluster aperture at which the final signal is measured.

\begin{figure}[h!]
\includegraphics[width=7in]{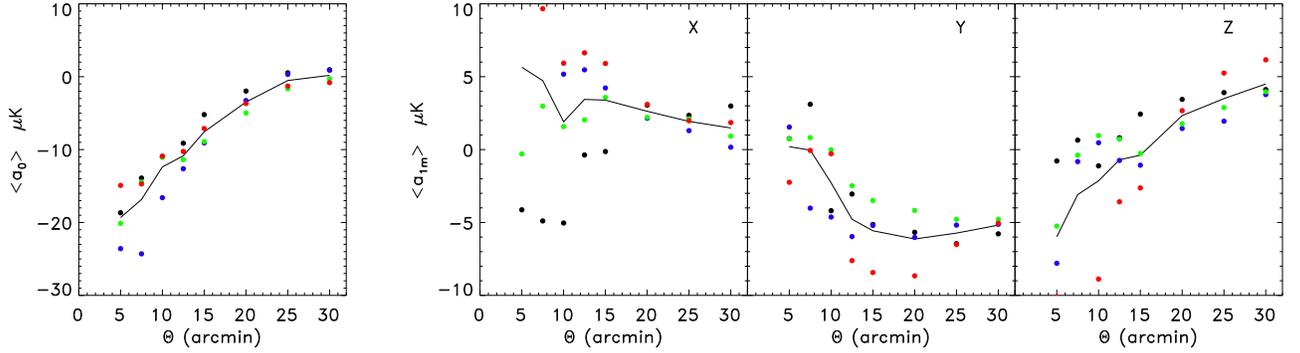}
 \caption[]{Monopole (left) and dipole components plotted vs the aperture radial size for the 7-year WMAP W band data. The numbers are evaluated at the
positions of $L_X\ge 2\times 10^{44}$ erg/sec and $z\leq 0.2$ clusters of the catalog used in KAEEK. Black, blue, green, red circles correspond W1, W2, W3, W4 DA channels respectively. Adapted from KAE.}
\label{fig:kae6}
\end{figure}
Do the data support this interpretation? The 7-yr WMAP data have sufficiently low instrument noise to probe the SZ signal in the inner parts of the (stacked) clusters. Using data restricted to the W channel four DAs, which provide the best angular resolution, we have evaluated the dipole in increasing cluster-centric apertures. The results are shown in Fig. \ref{fig:kae6} where we plot the monopole and the dipoles against the aperture size for each of the four W DAs. We used the same cluster catalog as KAEEK but limited to systems with $L_X\ge 2\times 10^{44}$ erg s$^{-1}$ which yield the highest S/N. Even though the TSZ contributions can not be subtracted until we have  recomputed fundamental cluster properties for an NFW model, the data, if noisy, indeed suggest a sign change as the aperture is increased. Note that the zero-crossing is consistent for the $Y, Z$ components where the measurement is statistically significant.

In this context we stress again (see also KABKE)  the importance of using the entire aperture containing the full extent of the X-ray emitting gas that gives rise to the SZ effect, if one is to measure a statistically significant signal.

While there is thus indeed evidence for a sign change in the KSZ signal which affects the direction of the KAEEK-measured dipole, we emphasize again that a definitive answer will have to await a more complete, expanded and recalibrated SCOUT catalog. Mathematically, the sign-change scale is very sensitive to the polytropic index (or its effective equivalent) of the compiled cluster sample and their distribution. The flow direction can, however, be determined from applications of the KA-B method to Planck data, taking advantage of Planck's angular resolution of 5$^\prime$ --- a good match to the inner parts of clusters out to the limit of the SCOUT catalog --- and of the mission's 217 GHz channel for which the TSZ component vanishes. Since Planck (as well as Chandra on the X-ray side) will resolve
SCOUT clusters all the way out to $z\sim 0.6$, modelling of cluster properties with NFW profiles will be possible specifically for the most X-ray luminous systems which contribute most strongly
to the dipole signal. This measurement was already proposed by us to the Planck mission and will be performed when the Planck data become public and the final SCOUT catalog is assembled.

With this caveat in mind, the tentative evidence of the sign flip from filtering, allows to constrain the direction along the axis of motion determined in KAEEK.

\subsubsection{Systematic bias in the current calibration}

In addition to the sign, there is a problem of systematic uncertainty induced by the calibration using the current cluster catalog based on the parameters evaluated for an isothermal $\beta$-model. In the calibration, when evaluating $C_{1,100}$ we
restricted our calculation to the central $1\theta_{\rm X-ray}$
where the $\beta$-model and NFW profiles differ by 10-30\% and
where the central values of the measured dipole are similar to the
values measured at the final aperture extent. In other words, we
assume that for each cluster all pixels measure the same velocity
(in modulus) across the sky, so the calibration constant, measured
from any subset of pixels is the same, irrespective of the signal
(in $\mu$K) measured at their location.

Filtering
reduces the effective $\tau$ by a factor of $\simeq 3$. As
mentioned above, since a $\beta$-model provides a poor fit to the
measured TSZ component outside the estimated values of
$\theta_{\rm X-ray}$, we computed $C_{1,100}$ with
the total extent assumed to be $\theta_{\rm X-ray}$. Owing to the large size of our cluster sample, the random uncertainties in the estimated values of $C_{1,100}$ should be small, but there would be a systematic
offset related to the difference between the filtering of NFW and $\beta$-modeled clusters. Since the filtering effectively removes the
profile outside, approximately, a few arcmin (typical value of $\theta_X$), it removes a more substantial amount of power
in the $\beta$-model when the cluster SZ extent is increased
beyond $\theta_{\rm X-ray}$, than in the steeper NFW profile measured
by us. Therefore, the effective $\tau$ is
{\it underestimated} by using a $\beta$-model. Nevertheless, the
calibration factor {\it cannot exceed} $\sqrt{C_{1,100}}\simeq 0.8
\mu$K given by that of the unfiltered clusters. The
above number for the calibration is {\it lowered} by filtering.
Filtering removes somewhat more power in the NFW clusters than in
the $\beta$-model, so the value of $\sqrt{C_{1,100}}=0.3 \mu$K for
filtered clusters appears to be a firm lower limit. We
also note that they only affect the accuracy of the determination
of the amplitude of the bulk flow, but cannot put its existence
into doubt which is established from the CMB dipole detected at
the cluster locations.

Table \ref{tab:calibration} illustrates further the systematic calibration uncertainty. Each of the quantities listed there (the mean central optical depth, $\langle \tau_0\rangle$, and the two sets of values of $C_{1,100}$ evaluated at $5^\prime$ and at $\theta_X$ using the $\beta$-modeling) can be used equally plausibly (at this stage) for calibrating the dipole in terms of the equivalent velocity. Yet the spread in the values of these quantities is such that, while the fact of the large scale coherent motion in excess of the prediction of gravitational instability remains, the exact value of that velocity is systematically uncertain by perhaps as much as $\sim 30\%$.

While the above already limits calibration to a relatively narrow
range, a more accurate determination of $C_{1,100}$ would require
an adequate knowledge, not yet available, of the NFW profile of
each individual cluster. It is not sufficient to know the average
profile of the cluster population (AKKE). Filtering acts
differently on the NFW-type clusters depending on their angular
extent and concentration parameter, i.e., the filtered mean
profile is not the same as the mean of all filtered profiles.
However, since $C_{1,100}$ was computed using the central pixels,
the region where the filter preserves the signal most  and where
both profiles differ less, we believe that our estimate of
$C_{1,100}\simeq 0.3\mu$K is fairly accurate, at least in the
sense that our overall cosmological interpretation holds within
the remaining uncertainties.

\subsection{CMB dipole and cluster motion}
\label{sec:results2df}

Bearing in mind the approximate nature of the current calibration scheme, we can now move to evaluating the velocity field implied by the measurement. The data strongly suggest a dipole flow as evident by the quadrupole (null) measurment shown in Sec. \ref{sec:quadrupole}.
The calibration results will follow the KAEEK presentation.
\begin{deluxetable}{c | c c c | c | c}
\tablewidth{0pt} \tabletypesize{\scriptsize}
\tablehead{ \colhead{Scale ($h_{70}^{-1}$Mpc)} & \colhead{$V_X$} & \colhead{$V_Y$} & \colhead{$V_Z$} & \colhead{$V_{\rm total}$} &
\colhead{$(l_{\rm Gal}, b_{\rm Gal})$ }}
\tablecaption{BULK FLOW RESULTS  from KAEEK. }
\startdata
$\sim 250-370$ & $174 \pm 407$ &  $-849 \pm 351$ & $348 \pm 342$ & $934 \pm 352$ & $(282\pm 34, 22 \pm 20)^\circ$ \\
$\sim 370-540$ & $410 \pm 379$ & $-1,012 \pm 326$ & $566 \pm 319$ & $1,230\pm 331$ & $(292\pm21, 27 \pm 15)^\circ$ \\
$\sim 380-650$ & $213 \pm 341$ & $-872\pm 294$ & $529 \pm 287$ & $1,042 \pm
  295$ & $(284\pm 24, 30 \pm 16)^\circ$ \\
$\sim 385-755$ & $313\pm 308$ & $-707\pm 265$ & $643\pm 259$ & $1,005\pm267$ & $(296 \pm 29, 39 \pm 15)^\circ$ \\
\enddata
\label{tab:vfromkaeek}
\tablecomments{Velocities are in km/sec assuming the motion in the direction of the dipole measured from the filtered maps and using calibration from KAEEK. }
\end{deluxetable}

Since the different cluster
subsamples probe different depths ($z_{\rm mean/median}$), KAEEK
proceeded as follows to isolate the overall flow across all
available scales: for the flow which extends from the smallest to
the largest $z$ in each $z$-bin, we model the dipole coefficients
(e.g. $y$) as $a_{1y}^n = \alpha_n V_y$, where $\alpha_n$ is the
calibration constant for cluster in $n$-th luminosity bin in Table \ref{tab:calibration}. Table \ref{tab:vfromkaeek} shows the bulk velocities recovered from this fit by KAEEK. The derived velocity field is consistent with a coherent flow pointing in the same direction out to the depth of $\sim 800$ Mpc.

\begin{figure}[h!]
\includegraphics[width=7in]{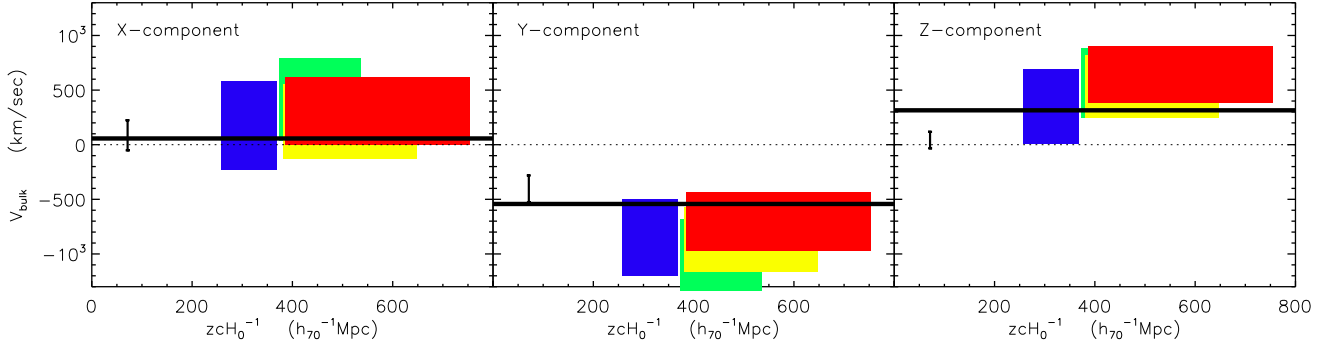}
 \caption[]{Three bulk velocity components from KAEEK vs CMB dipole. Error bar shows the 2-$\sigma$ range from Feldman et al (2010), thick horizontal line shows the CMB dipole velocity after correcting for local motions from Table 3 of Kogut et al (1993). Colors are from KAEEk as per Table \ref{tab:vfromkaeek}. Note that the Feldman et al (2010) and Watkins et al (2009) derived flow does not account fully for the overall CMB dipole.}
\label{fig:kaeek}
\end{figure}

Fig. \ref{fig:kaeek} plots the velocity components obtained from this calibration and compares the derived velocity field with that from the CMB dipole anistropy converted into equivalent velocity. The CMB dipole has been corrected for the Local Group motion as described in Kogut et al (1993). Even within the systematic uncertainties (overestimate) of the current calibration, the agreement between the two dipoles is remarkably good and is consistent with the CMB dipole not converging until at least $\sim 800$ Mpc. If the $\beta$-modeling of the clusters in the catalog indeed produced an upward bias in the derived velocities of $\sim 20-30\%$ the agreement with the overall CMB dipole will be even more remarkable.

\subsection{Further tests of "systematics"}
\label{sec:systematics}

We have shown that our signal (a) is present exclusively at cluster locations,
(b) is measured at zero monopole and (c) correlates with cluster X-ray luminosity
so clusters with the largest TSZ effect at the center show the largest dipole.
If our measurement is due to a systematic effect, it will have to
display the three properties. Any systematic coming from
instrumental noise (not necessarily white noise) or intrinsic CMB
will not correlate with cluster luminosity. This systematic
must necessarily come from the TSZ component.
Given that we evaluate the dipole at the aperture where the monopole
vanishes, there are three ways that could potentially confuse the measurement:
1) Systematic effects that could fold the Doppler shift of our local motion
into the tSZ contributions, 2) cross-talk effects between the
tSZ monopole and dipole terms in sparse/small samples (Watkins \& Feldman 1995);
and 3) inner motions of the intracluster medium (ICM) as opposed to the
coherent flow of the entire cluster sample. We now proceed to demonstrate
that any systematic that might arise due to either of this effect is negligible.
Foremost, let us remark that any TSZ induced dipole will be frequency dependent.
However, given our error bars and the small variation of the TSZ at
WMAP frequencies for the Differencing Assemblies (DA) Q, V and W
we can not use this frequency dependence to discriminate a TSZ origin
of the measured dipole.

\subsubsection{Dipole induced by a Doppler shift of the TSZ signal.}

The motion of the Sun with respect to the CMB frame has been accurately
determined to be: $u_\odot=370\pm 3$km/s in the direction $(l,b)=(264^0,48^0)$.
close to the direction $(276^\circ, 30^\circ)$ of the Local Group with respect to the same
reference frame (Kogut et al, 1993). It is also close to the direction
of the dark flow (see Table \ref{tab:vfromkaeek}). If a small systematic effect in
the construction of CMB maps from the Time Ordered Data (TOD) couples
the TSZ cluster and the kinematic solar dipole, it could produce
a residual dipole that would verify the three properties described above.
First, this residual dipole $(\Delta T)_{res}$ would be bound
by $(\Delta T)_{res} < (\Delta T)_{TSZ}(u_\odot/c)$. In AKKE we
measured $(\Delta T)_{TSZ}\sim -30\mu$K for our cluster sample and
this amplitude is reduced a factor $\sim 3$ due to filtering, then
$(\Delta T)_{res} < 10^{-2}\mu$K, more than two orders of magnitude
smaller than the measured sample.

A different kind of systematic could come from the calibration of
the time-ordered data (TOD). The CMB monopole and the velocity dipole resulting from
WMAP orbit around the solar system are used in the calibration algorithm
(Jarosik et al 2011).
After one year of integration, the velocity dipole will trace the
CMB dipole. Any undetected offset that couples with the measured
the measured CMB anisotropies could have a larger offset at cluster
locations, larger for the more massive clusters, that would be
dipolar in nature. If our dipole had this origin, then it will show
a dependence in frequency, that at present we can not established,
but also will increase in time. Any systematic effect will increase
when adding data for several years. Comparing the measured dipole
for our original analysis of WMAP 3, 5 and 7 years of data, our dipole
would be a factor $\sqrt{7/3}\approx 1.5$ times larger, going from
$\sim 3\mu$K to $\sim 4.5\mu$K. There was no increase in the
dipole we measured in the different WMAP data releases, so we can
also disregard that our measurement is due to a systematic of this type.

\subsubsection{Cross-talk between monopole and dipole terms.}

Since clusters are not randomly distributed on the sky, the TSZ
signal will give rise to a non-trivial dipole signature that, in
principle, may confuse the KSZ dipole. The TSZ dipole for a random
cluster distribution is given by $a_{1m}^{\rm tSZ}\sim \langle
(\Delta T)_{\rm tSZ}\rangle \large({3/N_{\rm cl}}\large)^{-1/2}$
decreasing with increasing $N_{\rm cl}$. This decrease could be
altered if clusters are not distributed randomly and there may be
some cross-talk between the monopole and dipole terms especially
for small/sparse samples (Watkins \& Feldman 1995). As discussed
in KABKE2, the dipole from the TSZ component varies with the
cluster sub-sample, contrary to measurements, and also has
negligible amplitude because it is bound from above by the
remaining monopole amplitude of $\langle (\Delta T)_{\rm
TSZ}\rangle\ll 1\mu$K measured at the final aperture.

In order to assess that there is no cross-talk between the
remaining monopole and dipole which may confuse the measured kSZ
dipole, KABKE2 and AKEKE conducted the following experiment: 1) The TSZ and KSZ
components from the catalog clusters were modeled using cluster
parameters derived for our current catalog. To exaggerate the
effect of the cross-talk from the TSZ component, the latter was
normalized to $\langle (\Delta T)_{\rm TSZ}\rangle=-1\mu$K, a
value significantly larger than the monopoles in Table 1 of KAEEK
at which the final dipole was measured; the results for even
larger monopoles were also computed and can be scaled as described
below. For the KSZ component each cluster was given a bulk
velocity, $V_{\rm bulk}$, in the direction specified in Table 1 of
KAEEK, whose amplitude varied from 0 to 2,000 km/sec in 21
increments of 100 km/sec. The resultant CMB map was then filtered
and the CMB dipole, $a_{1m}({\rm cat})$, over the cluster pixels
computed for each value of $V_{\rm bulk}$. 2) At the second stage
we randomized cluster positions with $(l,b)$ uniformly distributed
on celestial sphere over the {\it full} sky for a net of 500
realizations for each value of $V_{\rm bulk}$. This random catalog
keeps the same cluster parameters, but the cluster distribution
now occupies the full sky (there is now no mask) and on average
does not have the same levels of anisotropy as the original
catalog. We then assigned each cluster the same bulk flow and
computed the resultant CMB dipole, $a_{1m}({\rm sim})$, for each
realization. The final $a_{1m}({\rm sim})$  were averaged and
their standard deviation evaluated.

\begin{figure}[h!]
\includegraphics[width=6in]{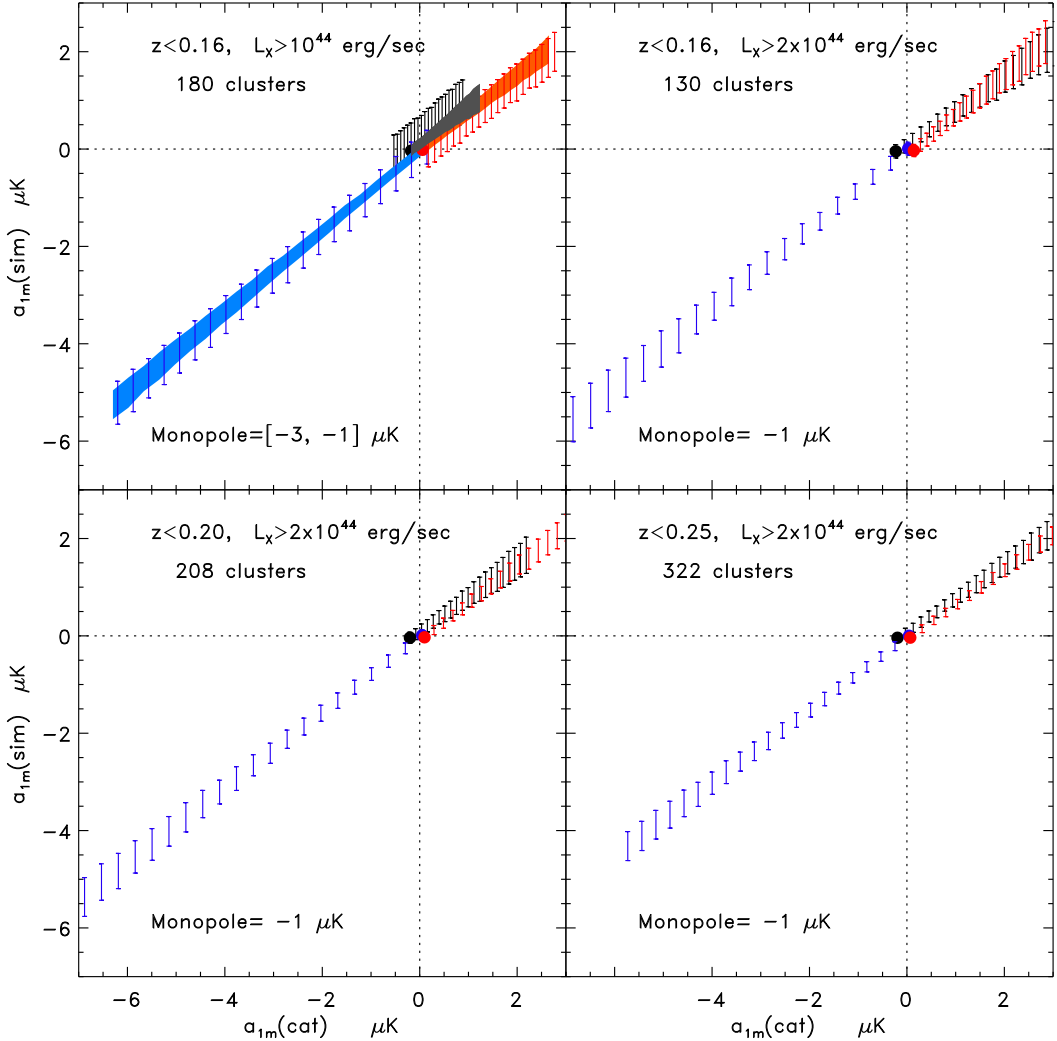}
 \caption[]{The dipole coefficients for
simulated cluster distribution (random and, on average, isotropic)
are compared to that from the true catalog. (See text for
details). Each cluster in each catalog is given bulk flow of
$V_{\rm bulk}$ from 0 to 2,000 km/sec in increments of 100 km/sec
towards the apex of the motion from Table 1 of KAEEK. The results
from 500 simulated catalog realizations were averaged and their
standard deviation is shown in the vertical axis. Dotted lines
mark the zero dipole axis of the panels. The four most sparse
samples from Table 1 of KAEEK are shown which correspond to the
largest $L_X$-bins giving the best measured S/N. Black/blue/red
colors show the x/y/z components of $a_{1m}$. Filled circles of
the corresponding colors show the dipole components due to the
modelled tSZ component. The upper left panel shows the results for
two values of the monopoles: in the case of $\langle \Delta T_{\rm
tSZ}\rangle=-3 \mu$K the results are shown as individual error
bars; the case of $\langle \Delta T_{\rm tSZ}\rangle=-1 \mu$K is
shown with filled contours. All other panels show the results for
$\langle \Delta T_{\rm tSZ}\rangle=-1 \mu$K and our simulations
find good scaling with higher monopole values as described in KABKE2 and AKEKE. Adapted from AKEKE.}
\label{fig:monleak}
\end{figure}

Fig. \ref{fig:monleak} shows the comparison between the two dipoles for
each value of $V_{\rm bulk}$ for the most sparse sub-samples from
Table 1 of KAEEK. In AKEKE we also made the computations at TSZ monopole
values still larger than above (see upper left panel for one such
example). The overall contribution from the TSZ component to the
dipole is $\propto \langle \Delta T_{\rm TSZ}\rangle$, so in the
absence of cross-talk effects the amplitude of the scatter in the
simulated dipoles is made of two components: 1) remaining TSZ
$\propto \langle \Delta T_{\rm TSZ}\rangle$ and 2) genuine KSZ
dipole with amplitude $\propto V_{\rm bulk}$ to within the
calibration. One can see that there is no significant offset in
the CMB dipole produced by either the mask or the cluster true sky
distribution. The two sets of dipole coefficients are both
linearly proportional to $V_{\rm bulk}$ and to each other; in the
absence of any bulk motion we recover to a good accuracy the small
value of the TSZ dipole marked with filled circles. As discussed
in KABKE2, since the bulk flow motion is fixed in direction and
the cluster distribution is random, one expects the calibration
parameterized by $C_{1,100}$ to be different from one realization
to the next, e.g. in some realizations certain clusters may be
more heavily concentrated in a plane perpendicular to the bulk
flow motion and the measured $C_{1,100}$ would be smaller. In our
case, the mean $C_{1,100}$ differs by $\la 10\%$ suggesting that
our catalog cluster distribution is close to the mean cluster
distribution in the simulations. This difference in the overall
normalization would only affect our translation of the dipole in
$\mu$K into $V_{\rm bulk}$ in km/sec, but we note again the
systematic bias in the calibration resulting from our current
catalog modeling clusters as isothermal $\beta$-model systems
rather than the NFW profiles required by our observations .
In addition we note also the diagonality of the dipole covariance matrix, ${\cal A}_{mn}$, in the filtered maps.
\clearpage
\clearpage

\subsubsection{Directly measured dipole over cluster pixels}

\begin{figure}[h!]
\includegraphics[width=7in]{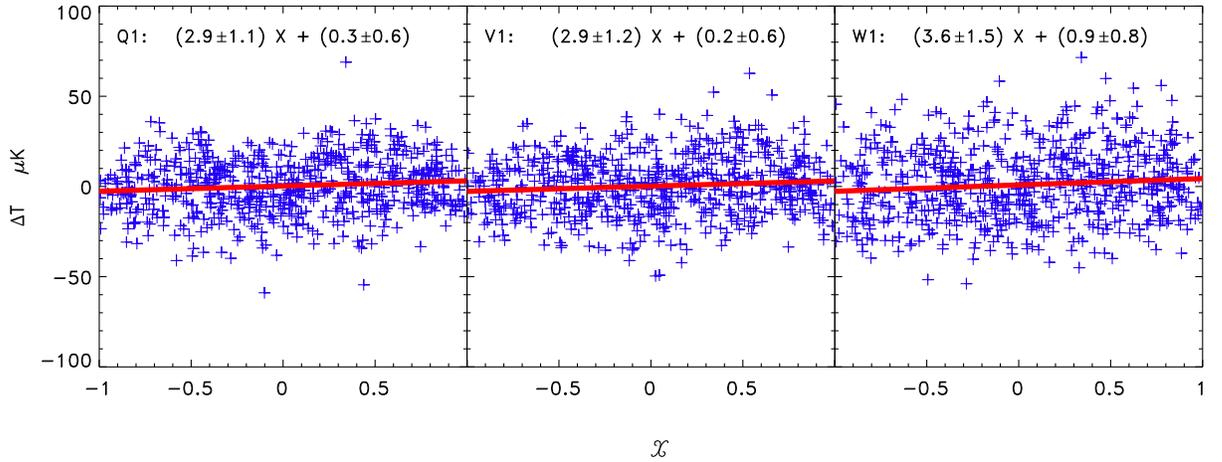}
 \caption[]{CMB temperature averaged over the final $30^\prime$ aperture (blue plus signs) for each of 674 clusters out $z\leq 0.3$ is plotted
 vs ${\cal X}$, the cosine of the angle between the dipole apex and each cluster. The plots
 are shown for one DA channel at each frequency. The linear fit to the data
 is shown with thick red line; its parameters and their uncertainties are displayed at the top
 of each panel. The uncertainties in the displayed fits were computed using uniform weighting. Adapted from KABKE2 who used WMAP 3yr data. }
\label{fig:kabke2_8}
\end{figure}

As shown in KABKE2, the reality of the measured dipole can also be seen in from the
following test: In Fig. \ref{fig:kabke2_8} we show present the measured signal
of the entire cluster sample ($z\leq 0.3$) plotted against $X$,
the cosine of the angle between the detected dipole and the
cluster itself for three channels at three different frequencies
(Q1, V1, W1). The figure is reproduced from KABKE2 who used WMAP 3yr data. For each cluster the CMB temperature was averaged
over the cluster pixels out to the fixed aperture corresponding to zero monopole.
Results from linear fits (thick solid lines) to the data and their
uncertainties are displayed in each panel. As expected there is a
statistically significant dipole component in the cluster CMB
temperatures. In each of the eight channels the significance is $>
2 \sigma$ already for WMAP 3 yr data leading to the overall result.

\subsubsection{Inner motions of the Intra-Cluster Medium.}

The intra-cluster medium (ICM) may not be at rest in
the cluster potential wells as a result of mergers during cluster formation
process. In principle, our measurement and interpretation then may be affected
by turbulent motions that give rise to a kSZ effect that
would be larger for the more massive clusters. However, since the motions are randomly
oriented with respect to the line of sight, they will not produce a
significant effect. In order, to reach the value comparable to
$V_{bulk}\sim 1,000$km/s. In a sample of $N_{cl}$, a typical cluster
would need to have thermal motions of an amplitude
$\sim V_{bulk}N^{1/2}_{cl}$, to produce an effect. This random
velocities are about one order of magnitude larger than the
velocity dispersion of Coma-type clusters and would preferentially
give rise to a monopole contribution.
Rather, these motions will enter the overall dispersion budget
(noise, gravitational instability and this component) around the
measured coherent bulk flow.

\subsection{Comparison with post-DF distance indicator results: an overall consistency or conflict?}
\label{sec:df_comparison}

We now turn to discuss comparison of our measurements with other post-DF studies which used alternative methods of velocity determination. We will try to follow the discussion chronologically, whenever possible.


Watkins, Feldman \& Hudson (2009) developed a method to minimize the variance between the measured {\it line-of-sight} flow velocity and the flow, $U_\alpha$, when combining data from many surveys of different selections and geometries and measuring the resultant bulk flow with minimal sensitivity to small scale power:
\begin{equation}
\langle [\sum w^\alpha_i(\vec{v}_{p,i}\cdot \hat{r}_i) - U^\alpha]^2\rangle=\min
\label{eq:mv}
\end{equation}
The weights for the bulk velocity determined in this way depend on the underlying power spectrum of the velocity flow (Watkins et al use $\Lambda$CDM model with standard cosmological parameters).
In their post-KABKE1, but independent, study Watkins et al (2009) combined the data from nine velocity surveys available to-date, demonstrated their mutual consistency (except for the Lauer \& Postman (1994) sample) and measured a bulk flow at the level $407\pm 81$ km/sec at scale $\sim 50h^{-1}$Mpc in the direction of only $\simeq 6^\circ$ away from the central dipole direction of KABKE1. They suggested that the flow they measured results ``from the local density field" and did not consider the tilt possibility.

Lavaux et al (2010) generate the peculiar velocity field for the 2MASS Redshift Survey (2MRS)
catalog using an orbit-reconstruction algorithm based on the Zeldovich (1970) approximation.
Their reconstructed velocities of individual objects are well correlated with the peculiar velocities
obtained from high-precision distance measurements within 3,000 km/s. The reconstructed motion
of the Local Group in the rest frame established by distances within 3,000 km/s agrees with
the observed motion and is generated by fluctuations within this volume. They also analyze
the velocity field in successively larger radii, to study its convergence towards the CMB dipole.
They found that while most of the amplitude of the CMB dipole seems to be recovered by 120 $h^{-1}$Mpc,
its direction is more than $50^\circ$ away the CMB dipole direction. The misalignment angles
are significantly larger than anticipated by the $\Lambda$CDM model on scales larger than
50-60 $h^{-1}$Mpc. Contrary to Erdogdu et al. (2006) but more in agreement with Pike \& Hudson (2005),
they show that the depth of the convergence in 2MRS lies beyond 120 $h^{-1}$Mpc. Due to severe
incompleteness of the 2MRS catalog beyond 120Mpc/h, they could only place a lower limit
on the value of the convergence depth. For instance, at 150 $h^{-1}$Mpc, the amplitude was
35\% smaller than the CMB dipole and direction is almost $60^0$ away, with no sign of
convergence. 
The Shapley concentration is not well sampled in 2MRS, so the authors
conclude that deeper redshift surveys are required to see if the Shapley concentration around
150 $h^{-1}$Mpc is enough to reach convergence or one has to go well beyond that distance.

Feldman et al (2010) extended the Watkins et al (2009) analysis to include measurement of higher moments of the velocity field they detected earlier, such as shear. They find from their method
that the flow is driven by mass concentration(s) more distant than
the Great Attractor, in the same general direction, and conclude that the absence of any detectable shear in their flow implies that the attractor is {\it at least} $300h^{-1}$Mpc away and the data are consistent with it being at infinity. The results of their bulk-flow measurement (which supersedes the earlier Watkins et al result from the same group) are shown in Fig. \ref{fig:kaeek}. Note that, unlike the KAEEK results,  the flow derived from Feldman et al (2010) (and Watkins et al 2009) does not account fully for the overall CMB dipole.


The result of Watkins et al. (2009) claiming a
bulk flow of amplitude $407\pm 81 {\rm km s^{-1}}$ within $R = 50 h^{-1}$ Mpc,
has been challenged by the re-analysis of the same data by Nusser \& Davis (2011).
Contrary to Watkins et al. (2009) who used all the available data sets,
Nusser \& Davis (2011) restrict their analysis to the SFI++ survey
of spiral galaxies in the $I$ photometric band (Springob et al 2007) with distances estimated using
the Tully-Fisher (TF) relation. The survey has an effective depth of only $\sim 30h^{-1}$Mpc. Galaxies fainter than $M = 20$ were removed
to ensure the linearity of the TF relation.
Nusser \& Davis (2011) find a bulk flow amplitude consistent with $\Lambda$CDM expectations and claim that their discrepancy with
Watkins et al (2009) comes from the latter combining several catalogs
because minor miscalibration errors between different catalogs could
lead to large artificial flows when these catalogs are combined. But is then hard to understand why  such a bias, if it exists, would be aligned with the directions of the CMB dipole and/or DF.

Lavaux \& Hudson (2011) have studied convergence of the velocity field i9mplied by the peculiar gravity reconstructed from a compilation of the data from three redhsift surveys (2MRS, SDSS and 6dFGRS). After modeling out the ZoA effects, they find that out to $\sim 200 h^{-1}$Mpc the most prominent attractor is the Shapley concentration at $\sim 150 h^{-1}$Mpc away in the direction of $(l, b)_{\rm Shapley} = (312^\circ, 30^\circ)$ and it can account for about 90 km/sec of the LG motion with respect to the CMB rest frame. The plan to continue with the compiled catalog ``in the hope that the distribution of density in [its] volume will account for the [observed] high amplitude bulk motions on scales of $100 h^{-1}$Mpc".

Ma, Gordon \& Feldman (2011) combined peculiar velocity from five different samples containing various galaxy surveys and a SN sample containing $\sim 10^2$ objects from the compilation by Tonry et al (2001). They subtracted CMB dipole from each galaxy and then modelled the residual line-of-sight velocity as a sum of the tilted - dark flow - velocity, which is invariant with distance, and a random component with dispersion given by the gravitational instability within the concordance $\Lambda$CDM model. They a statistically significant dark flow component which is consistent between the five samples, including from their SN-compilation. The dark flow (tilted) component for the samples used there is around $V_{\rm tilt}\sim 300-500$ km/sec with typical uncertainty of about 30\%. The direction of the flow evaluated in each sample is in good agreement with that in KABKE1 and KAEEK. Within their errors these numbers are consistent with our results within the systematic calibration uncertainties of the dark flow measurement.

On the other hand, results from studies using exclusive SN samples are less in agreement with the DF and several of these claim not to have measured any significant flow at very large distances ($\gsim 100 h^{-1}$Mpc for those samples). Because of the potential promise of SNIa samples in the DF study we devote a separate sec. \ref{sec:sn} to discussion of the current SN-based results in detail, but briefly summarize here what the various SN-based results are as of this writing: Colin et al (2011) analyzed the Union2 SN dataset and their results are consistent with the Watkins et al flow and with the DF on larger scales, where the errors are large. With a smaller sample Jha et al (2007), Haugb{\o}lle et al
(2007) and Weyant et al (2011) found bulk flows of amplitude $\sim 500$km/s in a direction consistent with that of the Dark Flow. Ma et al (2011) find a statistically significant tilt velocity of $\simeq 450\pm 180$ km/sec in the direction which coincides with that in Table \ref{tab:vfromkaeek}. On the other hand, Dai et al analyzing the same Union 2 dataset find no evidence of DF at $z\gsim 0.06$, although the Union 2 dataset is highly anisotropic and inhomogeneous on the sky for these redshifts. Turnbull et al (2011) use a larger compilation of SNIa sources, compare their bulk velocity to the gravitational acceleration from the IRAS PSCz source catalog and find only a small, but statistically significant, excess velocity of $\simeq 150\pm 43$ km/sec in the direction which is only marginally consistent with that of Table \ref{tab:vfromkaeek}. We note that such modeling involves reconstruction of the gravitational field from regions (and depths) which are not the same where the SNIa sources have been measured. The situation from the SNIa measurements is thus confusing  and the two recent results claim to be in disagreement with our measurement, although the methodology there is not without assumptions as we discuss in more detail in  sec. \ref{sec:sn}.

Abate \& Feldman (2011) used luminous red galaxies from the SDSS as probes of peculiar flows. They argued that, because these systems present a isolated set of passively evolving populations, one can such magnitude fluctuations in these objects as distance indicators. After calibrating to the nearby luminous red galaxies, Abate \& Feldman find a statistically significant residual flow at $z\gsim 0.08$ in the same direction as Table \ref{tab:vfromkaeek}. With the current calibration of this novel method the flow appears to be (unrealistically?) large at $\sim 4,000$ km/sec, although this may be a  result of the imperfection of their calibration or an artifact of systematics, as they discuss.

Also, the recent study of Wiltshire et al (2012) with over 4,500 galaxy distances out to $\sim 100h^{-1}$Mpc provides support for the DF by finding that the Hubble flow is significantly more uniform and isotropic in the reference frame of the Local Group motion as opposed to that of the CMB dipole (see more discussion in Sec. \ref{sec:tests}.

In summary, we note that the DF, if reflective of the primordial CMB dipole, does not generate the Kaiser (rocket) effect of distortions of the measured galaxy clustering in the $z$-space and cannot be probed analyzes such as Song et al (2011). Rather, these distortions would constrain the {\it gravitational instability component} of the peculiar flows and their low limits on this component could directly imply that the flow measured by Watkins et al has its origin in the tilt consistent with the measurement and origin in KABKE1.

\subsection{Summary: the "dark flow"}
\label{sec:df_summary}

 \begin{figure}[h!]
\includegraphics[width=6in]{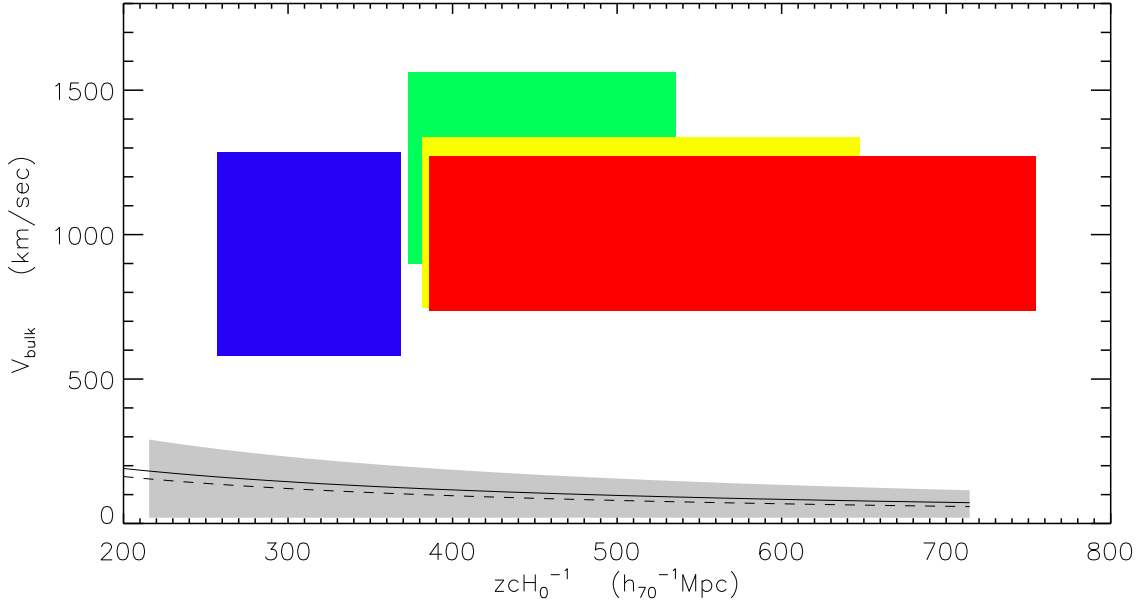}
 \caption[]{The bulk flows per Table \ref{tab:vfromkaeek}. Shaded grey region shows the velocity that 95\% of cosmic observers would measure in the standard $\Lambda$CDM model. Black and dashed solid lines correspond to the rms expectation for top-hta/Gaussian windows assuming the standard $\Lambda$CDM model. Reproduced from KAEEK. }
\label{fig:kaeek}
\end{figure}
We now summarize the evidence for a coherent flow of galaxy clusters with respect to the CMB dipole frame which extends to at least $\sim 800$ Mpc. The evidence is summed up below and is based on interpreting the measurement of a statistically significant dipole at cluster locations; no alternative explanation for the measurement has been suggested as of now. The reasons for this conclusion are as follows:
\begin{enumerate}
\item Dipole appears only at cluster positions at high significance ($\sim 4\sigma$ for the brightest clusters).
\item The CMB pixels within the identified clusters display a negative monopole, $\langle\delta T\rangle<0$ so their CMB signature arises from hot X-ray gas these clusters contain.
\item The final dipole signal is measured at zero monopole and hence the TSZ contribution to the dipole is small (and limited in magnitude by the residual monopole).
\item The amplitude of the measured dipole correlates well with the $L_X$-threshold of cluster sample which is consistent with its SZ origin and is highly unlikely to have been produced by some putative systematics.
\item The dipole disappears at larger apertures as shown in Fig. 10 of KABKE2.
\item An all-sky filter cannot imprint a dipole with the above properties whereas inappropriate filtering can reduce the KSZ component enough to make the measurement (with such a filter) statistically insignificant.
\item The direction/axis {\it and} amplitude of the flow agree remarkably well with the galaxy distance indicator measurements on much smaller scales.
\end{enumerate}

\begin{figure}[h!]
\includegraphics[width=3.5in, angle=+90]{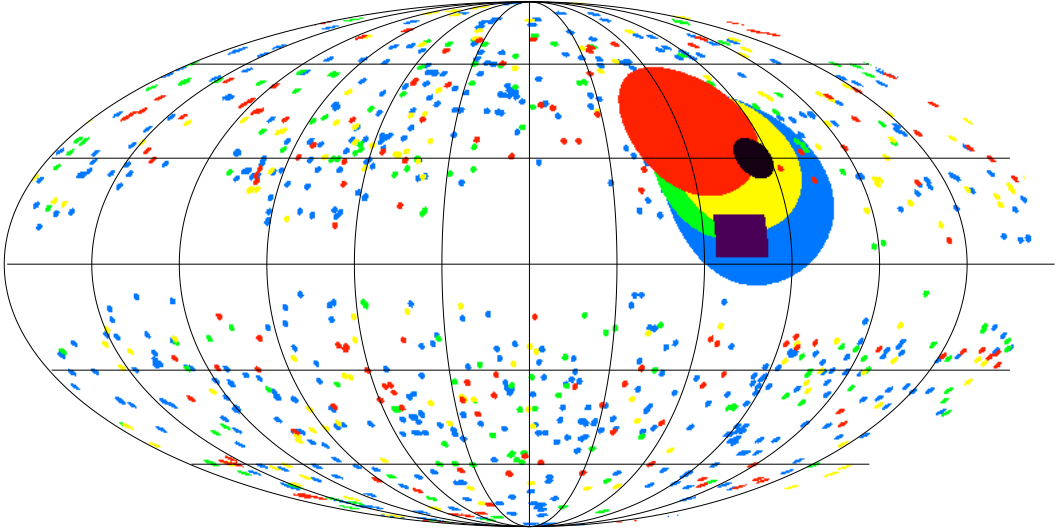}
 \caption[]{Filled colored circles correspond to the error contours on the direction of the flow from Table \ref{tab:vfromkaeek}. Filled dots (each encompassing a circle of 0.5$^\circ$ radius approximately corresponding the final aperture size) show with the same color the cluster samples that went into each measurement. Colors match Fig. \ref{fig:kaeek}.  The CMB dipole after correcting for local motion as per Table 3 of Kogut et al (1993) is shown with black circle of $6^\circ$ radius (for clarity) which corresponds to approximately 2-$\sigma$ uncertainty from Table 3 of Kogut et al (1993). The direction from Watkins et al (2009) and Feldman et al (2010) at $\sim 50h^{-1}$Mpc is shown in violet.}
\label{fig:directions}
\end{figure}

The overall probability the signal being produced by chance appears to be low when everything is considered: 1) the detection of the KSZ dipole for the brightest clusters is at $\sim (3.5-4)\sigma$ level with Gaussian probability of better than $P_1\sim 10^{-4}$, 2) the probability of the detected dipole coinciding so well with the independent measurements at small scales as per Fig. \ref{fig:directions} is $P_2 \sim 10^{-2}$ given the direction errors in Table \ref{tab:a1mfromkaeek}, 3) the probability of the found amplitude agreeing with that of the CMB dipole motion is harder to quantify, but is $P_3 <1$, and 4) the probability of the detected CMB dipole at cluster positions correlating with the cluster $L_X$-bin threshold as per Figs. \ref{fig:kaeek1c}, \ref{fig:kaeek_c1} is hard to quantify but is likewise $P_4<1$. This leads to the net probability of the detected signal being to chance of only $P_{\rm total} = P_1 P_2 P_3 P_4 \lsim 10^{-6}$.

With the sign tentatively measured in Fig. \ref{fig:kae6} the bulk flow motion is roughly in the direction of the detected CMB dipole at cluster positions; it is given in Table 1 of KAEEK  and the present data make little difference to it. Within the errors this direction will then coincide with the direction of the flow from Watkins et al at smaller scales ($\lsim 100$ Mpc) suggesting a coherent flow from the sub-100 Mpc scales to those probed in KAEEK ($\sim 800$Mpc).

As mentioned above Waktins et al (2009) and Feldman et al (2010) assign the motion to gravitational instability from structures within the observable Universe, although Feldman et al (2010) note that the absence of shear in their flow point to the attractor lying at least $\> 300h^{-1}$Mpc away. This conclusion is also reiterated by Abate \& Feldman (2011), although Feldman et al (2010) themselves admit that their results implies the attractor can be at infinity. On the other hand, if the Watkins flow was indeed due to matter inhomogeneities within the observable Universe, it should give appreciable redshift distortion. No such distortion has been observed by Song et al (2009), which then seems at odds with the gravitational instability interpretation of the Watkins et al result and would argue for the ``dark flow" interpretation. Still, two recent analyses of the SNIa samples fail to find such high amplitude motions, or any statistically significant motions at very large scales.  This does present a challenge to the ``dark flow" interpretation of the motions, although we propose the possible reasons for this discrepancy in the following sections. (We further note that two other SN-based analyses find results consistent with the DF as we discuss later in Sec. \ref{sec:sn}).

\subsection{Criticisms - addressed and resolved}
\label{sec:rebuttals}

Three challenges to the dark flow measurement have appeared as of this report writing: one as unrefereed posting, and two in peer-reviewed literature (Keisler 2009, Osborne et al 2011). The unrefereed posting was fully refuted elsewhere \\ (\url{http://www.kashlinsky.info/bulkflows/Wright\_is\_wrong}) and is not addressed below.

\begin{figure}[h!]
\plotone{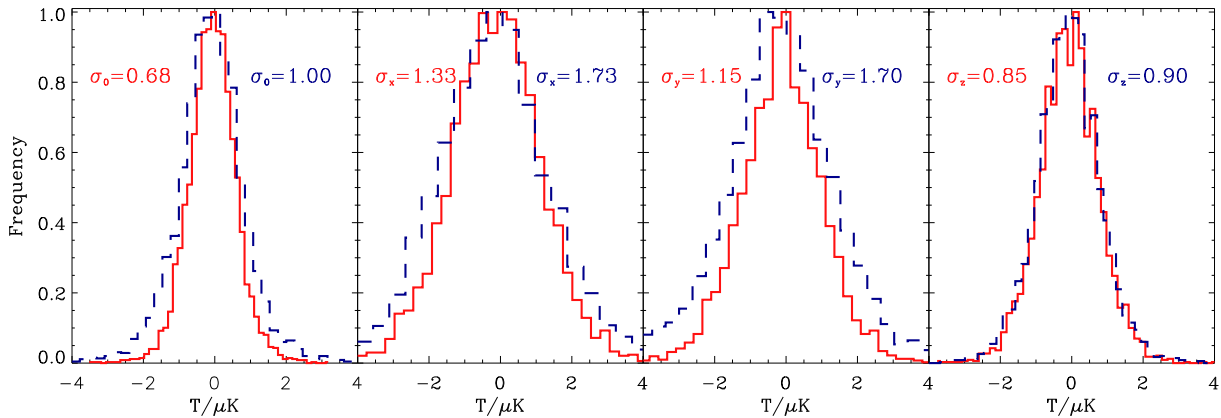} \caption{\small{Histograms  of 4,000
realizations of the CMB sky using Q1 DA parameters. Solid, dashed
lines correspond S1, S2 simulations; in S1 (S2) the monopole and
dipole outside the mask are (are not) removed. The left,
right rms dispersion corresponds to S1, S2, respectively. The figures
are given in micro Kelvin. Adapted from AKEKE.
 }} \label{fig:akeke5}
\end{figure}

Keisler (2009, ApJ, 707, L42) replicated the KABKE1,2 analysis using
a cluster catalog from public data. He confirmed the
central dipole values measured by KABKE2, but claimed larger errors after correctly identifying the correlations between the CMB remaining in the different DA's after the filtering which were addressed concurrently by KAEEK and AKEKE. However, Keisler (2009) claimed an increase
in errors by a factor of $>\sqrt{20}$ compared to KABKE2. Clearly,
the effect of residual CMB correlations between the $N_{DA}=8$
WMAP channels can at most increase the KABKE1,2 errors by a factor
of $\sqrt{N_{DA}}< \sqrt{8}$. (In reality, because the instrument
noise is also present, the errors on individual dipole components
in Table 2 of KABKE2 would be increased for 3-year WMAP data by a
factor of $\simeq \sqrt{6}$ to become $\la 1\mu$K at the largest
redshift bin.) A larger increase, as we have demonstrated above,
cannot happen. AKEKE have since shown that the latter are due to Keisler not having removed the monopole and dipole from the
CMB maps outside the mask; the AKEKE identification was confirmed
by Keisler in private correspondence. This is  shown in Fig. \ref{fig:akeke5} adopted from AKEKE.

\begin{figure}[h!]
\includegraphics[width=4in]{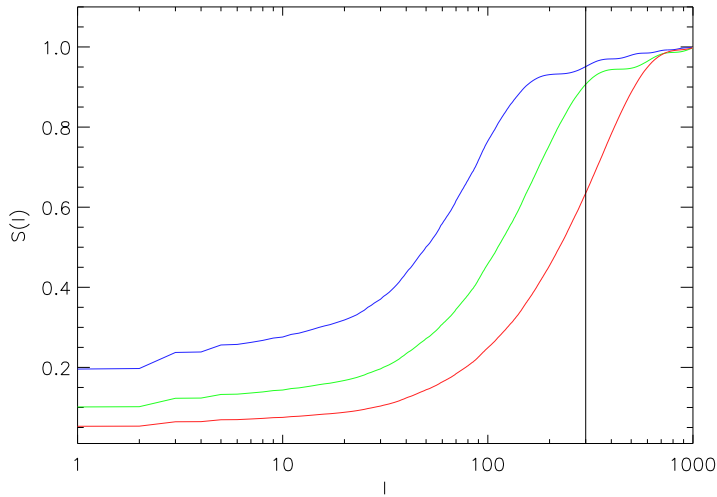} \caption{\small{We measure a dipole at 600 clusters randomly distributed on the sky outside the CMB Kp0 mask. The contribution of each $\ell$-multipole to the power of the dipole S in  $\ell$-space is $S^2(\ell)=\sum_{i=2}^\ell (2i+1)C_i$. In the plot we see how the signal builds up. Green, yellow and grey lines correspond to clusters having 60, 30 and 15 arcmin radius. In all cases, more than 50\% of the signal comes from multipoles $\ell<300$. The filters employed by Osborne et al (2011) are lower pass filters, effectively removing power at $\ell<300$, so these authors are dismissing a large fraction (most, in fact) of the signal.
 }} \label{fig:rebut_omcp}
\end{figure}

Osborne et al (2011) replicated the
KA-B approach using different filters, designed in other studies to detect
radio point sources rather than remove the primary CMB, the main contaminant in
the KA-B method.  Since the KSZ dipole is present at cluster locations, a
fraction of less than 1\% of the sky, the dipole in Fourier space is
spread over all multipoles.  The filters employed by Osborne et al reconstruct the KSZ signal
from $\ell\gsim 300$, neglecting the 85\% of the signal present at $\ell < 300$ (see their Fig. 3) and, at the same time, they boost the power at high $\ell$'s to account for the missing signal. This is demonstrated quantitatively in Fig. \ref{fig:rebut_omcp} and renders the KA-B method useless if instrument
noise and primary CMB are added since these two components, that dominate
the error bar, are also boosted. As a result, the Osborne et al filters appear to
measure a KSZ dipole only when their sample of $\sim 700$ clusters is moving almost at
relativistic velocities. Fig. 13 of their paper shows that the original input velocity is recovered only at $V\gsim 6,000$ or $\gsim 10,000$ km/s for their two filters. At such velocities, the KSZ signal of their clusters, with $\langle \tau\rangle =7\times 10^{-3}$ for their cluster sample, is very high at $\delta T_{\rm KSZ}\gsim 500-1,000\mu$K and so the {\it KSZ should have been measured without any filtering} whereas the Osborne et al filtered maps fail to show it.
[This is also true for their Wiener filter, which does not take into account
the actual realization of the noise as our filter does, and removes all
power below $\ell\simeq200$ - see their Fig 24]. Putting it differently, the Osborne et al filtering reduces the $S/N$ of the KSZ signal. In our studies the result is recovered at $S/N\sim 4$, which is highly statistically significant. But if filtering were to reduce the KSZ term by a mere factor of $\sim 2$, the $S/N$ would drop below the statistically significant threshold. Fig. \ref{fig:rebut_omcp} demonstrates how critical it is to preserve the contributions from low $\ell$'s in filtering. This shows the problems that might arise from  using filters unsuitable for this measurement, as planned by the same group in the context of the Planck data analysis (Mak et al 2011)\footnote{Note that the direction of the dipole measured in our analysis, and shown in Table \ref{tab:vfromkaeek} from KAEEK, is misrepresented by these authors in their Fig.1.}.

When using our Wiener-type filter in their final revision, Osborne et al now\footnote{Their original draft was not able to recover these results pointing to flaws in their constructed pipeline.} recover our results [see their Tables 9 \& 10] yet claim that their significance is, at most, ``at the
$2.9-\sigma$ confidence level". While clearly a ``2.9-$\sigma$" detected dipole constitutes a statistically significant result\footnote{Their error bars are evaluated using only 100 simulations, which leads to a $\gsim 10\%$ (at one standard deviation in their sigma's) extra uncertainty in their error bars. In this respect, what they call a "2.9-sigma detection" with our filter, is easily equivalent to $\sim (3-3.5)\sigma$. For comparison, our errors are based on 4,000 realizations in each simulation.}, other numbers point to further inconsistencies in their pipeline: in Tables 9 \& 10 the error on the X component is smaller than in the Y component, when WMAP masks remove
more solid angle in the direction of the Galactic center and anticenter. This
is an indication of a numerical artifact in their pipeline. Osborne et al never present their analysis of the errors based on eq. 1 of AKEKE given here as eq. \ref{eq:sigma_fil_map} making it impossible to pinpoint their pipeline artifacts. However, they compute
the dipole at 15$^\prime$ instead of at the zero monopole, increasing the error bar
and allowing the TSZ component to contaminate their measurement [see their Tables 3, 4] and their
errors are computed from simulations instead of from the actual realization
of the sky that, as was discussed in AKEKE and above, increases error bars by an extra 10-15\%. Moreover,
the X-ray properties of their clusters were derived by combining
different sources, increasing the uncertainty of their results
[see for instance Fig 2 of KAE].

Worth mentioning is also Table 8 of Osborne et al (2011) in which these authors present a comparison between the numbers of clusters in their and in our (KABKE1,2) sample within various redshift and X-ray luminosity ranges. The origin and validity of these numbers remains a mystery though. For one, the sample used by KABKE is for the moment proprietary and, with that in mind, we do not see how to derive Table 8 of Osborne et al (2011) from any of the figures in KABKE, let alone identify individual clusters in any redshift or X-ray luminosity bin for comparison with the Osborne et al sample\footnote{While the sample using public data used in KAE is available from us {\it upon request}, no such request has been made by Osborne et al, so we do not understand how to derive their Table 8 comparison with the KAE sample}. Second, the X-ray luminosity of the clusters used in KABKE1,2 are based on X-ray fluxes recomputed for KABKE1,2 directly from the RASS raw data in order to eliminate systematic biases from different measurement methods employed by the eBCS, REFLEX, and CIZA teams (see Section~\ref{sec:xraycat} for details); they thus differ from the X-ray luminosities published by these three teams. Third, KABKE1,2 apply a common flux limit to their homogenized, merged cluster sample to avoid systematic differences in the depth of their catalog in different regions of the sky. Osborne et al (2011) do not follow any of these procedures but, instead, simply merge the aforementioned three catalogs as published. A comparison of the number of clusters in different X-ray luminosity bins would thus be nonsensical, even if it were possible. However, with the KABKE1,2 sample remaining unpublished, even this comparison -- and thus Table 8 in Osborne et al (2011) -- becomes entirely fictitious.



%
%
%
%
%
%
%
%
%
%
%
%

\section{Alternative tests of ``dark flow"}
\label{sec:tests}

If the coherence scale of the measured ``dark flow'' is compatible
with the motion extending all the way to the last scattering surface the
implication is that at least a significant part of the CMB dipole is intrinsic
and is unconnected to the local motion induced by gravitational instability. One
can define two sets of observers: (1) at rest with
respect to average matter distribution (MRF, Matter Rest Frame) and
(b) observers that see an isotropic CMB frame defined by its intrinsic
dipole (ICF, Isotropic CMB Frame also known as cosmic rest frame).
In the standard cosmological model the ICF defines a preferred
cosmic rest frame and the inflationary paradigm predicts that temperature
fluctuations should be statistically isotropic in this frame.
It is conventionally assumed that both reference frames coincide,
and all peculiar motions are due to gravitational instability.
The ``dark flow'' measurement provides the first observational evidence
that both reference frames may differ and this may also be reflected
in a variety of other observed consequences which have been pointed out
in recent literature. In this section
we review the tests that are sensitive to our peculiar motion local - or global - origin, which have been discussed in the
literature as of this review writing.

In linear theory, the irrotational component of the velocity field
grows due to the gravitational field while the primeval (solenoidal) component
decays with expansion because of the angular momentum conservation. Then,
in the gravitational instability picture, the local peculiar velocity
field must align with the peculiar acceleration. Alternative tests
of the dark flow are usually sensitive to the local peculiar
velocity field and/or to the acceleration/gravity field. 

Measurements of peculiar velocities or peculiar acceleration
vectors offer alternative methods to test the dark flow motion.
In this section we discuss some of the methods which have appeared
in the literature. The subsections below are ordered according to our personal take on which tests can potentially provide the most discriminating measurement of the reality of the dark flow.

\subsection{Diffuse Cosmic Backgrounds from collapsed objects.}

Due to their origins, the various diffuse cosmic backgrounds at other wavelengths provide
important information about different aspects of the Universe's
structure and evolution. Unlike the CMB, the cosmic infrared background (CIB) and the cosmic X-ray
background (CXB) are produced by emissions from collapsed
structures and trace the evolution of the Universe that took
place at relatively low $z$ compared to that of last scattering (see reviews by Kashlinsky, 2005 and Boldt, 1987 for CIB and CXB respectively). Thus, if the DF extends all the way to the last scattering, the dipole of the CMB should have a different amplitude than and show a misalignment with that of the CIB and CXB backgrounds, whose dipole is expected to originate entirely from the local motion with respect to the distant galaxy frame. Note, however, that even in this case at least a substantial part of the CMB dipole
must originate from the local motions of the Sun and the Galaxy,
so there should still be an overlap between the CMB and CIB/CXB dipoles making isolation of the amplitude of the intrinsic CMB dipole more difficult. Nevertheless, this presents a potentially testable effect as was already appreciated in early propositions of the offset between the CMB and galaxy expansion frames by Matzner (1980).

The motion at velocity $V$ with respect to the background of specific intensity $I_\nu$ at frequency $\nu$ would induce a dipole of
\begin{equation}
\delta I_\nu/I_\nu = (3-\alpha_\nu) (V/c)
\cos\theta
\label{eq:cib_dipole1}
\end{equation}
with $\theta$ being the angle to the apex of the
motion and $\alpha_\nu=d\ln I_\nu/d\ln\nu$. Here we ignored the
quadrupole, $O(V^2/c^2)$, and higher contributions resulting from
the relativistic Doppler corrections (Peebles \& Wilkinson 1968).
If the background and CMB dipoles are perfectly aligned, the CIB dipole
must lie in the direction of $(l,b)_{\rm CMB}$ given in Table \ref{tab:localmotion} and have the
amplitude of:
\begin{equation}
\delta I_\nu^{\rm dipole} = 1.23\times
10^{-3}(3-\alpha_\nu)I_\nu
 \label{eq:cib_dipole2}
 \end{equation}
For the Rayleigh-Jeans part of the CMB spectrum $\alpha_{\rm CMB}\simeq 2$. However, both far-IR CIB and CXB have spectral energy distributions such that $\alpha \sim -3$ over a range of frequencies which can be probed leading to a significantly amplified relative dipoles (Kashlinsky 2005, Hogan, Kaiser \& Rees 1982). This potentially presents a way to probe the tilt by testing the alignment  between CMB and CXB/CIB dipoles. Two problems, however, make it difficult to accomplish this test conclusively with the currently available data: 1) the two sets dipoles (CMB and CIB/CXB) would have at least a partial overlap as both are subject to the local motion (Solar System, the Galaxy, LG and beyond), and 2) the uncertainties in the foreground contributions at the relevant frequencies remain very substantial to enable any discriminative measurement because the foregrounds are modelled from the (presently available) instruments with the relatively large noise, poor angular resolution and/or insufficient frequency span.

In summary, in the absence of DF, the dipole of other cosmic backgrounds measured in the Sun-rest frame should correspond with a very high accuracy to velocity of $\simeq 370$ km/sec in the direction of $(l,b)\simeq (264^\circ, 48^\circ)$. Any deviation from this could be indicative of the DF. Two other cosmological backgrounds may prove useful in this regard in the future: the cosmic X-ray background (CXB) and cosmic infrared background (CIB).

Galactic subtraction is difficult in the current datasets. The Galaxy in such measurements is usually modelled with templates which are determined from Galactic lines or other known Galaxy tracers (see Fixsen \& Kashlinsky 2011 and references therein for discussion). Given spectral templates of the CMB and Galaxy contributions (the Planck spectrum, $B^\prime_\nu(T_{\rm CMB})$ in the case of the CMB and $G_\nu$ for Galaxy template at each channel $\nu$) one can model any measurement with more than three frequency channels, $\nu$,
decomposing the measured dipole, ${\cal D}$,  into the following terms:
\begin{equation}
{\cal D}^{\rm model}_\nu = {\cal D}_\nu^{\rm noise} + a\; G_\nu + b\; B^\prime_\nu + d\; I_\nu^{\rm CIB/CXB}
\label{eq:dipole_cib_future}
\end{equation}
The CMB dipole term ($b$) is, of course, vanishingly small in CXB measurements. The last term describes the CIB (or CXB) dipole with a known from measurements spectral template. Given the Galaxy templates, one can evaluate the CIB (or CXB) dipole after marginalizing over $a$ and $b$ and summing over all the available channels. This is achieved by minimizing $\chi^2=\sum_\nu ({\cal D}^{\rm sky}_\nu-{\cal D}^{\rm model}_\nu)^2/\sigma_\nu^2$ with respect to $(a,b,d)$. The solution for $d$ and its uncertainty, $\sigma_d$, is then given by standard regression and error propagation and the signal-to-noise of the prospective measurement will be given by $S/N=d/\sigma_d$ (Fixsen \& Kashlinsky 2011).

The general rule-of-thumb for the accuracy required to make this test valuable to the present discussion is as follows: If a background dipole is measured with signal-to-noise of $S/N$, its directional accuracy would be $\Delta \theta \simeq \sqrt{2} (S/N)^{-1}$ radian. Thus in order to be able to distinguish between the, say, local peculiar acceleration direction vector and that of the CMB dipole (currently misaligned  by as much as $\sim (10-20)^\circ$) the dipole of the background in question should be measured with $S/N>15-20$. In addition, its amplitude should agree with motion of 370 km/sec in the CMB dipole direction.

The discussion below will show that the far-IR CIB, if measured with appropriate instrumentation, may well present the best alternative testing ground for the possible tilt.

\subsubsection{CXB dipole measurements}

CMB dipole in eq. \ref{eq:dipole_cib_future} is negligible in X-ray measurements. The chief, by far, impediments to accurate measurement here are 1) the Galactic foreground produced by hard X-ray binaries and hot halo gas and 2) that source confusion dominates large angular scales, which make the systematics and uncertainties in such measurement with CXB very high (Boldt 1987, Fabian \& Barcons 1992). The situation with CXB based on HEAO measurements is
inconclusive although the results are marginally consistent with
the CMB dipole (Shafer 1983, Scharf et al 2000, Boughn et al 2002). ROSAT has greater foreground problems in such a measurement given its energy bands (Snowden, private communication) leading at present to a similarly inconclusive situation (Plionis \& Georgantopulus 1999).

The overall situation with the CXB dipole measurements, using the HEAO and ROSAT instruments, is at present inconclusive as far as probing the tilt via the alignment between the dipole and that of the CMB is concerned. The reason is that the present analyses can detect the CXB dipole only at less than 2-$\sigma$ significance, or signal-to-noise $S/N<2$. Since the directional accuracy of the dipole is $\Delta\theta \simeq \sqrt{2}(S/N)^{-1}$ rad, the current measurements can only constrain the direction of the CXB dipole within $\sim 30^\circ-40^\circ$ of the CMB dipole which is not enough to probe the possible misalignment. In addition, the magnitude of the CXB dipole is uncertain by the seeming excess of X-ray emitters in the general direction roughly aligning with the CMB dipole (Fabian \& Barcons 1992 and references cited therein).

\subsubsection{CIB dipole measurements}

The far-IR CIB (Puget \etal\ 1996, Schlegel et al 1998, Hauser
\etal\ 1998, Fixsen \etal\ 1998) has been reliably measured, both
its amplitude and its spectral-energy distribution. It is produced
by emission by cold ($T_d\sim 20$~K) dust components in
galaxies and most of it seems to originate at early times, $z\ga
1$. Its spectral energy distribution is such
that the dipole component produced by local motion is amplified,
in relative terms, over that of the CMB (Kashlinsky 2005). This provides a potential way of isolating the
CIB dipole component from the local motion and probing its
alignment with that of the CMB. Since the dust temperature $T_d \approx 18.5$~K, the CIB does not
reach the Rayleigh-Jeans regime, where the spectral index
$\alpha_\nu\simeq 2$, until $\lambda \ga 400-500~\mu$m. This results
in the significantly negative spectral index of the CIB over much of
the FIRAS and DIRBE probed bands.

\begin{figure}[h!]
\includegraphics[angle=0,trim=30 10 10 10,width=4in]{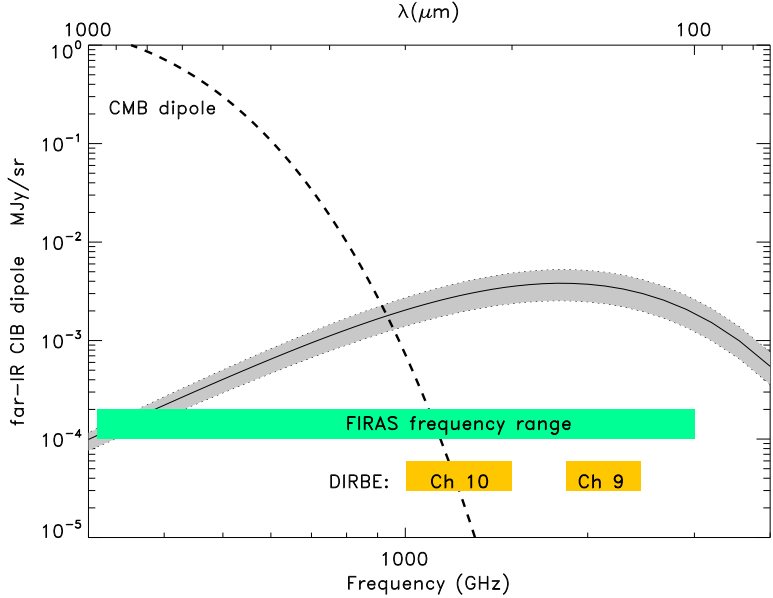}
\caption{Shaded region denotes the uncertainty contours of the spectrum the CIB dipole assuming {\it all} of the CMB dipole is produced by motion within the local volume; its mean value is shown with the black solid line. Dashes show the measured CMB dipole. The continous frequency coverage of FIRAS
and the two lowest frequency (broad) channels of DIRBE are marked. Adapted from Fixsen \& Kashlinsky (2011).}
\label{fig:cib_dipole}
\end{figure}

Fig. \ref{fig:cib_dipole} shows the predicted CIB dipole spectrum,
using the spectral energy distribution of the far-IR CIB from Fixsen et al (1998) and assuming perfect alignment with the
CMB; the dipole must have a peak value of $(3-5)\times
10^{-3}$~MJy/sr at 100-300$~\mu$m, or frequencies 1-3 THz. At
longer wavelengths the CMB dipole would overwhelm the signal and
at $\lambda \la 100~\mu$m the CIB dipole decreases, becoming
confused with Galactic and zodiacal light emission. In this
wavelength window, however, because of the slope of its spectral
energy distribution, the dipole in the CIB becomes $\sim10^{-2}$
of its mean level compared to $\sim 10^{-3}$ for the CMB.

Fixsen \& Kashlinsky (2011) have noted that this presents a potential way to constrain the possible tilt of the Universe and attempted (unsuccessfully) to measure the CIB dipole after noting that the FIRAS Pass 4 data noise levels could enable the measurement of the CIB dipole at S/N$\sim 2-3$ if it is perfectly aligned with that of the CMB. The reason for this potential S/N is that the FIRAS instrument onboard COBE had fine continuous frequency resolution of 15 GHz over a wide range of frequencies covering both sides of the prospective peak of the CIB dipole and the expected S/N was obtained after summing over all the channels and assuming negligible Galaxy foreground contribution. Accurate modeling of the foregrounds, however, proved crucial in this measurement.

In case of the CIB, the Galaxy emissions can be traced via its lines (C{\sc ii}, N{\sc ii}
and CO) which are resolved by FIRAS well enough to separate them from the dust continuum enabling the
use of them as templates. However, the key to any successful measurement of the CIB dipole is to break the degeneracy between the far-IR
CIB energy spectrum and that of the Galaxy over the wavelengths
where the CIB dipole is near its peak. CMB dipole dominates the
long-wavelength emissions, but its energy spectrum is very
accurately known and so it can be subtracted making the residual
small at wavelengths below $\sim 500 \mu$m. If one were able to
resolve galaxies out to sufficiently high $z$ and remove them from
the maps, the spectrum of the remaining CIB would potentially be
sufficiently different to allow robust removal of the Galactic
contribution to the dipole. Such experiment should be finely
tuned, since at the same time one would need to leave enough
sources in the confusion to generate sufficiently measurable
levels of the far-IR CIB. Or alternatively, the low-$z$ part of
the CIB can be removed together with the Galactic foreground, but
that too requires sufficient resolution to remove the
Galaxy accurately enough.

\begin{figure}[h!]
\includegraphics[trim=20 10 10 5,width=4in]{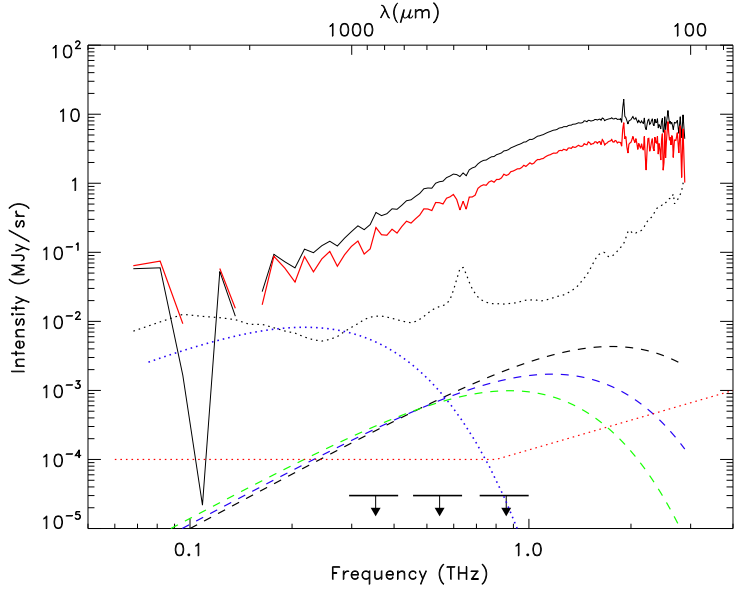}
\caption{The solid black line is the average Galaxy spectrum
for $|b|>10^\circ$  corresponding to the G10 template in the main text. The solid red line is
the average Galactic spectrum at $|b|>30^\circ$ corresponding to the G30 template in the text.
The dashed lines are the expected spectra of the CIB dipole; black, blue, green line
correspond to sources at $z\ga 0, 0.5, 1$ respectively. The dotted blue
line shows the CMB dipole residual uncertainty (at 0.017~mK rms
uncertainty). The FIRAS
noise is shown with the dotted black line. An estimate of
the Pixie noise is shown with dotted red line. Three horizontal
bars with arrows show the Planck noise limits (below the plot) at
its three highest frequency channels. Adapted from Fixsen \& Kashlinsky (2011).} \label{fig:cib_dipole_future}
\end{figure}

Thus the major difficulty in this measurement is confusion with the Galactic foreground; Fig. \ref{fig:cib_dipole_future} shows that the CIB spectral shape (from all galaxies) is quite similar to that of the Galaxy making the separation of the two components in eq. \ref{eq:dipole_cib_future} degenerate. The physical reason for this degeneracy is that the CIB as determined from maps of low resolution ($\sim 7^\circ$ for FIRAS) is dominated by low-$z$ galaxies whose dust energy spectrum resembles that of the Milky Way. A way to break the degeneracy would be to remove the resolved low-$z$ galaxies from the maps leaving behind enough CIB to be measured with a sufficiently redshifted spectrum as discussed in Fixsen \& Kashlinsky (2011). The figure shows the CIB dipole from galaxies remaining at $z>0.5,1$
(dashed blue and green lines respectively) can have a sufficiently different energy spectrum from the Galaxy and may have enough amplitude to be measurable in specially designed experiments (see below). Fig. \ref{fig:cib_dipole_future}
shows that CIB produced by high-$z$ sources becomes progressively more distinguishable in spectrum at increasing $z$. This still
leaves the CIB dipole significantly below the foreground Galactic
radiation but the spectral difference allows discrimination between
the bluer Galactic spectrum and the redder CIB spectrum.

Application of this methodology to the FIRAS Pass 4 data by Fixsen \& Kashlinsky produced only upper limits on the CIB dipole which are a factor of a few higher than the cosmologically interesting level given by eq. \ref{eq:dipole_cib_future} because Galaxy foregrounds cannot be isolated sufficiently accurately with the FIRAS maps. The currently flying {\it Planck} and {\it Herschel} missions will perform even worse in this respect because with only a few frequency bands and their broad channels, convincing separation of the Galactic foreground will be difficult, a point also realized by Piat \etal\ (2002).

The recently proposed Primordial Inflation Explorer ({\it PIXIE}) space mission, or a similarly designed future mission, (Kogut et al 2011) can, however, be significantly more successful in this measurement and may provide crucial test of the tilt's existence (Fixsen \& Kashlinsky 2011). Fig. \ref{fig:cib_dipole_future} shows the estimated noise levels over the continuous frequency coverage by {\it PIXIE}; they are over two orders of magnitude better than FIRAS.

As discussed earlier, the key requirement is to break the degeneracy between the far-IR
CIB energy spectrum and that of the Galaxy over the wavelengths
where CIB dipole is near its peak. Fixsen \& Kashlinsky (2011) evaluated the S/N of the CIB dipole measurements for {\it PIXIE} parameters and have shown that CIB dipole can be measured with S/N$\simeq 30-40$ there.
Importantly, the high S/N for the prospective CIB dipole measurement with {\it PIXIE}-type mission would allow to also measure the dipole direction with good accuracy. The accuracy of the measured direction for $S/N\gg 1$ would be $\Delta \theta \simeq \sqrt{2} (S/N)^{-1}$radian. Thus for {\it PIXIE} the accuracy of the CIB dipole direction would be
\begin{equation}
\Delta\theta_{PIXIE} \simeq 2^\circ \; \frac{40}{(S/N)_{PIXIE}}
\label{eq:pixie_dir}
\end{equation}
The current discrepancy between the local acceleration vector direction measured from galaxy surveys and the direction of the CMB dipole is about $\sim 15^\circ-20^\circ$ (Sec. \ref{sec:earlyresults} and references therein) presenting a challenge for the purely kinematic interpretation of the CMB dipole. Thus a measurement with such an instrument can settle the meaning of that discrepancy as the CIB is expected to be aligned with the true direction of the local motion.

\subsection{Supernovae-based velocity measurements}
\label{sec:sn}

Because of their narrow range of intrinsic luminosities, supernovae (SN) present a special class of distant indicators of potential use in peculiar flow determinations. For instance, SNIa based distances have statistical errors of only $\sim 6\%$ per object and can then probe deviations from the universal expansion to much greater distances than galaxy distance indicators. However, because their frequency in the Universe is determined by the star formations rates, their samples are currently sparse and their distribution on the sky is far from homogeneous. Their potential use in these and other cosmological probes has been discussed by Davis et al (2011). Below we discuss the theoretical aspects related to their application for peculiar velocity measurements followed by an overview of the currently available results.

Note, however, that SN-based distances are derived from their magnitudes which decrease rapidly at cosmological distances (and decay with time). Thus, even with the small uncertainty in the distance modulus this measurement becomes progressively more difficult toward higher $z$, where - in addition - would is biased toward brighter objects. This is very different from the situation with the KSZ-based measurements which are redshift-independent.

\subsubsection{Theoretical prelude}

Luminosity distances derived from SN type Ia (SNIa) provide an alternative useful
method of determining large scale velocity flows. At cosmological distances SN are unresolved sources, which leads to some significant differences in their application from galaxy distance indicators which use well resolved sources. Below we provide the minimal theoretical background needed in their application.

Despite significant recent progress, we feel that the technique has not yet come to fruition since discovering of high redshift SN requires
continuous monitoring of the same field. The few patches that are being
regularly observed by the same groups do not provide for homogenous
coverage of the sky and do not have enough statistical power to constraint
velocity fields beyond $\sim 100$ Mpc. Since this technique is totally
unrelated to the KSZ effect,
and has different systematics, as more fields are observed by the different
groups, the method may soon provide an independent test of the dark flow.

The effect of peculiar velocities on luminosity distance estimators of point sources has been studied
extensively in the literature. Sasaki (1987) derived the general expressions for
an Einstein-de-Sitter Universe followed by Pyne \& Birkinshaw (1996, 2004) who derived expressions for more general cases, including
non-flat models, as well as models with dark energy. Hui \& Greene (2006) analyzed
the effect of peculiar velocities and other systematics on luminosity distance
measurements on the dark energy equation of state. Following Pyne \&
Birkinshaw, Hui \& Greene gave a derivation of the luminosity distance fluctuation accurate to
first order. We now briefly summarize their discussion and refer the reader
to the original references for more details.

If $d\omega_e$ is the solid angle subtended by a
detector of projected area (along the LOS) $dA_0$ as it is seen by a source
located at redshift $z$, $F_0$ is the observed flux and
$L_e$ the source luminosity, then
\begin{equation}
F_0(z)=\frac{L_e}{4\pi(1+z)^2}\frac{d\omega_e}{dA_0}
\end{equation}
is an expression that is equally valid for a universe with
or without homogeneities (Weinberg 1976). The luminosity distance is defined
as
\begin{equation}
d_L(z)=(1+z)\sqrt{dA_0/d\omega_e}
\end{equation}
In an homogeneous universe, the redshift $z_H$ is proportional
to the recession velocity due to the Hubble expansion.
If emitter and observer are affected by a peculiar
velocity field the measured redshift $z$ would be the
Hubble value $z_H$ Doppler shifted to
\begin{equation}
1+z=(1+z_H)(1+\vec{v}_e\hat{x}-\vec{v}_0\hat{x}),
\end{equation}
where $\hat{x}$ is the unit vector along line of sight between the emitter
and the observer. The solid angle subtended by the detector
would also be affected at first order of the Lorentz transformation while the projected area along the line
of sight would be invariant, since Lorentz transforms leave coordinates
perpendicular to the motion invariant. If we use Lorentz boosts to transform the
solid angle and projected detector area from the homogeneous Hubble expansion
frame H to the observer/emitter frame,  then $dA_0=dA_H$ and
\begin{equation}
d\omega_e=\frac{d\omega_H}{\gamma^2(1-\vec{v}_e\hat{x})^2}
\end{equation}
The luminosity distance is now
\begin{equation}
d_L(z)=(1+z_H)\sqrt{dA_H/d\Omega_H}[(1+\vec{v}_e\hat{x}-\vec{v}_0\hat{x})
\gamma(1-\vec{v}_e\hat{x})]\\
\simeq d_L(z_H)(1+2\vec{v}_e\hat{x}-\vec{v}_0\hat{x})
\label{eq:dl}
\end{equation}
In the DF context, eq.~\ref{eq:dl} has a very clear interpretation.
If the observer and emitter are at rest with respect to the MRF, then
in the CIF they would have a peculiar velocity $\vec{v}_e=\vec{v}_0=\vec{v}_{bulk}$.
If we refer redshifts measured in the MRF $z$ to the CIF $z_H$, then
the luminosity distance of different
objects will show a dipole pattern: $d_L(z)=d_L(z_H)(1+v_{bulk}\cos\theta)$,
identical to the KSZ measurement.

\subsubsection{First results}

Four post-DF studies have appeared in the literature as of this review writing. We address them in chronological order:

Weyant et al (2011) measured the bulk velocity of the Local Group  out to a depth of $40h^{-1}Mpc$ from a sample of 112 SN with $z<0.028$ using luminosity distances from Hicken et al (2009).
Two non-parametric methods, Weighted Least Squares
and Coefficient Unbiased were used to compute the coefficients of
the spherical harmonic expansion of the radial component of the peculiar
velocity field. They measured velocity of $538\pm 86$ toward  $(258\pm 10,36\pm 11)$ km/sec for the first method and $446\pm 101$ km/sec in the direction of $(273\pm 11,46\pm 8)$ using the second method. Because of the shallow depth of the data, this result has only limited relevance for the DF measurement inasmuch as it is consistent with the latter.

\begin{figure}[h!]
\includegraphics[width=0.32\textwidth]{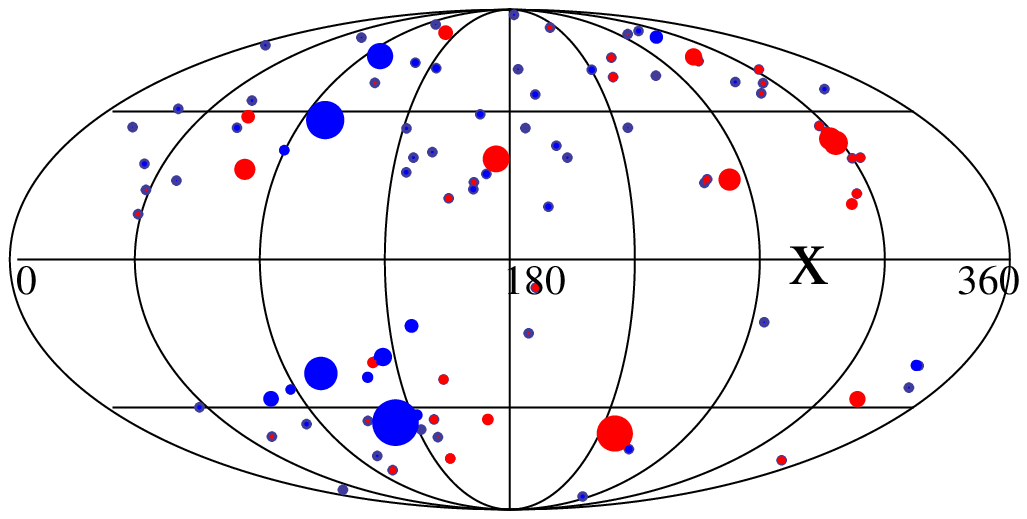}
\includegraphics[width=0.32\textwidth]{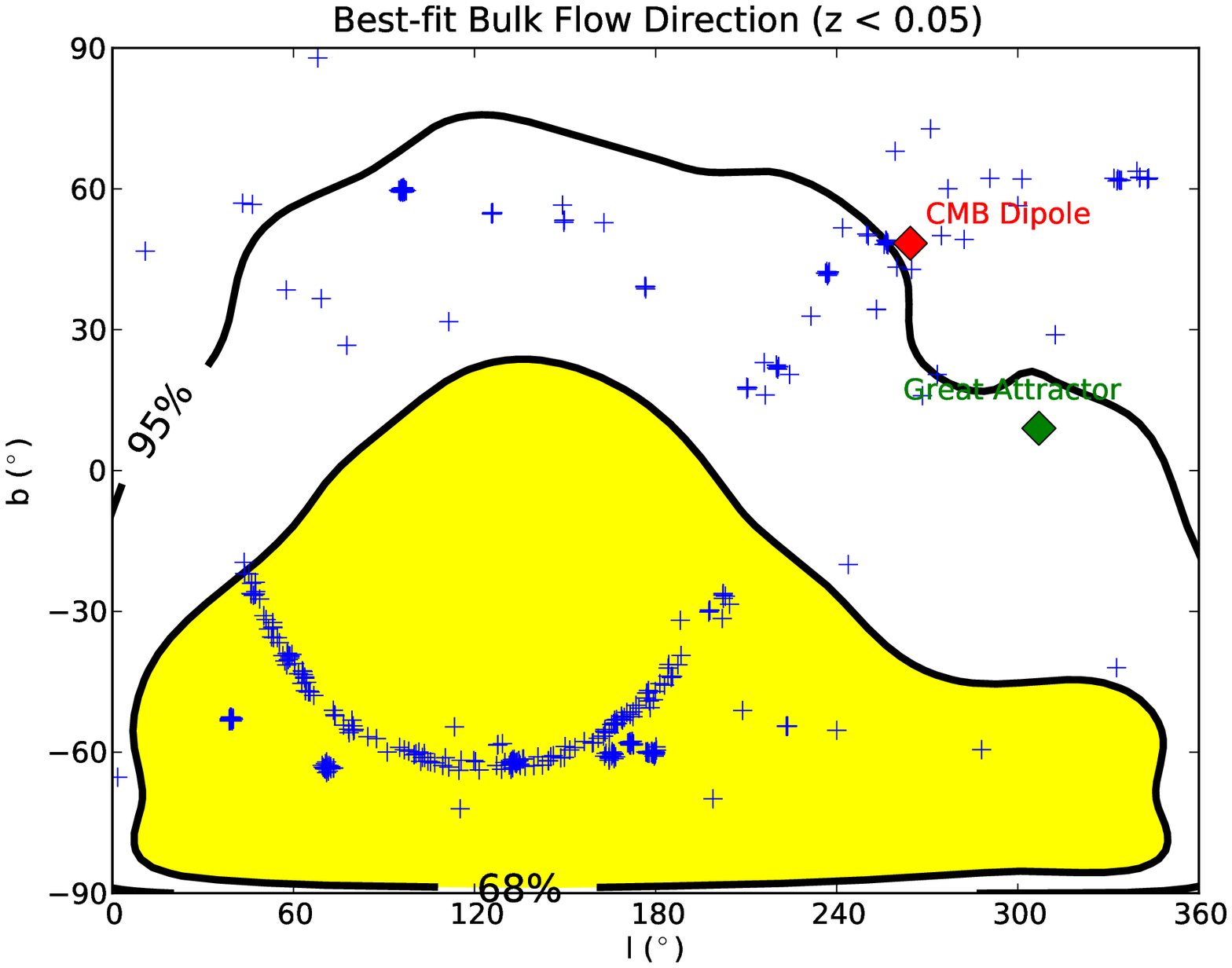}
\includegraphics[width=0.32\textwidth]{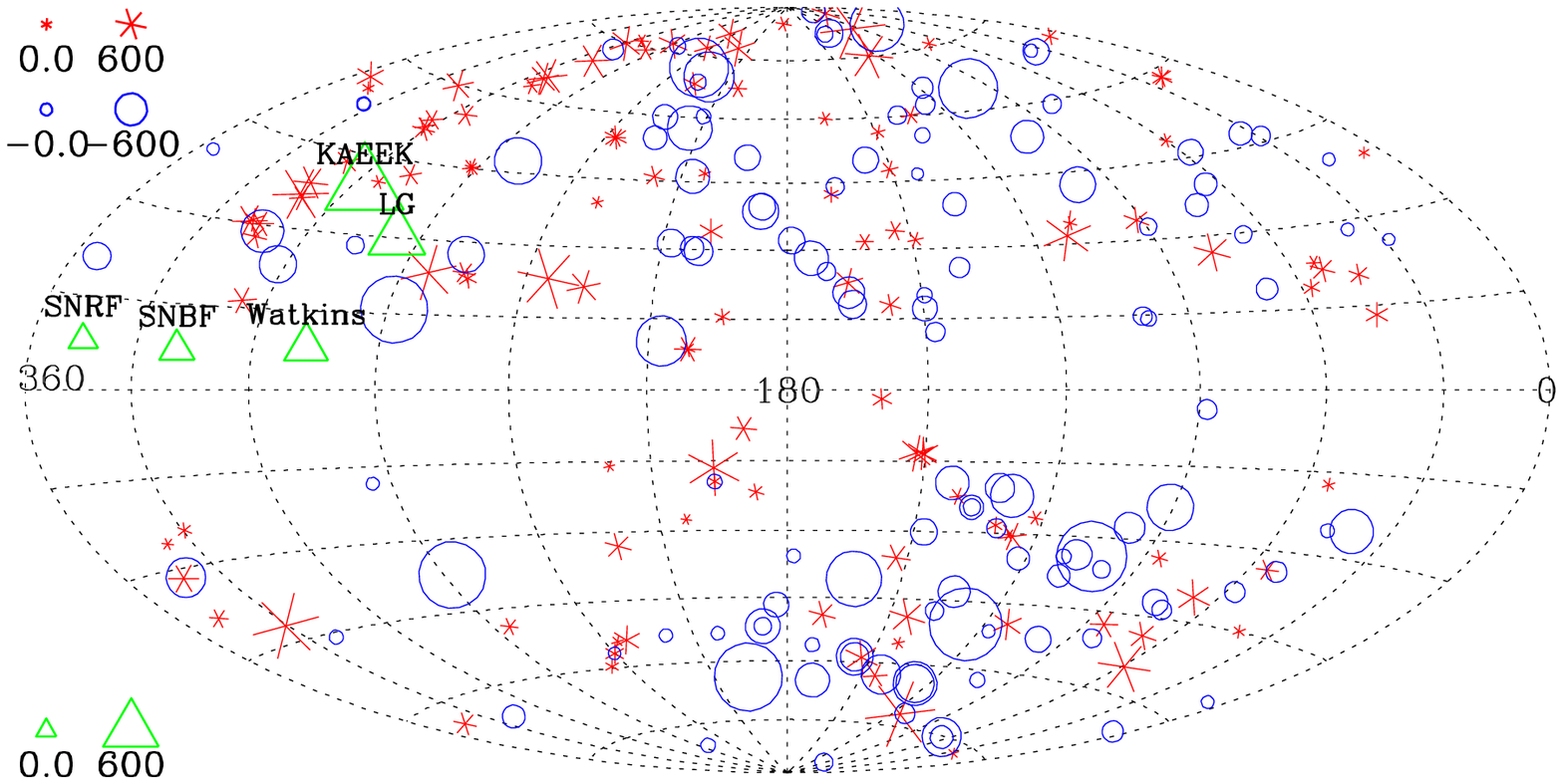}
\caption[]{Sky distribution of SNIa used in the different surveys. {\bf Left}: The sample of SN used in Ma et al (2011). Blue/red symbols correspond to outgoing/incoming sources. The cross shows the direction measured in Watkins et al (2009). {\bf Center}: The distribution of SN from the Union 2 compilation of SDSS survey used by Dai et al (2011). The arc shows the distribution of SN at $z>0.05$ which is restricted to a narrow strip at constant declination. {\bf Right}: From Turnbull et al, Fig.1 - blue symbols are approaching objects, red symbols denote receding sources. The size is proportional to the radial velocity component.}
\label{fig:sn_surveys}
\end{figure}
Colin et al (2011) have used a sample of Union 2 SNIa from Amanullah et al (2010). If there is a large scale flow extending to a given $z$, the luminosity distance should have a dipole distribution. They designed methods to probe for anisotropy in the sky distribution of the luminosity distance measured for these objects. They note that at $z\gsim 0.06$ the data become sparse and do not provide an isotropic coverage on the sky, but conclude that the $z<0.06$ subsample is suitable for the study. At distances $z\sim 0.05$, roughly coinciding with the Shapley concentration, Colin et al find anisotropy consistent with a large-scale flow of the amplitude comparable to the Watkins et al flow and approximately in the same direction. (We emphasize again that the direction can be probed only to $\Delta \theta \sim 20^\circ-30^\circ$ given the limited S/N of such measurements). On these scales their flow disagrees with the $\Lambda$CDM prediction at about $\sim (2-3)\sigma$. On large scales the results are inconclusive being at $\sim 1\sigma$ level of significance. The authors conclude that convergence of the Local Group velocity to the CMB rest-frame must occur ``well beyond Shapley" concentration (distance $\gsim 160h^{-1}$Mpc).

As mentioned earlier, Ma, Gordon \& Feldman (2011) used the data from $103$ SN from Tonry et al (2001). The distribution of their SN is shown in the left panel of Fig. \ref{fig:sn_surveys}. They provide a self-consistent model for probing the DF (which they called ``tilt velocity"): the velocity field is modeled as having contributions from 1) thermal motion of galaxies (including the noise), 2) component from gravitational instability with the dispersion given by eq. \ref{eq:vg_rms} with the power spectrum given by the observationally established $\Lambda$CDM model, and 3) a component due to tilt (DF) which is constant for all sources. Using the SNIa sample alone they found a bulk flow $v_{bulk}=450^{+180}_{-190}$ km/sec in the direction consistent with that of the DF motion from KAEEK.

Dai et al (2011) analyzed 557 SN from the Union2 sample and searched for the
effect of the dark flow on the luminosity distance estimates. The distribution of SN in their study is shown in the middle panel of Fig. \ref{fig:sn_surveys}; it is the same sample as used in Colin et al (2011). While Dai et al (2011) found
of a bulk motion for SN with $z<0.05$ of amplitude $v_{bulk}=188^{+119}_{-103}$km/s, significantly smaller than in Watkins et al (2009).
For their high redshift ($z>0.05$) they found no evidence of a dark flow and
concluded the dark flow measurement was inconsistent with SN data at the
99.7\% confidence level. However, it is worth pointing our that their high redshift data comes primarily from the Sloan slice, that is very localized in the sky, since the telescope scans at constant declination producing a narrow strip of $\sim 2^\circ\times 80^\circ$ centered approximately in the anti-dipole direction and much of its objects lie orthogonally to the DF plane. Second, their likelihood
function is rather flat, so a more conservative conclusion would be that
they see neither the evidence of a bulk flow nor its absence (Kinney 2011, private communication).

More recently, Turnbull et al (2011) used a compilation
of 245 SNe peculiar velocities and use a Maximum Likelihood and
a Minimum Variance methods to estimate bulk flow velocities.
Their sample was an extension of the Constitution data set of Hicken
et al (2009) and contains SN with redshifts $z\lsim 0.067$. The distribution of their SN is shown in the right panel of Fig. \ref{fig:sn_surveys}.For these
nearby SN, they confirm the results of Dai et al (2011) and find
a bulk flow amplitude consistent with the expectation of $\Lambda$CDM.
However, by comparing the SNe peculiar velocity field with the predictions
of the IRAS Point Source Catalog Redshift Survey (PSCz) galaxy density
field, they show the density PSCz density field fails to account for
$150\pm 43$km/s of the SNe bulk motion. We caution that the main conclusions of that interesting study as far as DF goes are based on their eq. 7 which models the tilt (DF) component in addition to the gravitational instability flow {\it as predicted by the distribution of light they adopt for the nearby Universe}. Hence their derivation of the tilt component depends on 1) the assumption that light traces mass very accurately, and 2) the distribution of the mass thus derived reflects the overall distribution over the distances they probe. In addition, Wilthsire et al (2012) raised another issue of whether these and other SNIa-based results can be affected by the reddening correction used.

These are all important tests of the DF phenomenon, precisely precisely the sort of science needed to test the latter. The current results results are still inconclusive as far as DF goes with some being inconsistent with it and some supportive. This may be due to the limitations of the current samples whose coverage we display in Fig. \ref{fig:sn_surveys}. While potentially very promising the problems with the current SN samples are the limited coverage due to ZoA from the Galaxy, incompleteness of the samples, inhomogeneous coverage at different depths, and the small number of sources available in the samples. Table \ref{tab:sn_results} illustrates the currently conflicting results with our comments on each study.

\newpage
\begin{table}[h!]
\caption{Summary of SN-based results as of this report writing.}
\begin{tabular}{p{1in} | p{1.2in} p{1.5in} | p{3.0in} }
\hline
Source & $V_{\rm tilt}$ or equivalent & $(l,b)$ & Comments\\
 & km/sec & deg & \\
\hline \hline
Weyant et al (2011) using a sample of 112 SN with $z<0.028$
out to a depth of $40h^{-1}Mpc$ & $538\pm 86$ &  $(258\pm 10,36\pm 11)$ & {\footnotesize Because of the shallow depth the analysis is not suitable for probing DF and the results are only meant to show consistency with the possible DF motion at much larges distances.} \\
\hline
Colin et al (2011) using Union 2 dataset & $260\!\pm\!130$ & $\sim(298, 8)\!\pm\! 40$ &  {\footnotesize The SN sample is subdivided into many $z$-bins. The uncertainties in the direction are shown as $\Delta \theta\simeq \sqrt{2}(S/N)^{-1}$radian. The direction is consistent with DF. The amplitude overlaps within the systematic uncertainties of the calibration in KAEEK and the statistical errors there.}\\
\hline
Ma et al (2011) using 103 SN from Tonry et al (2001) (Fig. \ref{fig:sn_surveys}, right) & $450^{+180}_{-190}$ & $(284.9^{+22.9}_{-22.1}, -1.0^{+18.8}_{-18.3})$  & {\footnotesize Tilt velocity determined by decomposing the velocity field into 1) thermal/noise component, 2) gravitational instability specified by the concordance $\Lambda$CDM model, and 3) tilt velocity. (Other surveys considered there give consistent results).}\\
\hline
Dai et al (2011) using Union 2 dataset (Fig. \ref{fig:sn_surveys}, center) & null result & N/A & {\footnotesize Divide data into low/high-$z$ bins ($z<\!/\!>0.05$). Report not finding measurable velocities in the high $z$ bin. The distribution of SN in the high-$z$ bin comes from a very narrow strip whose center is roughly in the anti-CMB dipole direction (see Fig. \ref{fig:sn_surveys}, middle panel).}\\
\hline
Turnbull et al (2011) using SN dataset in Fig. \ref{fig:sn_surveys}, right & $150\!\pm\! 43$ & $(345\!\pm\! 20, 8\!\pm\! 13)$ & {\footnotesize Tilt velocity determined as a large-scale residual from the actual velocity of the SN sample and that modeled by gravitational instability from light-tracing structures observed in the IRAS PSCz survey.}\\
\hline
\end{tabular}
\label{tab:sn_results}
\end{table}

\subsection{Uniformity/isotropy of the universal expansion}

Uniformity, isotropy and statistical properties of the uniform expansion and its parameters provide another testing ground of the DF reality. Indeed, if the DF is real it defines a preferred direction in our Universe. In that case, at least part of the CMB dipole has to be primordial and that component should not be subtracted out when evaluating the local motion hierarchy.

Very recently, Wiltshire et al (2012) have studied isotropy and uniformity of the Hubble flow using a massive compilation containing 4,534 galaxies with measured distances extending to about $\lsim 100h^{-1}$Mpc. Remarkably, they noted that the universal expansion is significantly more uniform, and the Hubble constant is closer to its global value, when measured in the rest frame of the Local Group rather than when referred to that defined by the CMB dipole. This is precisely what the DF would suggest. They conclude that their ``results suggest that, as far as observations are concerned, variance in the Hubble law over scales of tens of megaparsecs cannot be simply reduced to a boost at a point; space really is expanding, and by differential amounts".

Interestingly, a marginal detection of anisotropy in the acceleration parameter has been reported from the Union 2 SN data analysis in approximately the same direction (Cooke \& Lynden-Bell 2010, Antoniou \& Perivolaropoulos 2010).

\subsection{Unresolved KSZ induced by the Dark Flow.}

\subsubsection{Signature from the dark flow effect on ionized intergalactic matter}

If the CMB dipole is intrinsic, all observers at rest in the matter
rest frame should see the same dipole pattern. This is specially relevant
for ionized gas in clusters and elsewhere. In the cluster rest frame, photons interacting with the ionized gas would display a dipole pattern producing a kSZ effect contribution,
as discussed in \ref{sec:thomson}. The dark flow was measured
by computing the dipole at known cluster locations. But what would
be the effect of the unresolved hot gas population? The
dark flow should also induce CMB temperature fluctuations at
angular scales $\ll\sim 10^3-10^4$ (Zhang 2011). Their amplitude is about a
factor of 2 smaller than the KSZ due to the peculiar velocities
of clusters in the matter rest frame. Its existence worsens the low
SZ problem found by the South Pole Telescope (SPT) [Lueker et al 2009, VERIFY that
it is discussed in the corresponding section of tSZ measurements].
The SPT measured a combined TSZ+KSZ power spectrum at $l=3000$
of $4.2\pm 1.5 (\mu K)^2$ and the dark flow induced KSZ would
add a $0.3 (\mu K)^2$ to this measurement. At present,
the uncertainties in the modelling of the gas distribution
on clusters and in $\sigma_8$ did not allow Zhang (2011) to reach a conclusive constraint.

More importantly, Zhang (2011) has also pointed out that, due to the dark flow, the small
scale anisotropy of the CMB would correlate large scale structure\footnote{In this connection one should note the Ostriker-Vishniac effect (Ostriker \& Vishniac, 1986, Vishniac 1987)
resulting in the CMB anisotropy produced by scattering from ionized regions or
clouds with bulk motions. The temperature fluctuations induced along
the line of sight is the integrated contribution of eq.~(\ref{eq:ksz_def})
weighted by the probability of electron scattering off CMB photons
from some initial moment to the present time $\frac{\Delta T}{T_{CMB}}(\hat{x})=
-\sigma_T\int_0^{t_0}dt n_e (\hat{x}\cdot\vec{v}) e^{-\tau}$.
For standard reionization and gravitational instability paradigm,  Vishniac (1987) showed that these anisotropies would be
regenerated at arcminute scales due to the
bulk motions generated by gravitational clustering.}.
Assuming that galaxy catalogs, such as 2MASS, trace the distribution of
free electrons in the Universe, he showed that PLANCK
could measure the CMB-LSS correlation due to the dark flow, using
2MASS, with a signal to noise ratio of $S/N\sim 14 [V_{bulk}/10^3 (km/s)]^2$.

\subsubsection{Imprint on WHIM}

In a similar context, Atrio-Barandela et al (2009) discussed the contribution of the KSZ produced by the diffuse gas in the
Warm-Hot Intergalactic Medium (WHIM) to the cold spot detected
in the Corona Borealis supercluster by the
Very Small Array (G\'enova-Santos et al 2005). The cold spot at
$(l,b)\simeq (45^0,67^0)$ was not associated with any significant
X-ray emission in ROSAT data. If this contribution exists, it can not be due
to the dark flow since the cold spot is $\sim 80^0$ away from the apex of
the flow, almost perpendicular to it, and the resulting KSZ effect
is negligible. G\'enova-Santos et al, 2009 extended the formalism to
compute the contribution of the KSZ, produced by interaction of CMB photons with the WHIM, to WMAP temperature anisotropies.
They found a marginal evidence that there could exist a contribution of $3.1\%$
to the CMB power spectrum at $\ell 450$ from the KSZ effect in
the WHIM, but we note that at the $2\sigma$ confidence level, a null contribution
is also compatible with the data. If it is confirmed by Planck, such large
contribution would be an indirect indication of the existence of
the dark flow.

\subsection{Doppler shift of the CMB power spectrum.}

CMB defines a preferred reference frame and standard cosmological models based on strict uniformity and isotropy of the Universe predict it
should be statistically isotropic in this rest frame.
Kosowsky \& Kaniashvili (2011) pointed out that  due to the preferred direction
of motion for observers
moving with respect to the CMB frame, the CMB temperature fluctuations
would no longer be isotropic. In particular they noticed that velocity of the
observer with respect to the CMB frame will also induce changes in the angular correlation
function and lead to a potentially measurable among the different multipole $\ell$-moments.
If $C(\hat{x}^\prime_1,\hat{x}^\prime_2)$ is the correlation function
between two directions $(\hat{x}^\prime_1,\hat{x}^\prime_2)$ on the sky
in the CMB Isotropic Frame, an observer is
moving with velocity $\vec{v}$ it will measure
a correlation function $C(\hat{x}_1,\hat{x}_2)\simeq
(1+\vec{v}\hat{x}_1^\prime+\vec{v}\hat{x}_2^\prime)C(\hat{x}^\prime_1,\hat{x}^\prime_2)$
which is no longer isotropic. In this expression, the unprimed unit vectors
are the Lorentz transform of the primed vectors. The correlation function
is modified at $O(v/c)$ and varies with the angle between the boost direction
and the direction of observation.

This potentially provides an important test of the DF because any primordial component of the CMB dipole would not generate this coupling. Kosowsky \& Kaniashvili (2011) estimated that the PLANCK satellite could measure this effect at a signal to
noise level of $\sim 5$ if the entire CMB dipole arises from peculiar motion. Failure to detect the correlation due to our peculiar
motion at the level implied by the kinematic origin of the entire CMB dipole would lend independent support to the dark flow measurement.

Note, however, that here one would need to robustly distinguish the coupling introduced by the Lorentz boost from that generated by the CMB masking.


\subsection{Redshift distortions.}

Galaxy redshift surveys such as the SDSS have created 3-dimensional maps of the Universe
containing $\sim 10^6$ galaxies. Because of peculiar motions,
the redshift distribution of galaxies should  differ from its
spatial distribution as exemplified  by
the collapsed structures appearing elongated along the line-of-sight.
More importantly, in the linear regime
the coherent motions of galaxies also introduce redshift-space distortions resulting in 
an anisotropic features  of the measured clustering statistics.
Kaiser (1987) indicated that a distant observer should expect a
multiplicative enhancement of the overdensity field along the line-of-sight,
compared to the transverse direction, due to such coherent peculiar motion
of galaxies.  This ``rocket effect" can be seen as a
flattening of structures along the line of sight.
The two-dimensional two-point correlation function
$\xi(\sigma,\pi)$, can be decomposed into two vectors:
one parallel to the line-of-sight ($\pi$) and the other perpendicular to it ($\sigma$)
Information about the coherent velocities of galaxies in the component which correlates with the galaxy correlation function (or gravitational instability induced flows - see eq. \ref{eq:k92})
can then be extracted from the anisotropy of the two-dimensional correlation function via
careful theoretical modelling of these redshift-space distortion effects
(see Hamilton 1997 for a thorough review).

Song et al (2010) and (2011) investigated the likelihood of large scale flows using
redshift-space distortions and a sample of galaxy clusters selected
from the Sloan Digital Sky Survey (SDSS).
In contrast with the results of Watkins et al (2009), Feldmann et al (2010),
Lavaux et al (2010), that measure the bulk flow of the Local Group,
Song et al (2010) and (2011) provide statistical
estimates of coherent motions due to gravitational instability on scales up to $60h^{-1}$Mpc
derived from within a large volume of the Universe at
higher redshift, e.g.,  a $0.5h^{-1}$Gpc$^3$ at
redshifts $0.16 < z < 0.32$ and $1.1h^{-1}$Gpc$^3$ at $0.32 < z < 0.47$.
The measured velocities due to gravitational instability at the mean redshift
$z=0.25$ and $z=0.38$ have an amplitude of $270-320$km/s, compatible with the
$\Lambda$CDM cosmological model.
Taken at face value, these results indicate that if bulk flow motion
is due to ``gravitational instability'' our Galaxy would be occupying
an unusual part of the Universe, next to a highly overdense or underdense region.
However, since the  correlation function in
redshift space is distorted by the divergence of the peculiar velocity field,
redshift distortions are not sensitive to a constant (dark) flow, while the KSZ measurements are.
Then, the Song et al results are compatible  with the measured velocity
of the Local Group if a large fraction of the LG dipole is due to a constant
DF extending well beyond the SDSS sample ($z\sim 0.5$).
The scale of the flow could soon be tested with
the SDSS-III Baryon Oscillation Spectroscopic Survey (BOSS),
which will provide redshift-space distortion measurements both at higher
redshift and over larger volumes of the Universe (White et al 2010). Comaprison of the Song et al (2011) results with the Watkins et al flow make it difficult to reconcile if the latter is generated by the gravitational instability as the SN analysis of Dai et al (2011) and Turnbull et al (2011) suggest. In other words, the flow identified by Watkins et al (2011) is either an artifact or must reflect the intrinsic CMB dipole, or ``the dark flow".



\subsection{Aberration: Number density of galaxies.}

Together with a dipole pattern in the luminosity distance of SN type Ia,
redshift distortions and coupling of different multipole moments,
the peculiar motion of the Earth also causes a dipole anisotropy modulation in
the distant galaxy distribution due to the aberration effect.
If we define the matter frame of the present-day Universe as the system where
the galaxy distribution looks isotropic on a sufficiently large scale
the relative motion of the Earth to the matter rest-frame causes a dipole
anisotropy modulation in the observed galaxy distribution.
This aberration effect causes a dipole anisotropy
in the galaxy distribution due to two relativistic effects. First, the photons
are blue/red shifted for galaxies in the direction of the motion or opposite
to it, causing galaxies to be brighter/dimmer and therefore included/excluded
in a magnitude-limited sample. Second, the relativistic contraction of the solid
angle for a moving observer causes the surface number density of galaxies
to be enhanced/suppressed even if the intrinsic galaxy distribution is
perfectly homogeneous on the sky. 

There have been several attempts to
explore this dipole anisotropy from such surveys (Baleisis et al 1998,
Scharf et al 2000). In particular,
Blake \& Wall (2002) analyzed the radio source distribution based on the
NRAO VLA Sky Survey (NVSS) data of Condon et al (1998), and claimed a possible
detection of the dipole anisotropy consistent with the CMB dipole
in the amplitude and direction at the $2\sigma$ and $1\sigma$ levels, respectively.

Itoh, Yahata, Takada (2010)
suggested to use this dipole pattern in the galaxy distribution
to test the motion of the LG with respect to the matter rest-frame.
In their analysis, they accounted for the covariances due to the Poisson shot
noise, the intrinsic clustering contamination and partial sky coverage.
They applied their technique to the SDSS Data Release 6, but their results
were inconclusive. Their more robust sample
indicates no dipole anisotropy in the galaxy distribution,
but the error bars were large and their results are both compatible with
the DF result and with the LG motion being of local origin.

In a similar manner, Nusser, Branchini \& Davis (2011) suggested to use
the apparent dimming or brightening of galaxies due to their peculiar motion
in large redshift surveys for measuring cosmological bulk flows.
Constraints on the bulk flow are obtained by minimizing systematic variations
in galaxy luminosities with respect to a reference luminosity function measured
from the entire survey. The method requires galaxy magnitudes and redshifts
and the shape of the luminosity function, but does not require
error-prone distance indicators and is independent of the poorly
known galaxy bias. The main difficulty is that very large numbers of
galaxies are required. Nusser et al (2011) applied their
method to the 2MASS redshift survey to measure bulk flows of spherical shells
centered on the Local Group. At $R\simeq 60 h^{-1}$Mpc they found a bulk
velocity of $(v_x,v_y,v_z)=(100\pm 90, 240 \pm 90,0 \pm 90){\rm km s^{-1}}$,
consistent with the results of Nusser \& Davis (2011).
But to measure bulk motions of amplitude $\sim 200{\rm km s^{-1}}$
at the $3\sigma$ level at redshifts $z=0.15$ or $z=0.5$ with this
method, about $10^6$ galaxies with photometric redshifts are required.

\section{Implications: connection to global structure of space-time or modified physics?}
\label{sec:implications}

The statistically significant dipole at cluster positions for all samples with
mean/median depth out to $\sim 800$Mpc indicates that 1) either the
entire Hubble volume is moving with respect to the CMB rest frame, as measured by
its all-sky dipole, or 2) the flow extends to about $\sim 0.5-1$ Gpc.
The first possibility is equivalent to a primordial all-sky CMB dipole and
would require revising our understanding of the Universe,
placing the context of the DF into what is generally called {\it the Multiverse}.
The second proposition requires a suitable adjustment to the laws of physics
governing such motions. Gravitational instability based in the highly successful
(i.e. established) concordance $\Lambda$CDM model for structure formation predicts
very different velocity field, smaller in amplitude and declining with scale.
Simply, does the DF require revising the structure of the space-time
or does it imply modifying gravity in one form or another?

Before we delve into the implications of the "dark flow" measurement,
we emphasize again that the DF result very likely, if it is
equivalent to the existence of the primordial dipole, it is not in conflict
with the standard cosmological $\Lambda$CDM model. The fact that our filtering
procedure, which is based on the standard $\Lambda$CDM model, removes the
primary CMB so successfully down to the fundamental cosmic variance limit,
is by itself an explicit demonstration of the validity of that model.

\subsection{Primordial CMB dipole anisotropy?}

If the CMB dipole is of cosmological origin, any stochastic mechanism
would be require to produce an amplitude $d\sim 10^{-3}K$ while
the quadrupole must be constrained to the measured value of $Q\sim 10^{-5}K$.
Also, as discussed in Sec~\ref{sec:tests}, if the CMB dipole is due to
a local motion, temperature fluctuations would not longer be isotropic,
due to the preferred direction of motion (e.g. Kosowsky \& Kaniashvili 2011).
Within the context of the Friedmann-Robertson-Walker model, is it
possible to generate a dipole of cosmological origin while the
quadrupole and higher order multipoles remain consistent with observations?

Interestingly, an early - preinflationary - model was indeed proposed
for this possibility by Matzner (1980) right after the measurement of the
CMB dipole by Smoot, Gorenstein \& Muller (1977). Matzner (1980) showed how,
in the context of chaotic cosmologies, the dipole anisotropy (i.e. superhorizon
scale motion) can arise as a global feature of the CMB. If at recombination matter
and radiation shared the same velocity field on superhorizon scales,
the conservation of angular momentum of the volume
elements of matter and radiation with respect to an arbitrary observer,
could have generated an intrinsic CMB dipole of amplitude $\sim 600$ km/s with
respect to comoving observers while the matter would have an
undetectable velocity. As discussed in Sec. \ref{sec:vprimeval} this arises
since after decoupling the primeval component of the peculiar velocity
field of radiation remains constant whereas that of the matter decays
as $\propto (1+z)^{-1}$.

The advent of inflation (Kazanas 1980, Guth 1981) led by the significant
progress of the high-energy vacuum physics opened the possibility of
the existance of a primeval CMB dipole also in this context. The DF
could provide an observational window to the overall (preinflationary)
structure of the Universe beyond the reach of the concordance $\Lambda$CDM
model.

\subsubsection{Curvature vs isocurvature modes of CMB}
\label{sec:cmb_anisotropies}

We first describe the theoretical connection between the CMB dipole and super-horizon
sized fluctuations. Grischuk \& Zel'dovich (1978) were the first to describe the
effect of super-horizon modes on large angular scale anisotropies
of the CMB. They showed, in what is now known as the Grischuk-Zeldovich (GZ) effect, that in a matter dominated universe, the contribution
of long-wavelength curvature modes to the CMB dipole is strongly suppressed.
Turner (1991), as discussed below, argued that this suppression would not occur in models with
isocurvature modes. If the dipole is primordial, it would reflect the nature of super-horizon isocurvature density perturbations.

We start by reviewing the discussion of the evolution of superhorizon scale perturbations and their contribution
to the CMB temperature anisotropies and discuss under what conditions
the measured CMB dipole, as implied by the DF, could be generated by those
isocurvature modes without, at the same time, given rise to a
quadrupole in excess of what is observed.

The evolution of density perturbations on superhorizon scales is complicated
by the freedom to perform gauge transformations in general relativity.
A proper treatment of cosmological perturbation theory requires clear
separation between physical and gauge degrees of freedom.
Linear perturbation theory has been extensively studied
both using gauge invariant variables (for example, Bardeen 1980, Kodama \& Sasaki 1984,
Mukhanov, Feldman and Brandenberger 1992) and restricted to specific
gauges (Lifshitz 1946, Efstathiou 1990, Bond \& Efstathiou 1986, Ma \& Bertschinger 1995).
For the present purposes, we only need to consider the scalar degrees of
freedom of the metric. Physically, these scalar modes
correspond to the Newtonian (metric) gravity with relativistic
corrections. We use the gauge freedom to eliminate the non-diagonal
terms in the perturbed metric. In this gauge, the metric element can
be expressed in terms of two scalar potentials $\phi,\psi$
\begin{equation}
ds^2=a^2(\tau)[-(1+2\psi)d\tau^2+a^2(1-2\phi)dx^idx_i]\, .
\label{eq:metric}
\end{equation}
This gauge is termed conformal Newtonian gauge and is also known
as the longitudinal gauge. It is a very convenient gauge to use
when only the scalar mode of the metric is to be considered since
the vector and tensor degrees of freedom are eliminated from
the start. First, in this gauge the metric is diagonal and
consequently the geodesic equations are simplified.
Second, $\psi$ plays the role of the gravitational potential in
the Newtonian limit and has a simple physical interpretation.
Finally, both $\phi$ and $\psi$ are gauge invariant quantities.
A comprehensive description of perturbation theory in the conformal newtonian
gauge and CMB temperature anisotropies can be found in Dodelson (2003)
and Lyth \& Liddle (2009).

In a two (matter-radiation) fluid system, a general perturbation can be expressed
in terms on the density perturbation of each component $\delta_1,\delta_2$
or in terms of the total density perturbation $\bar{\rho}\delta=
\bar{\rho}_1\delta_1+\bar{\rho}_2\delta_2$ and the perturbation
on the number density ratio $S_{12}$. For uncoupled perfect fluids,
if $w_i$ is the equation of state parameter of the $i$ component, then
$\bar{\rho_i}=a^{-3(1+w_i)}$ and the perturbation in entropy is
\begin{equation}
S_{12}\equiv\frac{\delta(n_1/n_2)}{(n_1/n_2)}=\frac{\delta_1}{1+w_1}-\frac{\delta_1}{1+w_1} .
\label{eq:defS}
\end{equation}
An adiabatic perturbation satisfies $S=0$ and an isocurvature perturbation
satisfies $\delta \rho_{\rm total}=0$. {\it These conditions are not
invariant in time; for instance, a perturbation that begins as an isocurvature
perturbation generates an adiabatic component and vice versa}.

In the first order perturbation theory, we assume that
a fluid moving with a small peculiar velocity $v^i\equiv dx^i/d\tau$
can be treated as a perturbation of the same order as
the metric potentials, the energy density $\delta\rho=\rho(\vec{x})-\bar{\rho}$
or the pressure $\delta P=c_s^2\delta\rho$, where $c_s$ is the sound
speed of the fluid. We restrict this discussion to a 3-component fluid:
cold dark matter (c), baryons (b) and radiation (r)  and express
the equations of evolution in the Fourier space. In what follows, we will
not consider any period of accelerated expansion and we neglect any dark
energy contributions.

Cold dark matter interacts with other particles
only through gravity and it can be treated as a pressureless fluid
whose energy-momentum tensor is independently conserved.
Photons and baryons are tightly coupled before recombination,
interacting mainly via Thomson scattering. After recombination the photon
mean free path increases and both fluids decouple. Baryons fall into the
DM potential wells and in the subsequent evolution they follow the
DM distribution. If we define $\rho\delta=\rho_c\delta_c+\rho_b\delta_b$,
then, the perturbed part of the energy momentum tensor gives
(Ma \& Berschinger 1995)
\begin{eqnarray}
\dot{\delta}+ikv=-3\dot{\phi}, \label{eq:delta_m}\\
\dot{v}+\frac{\dot{a}}{a}v=-ik\psi . \label{eq:velocity_m}
\end{eqnarray}
The fluid limit for photons is valid when
no causal processes such as free streaming or diffusion can separate
the baryon-photon plasma. This occurs on superhorizon scales
and when the interaction between baryons and photons is negligible.
This fluid limit is valid on scales larger than the horizon at recombination
and the photon component can be described by only two moments of the Boltzman hierarchy,
the monopole and dipole (Kodama \& Sasaki 1984)
\begin{eqnarray}
\dot{\theta}_0+k\theta_1&=&-\dot{\phi}\\ \label{eq:delta_0}
\dot{\theta}_1-\frac{\theta_0}{3k}&=&\frac{k}{3}\psi \label{eq:delta_1}
\end{eqnarray}
In this limit, the photon density contrast is $\delta_r=4\theta_0$ and
the fluid velocity is $v_r=3i\theta_1$. These expressions are valid
any relativistic species like neutrinos, so the subindex (r) refers to
perturbations in all relativistic (radiation) components.
Using eqs.~(\ref{eq:delta_m}) and (\ref{eq:delta_0}) the
time evolution of the perturbation in the matter-radiation number density
can be expressed as
\begin{equation}
\dot{S}_{mr}=\dot{\delta}-\frac{3}{4}\dot{\delta}_r=ik(v-v_r) .
\label{eq:dotS}
\end{equation}
Since fluid velocities are identical on superhorizon scales, $\dot{S}_{mr}=0$,
i.e., {\it the entropy perturbation mode remains constant on superhorizon scales}.

The anisotropies generated on the CMB radiation when
it propagates on a weakly inhomogeneous Universe were first computed by
Sachs \& Wolfe (1967). The temperature fluctuations on the sky measured by
an observer today and induced solely by scalar modes can be written as
\begin{equation}
\frac{\delta T}{T_0}\equiv\frac{T(\hat{x})-T_0}{T_0}=
\left[\frac{1}{4}\delta_r+\psi\right]_e^0-[x_iv^i_b]_e^0-\int_e^0[\phi'-\psi']d\tau ,
\label{eq:sw_effect}
\end{equation}
where $e,0$ corresponds to the time of recombination and the present
time, respectively and $\hat{x}$ is the unit
vector in the direction of the line of sight. The different terms
in eq~(\ref{eq:sw_effect}) correspond to the intrinsic photon inhomogeneity
that is present at the last scattering surface, the inhomogeneites of the
metric (termed Sachs-Wolfe -SW- effect), the Doppler shifts induced
by the relative velocity of the emitter and the observer and the Integrated
Sachs-Wolfe (ISW) effect due to the time variation of the gravitational potentials
along the line of sight. The term $[\delta_r+\psi]$ evaluated today
contributes only to the monopole and is, therefore, unobservable.

In the context of superhorizon perturbations,
the velocity term in eq~(\ref{eq:sw_effect}) requires a more detailed discussion.
While the velocities are defined with respect to comoving observers
in the conformal newtonian gauge, those velocities are unobservable.
The motion of the Local Group $\vec{v}_{LG}$ is measured with respect to the
Last Scattering Surface and we have to add the motion of the Last Scattering
Surface with respect to comoving observers $\langle \vec{v}(\tau_e)\rangle$, i.e.,
$\vec{v}_{LG}=\vec{v}(\tau_0)-\langle \vec{v}(\tau_e)\rangle$.
If $\theta$ is the angle measured from the direction of the apex of the motion
and $\mu=\cos\theta$ then $x_iv^i(\tau_0)=\mu v_{LG}-\langle x_iv^i(\tau_e)\rangle$.
The darkflow measurement argues that the dipole due to the motion of
the LG is subdominant with respect to an intrisic, cosmological, dipole.

If we define
\begin{equation}
\frac{\delta T}{T_0}\equiv\frac{T(\hat{x})-T_0}{T_0}=
\int\frac{d^3k}{(2\pi)^3} \sum_{\ell=0} i^\ell (2\ell+1) P_\ell(\mu)
\Delta_\ell(k,\tau_e,\tau_0)
\end{equation}
then
\begin{equation}
\Delta_\ell = \Delta^I_\ell+\Delta^{II}_\ell+\Delta^{III}_\ell+\Delta^{ISW}_\ell
\label{eq:delta_l}
\end{equation}
with
\begin{eqnarray}
\Delta^{I,ad}_\ell&=&[\frac{1}{4}\delta_r+\psi](\tau_e)j_\ell(k(\tau_0-\tau_e)) ,
\label{eq:Delta_I}\\
\Delta^{II}_\ell&=& iv(\tau_e)j_\ell'(k(\tau_0-\tau_e)) ,\label{eq:Delta_II}\\
\Delta_\ell^{III}&=& i\frac{1}{3}\left[v(\tau_0)-v(\tau_e)j_0(k(\tau_0-\tau_e))\right] ,
\label{eq:Delta_III} \\
\Delta^{ISW}_\ell &=& -\int_{\tau_e}^{\tau_0}\tau
\frac{\partial (\phi-\psi)}{\partial\tau}(\vec{k},\tau)j_\ell(k(\tau_0-\tau)) .
\label{eq:Delta_ISW}
\end{eqnarray}
Velocities and potentials are related by
\begin{equation}
\dot{\phi}+\frac{\dot{a}}{a}\phi=-i\frac{3}{2}\left(\frac{\dot{a}}{a}\right)^2k^{-1}v .
\label{eq:potential_velocity}
\end{equation}
For the scales of interest and
in the matter dominated regime $\phi=-\psi\simeq const$,
$\dot{a}/a=2/\tau$ and
eq.~(\ref{eq:potential_velocity}) can be written as
\begin{equation}
v(\tau)=\frac{1}{3}ik\tau\phi(\tau) .
\label{eq:velocity_potential}
\end{equation}
Since the potentials are constant in the matter dominated regime, there ISW effect
is null. For adiabatic initial conditions (see Dodelson 2003) we have
\begin{eqnarray}
\Delta^{I,ad}_\ell&=&
\frac{1}{3}\phi(\tau_0)j_\ell(k(\tau_0-\tau_0)), \label{eq:Delta_Ib}\\
\Delta^{II}_\ell &=& -\frac{1}{3}k\tau_e\phi(\tau_0)j'_\ell(k(\tau_0-\tau_0),
\label{eq:Delta_IIb}\\
\Delta_\ell^{III}&=&-\frac{1}{9}\phi(\tau_0)k (\tau_0-\tau_e) .
\label{eq:Delta_IIIb}
\end{eqnarray}
In eq.~(\ref{eq:Delta_IIb}) and (\ref{eq:Delta_IIIb})
we have replaced $\phi(\tau_e)$ by $\phi(\tau_0)$ since recombination
occurs well within the matter dominated regime.

We can now compute the contribution to the CMB dipole and quadrupole
from perturbations that are outside the horizon today by taking the limit
of $k\tau_0\rightarrow 0$ in eq.~(\ref{eq:delta_l}). In this limit
$j_1(x)\approx (x/3)$, $j_2(x)\approx (x^2/15)$, $j_1'(0)=1/3$ and
\begin{eqnarray}
\Delta_1&=&\frac{1}{90}\phi(\tau_0)(k\tau_0)^3 + O[(k\tau_0)^5]
\label{eq:dipole_1}\\
\Delta_2&=& \frac{1}{45}\phi(\tau_0)(k\tau_0)^2+O[(k\tau_0)^4]
\end{eqnarray}
The leading order $(k\tau_0)$ of the dipole is canceled exactly,
as discussed by Turner (1991), and the first non negligible contribution
is of order $(k\tau)^3$, the same as in the octupole. Since
\begin{equation}
\frac{\Delta_1}{\Delta_2}=\frac{1}{2k\tau_0} ,
\end{equation}
the adiabatic modes that are outside the horizon today can not contribute
significantly to the dipole without generating a much larger quadrupole.
Then, the low value of the measured quadrupole sets an upper limit
to the contribution of large scale perturbations to the intrinsic dipole.
The measured dipole is $\Delta_1\sim 10^{-3}$ while the quadrupole
is $\Delta_2\sim 10^{-5}$ then, it could not have been generated
by adiabatic density perturbations.

The exact cancelation of the dipole at first order in $k\tau_0$ holds not only
during the matter dominated regime. Erickcek, Carroll and Kamionkowski (2008) and
Zibin and Scott (2008) have shown analytically and numerically, respectively,
that for adiabatic super-Hubble modes the contribution to the dipole is indeed
suppressed, regardless of the matter content, pointing out that this behavior
is a simple consequence of adiabaticity. This is not true when isocurvature
density perturbations are considered. These perturbations
add an extra contribution to the large scale temperature anisotropies
(Turner 1991, Langlois \& Piran 1996, Langlois 1997) that is different
for matter or neutrino isocurvature density perturbations (Gordon \& Lewis, 2003).
For matter isocurvature perturbations $S_{mr}$ this contribution is
\begin{equation}
\Delta_\ell^{I,iso}=-\frac{2}{5}S_{mr}(k,\tau_0)j_l(k\tau_0) ,
\end{equation}
and the dipole to quadrupole ratio induced by this contribution is
\begin{equation}
\frac{\Delta_1}{\Delta_2}=5(k\tau_0) .
\end{equation}
The observed dipole can be generated without perturbing the quadrupole
for sufficiently large isocurvature modes, i.e., $k\tau_0\gsim 100$.
This was the proposal of Turner (1991) discussed below.

\subsubsection{Preinflationary remnants}

In Sec.~\ref{sec:cmb_anisotropies} we have shown that,
as opposed to curvature perturbations, isocurvature
modes with wavelengths much larger than the current Hubble radius
can produce an intrinsic CMB dipole of the right amplitude and be
compatible with the low value of the quadrupole.
In the context of the DF, the $\Lambda$CDM model would contain
isocurvature modes that would be the origin of primordial CMB dipole
whereas curvature fluctuations such as the adiabatic density
perturbations, presumably generated during inflationary slow roll-over,
are responsible for gravitationally induced velocity components. We now
turn to some specific models that have been proposed to generate the
primordial CMB dipole component.

Turner (1991,1992) was the first to propose that an intrinsic
dipole or ``tilt'' across the observable Universe could be generated
by preinflationary structures. He noted
that any curvature component of the perturbations would lead to zero CMB
dipole because it is precisely cancelled by a corresponding dipole in the
gravitational potential. Simply put, the frame defined by the isotropy of
the CMB must coincide with that of the universal isotropic expansion.
However, a tilt will still arise if, in addition to the inflaton field, other
fields with negligible contribution to the total energy density also existed
(e.g. axion). The perturbation associated with the dynamically subdominant,
at the time, fields would be approximately isocurvature because it leaves the
net density (approximately) unperturbed. In that case, an intrinsic CMB dipole
could be generated by a superhorizon sized isocurvature perturbation
remnant of the preinflationary period.

Since the ``tilt'' requires isocurvature remnants  of amplitude $\delta \sim 1$
on scales $L\sim 100\tau_0$ to have survived inflation, it is useful to ask what
is the likelihood that inflation had extended $5-10$ e-foldings more than
necessary for solving the horizon problem. Freivogel et al (2006)
use a parametrization of the inflationary potential to evaluate the
probability distribution for the number of e-folds $N$, finding that it is
proportional to $1/N^4$. Given the observed bound of $N\sim 62$ e-foldings, they
found a probability of about 10\% that the actual number of e-foldings is
between 62 and 64 and argue that such a small number of e-folds
would have observable consequences. This probability distribution implies a non-negligible likelihood of other preinflationary remnants being close enough to cause a dark flow (or primordial CMB dipole).

In a different context, Paczynski \& Piran (1990) showed that a non-centered
observer in a spherical symmetric Tolman-Bondi dust model containing
a spherical distribution of radiation could measure a CMB dipole significantly
larger than the quadrupole due to a radially varying specific entropy.
Jaroszynski \& Paczynski (1995) claimed that it was impossible
to obtain a dipole far larger than the quadrupole from either
adiabatic or isocurvature perturbation.
Langlois \& Piran (1996) demonstrated that the latter claim was not
correct by examining the growth of density perturbations on scales
larger than the Hubble radius. They showed that a CMB dipole could
be the result of isocurvature perturbations on scales larger than the
current Hubble radius, but only with wavelengths several hundred
times greater than the Hubble radius in order to avoid the quadrupole
amplitude being inconsistent with observations. They pointed out that a
perturbation that begins initially as a purely isocurvature species would
remain so until enters the horizon. In order to explain the origin
of these superhorizon perturbations
Langlois (1996) proposed a double inflation scenario where the
transition between inflation driven by the heavy and light scalar fields
takes place at the scales far larger than the Hubble radius today.
Langlois (1997) examined the idea of a dipole generated
by ultralarge isocurvature modes in the context of open cosmologies and showed
that the isocurvature interpretation to be feasible only if the model
is very close to being flat, $|\Omega_0-1|\le 10^{-4}$ as required by
an earlier discussion in Kashlinsky et al (1994).
In another words, departures from spatial flatness also rule out
a primordial dipole, so the measured flatness indirectly support
the DF measurement as an intrinsic dipole. Zhang \& Stebbins (2011) discuss the problems the Tolma-Bondi type models face viz-a-viz the new observations of the CMB from ground-based telescopes with arcminute resolution.

\subsubsection{Primeval landscape}

In the context of string theory, our current observed Universe is just one
of a large ensemble $\sim 10^{500}$ of similar pocket Universes, termed
string landscape. Without a guiding principle that selects a subclass of
those vacua where the inflationary dynamics generates an homogeneous
and isotropic Universe as we observe, has not been possible to make
the string landscape predictive.
Interestingly, anthropic principle can provide a hint as to why our
Universe is homogeneous on scales encompassing at least $\gsim 10^{22}M_\odot$.
For carbon to form in stellar nucleosynthesis (and hence lead to carbon-based life), the present day Hubble constant
has to be $H_0^{-1}\lsim 10^{10}$ yrs. The mass contained within the observable
patch that has had the time to expand for that amount of time is then given
by $M_{\rm hor}\sim \frac{c^3}{G}H_0^{-1} \simeq 6\times 10^{22}M_\odot$
in agreement with what is observed.

Mersini-Houghton (2005), Holman \& Mersini-Houghton
(2006) and Holman, Mersini-Houghton \& Takahashi (2008a)
argue that the initial conditions for inflation in the surviving universes
could be selected through gravitational quantum dynamics in the landscape if
string theory gives the correct description of quantum gravity.
In their view, the landscape is the configuration space for the wave
function of the Universe and superhorizon inhomogeneities have induced non-local
quantum effects between our local volume and modes and domains beyond the
current horizon. Holman, Mersini-Houghton \& Takahashi (2008b) confront
their model of the string landscape with cosmological observations. They argue
that the decohering effects of long wavelength modes on the wavefunction of
the Universe leave unique signatures on the CMB spectra and Large Scale Structure.
In Holman \& Mersini-Houghton (2009) that the quantum entanglement of our Hubble
volume with those superhorizon modes will give naturally a bulk flow
with a correlation length at a horizon scale. Their model adscribes
the 'tilt' in the gravitational potential to preinflationary remnants of the
landscape. While the mechanism is similar to the one advocated by Turner (1991)
it presents a framework to explain the origin of the dark flow, predicting an
amplitude close to the measured value. Placed in this context, the bulk flow could
provide a probe of the preinflationay physics and a window onto the landscape multiverse.

\subsection{Alternative models.}

If the current DF measurements do not reflect a primordial CMB dipole, but simply are a consequence of the velocity field with a large, but finite and sub-horizon, coherence length, accommodating the data could imply certain modifications to the overall physics laws rather than to our understanding of the global world structure. This may also require fundamental departures from the inflationary paradigm based on the vacuum-like quantum physics. Those include modifications of the concordance model paradigm like higher dimensional gravity models, such as the Dvali et al (2000, DGP) gravity with extra spatial dimension within which the observed (3+1)-dimensional world is embedded as a brane.  Other modifications used models with peculiar motions between ordinary matter and dark energy to alternative models with no accelerated expansion or modified gravity designed as an alternative the dark matter.


Beltr\'an-Jim\'enez \& Maroto (2009a) argued that the DF is an indication
of the dark energy reference frame being different from the matter-radiation
frame at decoupling. The dark energy flow would be responsible for the
relative motion between the Matter Rest Frame and the CMB Isotropic Frame.
Like superhorizon isocurvature perturbations, this moving dark energy mechanism gives an
additional contribution to the quadrupole that could erase a fraction of the
inflation generated quadrupole, explaining the low measured value and its
alignment with the dipole.

Beltr\'an-Jim\'enez \& Maroto (2009b) considered the cosmic evolution in
general vector-tensor theories of gravity and found that vector perturbations
could be supported by perturbations of the vector field so that, unlike the
concordance $\Lambda$CDM model, one could obtain large peculiar velocities such as
the DF measurement. 

Carroll et al (2010) considered the observable effects on the
CMB that would have a small violation of translational
invariance during inflation, that could be achieved by means of a vector
field, but gave no specific mechanism as of how to generate a DF.

Wyman \& Khoury (2010) considered a class of modified gravity theories,
using DGP brane-induced gravity, that include additional scalar forces
that are inherited from the higher dimensional massless graviton.
Compared with the Newtonian gravity, these new degrees of freedom enhance
the effective gravitational attraction at late times and at large scales,
speeding up the growth of structure at low $z$. In these models,
bulk flows can be enhanced about 40\% relative to the $\Lambda$CDM model.
The predicted peculiar velocities alleviate the tension with the observed
bulk flow of Watkins et al (2009) but seems insufficient the amplitude
and coherence of the DF assuming the KAEEK measurement extending to $\sim 800$ Mpc.

Tsagas (2010, 2011) argues that peculiar velocities change the rate of expansion
of observers moving with respect to the smooth Hubble flow and observers
could experience accelerated expansion within a globally decelerating
Universe that would also have a dipole pattern fairly aligned with the
velocity dipole.


There were also the perhaps inevitable suggestions that DF support more radical revisions of the laws of gravity and motion. Before we describe this possibility, we note that the general relativity
theory remains remarkably accurate (Reyes et al 2010, Wojtak et al
2011)\footnote{Bekenstein \& Sanders (2011) noted, however, that the Wojtak et al (2011)
method of measuring the mean profile
of the redshift distortion in (stacked) clusters should lead to the same theoretical expectation the same for all metric theories of
gravity. Bekenstein \& Sanders then noted that the Wojtak et al result and
conclusion are valid  inasmuch as the average NFW profile of cluster mass extends to
the scales probed in the stacking ($\sim$several virial radii). We should
point out in this context that it has been demonstrated by us (AKKE) that
the true distribution  the hot gas clusters of galaxies is well described by
the NFW model and the isothermal $\beta$-model provide unacceptable fits to the measured
TSZ profiles.}

Lastly, we note that DF was cited by Milgrom (2010) to provide evidence for modification
of gravity designed to eliminate dark matter (known as MOND). This is based on
misreading - or neglecting - the entirety of our results. Indeed, the filtering we
employed was based on the standard $\Lambda$CDM model and the filtered maps have
been demonstrated to have removed the primary CMB down to the fundamental limit
imposed by the cosmic variance. This by itself supports the $\Lambda$CDM
paradigm, {\it including the existence of dark matter as required by this model}.

Should the motions at such large scales reflect deviations
from the Newtonian-Einsteinian laws of motion they would constrain the possible modifications of the latter as follows. The linear
continuity equation requires via the mass conservation
$\vec{\nabla}\cdot\vec{v} = - \dot{\delta}$. Assuming that gravity
is irrotational within the modified theory, gives $v_k \propto
\dot{\delta}_k/k$ in the direction of $\vec{k}$, so any such
modification should thus require that the evolution/growth for
each mode depends on the scale/mass it contains and that harmonics
containing larger mass grow at a faster rate.  Specifically, since
the velocity field with an assymptotically flat bulk flow is
recovered when $|v_k|^2\propto k^{-3}$, such modified theory would
have to require a ``generic law" of $\dot{\delta}_k \propto
k^{-1/2} \propto M^{1/6}$. Assuming that the measured velocity in
the local part of the Universe is representative of a typical
(rms) velocity field and that the power spectrum is approximately
Harrison-Zeldovich ($\langle |\delta_k|^2\rangle \propto k$) on
the relevant scales ($\gsim 30 h^{-1}$Mpc), the relative
growth-rate of fluctuations would be required to increase with
mass as $|d\ln \delta_k /dt | \propto k^{-1} \propto M^{1/3}$.
Either way, it is unclear, however, whether such peculiar growth
of structure with large scales growing at a faster rate can at the same time 
account for the observed cosmogonical structure formation
sequence.

\section{Future prospects: SCOUT and beyond}
\label{sec:future}

\subsection{Currently unresolved issues}
\label{sec:unresolved}

While there is evidence of the statistically significant CMB dipole associated exclusively with clusters of galaxies, which has the properties of the KSZ component, the current methodology developed by us still has not resolved several outstanding issues. While their status does not affect the fact of the detection of the highly coherent motion out to at least $\sim 800$Mpc which is well above that predicted for the gravitational instability component, their accurate resolution is important for proper interpretation of the detected flow and the underlying physics behind this phenomenon.

Here we summarize the most important of these issues where we anticipate progress in the coming time, which we list in order of their acuteness:

\begin{itemize}
\item {\bf Calibration with sign}.
Fig. \ref{fig:kabke2_9} shows the inadequacy of the current calibration at large apertures. Whereas our current catalog (open circles) gives TSZ which matches well the measurements around $\theta_X$, it fails at larger apertures where the dipole is measured. The reason is that the clusters are not described by isothermal $\beta$-modeling which formed of the cornerstone of our initial catalog parameter calculation. Instead, it appears that clusters should be described by NFW (and/or Arnaud et al) type profiles. This requires significant re-calibration of the catalog, which is currently in progress. The necessity of highly accurate is exacerbated by the tentatively seen sign change, presumably arising from the wide side-lobes of the KABKE filter. The upcoming Planck data, with their finer angular resolution, lower noise and wide frequency coverage, will be particularly helpful to test the recalibrated catalog in resolving this issue. This is discussed in some detail below.

\item {\bf Understanding filtering}.  The Wiener filter employed by us  to remove
the primary CMB anisotropies is designed to minimize
$\langle(\delta T-\delta_{\rm noise})^2\rangle$ and, as shown by
AKEKE,  indeed
removes the primary CMB fluctuations down to the fundamental
limits imposed by cosmic variance.
We will explore alternative
filtering schemes in order to further increase the S/N of the
KSZ measurement and explore any
potentially remaining systematics.   In
addition, we will analyze the effect of filters in real space that
subtract the CMB locally (e.g., multifrequency-matched filters)
and have been used to search for point sources in CMB data, while
paying close attention to attenuation effects. If the filtering
suppresses the KSZ contribution as much as that of the primary
CMB fluctuations or more, bulk-flow measurements using the KA-B method
are rendered impossible. (One example of an ill-suited
filter would be filters designed to detect radio point sources
rather than remove the primary CMB, the main contaminant in the
KA-B method).
Our investigation  will thus be
restricted to filters that preserve or increase the S/N of the
bulk flow. To this end we will analyze contributions in
$\ell$-space to the measured dipole in order to eliminate possible
systematics that may arise from differences in sampling of
different parity multipoles. Modifications of our current
filter may also become necessary in order to account for
the specifics of the scanning strategy used by the Planck
satellite which lead to a larger 1/f component in the instrument
noise than for WMAP data.
When filtered, data are Fourier transformed to $\ell$-space and back into
real space. In this process, the power in the TSZ monopole and KSZ
dipole signals leaks to higher even and odd multipoles, respectively.
Since filtering preserves phases, the dipole is, in the
filtered map, recovered only at cluster locations; by
contrast, leakage from the Galactic cut will be seen in {\it all} pixels
in the filtered data. The
different behavior of the SZ components, thermal and kinematic,
under filtering can be exploited to devise better filtering
schemes. We will use numerical
simulations of the impact of filtering on the KSZ signal from
random NFW profiles to compute the calibration more accurately.
Filters suited to detect point sources or to remove the TSZ component could
remove the CMB on large scales but boost it on small scales, reducing the
S/N measurement of the KSZ dipole at cluster locations. The formalism described in AKEKE enables to determine the efficiency of any filter as discussed above.

Fig.~\ref{fig:akeke1} indicates that our filter removes the
intrinsic CMB down to the fundamental limit imposed by cosmic
variance. In this sense the filter is close to optimal, since it
minimizes the errors contributed to our measurements by primary
CMB. In principle, one can define a more aggressive filter that,
together with the intrinsic CMB, also removes the noise leaving
only the SZ signal. But filtering is not a unitary operation and
does not preserve power. Such a filter would then remove an
important fraction of the SZ component and would probably reduce
the overall S/N. In general, a different filter would give
different dipole (measured in units of temperature) and would
require a different calibration. Alternative filtering schemes
that maximize the S/N ratio and minimize the systematic error on
the calibration and their application to DF measurement will be
investigated in the course of this experiment.

\item {\bf Window function and coherence}. The current estimates of the flow scale and coherence are crudely taken as median/mean depths of the clusters samples. This is clearly inadequate for modeling any possible flow in the presence of components arising from 1) thermal motions in clusters and superclusters on smallest scales, 2) gravitational instability component described by the linear approximation of the $\Lambda$CDM model on the intermediate scales, and 3) possible dark flow on the largest scales. Thus proper modelling should include computation of the window function of the recalibrated catalog and evaluation of the coherence of the net velocity field decsribed by the above three components in the presence of noise.

\item {\bf Shear and higher moments}. Shear is an important test of the nature of the measured flow. It is reflected in the higher moments of the CMB signal measured at cluster pixels. However, as discussed above, once these moments are measured one needs to know cluster catalog parameters to high accuracy in order to subtract out the TSZ (and other cluster foreground) contributions to the accuracy necessary to identify/constrain the shear with any accuracy. Thus expanded and extended recalibrated catalog will be critical in this task as will the upcoming Planck data.

\end{itemize}

\subsection{SCOUT}

We have designed and are conducting an experiment, dubbed SCOUT ({\bf S}unyaev-Zeldovich {\bf O}bservations of the {\bf U}Universe's {\bf T}ilt), which builds upon the Dark Flow study and seeks to
confirm, improve upon, and extend our earlier measurement. SCOUT goals are to measure the bulk flow more accurately, extend the measurement to still greater scales and determine the flow properties with reasonable accuracy. The central elements of the
experiment, are divided into the two main components: the compilation
of a much-improved X-ray cluster database comprising over 1,500
clusters at redshifts extending to $z\simeq 0.7$, and the improved
KSZ measurement attainable with it, using CMB data from the WMAP
and Planck satellite missions.

SCOUT has four immediate experimental objectives:
\begin{enumerate}
\item A more accurate measurement of the amplitude of the flow. We
aim to achieve
  this goal by using a larger cluster catalog and less noisy CMB data, and by
  performing a more accurate calibration of the conversion from CMB temperature decrement to bulk-flow velocity.
\item An improved measurement of the extent of the flow. We will
probe bulk motions out to much larger scales (up to 1-2 Gpc) with
the help of a newly compiled cluster catalog containing systems
out  to redshifts of $z\simeq 0.7$.
 \item An accurate determination of the coherence of the flow. We will achieve this objective via detailed
evaluations of the final cluster sample using window functions for
various configurations.
 \item A measurement of the magnitude of the shear of the flow. This diagnostic measurement as well as
 improved assessments of possible systematic biases will become feasible by applying our
 extended cluster catalog to yet better CMB data from WMAP and Planck, and by taking advantage
 of Planck's greatly improved frequency coverage as well as of differences in the scanning strategies
 between the two missions.
\end{enumerate}

Via these experimental goals, SCOUT will enable us to
identify the origin of the flow and its implications for
fundamental properties of the pre-inflationary Universe.
We here describe in greater detail the central elements of the
SCOUT experiment, divided into the two main components: the compilation
of a much-improved X-ray cluster database comprising over 1,500
clusters at redshifts extending to $z\simeq 0.7$, and the improved
KSZ measurement attainable with it, using CMB data from the WMAP
and Planck satellite missions.

As outlined before in Section \ref{sec:xraycat}, the SCOUT cluster catalog is being compiled by building upon the identifications made in the course of the MACS and CIZA cluster surveys. It goes beyond these previous surveys by applying no flux limit to the initial RASS X-ray source catalogs and no redshift limit to the cluster candidates identified. Tentative identifications are obtained in three steps. The first level consists of visual scrutiny of Digitized Sky Survey images, a process that eliminates obvious non-cluster RASS sources and allows a reliable identification of X-ray luminous clusters at $z<0.2$. For RASS sources unidentified after the first step, Sloan Digital Sky Survey images will be examined, thus adding color information and sufficient depth to permit robust identifications at $z<0.4$ in about 1/3 of the sky. As the "3$\pi$" survey conducted by the PanSTARRS PS1 telescope progresses, SCOUT cluster  identifications will be enhanced and expanded to cover all X-ray luminous clusters within $z<0.6$ for the 3/4 of the sky surveyed by PS1. With no similar optical imaging database available at declinations below $\delta=-30$ deg, the resulting catalog will suffer from large-scale anisotropies. Modeling and, if possible, correcting the impact of any such anisotropy will be a priority for SCOUT. As the catalog is being compiled (a first version encompassing about 1,000 clusters was successfully used by KAEEK), dedicated follow-up observations are conducted from Mauna Kea (Hawai'i, USA) in the northern, and from La Silla (Chile) in the southern hemisphere, to ensure that all SCOUT clusters have spectroscopic redshifts. At redshifts exceeding $\sim 0.3$ we will complement this RASS-based catalog by adding the most X-ray luminous clusters from serendipitous cluster surveys, such as the 400 sq.deg. project or WARPS (cf.\ Fig.~\ref{fig:lx-z}). Again, the isotropy of these supporting catalogs will be examined closely and modeled if necessary.

\begin{figure}[b!]
\includegraphics[]{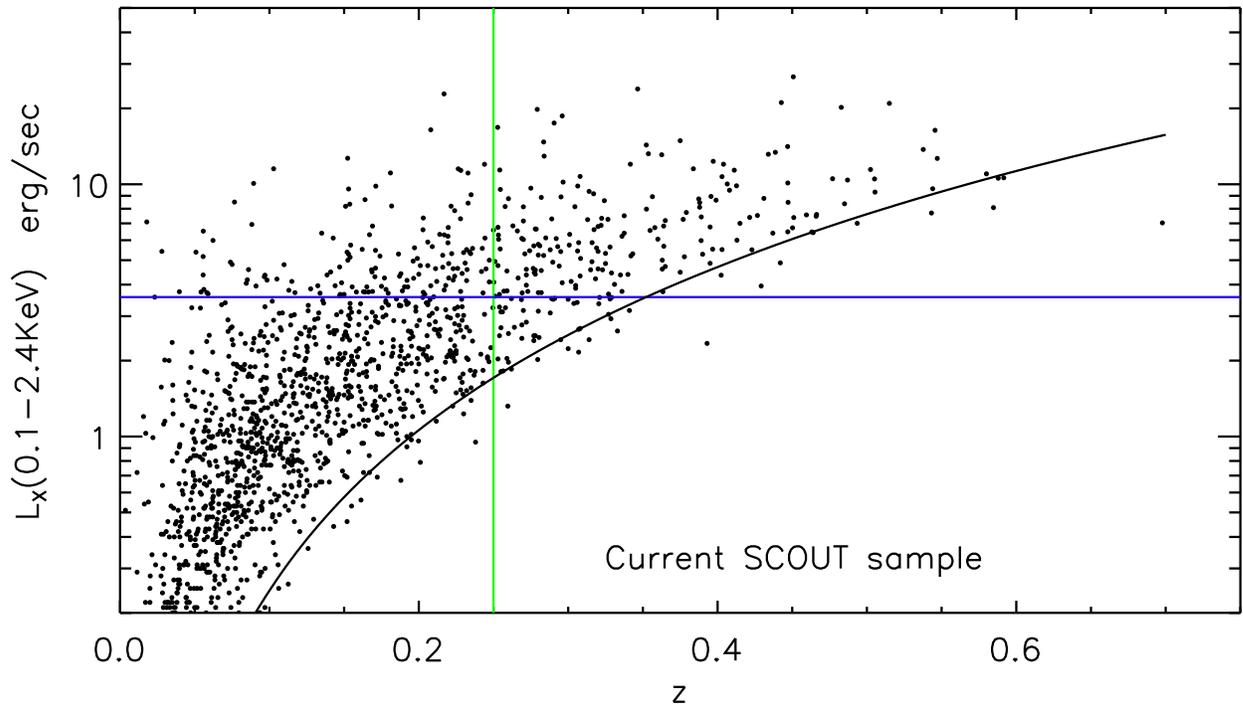}
\caption{\small{ X-ray
luminosity vs redshift for the clusters in the current SCOUT sample. The
shown luminosities represent the total emission detected during
the RASS. By contrast, the nominal flux limit  (black solid line)
relates to the flux measured in the original RASS ``detect cell".
The scatter around the flux limit is thus either caused by the
detect cell being too small to capture the entire flux, or by
contaminating point sources that erroneously inflated the
detect-cell flux. The blue line marks the X-ray luminosity of the
Coma cluster.} \label{fig:future_scout}}

\end{figure}
Fig.~\ref{fig:future_scout} illustrates why and how SCOUT will take the
Dark Flow measurement to the next level. The upper panel shows
that, already at the redshift limit of our current
study, even massive clusters are
unresolved in the WMAP channels. By contrast, the Planck mission
will resolve Coma-type clusters out to the redshift limit of the
SCOUT catalog ($z{\sim}0.7$) which will enable us to calibrate the
measurements, i.e., to convert the measured dipole CMB amplitude
into a flow velocity. As importantly,  Planck provides data at
217\,GHz where the TSZ signal vanishes, thus allowing us to probe
directly the KSZ component and its dependence on cluster-centric
radius.

The lower panel of Fig.~\ref{fig:future_scout} shows the $L_X-z$
distribution of our current cluster sample. As part of the SCOUT
experiment we will progressively lower the X-ray flux limit and
focus our survey work on distant systems, in order to extend our
present sample to well over 1,500 clusters out to $z\ga0.6$. All
clusters added at $z\ga0.2$ will be intrinsically X-ray luminous
by virtue of the flux-limited nature of the sample. Hence, they
will also feature higher values of $\tau$, leading to a larger KSZ
contribution and a stronger bulk-flow signal out to larger
distances (redshift).

\subsection{Application to Planck}

The Planck mission has been designed to produce a full-sky survey
of the CMB with unprecedented accuracy in temperature and polarization
(Ade et al 2011a). The data are projected to be released in January 2013.
The instrument operates at nine frequencies
logarithmically spaced in the range 30-857GHz. The in-flight
performance of the High and Low Frequency Instruments (HFI and LFI) have been
described by Ade et al (2011b) and Mennella et al (2011).
These particular data sets will enable several improvements.
First and foremost, at the Planck 217GHz channel the TSZ
signal is small down to the relativistic corrections to the TSZ terms, eliminating one source of systematic error.
The channel has the angular resolution with
FWHM of $5^\prime$ which allows resolving inner parts clusters out to larger $z$ than WMAP. Second, the noise levels are small.
The measured HFI in-flight performance has shown that  in the first
year of operation to have reached better than expected\footnote{http://www.rssd.esa.int/SA/PLANCK/docs/Bluebook-ESA-SCI(2005)1\_V2.pdf} sensitivity
requirements. On the negative side, the noise is fairly inhomogeneous
largely dominated by a white noise component, but containing a $1/f$ contribution.
The sky is scanned in great circles, every circle collected in
about $\sim 1$ minute. This strategy generates $1/f$ noise terms
that affect preferentially the {\it low} multipoles, but even then the
noise levels are very small. For the 217GHz channel, white noise level
after one year of operation is $13\mu$K per pixel of $3.4'$ on the side.

The advantages of Planck data for resolving the DF issues are several: Its instruments have much finer angular resolution than WMAP (5 arcmin FWHM), and the High Frequency Instrument (HF) covers both sides of the zero crossing frequency of the TSZ component. Its 217 GHz channel, where the TSZ component is negligible in the absence of relativistic corrections, will likely prove important for calibration of the measured CMB dipole at cluster positions, although, the foreground contributions at that frequency are substantial (reference). Still, with proper methodology as outlined above, it could rule in on the tentatively detected sign change in the CMB dipole profile. Additional potential problems could arise due to the  $1/f$ noise resulting from the Planck scanning strategy as this may require some modifications to the KABKE filtering scheme.

To characterize the noise and foreground emission, the HFI and LFI core
teams have constructed maps with the CMB cosmological contribution
subtracted off the Time Ordered Information (TOI). They have used six
different component separation algorithms to obtain the primordial CMB signal.
The difference between the six methods provides an estimate of the CMB residual
(Ade et al 2011c, Zacchei et al 2011). These maps can be useful as an
alternative to the filtering schemes used with WMAP data. Also,
due to its wide frequency coverage, Planck will allow a detailed
investigation of the astrophysical contaminants.

The satellite has shown to be an optimal instrument for blind detection
of clusters using the TSZ effect. The first clusters detected by PLANCK
include 189 cluster candidates with signal-to-noise larger than 6
(Ade et al 2011c). These SZ clusters are mostly at moderate redshifts
(86\% are at $z<0.3$) and span over a decade in mass, up to the rarest and most
massive clusters with masses above $10^{15}$M$_\odot$. In combination
with the South Pole and Atacama Cosmology Telescopes it will allow the determination
of the pressure profile of clusters with good accuracy which would be
essential in order to solve for the calibration of the temperature dipole in terms of the equivalent km/sec. It will also clarify the origin of the sign change tentatively already seen (KAE) in the W band of WMAP.

Thus Planck data will be an important testing ground for the reality of DF. However, given that the signal has been detected by us at only $S/N \simeq 3.5-4$, we emphasize again that incorrectly applied methodologies can, as demonstrated in mathematical detail in this report, reduce the S/N of the measurement to below statistically significant levels ($S/N\lsim 2$) rendering the measurement impossible. As discussed in KAE we have proposed such a measurement to the Planck collaboration as early as 2005\footnote{\url{http://www.rssd.esa.int/SA/PLANCK/docs/Bluebook-ESA-SCI(2005)1\_V2.pdf}} and again in 2009. We plan to perform the necessary verification as soon as the first Planck data is publicly released (projected in early 2013). The data needed to verification of the WMAP-based DF results has been posted by us publicly at \url{http://www.kashlinsky.info/bulkflows/data\_public} and has indeed been verified by numerous colleagues. Should the signal not be found in the Planck analysis that would imply a systematic difference (presumably at cluster locations) between the WMAP and Planck maps, which will require an explanation of their respective data processing pipelines. We use this space to provide several comments concerning specifically Planck-related issues and its potential promise in resolving the issues described above in Sec. \ref{sec:unresolved} for which the upcoming SCOUT catalog will be critical.

Provided the Planck measurements over the selected $\sim30,000-40,000$ pixels identified with clusters coincide with WMAP data, the results of applying the KABKE filtering methodology should produce similar results. The converse is also true, so looking at the consistency (or not) of the Planck/WMAP datasets over the selected (relatively) few pixels (in the correspondingly filtered maps) is critical for this comparison. If there are differences between the two datasets, that - quite likely - would imply differences in the underlying cosmological model, considered established by WMAP. In that eventuality, the argument can be settled via a corresponding refinement of the argument given in AKEKE and \ref{fig:akeke1}, which show that - with the WMAP-established $\Lambda$CDM model - the primary CMB is removed by filtering down to the cosmic variance limit.

In this regard we emphasize again the importance of using an appropriate filtering scheme in Planck measurements. KABKE filter is proven to remove primary CMB down to the fundamental limit of cosmic variance without the correspondingly large removal of cluster KSZ. On the other hand, filters such as used by e.g. Osborne et al (2011) are designed to detect point sources, not remove primary CMB, and,  as discussed in detail in Sec. \ref{sec:rebuttals}, reduce S/N ratio of the measurement if using the KA-B method. With the KABKE filter, and the current cluster catalog, we achieve a highly - but not hugely - statistically significant measurement at $S/N \simeq 3.5-4$. Because filtering can aim to remove primary CMB down only to the cosmic variance limit (already achieved in KABKE) one needs to be careful with alternative filtering schemes since they may reduce the S/N of such measurement below the statistically significant level. This  is especially true when working with cluster compilation which are significantly complete that even the current SCOUT catalog.

Clearly, since it is the primary CMB that ought to be filtered out, the underlying cosmological model of the Planck data must be established with high fidelity {\it before} the DF measurement is attempted. Because the errors of the measurement are - for the KABKE filter - dominated by the cosmic variance component of primary CMB, the only way to increase the S/N of the measurement is {\it by increasing the number of clusters in the catalog}. This would be especially true if using smaller apertures in conjunction with the 217 GHz channel Planck CMB data, where 1) noise will also contribute because of the fewer pixels covered by the aperture and 2) foreground pollution is significant. The latter may force a larger CMB mask leading to larger errors on the $y$-component of the DF (see Fig. \ref{fig:akeke4}), which is the most dominant and the best measured thus far.

\begin{table}[h!]
   \centering
\begin{tabular}{c|c|c|c|c}
\multicolumn{5}{c}{SCOUT goals}\\
\hline\hline
 $z_0$ & $L_{\rm X}^{\rm min}$ (erg/sec) & $N_{\rm cl}(z<z_0, \; L_{\rm X}>L_{\rm X}^{\rm min})$ & $\sigma_{1m}\; (\mu$K) & Expected S/N \\
\hline
 0.4 & $4.7\times 10^{44}$ & $266 \pm 31$ & $\sim 1.5$  & $\ga 5$ \\
 0.5 & $7.6\times 10^{44}$ & $138\pm 30$ & $\sim 2.2$ & $\ga 4.5$ \\
 0.6 & $1.1\times 10^{45}$ & $73\pm 27$ & $\sim 3.0$ & $\ga 3$ \\
 \hline
   \end{tabular}
 \caption[]{{\small Estimated parameters expected at the conclusion of SCOUT}\label{tab:future}}
\end{table}

So what will we achieve at the conclusion of the SCOUT experiment with Planck data assuming the current WMAP-based DF results? We estimated this as follows: at some fixed $z_0$, which is taken to be an upper end of the $z$-bin distribution, we use the  [0.1--2.4 keV] flux limit of
$1\times10^{-12}$ erg cm$^{-2}$ s$^{-1}$ and evaluate the {\it lower} limit on the cluster X-ray luminosity, $L_X(z_0)$, from Fig. \ref{fig:future_scout} corresponding to these parameters. Then we evaluate the number of clusters, and its Poissonian uncertainty, expected to be found {\it out to} $z_0$ given these parameters. For it, we then evaluated the anticipated statistical uncertainties in this measurement using the analytical, and numerically verified, formalism of AKEKE and Sec. \ref{sec:errors}. Finally we have assumed a CMB signal corresponding to the measurement of KAEEK using the calibration adopted there.

Table~\ref{tab:future} shows the estimated parameters which we
expect to reach in the course of the SCOUT experiment. The first
column lists the redshift limit, followed by the limiting X-ray
luminosity at that redshift implied by the [0.1--2.4 keV] flux limit of
$1\times10^{-12}$ erg cm$^{-2}$ s$^{-1}$  of the
SCOUT cluster catalog. The third column lists the number of
clusters within the respective redshift limit in the SCOUT catalog
with Poissonian uncertainties using the data in
Fig.~\ref{fig:future_scout}. The fourth  column lists the 1$\sigma$
dipole error that corresponds to this sub-sample. Finally, the
last column lists the S/N with which the bulk flow will be
measured, assuming the current measurements.

\section{Conclusions: Eppur si muove?}
\label{sec:conclusions}

In this report we have presented a comprehensive review of the current state of measurements of the phenomenon dubbed the "dark flow". The current evidence in favor of a substantial part of the observable Universe seemingly moving with respect to the rest-frame defined by the CMB dipole is strong. Indeed, there appears to be a highly statistically significant dipole which appears 1) exclusively at cluster positions, 2) at zero CMB monopole, and 3) its amplitude correlates strongly with the cluster X-ray luminosity. All this is highly suggestive of the SZ-produced effect and, because the measurement is done at zero monopole, the KSZ component has to dominate the detected signal. The dipole we find is aligned, within the errors, with the all-sky CMB dipole, after correcting for the motion of the Local Group. There appears no statistically significant CMB quadrupole at the cluster locations for any of the $L_X$-bins. The equivalent motion is then dipolar and also agrees well, within the statistical and systematic calibration uncertainties, with many other independent velocity measurements, but disagrees with some other. Its amplitude is currently systematically uncertain, but that does not affect the fact of a highly coherent motion extending out to at least $\sim 800$Mpc with respect to the CMB rest frame.

We have discussed the still outstanding and/or unresolved issues of this measurement and its interpretation. These are related to calibrating the signal more accurately than done by us so far. It also requires understanding highly accurately the cluster properties of our catalog. Designing alternative filtering schemes could be useful, but only if they increase the S/N of this delicate measurement. The filtering designed by us removes primary CMB down to the fundamental limit imposed by the cosmic variance. On the other hand, alternative filtering schemes suggested thus far have been shown by us to reduce the S/N and they have not been demonstrated to work down to the cosmic variance levels of primary CMB, and removing the TSZ monopole while leaving the KSZ component sufficiently intact. The work on the DF continues and presents a challenge to both us and the sceptics of this phenomenon.

It is imperative to either explain this result or explain it away. The measurement is currently achieved at a highly, but not hugely, statistically significant level of $S/N \simeq 3.5-4$. Thus, with this statistical significance, a (slightly) inappropriate scheme of measuring may reduce the S/N to below what is statistically significant. We demonstrate with detailed mathematical discussion that this is what happened in two CMB challenges to the DF measurement. Because of this, we have discussed at great length the proper procedures that need to be followed in reconstructing the measurement and taking it further using data soon to be made public from the Planck mission. In order to make the verification available, we have provided the WMAP-based results and procedures by making them publicly available from \url{http://www.kashlinsky.info/bulkflows/data\_public}.

While there are many independent measurements supportive of and consistent with the DF existence, some other recent independent non-CMB-based analyses contradict it. It may, of course, be that our signal is a result of some other yet-unspecified effect. However, the probability of that happening by chance is very small and no alternative explanation of our measurement has appeared in peer-reviewed discussions as of this writing.

If the DF result turns out correct, it will require radical revision of either our current understanding of the global world structure and our Universe's place in it, or (far less likely, in our opinion, given the new measurements) of the basic physics connecting gravity to motion. These implications have been suggested in a wide body of literature and have been reviewed above.

It is thus absolutely critical to bring this project to a definitive and highly accurate conclusion which forms the basis of our efforts over the coming years.

\section{Acknowledgments}

The literature search for this review has ended in January 2012.

We would like to acknowledge our colleagues on the DF project, Dale Koveski and Alastair Edge, for their past and ongoing contributions to the compilation of the cluster samples used in this research. We also acknowledge fruitful exchanges, discussions and conversations over the course of this ongoing project with the following alphabetically listed colleagues Nico Cappelluti, Dale Fixsen, Carlos Hernandez-Monteagudo, Trac Hy, Ryan Keisler, Will Kinney, Eiichiro Komatsu, Arthur Kosowsky, Anthony Lasenby, Laura Mersini-Houghton, Igor Tkachev, David Wiltshire and many others.

We acknowledge support for this project from two NASA ADP grants, NNG04G089G and 09-ADP09-0050. In Spain the project was supported by the Ministerio de Ciencia e Innovaci\'on grants FIS2009-07238 and CSD 2007-00050.


\begin{thebibliography}{3}
\bibitem [Aaronson et al 1986]{aaronson}{Aaronson, M. et al 1986, ApJ, 302, 536}
\bibitem [Abate \& Ergogdu]{abate1}{Abate, A. \& Ergogdu, P. 2009, MNRAS, 400, 1541}
\bibitem [Abate \& Feldman 2011]{abate2}{Abate, A. \& Feldman, H. 2011, arxiv:1106.5791}
\bibitem [Abbott \& Wise 1984]{abbott-wise}{Abbott, L.F. \& Wise, M. B. 1984, 282, L47}
\bibitem[Abell et al 1989]{abell}{Abell, G.O., Corwin, H.G., Olowin, R.P., 1989, ApJ Supl Ser, 70, 1}
\bibitem[Ade et al 2011a]{ade-a}]{ade-a}{Ade, P.A.R., et al 2011a, arXiv:1101.2022}
\bibitem[Ade et al 2011b]{ade-b}{Ade, P.A.R., et al 2011b, arXiv:1101.2039}
\bibitem[Ade et al 2011c]{ade-c}{Ade, P.A.R., et al 2011c, arXiv:1101.2048}
\bibitem[Afshordi et al 2009]{afshordi}{Afshordi, N. Geshnizjahi, G. \& Khoury, J. 2009, JCAP, 8, 30}
\bibitem[Aghanim et al. 2001]{aghanim2001}{Aghanim, N.,  Gorski, K.M. \& Puget, J.-L. 2001, A\&A, 325, 9}
\bibitem[Aghanim et al. 2001]{aghanim2005}{Aghanim, N.,  Hansen, S.H.  \& Lagache, G.  2005, A\&A, 439, 901}
\bibitem [Aghanim et al 2008]{aghanim}{Aghanim, N., Majumdar, S. \& Silk, J. 2008, Rep. Prog. Phys., 71, 066902}
\bibitem [Albrecht \& Steinhardt 1992]{albrech-steinhart}{Albrecht, A, \& Steinhardt, P. 1992, Phys.Rev.Letters, 48, 1220}
\bibitem [Amanullah et al 2010]{union2}{Amanullah, R. et al. 2010, ApJ, 716, 712}
\bibitem [Antoniou \& Perivolaropoulos]{antoniou}{Antoniou, I. \& Perivolaropoulos, L. 2010, JCAP, 12,12}
\bibitem[Arnaud et al.2010]{arnaud}{Arnaud, M. et al. 2010, A\& A, 517, A92}
\bibitem [Arnaud et al 2010]{arnaud}{Arnaud, M. et al 2011, A\&A, 517, A92}
\bibitem[Atrio-Barandela et al 2008]{akke}{Atrio-Barandela, F.,
Kashlinsky, A., Kocevski, D. \& Ebeling, H.  2008, Ap.J.
(Letters), 675, L57 (AKKE)}
\bibitem[Atrio-Barandela \& Doroshkevich 1994]{atrio-dorosh} {Atrio-Barandela, F. \& Doroshkevich, A., 1994, ApJ, 420, 26}
\bibitem [Atrio-Barandela et al 2008]{akeke}{Atrio-Barandela, F., Kashlinsky,
A., Ebeling, H. \& Kocevski, D. 2010, Ap.J., 719, 77 (AKEKE)}
\bibitem [Atrio-Barandela et al 2004]{mach2}{Atrio-Barandela, F., Kashlinsky, A. \&
Mucket, J.  2004, Astrophys. J., 601, L111}
\bibitem[Atrio-Barandela \& Mucket 1999]{atrio-mucket}{Atrio-Barandela, F. \& M\"ucket, J.P., 1999, ApJ, 515, 465}
\bibitem[Afshordi et al 2007]{afshordi}{Afshordi, N., Lin, Y.-T., Nagai, D. \& Sanderson, A.J.R. 2007, MNRAS, 378, 293}
\bibitem[Audit \& Simmons 1999]{audit}{Audit, E. |\& Simmons, J.F.L. 1999, MNRAS, 305, L27}
\bibitem[Baleisis et al 1998]{baleisis}{ Baleisis, A., et al 1998, MNRAS 297, 545}
\bibitem[Bartlett \& Silk 1994]{bartlett}{Bartlett, J. G., \& Silk, J. 1994, ApJ, 423, 12}
\bibitem[Bataglia et al 2010]{bataglia}{Battaglia, N. et al, 2010, ApJ, 725, 91}
\bibitem [Bekenstein \& Sanders]{bekenstein}{Bekenstein, J. \& Sanders, R.H. 2011, arxiv:1110.5048}
\bibitem[Beltr\'an-Jim\'enez  \& Maroto 2009a]{beltram1}{Beltr\'an-Jim\'enez, J. \& Maroto, A. 2009a, JCAP, 3, 15}
\bibitem[Beltr\'an-Jim\'enez  \& Maroto 2009b]{beltram2}{Beltr\'an Jim\'enez, J. \& Maroto, A. 2009b, Phys Rev, D80, 063512}
\bibitem [Bennett et al 2003]{bennett-wmap3}{Bennett, C. et al 2003, ApJS, 148,1}
\bibitem[Benson et al 2003]{benson}{Benson, B.A. et al. 2003, ApJ, 592, 674}
\bibitem [Betschinger \& Zukin 2008]{bert}{Betschinger, E. \& Zukin, P. 2008, Phys.Rev. D, 78, 024015}
\bibitem [Birkinshaw 1999]{birkinshaw}{Birkinshaw, M. 1999,
Ph.Rep., 310, 97}
\bibitem [Birkinshaw \& Gull 1983]{birkinshaw-gull}{Birkinshaw, M. \& Gull, S. F. 1983, Nature, 302, 315}
\bibitem[Blake \& Wall]{blake}{Blake,C. \& Wall, J. 2002, Nature 416, 150}
\bibitem [B\"ohringer et al 2004]{bohringer}{B\"{o}hringer, et al. 2004, A\&A, 425,
367}
\bibitem [Boldt 1966]{boldt66}{Boldt, E., McDonald, F.B., Riegler, G., Serlemitsos, P. 1966, PhRvL, 17, 447}
\bibitem [Boldt 1987]{boldt}{Boldt~E 1987, Phys Rep, 146, 215}
\bibitem[Bond et al 2005]{bond}{Bond, J.R. et al, 2005, ApJ, 626, 12}
 \bibitem [Borgani et al 2004]{borgani}{Borgani, S. et al 2004, MNRAS, 348,
1078}
\bibitem [de Bernardis et al 2000]{boomerang}{de Bernardis, P. et al 2000, Nature, 404, 955}
\bibitem [Bond \& Eftstathiou 1986]{be86}{Bond, J. R. \& Efstathiou, G. 1986, MNRAS, 218, 103}
\bibitem [Bond \& Eftstathiou 1984]{be-cdm}{Bond, J. R. \& Efstathiou, G. 1984, ApJ, 285, L45}
\bibitem [Borgani et al 2004]{simulations_temp}{Borgani, S. et al  2004, Mon. Not. R.
Astron. Soc., 348, 1078}
\bibitem [Boughn et al 2002]{boughn-cxb}{Boughn, S., Crittenden, R.G. \& Koehrsen, G. P. 2002,
ApJ, 580, 672}
\bibitem [Bowyer et al 1965]{bowyer}{Bowyer, S., Byram, E.T., Chubb, T.A., Friedman, H. 1965, Sci, 147, 394}
\bibitem [Byram et al 1966]{byram}{Byram, E.T., Chubb, T.A., Friedman, H. 1966, Sci, 152, 66}
\bibitem [Cai \& Tuo 2011]{cai}{Cai, R. \& Tuo, Z. 2011, arXiv:1109.0941}
\bibitem [Caldwell \& Kamionkowski 2009]{caldwell-kamionkowski}{Caldwell, R. \& Kamionkowski 2009, Ann. Rev. Nulcear and Particle Science, 59, 397}
 \bibitem[Carroll et al 2010]{carroll}{Carroll, S.M., Tseng, C.-Y., Wise, M.B., 2010, Phys Rev, D81, 083501}
\bibitem [Castro et al 2003]{castro}{Castro, P. et al 2003, Phys. Rev. D., 68, 127301}
 \bibitem [Carlstrom et al 2002]{carlstrom}{Carlstrom, J.E., Holder, G.P. \& Reese, E.D.  2002, ARA\&A, 40, 643}
\bibitem [Cavaliere et al 1971]{cavaliere}{Cavaliere, A., Gursky, H., Tucker, W.H. 1971, Nature, 231, 437}
\bibitem [Cavaliere \& Fusco-Femiano 1976]{beta-model}{Cavaliere, A. \& Fusco-Femiano, R. 1976, A\&A, 49, 137}
\bibitem[Cavaliere \& Fusco-Femiano 1978]{cavaliere78}{Cavaliere, A. \& Fusco-Femiano, R. 1978, A\&A, 70, 677}
\bibitem [Clark 1965] {clark}{Clark, G.W. 1965, PhRvL, 14, 91}
\bibitem[Cole \& Kaiser 1988]{cole-kaiser}{Cole, S., \& Kaiser, N. 1988, MNRAS, 233, 637}
\bibitem [Colin et al 2011]{collin}{Colin, J., Mohayaee, R, Sarkar, S., \& Shafieloo, A. 2011, MNRAS, 414, 264}
\bibitem [Collin \& Hawking 1973]{collins-hawking}{Collins, C. B. \& Hawking, S. W. 1973, MNRAS, 162, 307}
\bibitem[Condon et al 1998]{condon}{Condon, J.J. et al. 1998, Astronomical Journal 115, 1693}
\bibitem [Cooke \& Lynden-Bell 2010]{cooke}{Cooke, R. \& Lynden-Bell 2010, MNRAS, 401, 1409}
\bibitem [Courteau et al 2000]{courteau}{Courteau, S. et al 2000, Astrophys. J., 544, 636}
\bibitem [Dai et al 2011]{dai}{Dai, D-C, Kinney, W. \& Stojkovic, D. 2011, JCAP, 4, 15}
\bibitem [Davis \& Huchra 1982]{davis-huchra}{Davis, M. \& Huchra, J. 1982, ApJ, 254, 437}
\bibitem [Davis \& Peebles 1983]{davis-peebles}{Davis, M. \& Peebles, P.J.E. 1983, ARA\&A, 21, 109}
\bibitem [Davis et al 2011]{davistamara}{Davis, T. et al 2011, ApJ, arxiv:1012.2912}
\bibitem [Davis et al 2010]{tamara_davis} {Davis, T.M. 2010, arXiv:1012.2912}
\bibitem [Devlin et al 2009]{blast}{Devlin, M.J. et al 2009, Nature, 458, 737}
\bibitem [Djorgovski \& Davis 1987]{djorgovski}{Djorgovski, S. \&
Davis, M. 1987, Ap.J., 313, 59}
\bibitem [Dodelson 2003]{dodelson}{Dodelson, S. 2003, Modern Cosmology, Amsterdam (Netherlands): Academic Press.}
\bibitem [Dressler et al 1987]{7s-di}{ Dressler et al 1987, Ap.J., 313,
42}
\bibitem[Duffy et al.2008]{duffy}{Duffy, A.R., Schaye, J., Kay, S.T. \& Dalla Vecchia, C. 2008, MNRAS, 390, L64}
\bibitem[Dunkley et al 2011]{dunkley}{Dunkley, J. et al, 2011, ApJ, 739, 52}
\bibitem [Dvali et al 2000]{dgp}{Dvali, G., Gabadadze, G. \& Porrati, M. 2000, Phys. Rev.Lett., 485, 208}
\bibitem [Ebeling et al 1998]{ebeling1}{Ebeling, H., Edge, A.C., B\"{o}hringer, H., Allen, S.W., Crawford,
C.S., Fabian, A.C., Voges, W., \& Huchra, J.P. 1998, MNRAS,  301,
881}
\bibitem [Ebeling et al 2000]{ebeling2}{Ebeling, H., Edge A.C., Allen S.W., Crawford C.S., Fabian A.C., \&
Huchra J.P. 2000, MNRAS, 318 333}
\bibitem [Ebeling et al 2001]{ebeling2001}{Ebeling, H. et al 2001, 553, 668}
\bibitem [Ebeling et al 2002]{ebeling3}{Ebeling, H.,  Mullis, C.R., \& Tully R.B. 2002, ApJ, 580,
774}
\bibitem [Ebeling et al 2007]{ebeling2001}{Ebeling, H. et al 2007, ApJ, 661, L33}
\bibitem [Efstathiou 1990]{efstathiou-review}{Efstathiou, G. 1990, in Physics of the Early Universe, eds. J. Peacock, A. F. Heavens \& A. T. Davies}
\bibitem [Efstathiou \& Bond 1986]{efstathiou-bond}{Efstathiou, G. \& Bond, J.R. 1986, MNRAS, 218, 103}
\bibitem [Efstathiou, G et al 1992]{efstathiou}{Efstathiou, G., Sutherland, W. \& Maddox, S. 1992, Nature, 348, 705}
 \bibitem [Ergogdu et al 2006]{ergogdu1}{Erdogdu, P. et al 2006, MNRAS, 368, 1515}
\bibitem [Erdogdu \& Lahav 2009]{erdogdu2}{Erdogdu, P. \& Lahav, O. 2009, Phys.Rev.D, 80, 043005}
\bibitem [Erickcek et al 2009]{erickcek}{Erickcek, A., Hirata, C. \& Kamionkowski, M. 2009, Phys. Rev.  D, 80, 083507}
\bibitem [Fabian \& Barcons 1992]{fabian}{Fabian, A. \& Barcons, X. 1992, ARA\&A, 30, 429}
\bibitem [Feldman et al 2010]{feldman}{Feldman, H. Watkins, R. Hudson, M. 2010, MNRAS, 407, 2328}
\bibitem [Felten et al 1966]{felten}{Felten, J.E., Gould, R.J., Stein, W.A., Woolf, N.J., 1966, ApJ, 146, 955}
\bibitem [Fixsen 2009]{fixsen09}{Fixsen, D. 2009, ApJ, 707, 916}
\bibitem [Fixsen et al 1998]{fixsen}{Fixsen, D. J.  et al, 1998, Ap. J., 508, 123}
\bibitem [Fixsen \& Kashlinsky 2011]{cibdipole}{Fixsen, D. \& Kashlinsky, A. 2011, ApJ, 734, 61}
\bibitem [Forman et al 1978]{forman}{Forman, W., Jones, C., Cominsky, L., Julien, P., Murray, S., Peters, G., Tananbaum, H., Giacconi, R. 1978, ApJS, 38, 357}
\bibitem[Fosalba et al 2003]{fosalba}{Fosalba, P., Gazta\~naga, E., \& Castander, F.J., 2003, ApJ, 597, L89}
\bibitem[Freivogel et al]{freivogel}{Freivogel, B. et al. 2006, JHEP, 0603, 39}
\bibitem [Friedman \& Byram]{friedman}{Friedman, H.\ \& Byram E.T. 1967, ApJ, 147, 399}
\bibitem [Genova-Santos et al]{genova}{	
	G�nova-Santos, R., Atrio-Barandela, F., M�cket, J. P. \& Klar, J. S. 2009, ApJ, 700, 447}
\bibitem [Giacconi et al 1962]{giacconi}{
	Giacconi, R., Gursky, H., Paolini, F.R., Rossi, B.B. 1962, PhRvL, 9, 439}
\bibitem [Giovanelli et al 1998a]{giovanelli1}{Giovanelli, R. et al 1998a, ApJ, 505, L91}
\bibitem [Giovanelli et al 1998b]{giovanelli2}{Giovanelli, R. et al 1998b, Astron. J., 116, 2632}
\bibitem [Gordon et al 2008]{gordon}{Gordon, C., Land, K. \& Slosar, A. 2008, MNRAS, 387, 371}
\bibitem [Gorski et al 1994]{gorski-dmr}{Gorski, K. et al 1994, ApJ, 430, L89}
\bibitem [Gorski et al 2005]{healpix}{Gorski, K. et al 2005,
Astrophys. J., 622, 759}
\bibitem [Gorski 1988]{gorski88}{Gorski, K. 1988, ApJ, 332, L7}
\bibitem [Gorski 1991]{gorski91}{Gorski, K. 1991, Ap.J., 370, L5}
\bibitem [Gott 1982]{gott}{Gott, J. R. 1982, Nature, 295, 304}
\bibitem[Grischuk 1992]{grischuk}{Grishchuk, L. P. 1992, Phys. Rev. D 45, 4717}
\bibitem [Grischuk \& Zeldovich 1978]{gz}{Grishchuk, L. \&
Zeldovich, Ya.B. 1978, Sov. Astron., 22, 125}
\bibitem [Gunn 1988]{gunn}{Gunn, J. 1988, In ``The extragalactic distance
scale", ASP 100th Anniversary Symposium, p. 344}
\bibitem [Gursky et al 1971]{gursky71}{Gursky, H., Kellogg, E., Murray, S., Leong, C., Tananbaum, H., Giacconi, R. 1971, ApJ, 167, L81}
\bibitem [Gursky et al 1972]{gursky72}{Gursky, H., Levinson, R., Kellogg, E., Murray, S., Tananbaum, H., Giacconi, R., Cavaliere, A. 1971, ApJ, 167, L81}
\bibitem [Gurvits \& Mitrofanov 1986]{gurvits}{Gurvits, L. \& Mitrofanov, I. 1983, Nature, 324, 349}
\bibitem [Guth 1981]{guth}{Guth, A. 1981, Phys. Rev. D., 23, 347}
\bibitem [Guth \& Pi 1982]{guthpi}{Guth, A. \& Pi, S-Y. 1982, Phys. Rev. Lett., 49, 1110}
\bibitem[Haenhelt \& Tegmark 1996]{haenhelt-tegmark}{Haenhelt, M.G.  \& Tegmark, M. 1996, MNRAS, 279, 545}
\bibitem[Hallman 2007]{hallman}{Hallman, E. J., Burns, J. O., Motl, P. M., \& Norman, M. L. 2007, ApJ, 665, 911}
\bibitem [Hamilton et al 1991]{hamilton}{Hamilton, A. et al 1991, ApJ, 374, L1}
\bibitem[Hamilton 1997]{hamilton97}{Hamilton, A. J. S. 1997, astro-ph/9708102}
\bibitem [Hanany et al 2001]{maxioma}{Hanany, Sh. et al 2000, ApJ, L5}
\bibitem [Hansen \& Lilje 1999]{hansen}{Hansen, F. \& Lilje, P. 1999, MNRAS, 306, 153}
\bibitem [Harmon et al 1987]{harmon}{Harmon R. T., Lahav O., Meurs E. J. A., 1987, MNRAS, 228, 5}
\bibitem [Harrizon 1967]{harrison}{Harrison, E.R. 1967, Rev. Mod. Phys., 39, 862}\
\bibitem [Haugboelle et al 2007]{haugboelle}{Haugb{\o}lle, T., Hannestad, S., Thomsen, B.,
Fynbo, J., Sollerman, J., \& Jha, S. 2007, ApJ, 661, 650}
\bibitem [Hauser et al 1998]{hauser}{Hauser, M. et al 1998, ApJ., 508, 44}
\bibitem[Hernadez-Moteagudo et al 2000]{chm-fab}{Hern\'andez-Monteagudo, C., Atrio-Barandela, F. \& M\"ucket, J.P.,	2000, ApJ, 528, L69}
\bibitem [Hernandez-Monteagudo]{chm}{Hern\'andez-Monteagudo, C., Verde, L., Jim\'enez, R. \& Spergel, D. 2006, ApJ, 643, 598}
\bibitem[Herandez-Moteagudo et al 2004]{chm2004}{Hern\'andez-Monteagudo, C., G\'enova-Santos, R. \&
	Atrio-Barandela, F., 2004, ApJ,  613, L89}
   \bibitem [Hinshaw et al 2009]{hinshaw09}{Hinshaw~G  \etal, 2009, ApJSuppl, 180, 225}
 \bibitem [Henry \& Hendriksen 1986]{henry}{Henry, J.P. \& Henriksen, M.J. 1986, Ap.J., 301,
 689}
\bibitem [Hicken et al 2009]{}{Hicken, M., et al. 2009, ApJ, 700, 331}
\bibitem [Hogan, Kaiser \& Rees 1982]{hogan-kaiser-rees}{Hogan, C. J., Kaiser, N. \& Rees, M. J. 1982, Royal Society Philosophical Transactions, Series A, 307, 97}
\bibitem[Holman \& Mersini-Houghton 2006]{mersini-06}{Holman, R. \& Mersini-Houghton, L., 2006, Phys. Rev., D74, 123510}
\bibitem[Holman et al 2008a]{mersini-08a}{Holman, R. \& Mersini-Houghton, L. \& Takahashi, T. 2008a, Phys Rev, D77, 063510}
\bibitem[Holman et al 2008b]{mersini-08b}{Holman, R. \& Mersini-Houghton, L. \& Takahashi, T. 2008a, Phys Rev, D77, 063511}
\bibitem [Holzapfel et al 1997]{holzapfel}{Holzapfel, W.L. et al  1997, Astrophys. J., 479, 17}
\bibitem [Hudson \& Ebeling 1997]{hudson-ebeling}{Hudson, M.J. \& Ebeling, H.
 1997, Astrophys. J., 479, 621}
\bibitem [Hudson et al 1999]{hudson}{Hudson, M.J. et al 1999, Astrophys. J., 512, L79}
\bibitem [Hui \& Greene 2006]{hui}{Hui, L. \& Greene, P.B. 2006, PRD, 73, 123526}
\bibitem[Itoh et al 1998]{itoh1} {Itoh, N., Kohyama, Y., \& Nozawa, S. 1998, ApJ, 502, 7}
\bibitem [Itoh et al 2010]{itoh2}{Itoh, Y., Yahata, K. and Takada, M. 2010, Phys Rev, D 82, 043530}
\bibitem [Jaffe \& Kasier 1995]{jaffe-kaiser}{Jaffe, A. \& Kaiser, N. 1995, ApJ, 455, 26}
\bibitem [Jarosik et al 2011]{wmap7}{Jarosik, N. et al 2011, ApJSuppl, 192, 14}
\bibitem [Jha et al 2007]{jha}{Jha, S., Riess, A. G., \& Kirshner, R. P. 2007, ApJ, 659, 122}
\bibitem [Jones et al 1979]{jones}{Jones, C., Mandel, E., Schwarz, J., Forman, W., Murray, S.S., Harnden, F.R.\ Jr.}
 \bibitem [Jones \& Forman 1984]{jones-forman}{Jones, C. \& Forman, W. 1984, ApJ, 276,
 38}
\bibitem [Juszkiewicz et al 1990]{juszk1}{Jusziewicz, R., Vittorio, N. \& Wyse, R. 1990, ApJ, 349, 408}
\bibitem [Juszkiewicz \& Yahil 1989]{juszkiewicz-yahil}{Juszkiewicz, R. \& Yahil, A. 1989, ApJ, 346, L49}
\bibitem [Kaiser 1988]{kaiser-7s}{Kaiser, N. 1988, MNRAS, 231, 149}
\bibitem [Kaiser 1987]{kaiser-rocket}{Kaiser, N. 1987, MNRAS, 227, 1}
\bibitem [Kashlinsky 1988]{k88}{Kashlinsky, A. 1988, ApJ, 331, L1}
\bibitem [Kashlinsky 1992]{k92}{Kashlinsky, A. 1992, ApJ, 386, L37}
\bibitem [Kashlinsky 2005]{k-cib}{Kashlinsky~A. 2005, Physics Reports, 409, 361}
\bibitem [Kashlinsky et al 1994]{ktf}{Kashlinsky, A., Tkachev, I., Frieman, J.  1994, Phys. Rev. Lett., 73, 1582}
\bibitem [Kashlinsky \& Atrio-Barandela 2000]{kab}{Kashlinsky, A. \& Atrio-Barandela, F.  2000, Astrophys. J., 536, L67 (KA-B)}
\bibitem [Kashlinsky et al 2008]{kabke1}{Kashlinsky, A.,
Atrio-Barandela, F., Kocevski, D. \& Ebeling, H. 2008, Ap.J., 686,
L49 (KABKE1)}
\bibitem [Kashlinsky et al 20089]{kabke2}{Kashlinsky, A.,
Atrio-Barandela, F., Kocevski, D. \& Ebeling, H. 2009, Ap.J., 691,
1479 (KABKE2)}
\bibitem [Kashlinsky et al 2010]{kaeek}{Kashlinsky, A.,
Atrio-Barandela, F., Ebeling, H., Edge, A. \& Kocevski, D.  2010, Ap.J., 712, L81 (KAEEK)}
\bibitem [Kashlinsky et al 2011]{kae}{Kashlinsky, A., Atrio-Barandela, F. \& Ebeling, H. 2011, 731, 1}
\bibitem [Kashlinsky \& Jones 1991]{kashlinsky-jones}{Kashlinsky,
A. \& Jones, B.J.T. 1991, Nature, 349, 753}
\bibitem [Kazanas 1980]{kazanas}{Kazanas, D. 1980, ApJ, 241, L59}
\bibitem [Keisler 2009]{keisler}{Keisler, R. 2009, ApJ, 707, L42}
\bibitem [Keisler et al 2011]{keisler-spt}{Keisler, R. et al 2011, 743, 28}
\bibitem[Khoury \& Wyman 2009]{wyman}{Khoury, J. \& Wyman, M. 2009, PhRevD, 80, 064023}
\bibitem [Kocevski et al 2004]{kocevski3}{Kocevski, D.D., Mullis, C.R., \& Ebeling, H.
2004, Astrophys. J., 608, 721}
\bibitem [Kocevski et al 2006]{kocevski2}{Kocevski, D.D. \& Ebeling, H.  2006, Astrophys. J., 645, 1043}
\bibitem [Kocevski et al 2007]{kocevski1}{Kocevski, D.D., Ebeling, H., Mullis, C.R., \& Tully, R.B.  2007, Astrophys. J., 662, 224}
\bibitem [Kogut et al 1993]{kogut}{Kogut, A. et al 1993, ApJ, 419, 1}
\bibitem [Kogut et al 2011]{pixie}{Kogut, A. et al 2011, JCAP, 7, 25}
\bibitem [Komatsu \& Seljak 2001]{komatsu2001}{Komatsu, E. \& Seljak, U. 2001, Mon. Not. R. Astron. Soc., 327, 1353}
\bibitem [Komatsu \& Seljak 2001]{komatsu2002}{Komatsu, E. \& Seljak, U. 2002, Mon. Not. R. Astron. Soc., 336, 1256}
\bibitem [Komatsu et al 2011]{komatsu-wmap7}{Komatsu, E. et al 2011, ApJ, 192, 18}
\bibitem [Kosowsky \& Kahniashvili 2010]{kosowsky}{Kosowsky, A. \& Kahniashvili, T. 2010, Phys Rev Lett., 106, 1301}
\bibitem [Lancaster et al 2005]{lancaster}{Lancaster, K. et al 2005, MNRAS, 359, 16}
\bibitem [Lange et al 2001]{boomerang}{Lange, A. et al 2001, Phys. Rev. D, 63, 042001}
\bibitem [Langlois \& Piran 1996]{langlois-piran}{Langlois, D. \& Piran, T. 1996, Phys. Rev. D, 53(6), 2908}
\bibitem[Larson et al 2011]{larson}{Larson, D. et al. 2011, ApJSS, 192, 16}
\bibitem [Lauer \& Postman 1994]{lauer-postman}{Lauer, T. R. \& Postman, M. 1994, Astrophys. J., 425, 418}
\bibitem [Lavuax et al 2010]{lavaux1}{Lavaux, G. et al 2010, ApJ, 709, 483}
\bibitem [Lavaux \& Hudson 2011]{lavaux-hudson}{Lavaux, G. \& Hudson, M. 2010, MNRAS, 416, 2840}
\bibitem[Lewis et al.2003]{lewis}{Lewis, A. D., Buote, D. A., \& Stocke, J. T. 2003, ApJ, 586, 135}
\bibitem [Liddle \& Lyth 1993]{liddle-lyth}{Liddle, A. \& Lyth, D. 1993, Phys. Rep., 231, 1}
\bibitem [Lilje et al 1986]{lilje}{Lilje, P., Yahil, A. \& Jones, B.J.E 1986, 307, 91}
\bibitem [Linde 1983]{linde}{Linde, A. 1983, Physics Letters B, 129, 177}
\bibitem [Lueker 2010]{lueker}{Lueker, M. et al  2010, ApJ, 719, 1045}
\bibitem [Lynden-Bell et al 1987]{7s-motion}{Lynden-Bell, D. et al 1988,
Astrophys. J., 326, 19}
\bibitem [Lynden-Bell et al 1989]{lynden-bell2}{Lynden-Bell, D., Lahav, O. \& Burstein, D. 1989, MNRAS, 241, 325}
\bibitem [Ma et al 2010]{ma}{Ma et al 2011, Phys. Rev. D, 83, 103002}
\bibitem [Maddox et al 1990]{mesl}{Maddox, S. et al 1990, MNRAS, 242, P43}
\bibitem [Mak et al 2011]{mak}{Mak, D.S.Y., Pierpaoli, E. \& Osborne, S. J. 2011, ApJ, 736, 116}
\bibitem[Markevitch et al.1998]{markevitch}{Markevitch, M., Forman, W.R., Sarazin, C.L., \& Vikhlinin, A. 1998, ApJ, 503, 77}
\bibitem [Mather et al 1990]{firas}{Mather, J. et al 1990, ApJ, 354, L37}
\bibitem [Mathewson et al 1992a]{mathewson1}{Mathewson, D.S., Ford,
V.L. \& Buchhorn, M.  1992a, Astrophys. J., 389, L5}
\bibitem [Mathewson et al 1992b]{mathewson2}{Mathewson, D.S., Ford,
V.L. \& Buchhorn, M.  1992b, Astrophys. J. Suppl., 81, 413}
\bibitem [Matzner, 1980]{matzner}{Matzner, R. 1980, ApJ, 241, 851}
\bibitem [Meekins et al 1971]{meekins}{Meekins, J.F., Fritz, G., Chubb, T.A.\ \& Friedman, H. 1971, Nature, 231, 107}
\bibitem [Meiksin \& Davis 1986]{meiksin}{Meiksin, A. \& Davis, M.  1986 , AJ, 91, 191}
\bibitem[Mennela et al. 2011]{mennella}{Mennella, A. et al, 2011, A\& A, 536, A3}
\bibitem[Mersini-Houghton \& Holman 2009]{laura}{Mersini-Houghton, L. \& Holman,
R. 2009, JCAP,  2, 6}
\bibitem[]{mersini}{Mersini-Houghton, L., 2005, Class. Quantum Gravity, 22, 3481}
\bibitem [Milgrom 2010]{milgrom}{Milgrom, M. 2010, Phys. Rev. D, 82, 043523}
\bibitem[Molnar \& Birkinshaw 2000]{molnar}{Molnar, S. M., \& Birkinshaw, M. 2000, ApJ, 537, 542}
\bibitem[Mroczkowski et al 2009]{mroczkowski}{Mroczkowski, T., et al.  2009, ApJ, 694, 1034}
\bibitem[Myers et al 2004]{myers}{Myers, A.D. et al, 2004, MNRAS, 347, L67}
\bibitem [Nagai et al 2007]{nagai}{Nagai, D., Kravtsov, A., \& Vikhilin, A. 2007, 668, 1}
\bibitem [Navarro et al 1996]{nfw}{Navarro, J.F., Frenk, C.S. \& White, S.D.M. 1996, Astrophys. J., 462, 563}
\bibitem [Notari \& Quartin 2011]{notari}{Notari, A. \& Quartin, M. 2011, arxiv:1112.1400}
\bibitem [Nozawa et al 1998]{nozawa1} Nozawa, S., Itoh, N., \& Kohyama, Y. 1998, ApJ, 508, 17
\bibitem[Nozawa et al 2000]{nozawa2} Nozawa, S., Itoh, N., Kawana, Y., \& Kohyama, Y. 2000, ApJ, 536, 31
\bibitem [Nusser \& Davis 2011]{nusser1}{Nusser, A. \& Davis, M. 2011, ApJ, 736, 93}
\bibitem [Nusser et al 2011]{nusser2}{Nusser, A., Brachini, E. \& Davis, M. 2011, ApJ, 735, 77}
\bibitem [Olive, K. 1990]{olive}{Olive, K. 1990, Phys. Rep., 190, 307}
\bibitem [Osborne et al 2010]{omcp}{Osborne, S. J., Mak, D.S.Y, Church, S.E. \& Pierpaoli, E. 2010, arXiv:1011.2781}
\bibitem [Ostriker \& Suto 1990]{mach1}{Ostriker, J. \& Suto, Y.  1990, Astrophys. J., 348, 378}
\bibitem[Ostriker\& Vishniac]{ostriker-vishniac}{Ostriker, J.P. \& Vishniac, E.T., 1986, ApJ, 306, L51}
\bibitem [Peacock \& Dodds 1996]{pecock-dodds}{Peacock, J. \& Dodds, 1996, MNRAS, 280, L19}
\bibitem [Peebles 1980]{peebles-lss}{Peebles, P.J.E. 1980, {\it The Large Scale Structure of the Universe}, Princeton University Press}
\bibitem [Peebles \& Wilkinson 1968]{peebles-wilkinson}{Peebles, P.J.E. \& Wilkinson, D. 1968, Phys.Rev. D, 174, 2168}
\bibitem [Perlmutter et al 1998]{perlmutter}{Perlmutter, S. et al 1998, ApJ, 517, 565}
\bibitem [Phillips 1995]{phillips}{Phillips, P. R. 1995, ApJ, 455, 419}
\bibitem [Pike \& Hudson 2005]{pike-hudson}{Pike, R.W. \& Hudson, M. 2005, MNRAS, 635, 11}
\bibitem[Plagge et al.2010]{plagge}{Plagge, T. et al. 2010, ApJ, 716, 1118}
\bibitem [Plionis \& Georgantopoulos 1999]{rosat-dipole}{Plionis, M. \& Georgantopoulos, I. 1999, MNRAS, 306, 112}
\bibitem [Pratt et al 2007]{pratt}{Pratt, G.  et al 2007, Astron. Astrophys. 461, 71}
\bibitem[Pratt \& Arnaud.2002]{pratt-arnaud}{Pratt, G. W., \& Arnaud, M. 2002, A\&A, 394, 375}
\bibitem[Pratt et al.2007]{pratt}{Pratt, G. W., B\"ohringer, H., Croston, J. H., Arnaud, M., Borgani, S.,
Finoguenov, A., \& Temple, R. F. 2007, A\&A, 461, 71}
\bibitem [Priatt et al]{priatt}{Priat,~M et al 2002, A\&A, 393, 359}
\bibitem [Puget et al 1996]{puget}{Puget, J-P 1996, A\&A, 308, L5}
\bibitem [Pyne \& Birkinshaw 1996]{pyne96}{Pyne, T. \& Birkinshaw, M. 1996, ApJ, 458, 46}
\bibitem [Pyne \& Birkinshaw 2004]{pyne04}{Pyne, T. \& Birkinshaw, M. 2004, MNRAS, 348, 581}
\bibitem[Ryachaudhury 1989]{raychaudhury}{Raychaudhury S., 1989, Nature, 342, 251}
 \bibitem [Raymond \& Smith 1977]{raymond-smith}{Raymond, J.C. \& Smith, B.W. 1977, Ap. J. Suppl., 35,
 419}
\bibitem [Rees \& Sciama 1968]{rees-sciama}{Rees, M. J. \& Sciama, D. 1968, Nature, 217, 511}
\bibitem[Rephaeli 1995]{rephaeli} Rephaeli, Y. 1995, ApJ, 445, 33
\bibitem [Reyes et al 2010]{reyes}{Reyes, R. et al 2010, Nature, 464, 256}
\bibitem [Riess et al 1997]{riess}{Riess, A., Davis, M., Baker, J. \& Kirshner, R.P. 1997, Astrophys. J., 488, L1}
\bibitem [Riess et al 1998]{riess_lambda}{Riess, A. et al 1998, AJ, 116, 1009}
\bibitem [Rowan-Robinson et al 1990]{rowan-robinson}{Rowan-Robinson, M. et al. 1990, MNRAS, 247, 1}
\bibitem [Rubin et al 19763]{rubin-73}{Rubin, V., Ford, W.K, \& Rubin, J. 1973, ApJ, 183, L111}
\bibitem [Rubin et al 1976]{rubin-76}{Rubin, V. et al.\ 1976, AJ, 81, 719}
\bibitem [Sasaki 1987]{sasaki}{Sasaki, M. 1987, MNRAS, 228, 653}
\bibitem [Sachs \& Wolfe 196?]{savchs-wolfe}{Sachs, R.K. \& Wolfe, A. M. 1967, ApJ, 147, 73}
\bibitem [Sazonov \& Sunyaev 1998]{sazonov} {Sazonov, S. Y., \& Sunyaev, R. A. 1998, ApJ, 508, 1}
\bibitem [Scaramella et al 1991]{scaramella}{Scaramella, R., Vettolani, G., \& Zamorani, G.  1991, Astrophys. J., 376, L1}
\bibitem [Scharf et al 2000]{scharf-cxbdip}{Scharf, C. et al 2000, ApJ, 544, 49}
\bibitem [Schmidt et al 1998]{schmidt}{Schmidt, B. et al 1998, 507, 46}
\bibitem [Shafer 1983]{shafer}{Shafer, R. 1983, PhD thesis, University of Maryland/NASA's GSFC}
\bibitem [Shirokoff et al 2011]{shirokoff}{Shirokoff, E. et al, 2011, ApJ, 736, 61}
\bibitem [Smoot et al 1992]{dmr}{Smoot, G. et al 1992, ApJ, 396, L1}
\bibitem [Song et al 2011]{song}{Song, Y-S, Sabiu, C. G., Kayo, I. \& Nichol, R.C. 2011, JCAP, 5, 20}
\bibitem[Song 2011]{Song}{Song, Y.-S. 2011, PRD, 83, 103009; arXiv:1009.2753}
\bibitem [Spergel et al 2003]{spergel-wmap3}{Spergel, D. et al 2003, ApJS, 148, 175}
\bibitem [Springob et al 2007]{sfi}{Springob, C. M. et al. 2007, ApJSuppl, 172, 599}
 \bibitem [Stebbins 1997]{stebbins}{Stebbins, A. 1997,
 astro-ph/9705178}
 \bibitem [Starobinsky 1982]{starobinsky}{Starobinsky, A. 1982, Physics Letters B, 117, 175}
\bibitem[Strauss et al 1992]{strauss}{Strauss, M.A. et al 1982, ApJ, 397, 395}
\bibitem [Strauss \& Willick 1995]{strauss-willick}{Strauss, M. \&
Willick, J.A. 1995, Phys. Rep., 261, 271}
\bibitem [Sunyaev \& Zeldovich 1972]{sz1972}{Sunyaev, R.A. \& Zeldovich, Ya.B. 1972, Comments on Astrophysics and Space Physics, 4, 173}
\bibitem [Sunyaev \& Zeldovich 1980a]{sz1980}{Sunyaev, R.A. \& Zeldovich, Ya.B. 1980a, MNRAS, 190, 413}
\bibitem [Sunyaev \& Zeldovich 1980b]{sz1980}{Sunyaev, R.A. \& Zeldovich, Ya.B. 1980b,  Ann Rev A A, 18, 537}
\bibitem [Takey et al 2011]{takey}{Takey, A., Schwope, A. \& Lamer, G. 2011, A\&A, 534, 120}
\bibitem [Tonry et al 2001]{tonry}{Tonry, J. et al 2001, ApJ, 546, 681}
\bibitem [Trac et al 2011]{trac}{Trac, H., Bode, P., \& Ostriker, J.P, 2011, ApJ, 727, 94}
\bibitem[Tsagas 2010]{tsagas1}{Tsagas, C.G., 2010, MNRAS, 405, 503}
\bibitem [Tsagas 2011]{tsagas2}{Tsagas, C. 2011, Phys. Rev. D. 84, 063503}
\bibitem [Tully \& Fisher 1977]{tully-fisher}{Tully, R. B. \& Fisher, J.R. 1977, Astron. Astrophys., 54, 661}
\bibitem [Turnbull et al 2011]{turnbull}{Turnbull, S. J.; Hudson, M. J., Feldman, H. A., Hicken, M., Kirshner, R. P.. Watkins, R. MNRAS, in press}
\bibitem [Turner 1991]{turner}{Turner, M. S. 1991, Phys.Rev., 44, 3737}
\bibitem [Turner 1992]{turner2}{Turner, M. S. 1992, General Relativity and Gravitation,, 24, 1}
\bibitem[Vikhlinin et al.2006]{vikhilinin}{Vikhlinin, A., Kravtsov, A., Forman, W., Jones, C., Markevitch, M., Murray, S. S. \& Van Speybroeck, L. 2006, ApJ, 640, 691}
\bibitem [Villumsen \& Strauss 1987]{jens}{Villumsen J. V., Strauss M. A., 1987, ApJ, 322, 37}
\bibitem [Vishniac 1987]{vishniac}{Vishniac, E. 1987, ApJ, 322, 597}
\bibitem [Vittorio et al 1986]{vittorio}{Vittorio, N., Juszkiewicz, R. \& Davis, M. 1986, Nature, 323, 132-133}
\bibitem [Voges et al 1999]{rosat}{Voges et al.\
    1999, A\&A, 349, 389}
\bibitem [Watkins \& Feldman 1995]{watkins-feldman}{Watkins, R. \& Feldman, H. A.  1995, ApJ, 453, L73-L76}
\bibitem [Watkins, Feldman \& Hudson 2009]{wfh}{Watkins, R., Feldman, H. A. \& Hudson, M. J. 2009, MNRAS, 392, 743}
\bibitem [Way \& Nussbaumer 2011]{p-today}{Way, M. \& Nussbaumer, H. 2011, Physics Today (Letters), August, p. 8}
\bibitem [Wojtak et al 2011]{wojtak}{Wojtak, R., Hansen, S. \& Hjorth, J. 2011, Nature, 477, 567}
\bibitem [Webster et al 1997]{webster}{Webster, M., Lahav, O. \& Fisher, K. 1997, MNRAS, 287, 425}
\bibitem [Weinberg 1972]{weinberg}{Weinberg, S. 1972, {\it Gravitation and Cosmology}}
\bibitem [Weinberg 1976]{weinberg76}{Weinberg, S. 1976, ApJ, 208, L1}
\bibitem [Weyant et al 2011]{weyant}{Weyant, A. 2011, ApJ, 732, 65}
\bibitem[White et al.1997]{white-scaling-relation}{White, D. A., Jones, C., \& Forman, W. 1997, MNRAS, 292, 419}
\bibitem[White et al 2010]{boss}{M. White et al. (2010), 1010.4915}
\bibitem [Willick 1990]{willick90}{Willick, J.A. 1990, ApJ, 355, L5}
\bibitem [Willick et al 1999]{willick99}{Willick, J.A.  1999, Astrophys. J., 522, 647}
\bibitem [Willick 2000]{willick}{ Willick, J.A.  2000,
astro-ph/0003232, in Proceedings of the XXXVth Rencontres de
Moriond: Energy Densities in the Universe}
\bibitem [Wiltshire et al 2012]{wiltshire}{Wiltshire, D., Smales, P., Mattsson, T. \& Watkins, R. 2012, arxiv:1201.5371}
\bibitem [Wolter]{wolter}{Wolter, H. 1952, Ann.~Physik, 10, 94}
\bibitem[Wright 1979]{wright} {Wright, E. L. 1979, ApJ, 232, 348}
\bibitem[Wyman \& Khoury 2010]{wyman}{Wyman, M. \& Khoury, J., 2010, Phys Rev, D82, 044032}
\bibitem[Yahil et al 1980]{yst}{Yahil A., Sandage A., Tammann G. A., 1980, ApJ, 242, 448}
\bibitem[Yahil et al 1986]{yahil-iras}{Yahil, A., Walker, D. \& Rowan-Robinson, M. 1986, ApJ, 301, L1}
\bibitem[Zachei et al 2011]{zacchei}{Zacchei, A. et al. 2011, A\&, 536, 5}
\bibitem [Zeldovich 1972]{zeldovich}{Zeldovich, Ya. B. 1972, MNRAS, 160, P1}
\bibitem [Zhang 2010]{zhang}{Zhang, P. 2010, MNRAS, 407, L36-L40}
\bibitem [Zhang \& Stebbins 2011]{zhang-stebbins}{Zhang, P. \& Stebbins, A. 2011, Phys. Rev. Letters, 107, 041301}
\end{thebibliography}
\end{document}